# Twisted van der Waals Quantum Materials: Fundamentals, Tunability and Applications

Xueqian Sun[1†], Manuka Suriyage[1†], Ahmed Khan[2], Mingyuan Gao[1,3], Jie Zhao[4,8], Boqing Liu[1], Mehedi Hasan[1], Sharidya Rahman[5,6], Ruosi Chen[1], Ping Koy Lam[4,7,8] and Yuerui Lu[1,8*]

[1]School of Engineering, College of Engineering and Computer Science, the Australian National University, Canberra, ACT, 2601, Australia

[2]Department of Industrial and Manufacturing Engineering, University of Engineering and Technology Lahore (Rachna College Campus), Gujranwala, Pakistan

[3]College of Engineering and Technology, Southwest University, Chongqing 400716, China

[4]Department of Quantum Science & Technology, Research School of Physics, The Australian National University, Canberra ACT 2601 Australia

[5]Department of Materials Science and Engineering, Monash University, Clayton, Victoria, 3800 Australia

[6]ARC Centre of Excellence in Exciton Science, Monash University, Clayton, Victoria, 3800 Australia

[7]Institute of Materials Research and Engineering (IMRE), Agency for Science, Technology and Research (A*STAR) , 2 Fusionopolis Way, Innovis #08-03, Singapore 138634, Republic of Singapore

[8]Australian Research Council Centre of Excellence for Quantum Computation and Communication Technology, the Australian National University, Canberra, ACT, 2601 Australia

† These authors contributed equally to this work.

* To whom correspondence should be addressed: Yuerui Lu (yuerui.lu@anu.edu.au)

## Abstract

Twisted van der Waals (vdW) quantum materials have emerged as a rapidly developing field of two-dimensional (2D) semiconductors. These materials establish a new central research area and provide a promising platform for studying quantum phenomena and investigating the engineering of novel optoelectronic properties such as single-photon emission, non-linear optical response, magnon physics, and topological superconductivity. These captivating electronic and optical properties result from, and can be tailored by, the interlayer coupling using moiré patterns formed by vertically stacking atomic layers with controlled angle misorientation or lattice mismatch. Their outstanding properties and the high degree of tunability position them as compelling building blocks for both compact quantum-enabled devices and classical optoelectronics. This article offers a comprehensive review of recent advancements in the understanding and manipulation of twisted van der Waals structures and presents a survey of the state-of-the-art research on moiré superlattices, encompassing interdisciplinary interests. It delves into fundamental theories, synthesis and fabrication, and visualization techniques, and the wide range of novel physical phenomena exhibited by these structures, with a focus on their potential for practical device integration in applications ranging from quantum information to biosensors, and including classical optoelectronics such as modulators, light emitting diodes (LEDs), lasers, and photodetectors. It highlights the unique ability of moiré superlattices to connect multiple disciplines, covering chemistry, electronics, optics, photonics, magnetism, topological and quantum physics. This comprehensive review provides a valuable resource for researchers interested in moiré superlattices, shedding light on their fundamental characteristics and their potential for transformative applications in various fields.

CONTENTS

# 1. Introduction

2D materials have garnered significant interest and experienced remarkable growth in recent years because of their exceptional properties and wide range of applications. Their layer-dependent band structures and unique optical properties[1-6] have inspired deep research in this field and revealed their sensitivity to external environmental stimuli. These materials exhibit strong linear and nonlinear optical performance, surpassing those of conventional bulk materials, making them suitable for various optoelectronic device applications.[7-13] The twisted 2D materials are generated by vertically stacking an atomically thin layer on top of another, which are connected by the vdW interactions[14-15]. Vertical stacking enables the formation of both homo- and heterostructures, which inherit the properties of individual monolayers while expanding new novel characteristics by combing the advantages of disparate materials.[16] However, if the interlayer coupling is very weak, each layer will retain its intrinsic features with negligible modulation and interactions between layers brought about by stacking.[17-18]

Therefore, twist angle-induced coupling plays a critical role in shaping the permanence of twisted 2D materials.[19-22] This vdW stacked structure has introduced new avenues of exploration, leading to the development of a subfield known as "Twistronics". Twistronics focuses on the study of the influence of twist angles in stacked 2D materials, on their properties and functionalities, offering unprecedented opportunities for tailoring and tuning material characteristics. The concept of moiré superlattice was subsequently introduced owning to the in-plane periodic variations resulting from the atomic alignment between top and bottom layers.[23-27] It emerges as a result of rotational misalignment or lattice mismatch between the stacked layers, providing a promising platform for exploring novel physical phenomena and engineering material properties in a controlled manner.

Graphene and transition metal dichalcogenides (TMDs) have been extensively studied as representative vdW structures in terms of moiré patterns, but other family materials have also been expanded for investigation due to the success of the former. Although the periodic moiré pattern in bilayer graphene was observed a decade ago,[26] a plethora of intriguing effects resulted from the moiré structure were found and focused on recently, including the modulations in the electronic band structures,[28-30] formation of localized states,[31-33] emergence of long-range periodic potentials,[34] modulation of coupling strength between layers,[35-36] and the unconventional superconductivity[37-39] and topological conducting channels[40-41] in the highly correlated regimes.[42] The high tunability of twisted vdW materials is a defining characteristic that distinguishes them from their pristine counterparts. Even if there are no strong moiré effects, the properties of the stacked structure can be dramatically tailored by manipulating the twist angle, influencing the charge transfer,[21] interlayer coupling,[43] charge redistribution[44], lattice reconstruction,[45-46] and band hybridization.[37, 47] It offers an additional path for precisely modulating various material properties, such as optical, electronic, and even magnetic characteristics. With this high degree of tunability through variations in stacking configurations, they have also established themselves as compelling building blocks for quantum-enabled devices with a small footprint. On the other hand, the properties of multilayer phases of 2D materials can vary greatly depending on their stacking configurations. For example, even-numbered layers of 3R $MoS_2$ show a strong second harmonic generation (SHG) signal, while even layers in 2H $MoS_2$ usually do not exhibit an SHG signal.[48-49] This disparity has again led to the interest in vertically stacked structures formed by stacking 2D materials in different configurations.[50]

This review will survey moiré superlattices through the lens of cross-disciplinary interest. It will establish an in-depth fundamental understanding spanning from basic theory and fabrication methods to the full range of novel physical phenomena and their readiness for device integration. Figure 1 provides an overview of the topics covered in this article. After the inspiration of moiré superconductivity in the twisted bilayer and trilayer graphene at a magic angle,[37, 51] moiré superlattices in TMDs have recently shown the strong potential to trap excitons exhibiting sharp emission lines,[27] and distinctive valleytronic properties[24] which can be used to create arrays of quantum emitters.[32] Intriguingly, optical excitation of moiré trapped excitons results in high ferromagnetism in semiconductor heterostructures. We will provide a complete picture of the moiré superlattices starting from fundamental theory, fabrication and visualization techniques, outstanding physical properties to device integration and also emphasize the unique potential of moiré superlattices to bridge multiple disciplines. This will serve as a convergence point for emerging moiré related topics, providing a helpful guide to researchers in the fields of materials science, computational chemistry, nanotechnology, optoelectronics, nanophotonics, and nonlinear optics.

**Figure 1: Overview of the review.** Moiré superlattices, a prominent research domain within

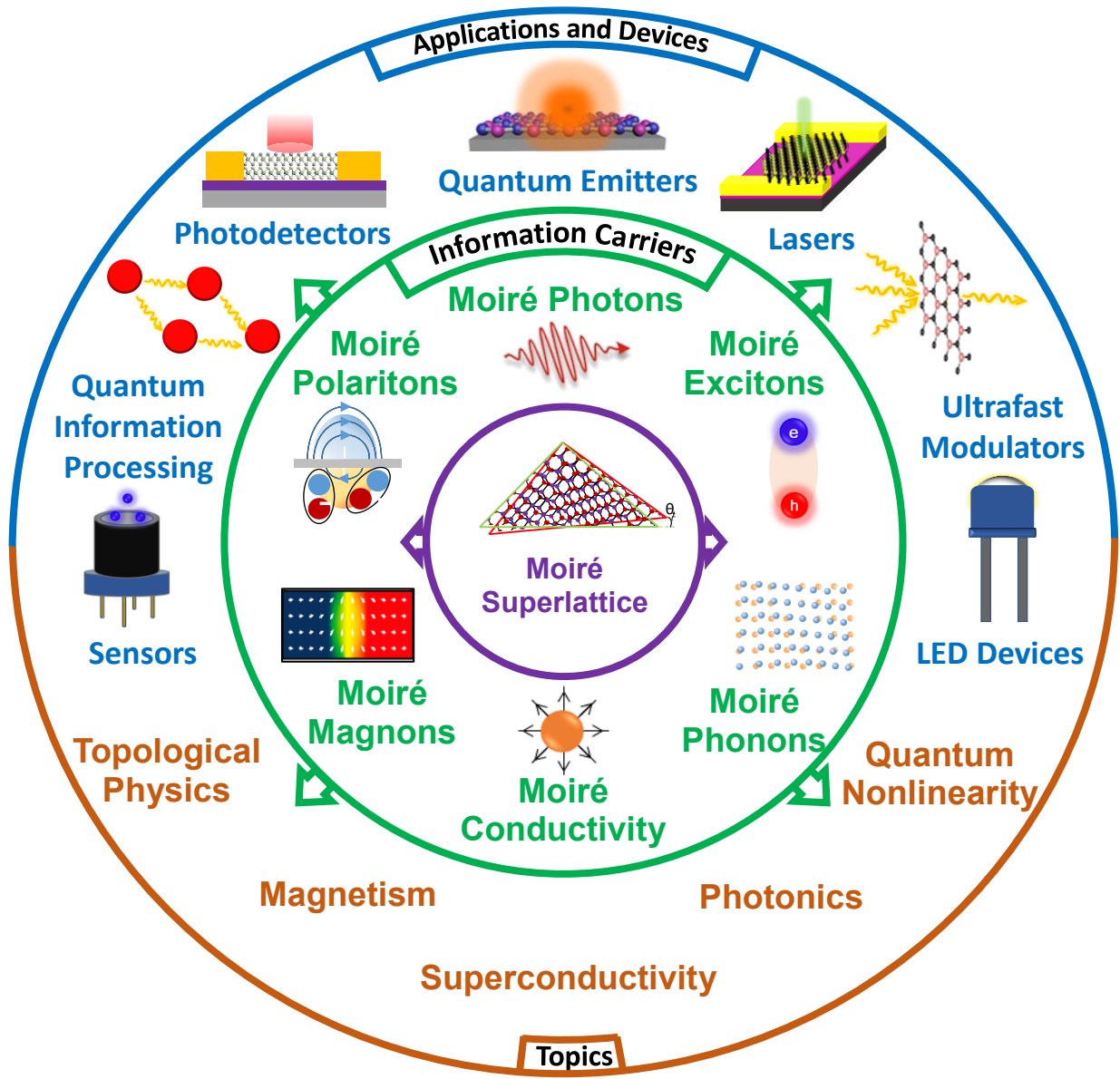

the realm of twisted vdW materials, offer a paradigm for effectively manipulating and engineering light-matter interactions through coupling, mismatch, or external means. They facilitate a spectrum of emerging photoelectric phenomena and instigate the emergence of moiré effects, adding a valuable layer of intricacy to the field and showcasing a unique cross-disciplinary convergence. Their high-degree of tunability establishes them as compelling building blocks for quantum devices and optoelectronic engineering.

## 2. Fundamentals of moiré superlattices

### 2.1 Basic theory

#### 2.1.1   Definition and basic description

Moiré superlattices are a captivating phenomenon that arises when two periodic entities are overlaid with a mismatch, giving birth to a new distinct periodicity known as the moiré pattern.[52-54] It produces a periodic modulation to the energy landscape of materials and the concept plays a crucial role in understanding quantum behaviours within twisted 2D materials. By adjusting lattice constants,[55] translational sliding,[56] and twist angles[45, 57] between stacked layers, the periodicity of the moiré pattern or superlattice can be modified, providing a powerful tool for tuning the physical characteristics of materials. The period length of this moiré unit cell can be calculated based on the underlying formation mechanism in different structures. For example, the moiré superlattice in a TMD heterostructure arises primarily from rotational misalignment and lattice mismatch between the two layers, thus these two factors play a pivotal role in deciding the moiré period in TMD heterostructures.[23, 58] But for homostructures, where the lattice constants of the top and bottom layers are same, the calculation is subsequently simplified,[59-60] and the rotation angle becomes the sole decisive factor governing the moiré period. The significance of rotation misalignment cannot be overstated, as even slight shifts like the magic angle (1.1 degrees) in bilayer graphene[37] (Figure 2a) can trigger remarkable changes in material behaviour. This sensitivity to twist angles opens exciting possibilities for manipulating material properties for diverse applications. Figure 2b demonstrates the impact of varying twist angles,[61] illustrating how different periodic functions of the moiré superlattice are generated.  The phenomenon of moiré effect is intimately linked to the mismatch in lattice constants and rotational twisting in stacked 2D atomic layers, bringing about a revolution in both fundamental research and technological innovation. With more than 2000 atomically thin 2D materials discovered, including insulators, semiconductors, superconductors and quantum spin liquids, a wide range of possibilities has opened up. By forming vdW heterostructures containing different 2D layers, a multitude of electronic,[62-63] photonic,[64] magnetic,[65-66] and topological functionalities can be synthesized even within a single quantum material. The geometrical moiré superlattice brings about fresh length and energy scales, creating a promising opportunity to design band structures and light-matter interactions. This leads to the discovery of exotic quantum phenomena, as seen in twisted bilayer graphene (t-BLG),[67] where correlated insulating states and unconventional superconductivity were detected. Despite the progress made, challenges remain in this field. Although some reports have indicated the possibility of controlling the twist angle within 0.1°,[68] achieving precise control of twist angles in practical device fabrication remains challenging. The absence of highly refined synthesis methods for twisted 2D structures necessitates extensive experimental efforts encompassing fabrication and characterization. On the other hand, the ongoing development of mathematical models will also advance research in this field. Some mathematical approaches/simulations can extract an accurate value of the twist angle by analysing the corresponding properties of twisted structures.[69]

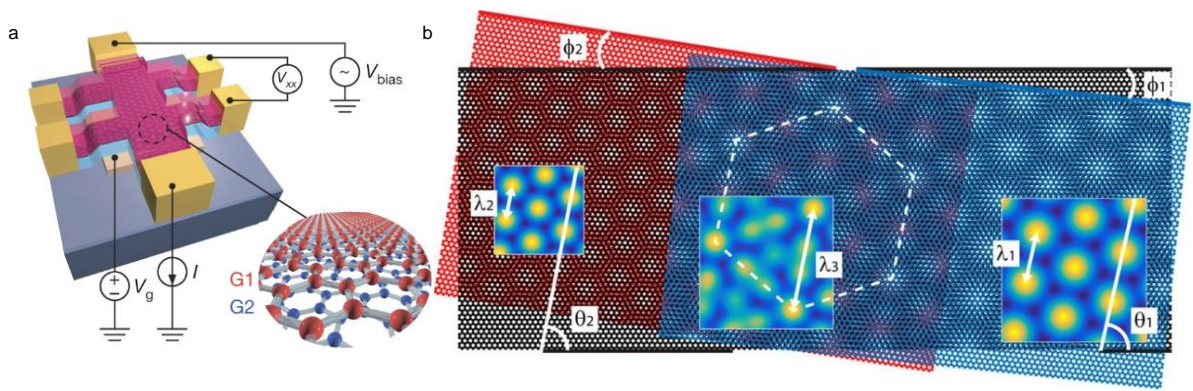

**Figure 2 | Description of moiré superlattices.** (a) Schematic representation of a standard t-BLG device with a four-probe measurement setup. The device is composed of hBN layers on the top and bottom, sandwiching two graphene bilayers (G1, G2) with a twist angle between them. Reprint with permission from ref.[37] Copyright 2018 Springer Nature. (b) Illustration of three distinct moiré superlattices (MSLs) formed in a hBN/graphene/hBN heterostructure. The top hBN lattice is depicted in blue, graphene in black, and the bottom hBN lattice in red. $\phi1$ ($\phi2$) represents the twist angle between the top (bottom) hBN and graphene layers, while

θ1 (θ2) indicates the orientation of the corresponding MSL relative to graphene. The resulting moiré periods are denoted as λ1, λ2, and λ3. The 3L-MSL (middle part) exhibits a larger period compared to both 2L-MSLs (left and right parts). Insets show moiré potential calculations. Reprint with permission from ref.[61] Copyright 2019 American Chemical Society.

### 2.1.2   Theoretical analysis

Moiré superlattice induces an in-plane periodic pattern that occurs when two periodic entities, such as atomic layers or crystal lattices, are stacked or twisted relative to each other. Understanding the underlying theoretical concepts[70] and mathematical models that explain moiré superlattices is of paramount importance to fully grasp their unique properties and explore their diverse applications. At the heart of explaining moiré superlattice formation lies the spectral theory[58-59, 71-72] based on Fourier analysis. This simple yet powerful mathematical approach enables a geometric interpretation of the twist angle-dependent behaviour of moiré patterns [73-76]. Assuming two stacked layers $L_1$ and $L_2$ are twisted by an angle $\phi$ (techniques used for the determination of this twist angle is explained in section 2.3) to generate a moiré pattern. The periodicity D according to the spectral theory of moiré patterns is given as;[58]

$$D = \frac{d_1 d_2}{\sqrt{d_1{}^2 + d_2{}^2 - 2d_1 d_2 \cos \phi}} \qquad (1)$$

Assuming $d_1$ and $d_2$ are the lattice constants of layer 1 and layer 2 The above equation represents the case of a heterostructure where lattice constants are different. However, for the case of homo-structures[84,85] and folds[61] where $d_1 = d_2$, Equation 1 simplifies into;

$$D = \frac{d}{2 \sin(\phi/2)} \qquad (2)$$

The periodicity of a moiré pattern is usually represented through a commensurate unit cell. The direction and length of the unit cell are represented through its own vectors. As the twist angle varies, the periodicity of moiré pattern varies, and new moiré pattern can be represented by its own vectors. This periodicity is often considered as the moiré supperlattice wavelength ($\lambda$) or the moiré length. This periodicity is often considered as the moiré supperlattice wavelength ($\lambda$) or the moiré length. Based on this moiré period and the corresponding shape of the moiré unit cell, the moiré surface area can be easily extracted using the formula for the area of a parallelogram.[77-78] If the lattice constant of the bottom layer of the moiré structure is taken as "a" and the lattice mismatch of the two layers is considered as "$\delta$" the moiré length can be written as in equation 3;

$$\lambda = \frac{(1 + \delta)a}{\sqrt{2(1 + \delta)(1 - \cos\phi) + \delta^2}} \qquad (3)$$

The relative rotational angle $\theta$ of the moiré pattern is extremely important. As illustrated in figure 2b the orientation of the moiré supperlattice $(\theta)$ of the moiré pattern with respect to the graphene lattice can be found using equation 4.[61]

$$\tan(\theta) = \frac{\sin\phi}{(1 + \delta) - \cos\phi} \qquad (4)$$

Another fundamental concept in explaining moiré superlattices is quantum mechanics,[58] the cornerstone of understanding electronic structures. When the periodic atomic layers are stacked or twisted, quantum mechanics comes into play, describing how the electronic states in the constituent 2D materials interact to create new electronic bands and energy levels within the moiré pattern.[38, 79-80] This phenomenon significantly influences the electronic properties of moiré superlattices, and theoretical approaches such as effective mass approximation and tight-binding model are often employed to simplify the description of charge carriers and electronic band structures, respectively. The interesting electronic phenomena[79, 81-83] generated from moiré superlattices such as Dirac fermions and Van Hove singularities have garnered significant attention in the study of these systems. The presence of Dirac fermions, with linear dispersion near the Fermi level, contributes to unique electronic transport properties. Van Hove singularities, conversely, occur when there are singularities in the density of states (DOS), leading to enhanced electronic interactions and novel electronic states. Moreover, the topological band theory comes into play when moiré superlattices possess topological properties.[84-86] This theoretical framework helps elucidate the emergence of nontrivial topological phases and robust edge states in the electronic band structure, opening new possibilities for harnessing topological effects in quantum materials.

Many-body interactions and correlations are significant factors in certain moiré superlattices[66, 87-90] where strong interactions shape the electronic or optical properties. The Hubbard model and other many-body theoretical methods are employed to explore the influence of interactions between electrons, shedding light on the emergence of correlated electronic states and their collective behaviour. To describe the long-wavelength behaviour and low-energy excitations of moiré superlattices, continuum models and effective Hamiltonians are often used.[91-92] These models capture essential physical aspects while simplifying the complexities of the full electronic structure, providing valuable insights into the collective excitations and quantum phenomena in these systems. Complementing theoretical approaches, ab initio simulations, such as density functional theory (DFT)[19, 69, 93-95] calculations, play a pivotal role in providing detailed information about the electronic and structural properties of moiré superlattices. DFT simulations enable researchers to predict and understand the ground state properties, electronic band structures, and electron-phonon interactions in moiré patterns.

Therefore, the theoretical concepts and mathematical models used to explain moiré superlattices encompass a broad range of tools from quantum mechanics to topological band theory and many-body physics. These approaches provide a comprehensive understanding of the formation and unique characteristics of moiré superlattices, laying the foundation for exploring their prospective applications in quantum technologies and materials science. The synergistic interplay of theoretical investigations and experimental studies contributes to unlocking the full potential of these intriguing systems and propelling the advancement of modern science and technology.

## 2.2 Modelling and simulation approaches

In the quest to understand and predict the behaviour of moiré patterns, researchers have embraced diverse mathematical models, which are tailored to varying length scales and complexities. These models span from atomistic approaches, including tight-binding Hamiltonians and DFT, to continuum models, offering unique insights into the fascinating world of moiré superlattices. Atomistic models exhibit remarkabe versatility, encompassing both the extensive length scale of the moiré pattern[69] and the smaller length scale of the basic unit cells. They typically commence with a moiré supercell as a basis and are particularly effective in dealing with commensurate systems. These models embrace Bloch theory and traditional electronic-bands approaches,[96-98] allowing for precise computations of electronic structures. Among atomistic methods, DFT stands out as a powerful tool due to its capacity to consider the local electronic environment of crystalline solids. By calculating the system's energy with electron density as a parameter, DFT offers flexibility and applicability to various materials.[99] It has proven successful in accurately determining electronic and atomic structures of moiré bilayer systems, positioning it as a preferred technique for many calculations involving twisted bilayer structures.[100-101] However, DFT calculations can become computationally demanding and costly when dealing with systems containing hundreds of atoms. An alternative to DFT is the tight-binding approximation,[69, 102-103] which also falls under atomistic models. This approach considers the local environment by assuming tight binding of electronic orbitals, resulting in reduced computational demands and cost-effectiveness compared to DFT. It is particularly effective near the Fermi energy for accurately reproducing electronic structures, making it a favored choice for larger systems where accuracy and computational efficiency are crucial. Both DFT and tight-binding approximation ensure that quantum behaviour is accounted for at all length scales, providing precise computational results for both finite and periodic structures. Stepping beyond atomistic models, continuum models[104] present another mathematical approach utilized in stacked bilayer graphene for calculating its electronic band structures and electrical transport behaviour. These models are computationally less demanding and cost-effective, making them suitable for larger moiré length scales, typically exceeding 10 nm. However, their versatility and precision are lower compared to atomistic models, especially when dealing with intricate moiré patterns involving a substantial numerous atoms and/or multiple layers. Therefore, the development of efficient and effective mathematical modelling methods for twisted structures remains an ongoing endeavor to tackle the challenges posed by complex moiré systems. As described, the study of moiré superlattices involves a diverse array of mathematical models, each offering unique insights into the behaviour of these captivating structures. Atomistic models, including DFT and tight-binding approximation, provide detailed understanding and accurate calculations across various systems. On the other hand, continuum models offer computational advantages for larger length scales but come with trade-offs in accuracy and flexibility. As research advances, achieving a harmonious equilibrium between different modelling approaches will remain pivotal fully unlocking the potential of moiré superlattices and exploring their applications in materials science and beyond.

### 2.3 Determination of Twist-angle

In the exploration of moiré superlattices, the precise determination of twist angles has become crucial for physical analysis,[105-108] requiring a fast, reliable, and non-destructive method. The following techniques are currently usually employed to measure the twist angle between stacked layers. Scanning Transmission Electron Microscopy (STEM) is one of the extensively used techniques,[109-110] which excels in discerning lattice structures in ultrathin materials and is particularly well-suited for gauging twist angles in 2D materials. STEM involves directing a convergent electron beam with lateral dimensions of approximately 0.1 nm across an ultrathin sample to probe its crystal lattice arrangement. A wide-angle detector placed opposite the sample collects the scattered electrons to determine the positions of nuclei.[111-113] When the electron beam encounters the nuclei, it scatters, generating dark signals on the detector that indicate atomic positions. Figure 3a presents a TEM image of a single-layer $MoS_2$ in high-resolution, displaying a regular atomic arrangement characterized by evident hexagonal symmetry,[114] where white spots represent gaps in the atomic structure while the sulfur and molybdenum atoms are indicated by the dark spots. The intensity of the dark signal varies based on the atomic number and the number of atoms encountered along the beam path.[115] Thus higher intensity is achieved for heavier atoms and thicker samples. Different stacking configurations can influence the number of scattered electrons collected by the detector, leading to various scattering patterns or electron diffraction patterns in STEM imaging. By analysing the symmetry of the scattering patterns which corresponds to different twist angles, researchers can differentiate between various stacking configurations,[116] to accurately determine the twist angles in stacked layers[59] (Figure 3b, c). Another potent tool is Polarization-dependent Second Harmonic Generation (PSHG), which has convincingly demonstrated its effectiveness as a swift and non-intrusive method for twist angle determination across a range of stacking configurations. SHG is highly sensitive to lattice inversion symmetry which makes it a precise technique for ascertaining lattice orientation in in materials lacking inversion symmetry such as monolayer TMDs.[117] By inserting polarizers into the system, SHG intensity is gauged at varying sample rotations to obtain a polarization angle-dependent SHG response, directly furnishing insights into the lattice orientation of monolayer TMDs.[118] For example, in case of 1L $MoS_2$, it shows 6-fold rotational SHG symmetry due to its $D_{3h}$ symmetry structure, and the maximal (minimal) SHG response is detected along the armchair (zigzag) direction for parallel (perpendicular) polarization modes (Figure 3d), formulated as explained in[119]. When two layers are twisted by an angle θ (θ represents the twist angle between armchair directions of layer 1 and layer 2), the resultant net SHG intensity is proportional to the vector summation of the SH intensities of the two layers, enabling the determination of twist angles in various stacking arrangements.[50] In addition, Raman spectroscopy constitutes an additional robust optical technique, capable of noninvasive determination of twist angles.[120-121] Variations in twist angles within stacked structures induce corresponding changes in the Raman spectrum, such as Raman shift and intensity, offering a means to gather a collection of twist angle-dependent Raman data. This enables the application of machine learning analysis technique to automatically identify the twist angles from the Raman spectrum of new samples, based on a robust database known as the training dataset (Figure 3e).[105] This approach has been demonstrated in bilayer graphene with low computational requirements but can provide fast and accurate results, showcasing the effectiveness of integrating machine learning and artificial research.

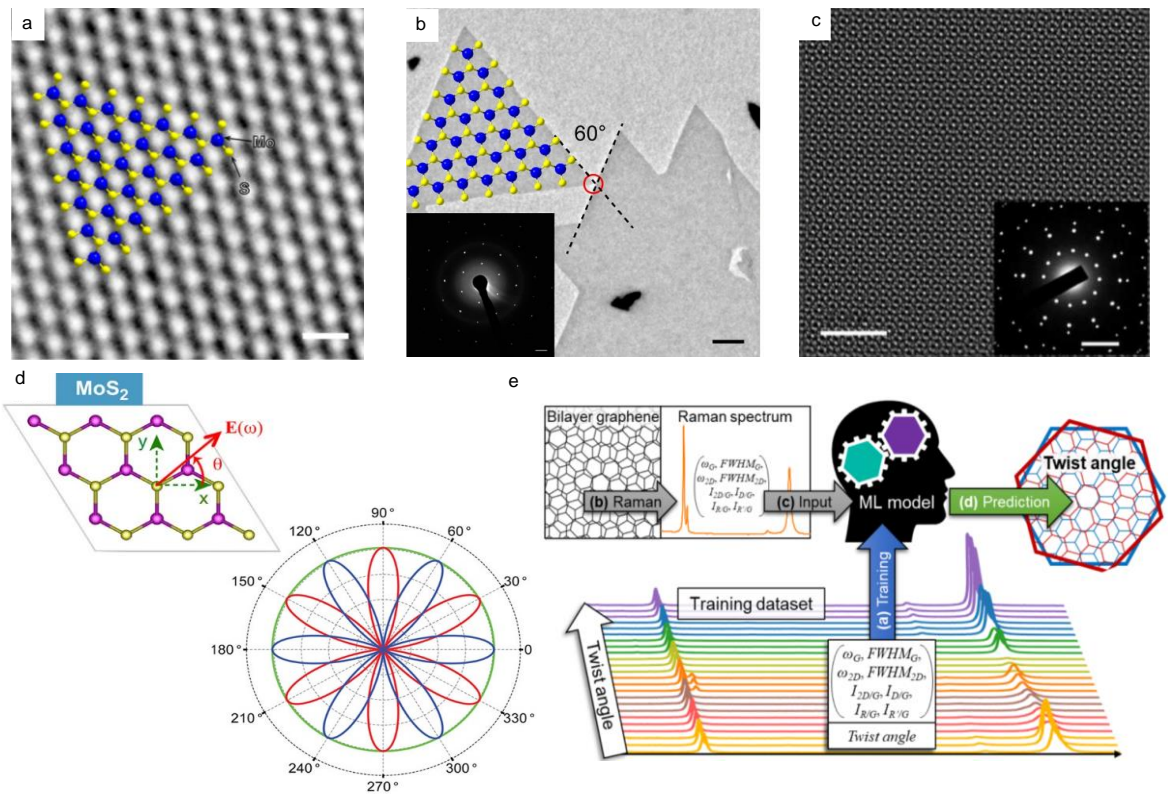

**Figure 3 | Determination of twist angle.** (a) High Resolution TEM image of a $MoS_2$ film showing the crystallinity of sample. (b) Low-magnification TEM image of $MoS_2$ islands. The inset shows the diffraction pattern obtained from the area marked with a red circle. Only one

set of diffraction spots is detected, indicating aligned crystalline lattices within the single-crystal island. The MoS$_2$ structure drawing illustrates the lattice orientation within the single-crystal island and is not drawn to scale. (a,b) Adapted with permission from ref.[114] Copyright 2015 American Chemical Society. (c) Scanning Transmission Electron Microscope (STEM) image of stacked bilayer MoS$_2$ film with 30° twist angle, scale bar 3 nm; insert is electron diffraction pattern of 30° stacked bilayer MoS$_2$ film, scale bar 5 nm$^{-1}$. Adapted with permission from ref.[59] Copyright 2020 Springer Nature. (d) Polarization dependent SHG susceptibilities in MoS$_2$. The red (blue) line indicates the polarization component of the SHG response parallel (perpendicular) to the polarization $E(\omega)$ of the incident electric field. $\theta$ is the rotation angle between $E(\omega)$ and the crystal lattice (i.e., the $x$-axis here for bar three 2D materials). Reproduced with permission from ref.[119] Copyright 2017 American Chemical Society. (e) Determination of Twist Angle through Machine Learning Analysis of its Raman Spectrum. Schematic of the process involved in the ML determination of the twist angle of BLG. Adapted with permission from ref.[105] Copyright 2022 American Chemical Society.

## 3. Moiré pattern from 2D materials

The moiré superlattice has been discovered in various layered structures of 2D materials. In this discussion, we will primarily focus on the moiré patterns observed in homo- and hetero-structures, respectively, due to their different mechanisms in the formation of periodic supercells. A homostructure refers to a layered material system composed of identical atomic layers stacked on top of each other. The layers possess the same crystal structures, lattice constants, and atomic arrangements, which create the regular lattice mainly determined by the twisting angle in the stacking. On the other hand, a heterostructure refers to a stacked structure composed of distinct atomic layers. The constituent layers have different crystal structures, lattice constants or compositions. It introduces additional degrees of freedom, leading to the formation of a moiré pattern through misalignment in angle and/or mismatch in lattice constants. These deviations from perfectly commensurate structures give rise to unique characteristics of moiré superlattices in different material systems.

### 3.1 Homo-structures

### 3.1.1   Graphene

Moiré patterns are formed especially when two identical structures are stacked with a small rotational angle,[122-124] as reported in the graphene/hBN system that will explain below. As a pioneer material of 2D science, graphene has been involved in a great deal of research. The occurrence of a moiré pattern in twisted graphene bilayers due to the mismatch between two identical graphene lattices was proposed with an outstanding electronic structure of "flat bands" at a magic angle of around 1.1° more than a decade ago[125] and the super-pattern was first observed in 2013.[26] It aroused great interest for researchers to experimentally demonstrate the realization and accessibility of fascinating properties. Some correlated quantum states including superconductivity and insulator behaviour have been raised in the twisted graphene bilayers.[37, 126] The technique of "tear and stack" was generally used for controlling the rotational angle between two graphene layers, forming a high-quality moiré superlattice (Figure 4a).[77, 127] It was generated because of the twist-induced in-plane periodic variations in the crystallographic alignment that exhibits alternating regions with different configurations of AA, AB/BA and SP (saddle-point), as shown in Figure 4b.[26, 128] This simultaneously brings about different energies to bilayer graphene because of the varying in-plane relative displacements between two layers.[26, 129] Figure 4c illustrates the energy landscape of bilayer moiré graphene structure, indicating the highest energy at AA stacking where the first layer is directly on the top of the second layer, and there are six energy minima around this central point, which are named as AB or BA stacking where the sublattice atom (A or B) from the top layer graphene is overlapped with the opposite atom (B or A) of the bottom layer graphene. These are also known collectively as Bernal stacking. A cross-cut line of the energy profile along an armchair direction of graphene layers in Figure 4d reveals more directly that the Bernal stacking of AB and BA are energetically favourable and connected by a saddle point (SP) with a slightly higher energy of 2.1 meV/atom but 10 times lower than the energy of unfavourable AA-stacked sites.[26] The rotation with a twist angle θ between two layers introduces a displacement between the corners of two Brillouin zones (BZ) of graphene layers, which results in a mini-Brillouin zone (Figure 4e) with the reciprocal lattice vectors and the size of this moiré BZ is strongly dependent on the rotational misalignment.[125-127] The Dirac cones near K or K' valley from the top and bottom layers experience mixing through interlayer hybridization and this hybridization between neighbouring Dirac cones leads to a significant modification of the electronic band structure.[125-126, 130] Especially at the small twist angle of around 1.05° (predicted magic angle), the electronic band structure shows a flattening feature around the charge neutral point (0 eV), owing to the strongly reduced or vanished Fermi velocity at Dirac-pints (Figure 4f). This creates a massive enhancement of the corresponding peak in the DOS spectrum with a narrow width because of the dense population of states within the energy range (Figure 4g).[125-127, 131-132] The visualization of AA stacking sites as bright spots in the SPM (scanning probe microscope) images of the graphene moiré pattern has been reported as shown in Figure 4h[77] and other STM (scanning tunnelling microscope) images,[67, 128] which further demonstrated that the enhanced tunnelling DOS makes them become elevated spots, while Bernal stacked regions appear at lower positions in STM topography. However, when two graphene layers are rotated relative to each other further to a large angle, the SPM images can only resolve AB and BA sites instead because the absence of flat bands makes the AA sites invisible. A dark-field TEM image of a bilayer-bilayer graphene structure with a twist angle of 16° is shown in Figure 4i,[26] revealing observed AB and BA stacked regions. This observation has garnered significant attention towards multilayer stacking with graphene, which can generate various moiré patterns and allow for their coexistence within the system.[52, 133] On the other hand, at a small twist angle (<1°), the vdW interaction between two graphene layers was found to induce a substantial structural reconstruction at the interface, although it is weak and the dangling bonds are absent there, which results in alterations to the lattice symmetry and electronic structure, as well as the deformation of the original moiré pattern, as displayed in Figure 4j and k.[134] Following the creation of the moiré supercell, three alternating stacking configurations explained above emerge within the superlattice (Figure 4j), however, a rapid increase of energetically unfavorable AA stacking regions in small t-BLG triggers the relaxation of moiré superlattice,[135] thus the atomic reconstruction is anticipated to take place with the rotation of lattice locally along the arrows (Figure 4j), forming an array of triangular domains consisting of Bernal stacking and demarcated by sharp mirror boundaries (Figure 4k).[134] The TEM dark-field images of bilayer graphene with a series of different rotational angles are presented in Figure 4l, experimentally demonstrating the interfacial atomic reconstruction in the bilayer, and the reconstruction intensity diminishes as misorientation increases.[134-135] It is evidenced by the weakened triangular domain contrast and nearly one-directional fringes at a large twist angle, which typically happens for the structure without atomic reconstruction.[136]

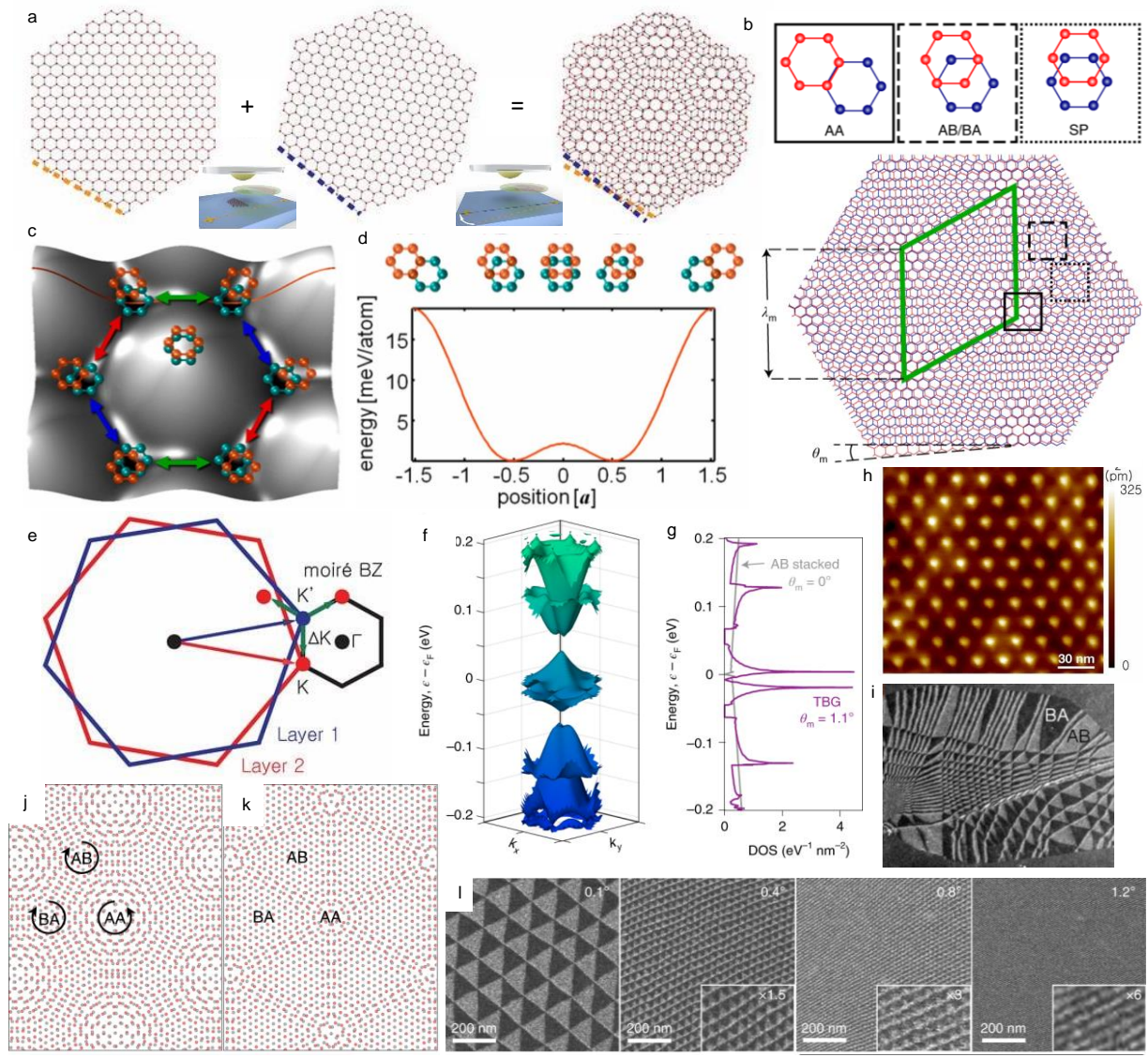

**Figure 4 | Moiré superlattice in t-BLG.** (a) Schematic illustration of the moiré pattern formation as a result of the twist angle between the two layers. The insets show the "tear and stack" fabrication process. Adapted with permission from ref.[77] Copyright 2017 Proceedings of the National Academy of Sciences. (b) Schematic of the three stacking configurations (AA, AB/BA and SP) and a moiré pattern in t-BLG with a moiré wavelength (green line), $\lambda_m = a/[2\sin(\theta_m/2)]$, where a = 2.46 Å is the lattice constant of graphene. Adapted with permission from ref.[128] Copyright 2022 Springer Nature. (c) vdW energy landscape for bilayer graphene with the corresponding orientations of the two layers shown schematically in orange and teal. The central location corresponds to AA stacking, six energy minima around correspond to Bernal-stacking. (d) A horizontal line cut through the energy landscape in (c) along an armchair direction, revealing the energy potential depending on the stacking orders and suggesting that AB is connected to BA through a saddle point (SP). (c,d) Adapted with permission from ref.[26] Copyright 2013 Proceedings of the National Academy of Sciences. (e) The mini-Brillouin zone of a TBG superlattice constructed from the difference between two K (or K′) wavevectors. Adapted with permission from ref.[77] Copyright 2017 Proceedings of the

National Academy of Sciences. (f) Calculated moiré band energy of 1.1° TBG. (g) The DOS corresponding to the bands shown in (f). (f,g) Adapted with permission from ref.[128] Copyright 2022 Springer Nature. (h) SPM topography image showing a TBG moiré pattern. The sample voltage is 0.3 V and the tunnel current is 100 pA. Adapted with permission from ref.[77] Copyright 2017 Proceedings of the National Academy of Sciences. (i) Dark-field TEM images of bilayer/bilayer graphene with a twist angle of 16°, image obtained by selecting the graphene diffraction peak of [-1010], enabling the observation of AB and BA domains. Adapted with permission from ref.[26] Copyright 2013 Proceedings of the National Academy of Sciences. (j, k) Schematics of t-BLG before (j) and after (k) the atomic reconstruction, the black arrows represent the rotational modulation of the lattice. (l) Dark-field TEM images in t-BLG acquired by selecting electrons from the [1010] family of diffraction angles with different controlled twist rotations. The alternating contrast of AB/BA domains is associated with the antisymmetric shift of the lattice period in these triangular domains. (j-l) Adapted with permission from ref.[134] Copyright 2019 Springer Nature.

### 3.1.2 Transition metal dichalcogenides (TMDs)

Similar to all other moiré superlattices that are produced from the twist-stacked structures, the angle between two individual layers of homostructure can be controlled to tune the moiré potential depth, which enhanced the formation of hybridized minibands at the same time, and the period of the moiré pattern will not be limited because the lattice and energy mismatch is absent in this structure due to the same constitutive monolayers.[137-138] In this case, the experimental access to a broader spectrum of superlattice length scales is enabled and the relative twist angle rather than lattice mismatch is the tool to govern the moiré lattice. The formation of a moiré pattern (Figure 5a) from twisted homobilayers (slight twist angle near 1°) will construct moiré potentials to trap the intralayer excitons(Figure 5b and c), enabling an array of two intralayer excitons that emerge at alternating positions in the supercell.[139] The high energy level of moiré potential is defined by the in-plane momentum of the system, and the spatial periodic potential can create some flat electron minibands by modulating the unperturbed electron band of the constitute single layer, leading to a change of the local energy band structure of twisted bilayers.[140] Consequently, depending on the size and depth of moiré potential, quasi-particles like electrons or holes and light-excited states including excitons, trions and other exciton complexes could get trapped in the potential. The moiré unit supercell will also result in the generation of the moiré Brillouin zone, the size of which in turn ascertains the number of occupied moiré cells and the strength of the optical dipole.[140] The twisted TMD homo- and hetero-structures provide a wider diversity of physical properties than graphene because of the absent inversion symmetry in the constituent layers.[46] The local-to-local variation of the interlayer registry produces different local atomic stacking configurations as shown in Figure 5d. The coupling between neighbouring monolayers in moiré pattern induces an atomic lattice reconstruction,[46, 139] where the lattices were locally strained to achieve the maximum size of energetically favored AB and BA stacking regions. It led to a triangular pattern of alternating AB/BA domains (top right, Figure 5d), which is different from the smooth transitions between the two domains normally observed in a typical moiré pattern. However, the gradual transitions are more commonly observed in heterobilayers with inherent lattice mismatch and at larger rotational angles, as shown in Figure 5d (top left). The detailed corresponding configurations and stacking orders for different atomic registries are illustrated in the bottom panel of Figure 5d. The tear-and-stack technique has been widely used in the fabrication of approaching zero-degree twisted homobilayers,[139, 141-142] and the reconstructed moiré pattern with triangular domains of AB and BA can be imaged via the Scanning electron microscope (SEM ) technique (Figure 5e). The reconstruction is created when the local twist is reduced near the AB and BA sites and the proposed critical angle was theoretically predicated as around 2.5°.[46, 143] An hBN-encapsulated $WSe_2/WSe_2$ homobilayer with a twist angle of 2.5° was studied to validate the prediction and it highlighted the necessity of high-resolution imaging in the investigation of the superlattice.[139] The direct observation that the group had from SEM imaging demonstrated the inhomogeneity of the sample, which provides direct evidence that there is significant variation in the local moiré periodicity both among and within the devices. Because of the identical constitute layers, the exciton hybridization of intralayer and interlayer excitons can be observed with the spatial variation of layer configurations in the moiré pattern.[144] The intralayer excitons are formed within the individual layers, consisting of electrons and holes within the same layer, while the electrons from one layer can also recombine with the holes from the other layer to form the interlayer excitons due to their overlapped wavefunctions (Figure 5f). Overlapping wavefunctions of the adjacent layer allow for efficient interlayer hopping and facilitate the hybridization of excitons. In addition, in TMD homobilayers, direct excitons are firstly created at the K point and the electrons from the conduction band can scatter into the energetic local minima at the Λ point through phonons. If we assume the holes are fixed in one layer, the electrons are located in Λ valleys of either the top or the bottom layer, which will subsequently form the intra- or interlayer excitons (Figure 5g).[145] Both intra- and interlayer excitons will experience energetic modulation due to the moiré potential induced band variations.[138] The potentials are visualized for different exciton species in Figure 5h. The interlayer excitons with opposite electric dipole orientations ($E_{cv'}$ and $E_{c'v}$ in Figure 5f) are confined to two high-symmetry sites within a moiré supercell and these points will create a honeycomb pattern with the dipolar attraction between adjacent points.[144] The twist angle between two layers can effectively affect the hybridization effects in the homostructure, which is supported by the angle-dependent PL spectra of both direct and indirect excitons (Figure 5i). For small twist angles around 0° to 5° and 55° to 60°, the intensity of PL emissions from momentum indirect states (orange circles) at lower energy is increased, which is accompanied by a drop in the intensity of direct exciton emissions (green circles). This implies a down-conversion of PL intensities from direct K-K excitons to indirect excitons when the twist angle is near 0° or 60°.[138]

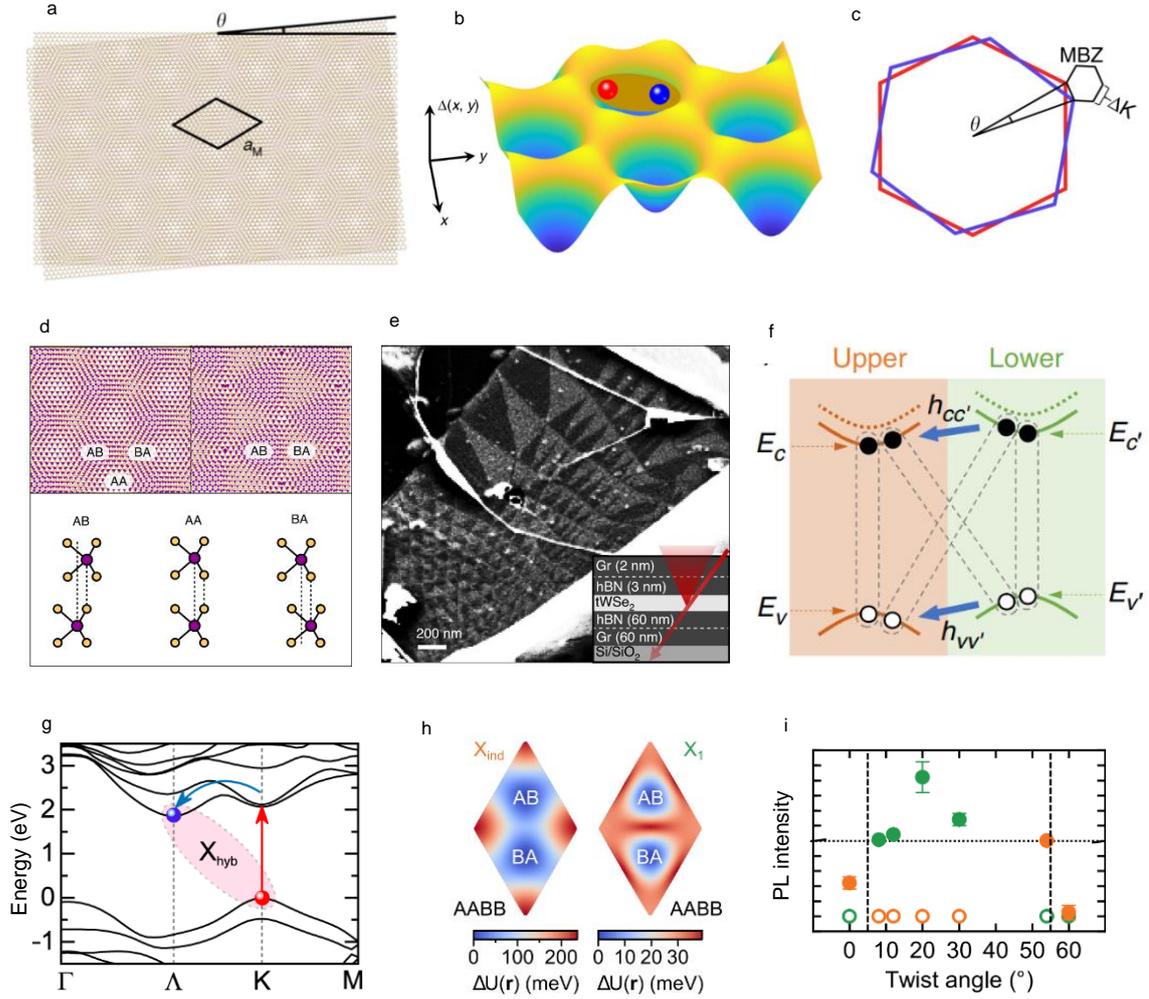

**Figure 5 | Moiré superlattice in twisted TMD/TMD homobilayer.** (a) Schematic illustration of the generated moiré pattern in TMD homobilayer with a rotational misorientation. (b) The moiré superlattice results in periodic modulation of the electrostatic potential in the twisted homobilayer, which can capture the excitons. (c) The moiré Brillouin zone that formed due to the formation of the moiré unit cell. (a-c) Adapted with permission from ref.[140] Copyright 2022 Springer Nature. (d) Top: top view schematic of the moiré pattern in twisted $WSe_2$ with twist angle of 5°with (right) and without (left) reconstruction. The reconstruction was created by reducing the local twist angle near the AB and BA sites. Bottom: side view of AB, AA and BA stacking orders. Purple circle presents W and yellow circle is Se. (e) SEM image of twisted $WSe_2$ showing a reconstructed moiré pattern with triangular AB and BA (3R) stacking domains ($\theta = 40°$). Inset shows the device schematic. (d,e) Adapted with permission from ref.[139] Copyright 2021 Springer Nature. (f) Excitons of intralayer and interlayer configurations, indicating the interlayer hopping between layers and the coexistence of the formation of these two excitons. Adapted with permission from ref.[144] Copyright 2021 American Physical Society. (g) Illustration of hybridized excitons though calculated single-particle band structure of the $WSe_2$ BLs for θ = 60°. After resonant excitation (red arrow) at the K point, electrons scatter via phonons (blue arrow) into the energetically favorable Λ valley to form hybrid excitons. Adapted with permission from ref.[145] Copyright 2020 Springer Nature. (h) Moiré potentials for interlayer exciton ($X_{ind}$) and intralayer exciton ($X_1$) within the moiré unit cell. (i) Normalized integrated PL intensity for interlayer exciton (orange) and intralayer exciton (green) as a

function of twist angle. (h,i) Adapted with permission from ref.[138] Copyright 2023 American Physical Society.

### 3.1.3 Hexagonal boron nitride (hBN)

The hBN layer is a wide bandgap insulator, featuring robust polar covalent bonding between the boron and nitrogen atoms. It possesses the same atomic structure as that of a graphene single layer and serves as the most commonly used material for engineering vdW heterostructures.[24, 146-147] As expected, the twisted double layer hBN gives rise to a similar moiré periodic triangular pattern with twisted graphene bilayers with alternating AB and BA stacked regions (Figure 6a). However, there are six high-symmetry configurations in the structure due to the non-identical atoms of boron and nitrogen in the unit cell, and they are divided into two different groups, parallel and antiparallel, based on the stacking orientations.[148] Figure 6b depicts a schematic representation of different stacking orders for the interface constructed in bilayer hBN, showing the AA, AB and BA stacked domains for parallel twist orientation, which is similar to the t-BLG, where the top layer of hBN fully overlaps with the bottom layer, aligning boron atoms with boron and nitrogen atoms with nitrogen, and the boron (nitrogen) atoms overlaid with the nitrogen (boron) atoms residing in opposite layers, respectively. Notably, the AB and BA stacking are essentially the same but with the only distinction of layer inversion[149] (left panel, Figure 6b). Because AB and BA stackings are energetically favourable, the triangular domains of AB and BA stacking are expected to be predominantly observed when the system is relaxed from the moiré induced local strain, as Figure 6a illustrates.[148, 150-151] The anti-parallel stacking configuration follows similar stacking rules as parallel ones but is symmetric under spatial inversion, which can form AA′, AB′$^{NN}$ and BA′$^{BB}$ stacked domains[149] (right panel, Figure 6b). AA′ configuration where the interfacial boron atoms of one layer are aligned to nitrogen atoms of the adjacent layer is more energetically favourable than the other two with the overlap between boron (nitrogen) and boron (nitrogen) of the neighboring layer. The hBN crystal typically grows in the AA′ stacking orientations thus the bulk form of the material naturally has AA′ configuration.[152] The technique of electrostatic force microscopy (EFM) was employed for imaging the moiré pattern produced between the hBN layers by detecting the variations in the surface potential (Figure 6c). Fumagalli *et al*. suggested that the observation of moiré pattern is attributed to the permanent out-of-plane dipoles generated by pairs of different atoms (boron and nitrogen) in the opposite layers, which creates ferroelectric-like domains with oppositely polarized dipoles (B-N or N-B) in the adjacent regions.[153] In addition to the dc-EFM curves shown in the inset of Figure 6c as a function of applied dc bias, a more direct indication is displayed in Figure 6d, the contrast of different domains shown in the EFM image will be inverted when the sign of the applied bias is changed from negative to positive. They both further confirmed the origination of the observed periodic pattern and this only occurs in parallel stacking orientation[148] because of the removal of restrictions imposed by inversion and mirror symmetries.[154-155] The dc-EFM images of two marginally twisted hBN devices with the presence of parallel and antiparallel and with parallel only are presented in Figure 6e, showing the periodic signal on one side but no regular pattern on the other side for the device with both domains (left panel, Figure 6e), in contrast, the representative triangular pattern was found on both sides with the same contrast for the other device (right panel, Figure 6e). The study demonstrated that the twisted vdW stacking not only alters the electronic band structure of materials such as graphene but also impacts the crystal symmetry, which allows for the creation of ferroelectric switching from parent compounds that are non-ferroelectric.[148-149, 153] Yao *et al*. exploited the electrical polarization at the interface with parallel stacking configuration to generate a superlattice potential that can act on general 2D materials. It is a non-invasive approach and was realized by simply aligning an hBN film with another as the substrate to support the 2D materials (Figure 6f). The top material will experience the modulation under the superlattice energy landscape determined by twisted hBN, regardless of the crystal orientation and lattice type of materials.[154] The first principle DFT calculated band-edge energies of monolayer $MoSe_2$ on hBN bilayer structure as functions of stacking registry positions are demonstrated in Figure 6g and it confirms the above statement. The stacking model used in the calculation was illustrated in insets of Figure 6g, where $r_t$ describes to the alignment between $MoSe_2$ and hBN, $r_{bn}$ characterizes the registry between two hBN layers with parallel stacking. The energies of CBM (conduction band minima) and VBM (valence band maxima) show anticipated modulation as a function of $r_{bn}$ caused by the $r_{bn}$-dependent electrical polarization in h-BN. When a mall twist is introduced at the h-BN interface, a superlattice energy landscape for electrons of $MoSe_2$ will be produced. However, there is no observed dependence of the band-edge energies on the stacking alignment $r_t$. The recent observation of Rydberg moiré excitons in $WSe_2$ utilized a similar approach[156] to modulate the properties of top layer TMD by adjusting the moiré period generated from a twisted graphene bilayer. This provides a new control parameter for tailoring the minibands and is distinct from the typical moiré superlattices.[154] Additionally, it can be further extended to combine with the moiré structure, enabling the coexistence of primary and secondary modulations on the target materials.

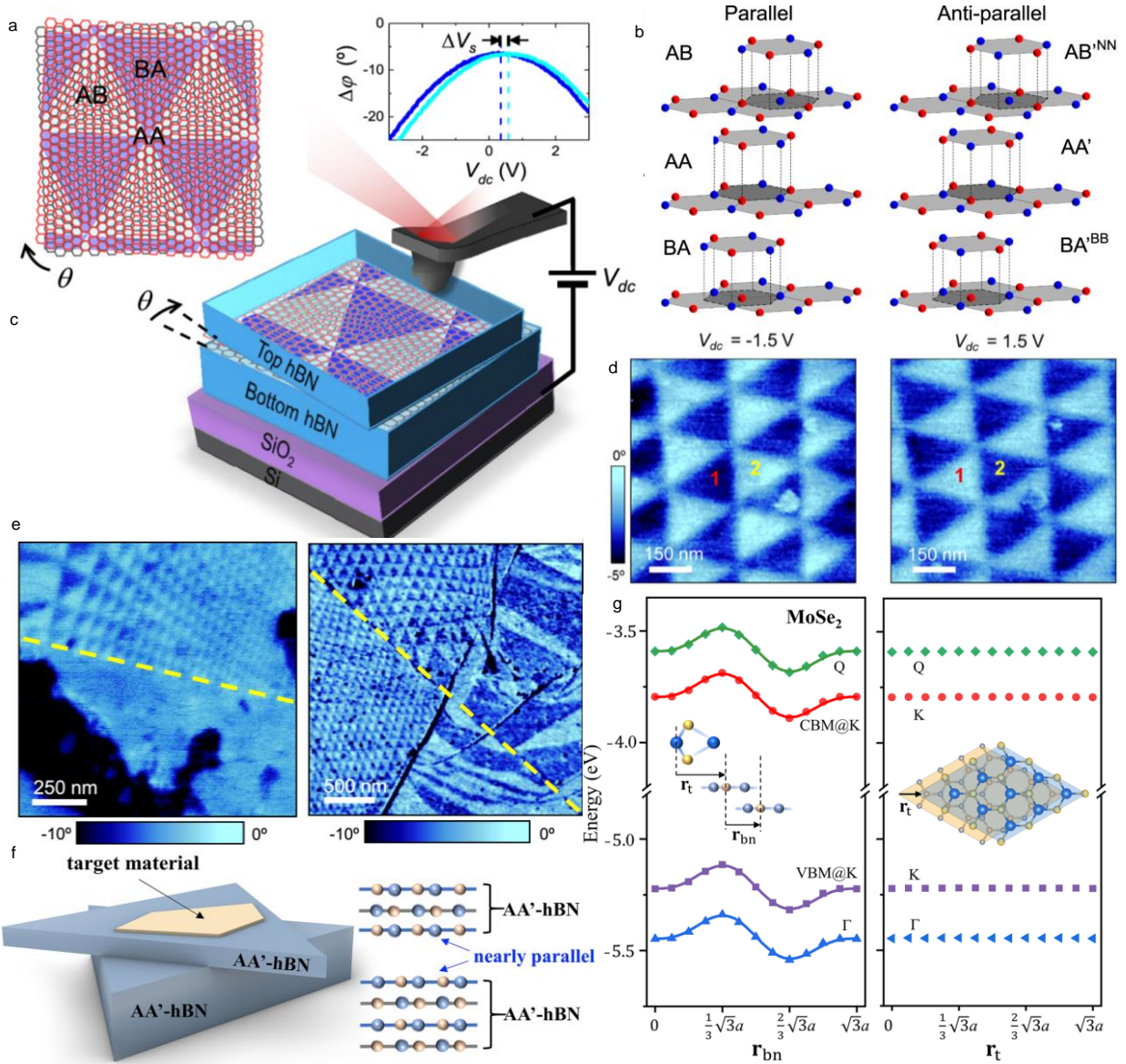

**Figure 6 | Moiré superlattice in twisted hBN bilayer.** (a) A schematic representation of the moiré pattern created in the hBN bilayer by misaligning two layers by a small angle, giving rise to predominantly AB and BA regions. (b) Illustration of six high-symmetry stacking configurations for the hBN–hBN interface. Nitrogen atoms are shown in red and boron atoms in blue. (c) Schematic of EFM experimental setup. A voltage bias is applied between the AFM probe and the silicon substrate. Inset shows representative dc-EFM curves as a function of the applied dc bias in two adjacent triangular domains. The horizontal shift of the maximum of the curves yields the variation in surface potential between the domains. (d) Representative dc-EFM images on marginally twisted hBN with -1.5 V and +1.5 V dc bias applied between the tip and the sample substrate. The triangular contrast reverses upon changing the sign of the dc bias. (e) The dc-EFM images of the hBN/hBN homo-bilayers with a transition from parallel alignments on one side to antiparallel on another side (left) and with only parallel alignment (right), the triangular potential modulations are visible on one side and both sides, respectively. (a-e) Adapted with permission from ref.[148] Copyright 2021 Springer Nature. (f) A moiré interface of nearly parallel alignment between two AA′-h-BN films is created as the substrate to modulate the performance of target material. (g) DFT-calculated band structures of monolayer MoSe$_2$ on a bilayer h-BN substrate shown in (f) as functions of $r_{bn}$ (registry between

the two parallel BN layers, inset of left panel), and $r_t$ (stacking registry between $MoSe_2$ and h-BN, inset of right panel). (f, g) Adapted with permission from ref.[154] Copyright 2021 Springer Nature.

### 3.1.4   Black phosphorus (BP)

The monolayer BP emerged among the various 2D materials as a prominent electronic material with distinctive characteristics,[157] especially possessing a tunable direct bandgap[158] and high mobility of carriers.[159] Similar to other hexagonal isotropic lattices demonstrated above in t-BLG or TMD materials, twisted phosphorene also shows the formation of moiré superlattice but it presents another intriguing system for investigating moiré physics with an anisotropic rectangular lattice.[160-161] As shown in Figure 7a, the rotational angle between two BP monolayer sheets leads to the emergence of various stacking orders, exhibiting four different high-symmetry local registries known as AA, AB, AA' and AB' (Figure 7b) to therefore give rise to a moiré pattern.[73, 162] The size of this moiré unit cell strongly depends on the twist orientation and is proportional to $1/\theta$. The different interlayer vdW interactions between atoms in two BPs across those different local stacking configurations lead to the variation in potentials and interlayer distances at different registry points, which work as impurities and provide a 'rugged land' of the electronic potential in the system, it is expected to have a remarkable impact on carrier transport.[162] The band structures of multiple twisted bilayer BPs with different twist angles ranging from1.8° to 5.4° are calculated as shown in Figure 7c. The increased band gaps were observed with the increasing angle misorientation, and the almost flat bands appear at both the conduction band minimum (CBM) and valence band maximum (VBM) of the bilayer structure with a small relative angle (first panel, Figure 7c) due to the localization of electronic states, which significantly affects the carrier mobility in the twisted BPs.[162] The dispersive bands are also observed at energy bands away from CBM and VBM. Figure 7d presents the electron and hole mobility along the armchair direction in twisted BPs as a function of the rotational angle between two monolayers. It demonstrates limited electron transport due to the deformation in potentials, while the hole mobility shows minimal changes with respect to the twist angle. When the rotation angle is twisted from 0º to 1.8º, the formation of a moiré supercell leads to a significant reduction of nearly 20 times in electron mobility, reaching the minimum, which is attributed to the strong localization of the electron wave function along the armchair direction at the small twist angle.[162] However, beyond the threshold twist angle of 1.8º, the electron mobility increases as the size of the moiré pattern shrinks with further rotation, eventually blurring. Notably, at a larger twist angle of around 5.4º, the predicted electron mobility is expected to exceed that in the zero-twist bilayer, where the moiré supercell is absent, owing to the anticipated variation of other parameters with respect to θ.[162] In addition, the optical anisotropic moiré transitions have also been experimentally reported in twisted 1L/2L BP, suggesting the distinct moiré resonances in phosphorene moiré supercell even for the structures with a large rotation of around 19º. Figure 7e and f present the high-angle annular dark-field scanning transmission electron microscopy (HAADF-STEM) image and schematic illustration of twisted monolayer/bilayer phosphorene with twist angles of approximately 5.7° and 19°, respectively, both demonstrating clear rectangular moiré patterns and are indicated by orthogonal vectors and grey dashed lines. This supports the preservation of moiré superlattices in twisted 1L/2L BP structures across a wider range of twist angles, differing from other systems where the moiré pattern typically exists only for rotational angles smaller than a few degrees.[161] The PLE measurement of twisted monolayer/bilayer phosphorene with varying degrees of twist angles (Figure 7g-i) demonstrates the distinct difference of optical transitions for various rotation alignments, showing highly angle-dependent optical properties. Although all of them exhibit PL emission near 0.83 eV, a small energy shift can be still observed among different samples. The PL emission at this low energy in structures with large twists such as 19° (Figure 7g) and 6° (Figure 7h) is much broader, but one with no clear resonances and another exhibits two absorption peaks with weak and strong intensities respectively. Whereas, the sample with 2° (Figure 7i) and 0° both show only a single prominent peak at the excitation energy of 1.92 eV. This twist angle-dependent observation suggests that the electronic band structures of twisted phosphorene can be modulated by the moiré potentials with the changing of twist orientations, and the effect remains pronounced for large twist angles up to 20°.[161] The calculated electronic structure of CBM and VBM in 20° twisted 1L/2L BP was examined, which displayed a similar VBM energy to that of nature bilayer phosphorene but significantly lower CBM energy compared to both bilayer and trilayer black phosphorene (Figure 7j). The charge density distributions of CBM and VBM were further illustrated in Figure 7k and l, the conduction band shows strong hybridization between the monolayer and double layer, with electrons at the CBM predominantly localized at the interface between them and exhibiting periodic modulation (Figure 7i). However, the VBM electrons are primarily localized in the bottom double layer and exhibit negligible coupling with the top monolayer (Figure 7k). The calculations identify that although this twisted structure has a similar bandgap size to the trilayer phosphorene (Figure 7j), their underlying conduction and valence bands are entirely distinct from each other.[161]

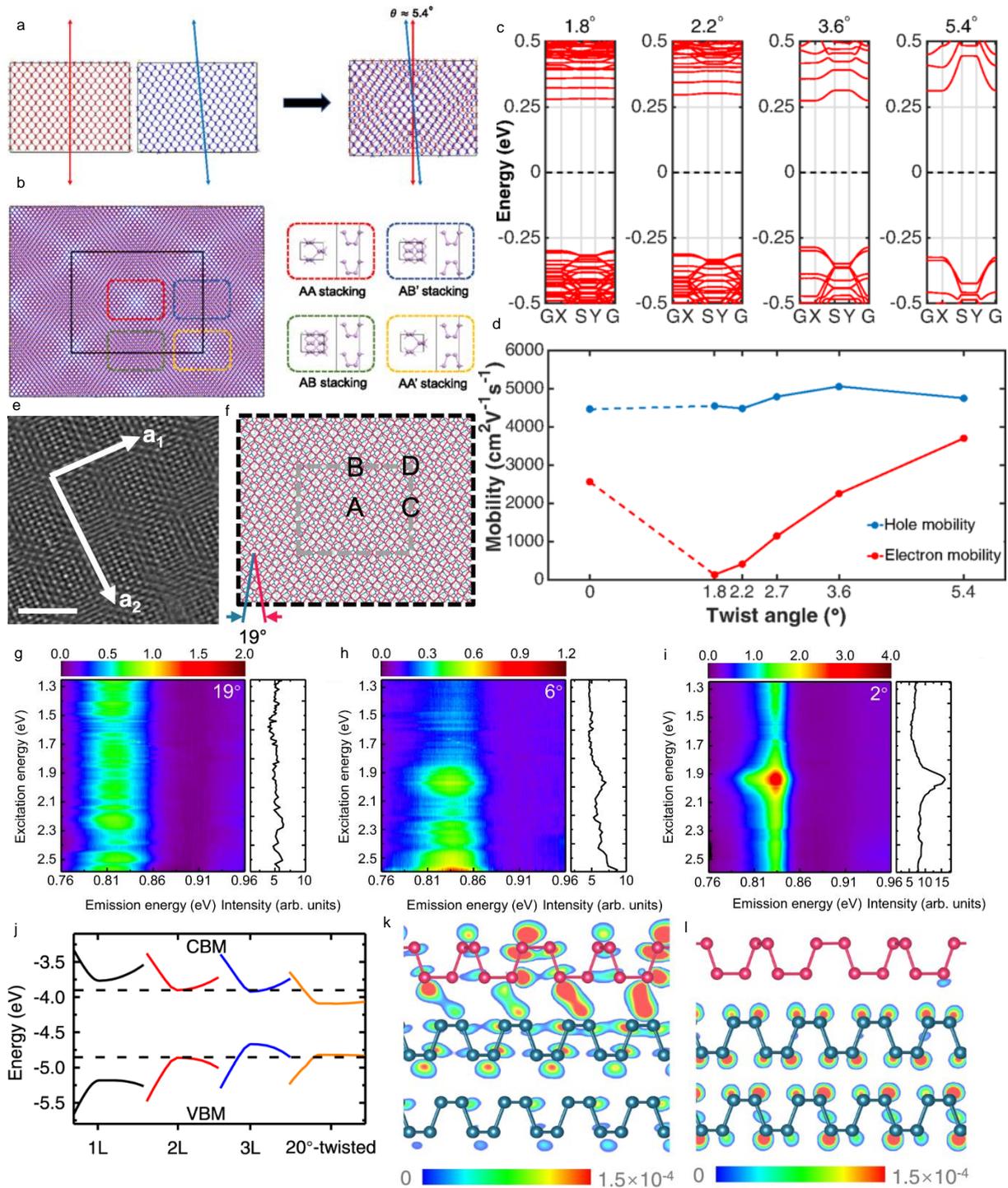

**Figure 7 | Moiré superlattice in twisted BP bilayer.** (a) Schematic construction of a twisted bilayer BP with a twist angle θ which is defined by the orientation between the two constituent BP lattices. The zero angle corresponds to the AA stacking of bilayer BP. (b) Moiré pattern in a twisted bilayer BP with a twist angle of 2.7∘. The coloured rectangles indicate the distribution of the high-symmetry local stacking configurations in twisted BP bilayers and the black rectangle indicates the simulated supercell containing 3588 atoms. The atomic structures of the four high-symmetry local stacking configurations are named AA, AB, AA', and AB', respectively and are shown in the right panels. (c) The calculated band structures of four twisted bilayer BP structures with twist angle of 1.8°, 2.2°, 3.6°, and 5.4°, respectively. (d) Deformation limited carrier mobility along the armchair direction of twisted BP bilayers versus the twist

angle θ. The dash and solid lines are guides to the eye. (a-d) Adapted with permission from ref.[162] Copyright 2017 American Physical Society. (e) HAADF-STEM image of an approximately 5.7° twisted monolayer/bilayer phosphorene structure. It shows rectangular shape moiré superlattices, as indicated by two orthogonal superlattices vectors labelled $a_1$ and $a_2$. (f) A top-view illustration of the twisted monolayer/bilayer phosphorene with a large angle of 19°. The top monolayer (red) is rotated by 19° respected to the bilayer phosphorene (light cyan for the middle layer and blue for the bottom layer). The gray dashed rectangle indicates the supercell that contains four special stacking configurations. (g-i) PLE spectra of twisted monolayer/bilayer phosphorene structures with the angle of 19° (g), 6° (h) and 2° (i). (j) Band structure of the monolayer (black solid line), bilayer (red solid line), trilayer (blue solid line), and 20° twisted 1L/2L phosphorene (orange solid line) near Γ points. The two black dashed lines indicate the CBM and VBM of bilayer phosphorene. The vacuum energy is set to be at 0 eV. (k, l) The sectional partial charge density distribution at the CBM (k) and VBM (l) of the 20° twisted phosphorene heterostructure along the armchair direction of the constituent bilayer phosphorene. The wavefunctions of the monolayer (red) and bilayer phosphorene (blue) are strongly coupled at the CBM, whereas, they are mainly localized in the bottom bilayer at the VBM. (e-l) Adapted with permission from ref.[161] Copyright 2021 American Physical Society.

### 3.1.5  Magnetic materials

The twist angle between layers in a magnetic homostructure can be adjusted to allow for a precise control over the depth of the moiré potential, offering a means to enhance the magnetic properties of layered twisted structure. Hongchao Xie et al.[163] fabricated a rotational stacked double bilayer of a 2D magnet using chromium triiodide ($CrI_3$) and demonstrated skilfully engineering of 2D magnetism through twisting, as illustrated in Figure 8a and b. They observed a novel magnetic ground state that differs from the intrinsic arrangements of $CrI_3$ in two-layer and four-layer structures (Figure 8c, d). When the twist angle is small, the emergent magnetism seems to be a combination of the magnetism of two- and four-layer $CrI_3$. However, with a larger twist angle, it bears a strong resemblance to the independent two-layer $CrI_3$ configuration. At an intermediate twist angle, they observed a finite net magnetization that defies explanation through a homogeneous stacking configuration (Figure 8e-g). This net magnetization arises due to spin frustrations caused by interplay between ferromagnetic and antiferromagnetic exchange couplings in moiré supercells. Similarly, researchers discovered a layered vdW magnet $Fe_3GeTe_2$ (FGT)[164], which exhibited a distinct two-step magnetic hysteresis loop in its homostructure (Figure 8h, i). This unique behaviour indicated an interaction between the two constituent FGT layers without their fusion into a thicker, homogeneous material. The coercivities of the homostructure differed from those of the individual FGT layers, indicating a mutual interaction between the constituents. Additionally, the homostructure displayed a Curie temperature intermediate to that of the thinner and thicker constituent layers, suggesting an interaction between the layers without the formation of a naturally occurring thicker layer. Weijong et al.[165] discovered that the interlayer coupling in bilayer $CrBr_3$ is highly influenced by the stacking order, leading to either ferromagnetic or antiferromagnetic interactions. Interestingly, while the individual monolayer of $CrBr_3$ demonstrated ferromagnetic properties, the bilayer configurations exhibited distinct magnetic behaviors dependent on the specific stacking arrangement. This intriguing stacking-dependent interlayer magnetism opens new possibilities for manipulating 2D magnetism through precise control of the rotational angles between neighboring layers. when two layers of graphene are stacked with a small twist rotation,[166] the resulting flat minibands from superlattice greatly enhance electron-electron interactions(Figure 8j). The study provides evidence that the improved interactions induce a ferromagnetic state in t-BLG. Within a specific density range, approximately three-quarters filling of the conduction miniband, the researchers observed the appearance of ferromagnetic hysteresis and a significant anomalous Hall effect, reaching up to 10.4 kilohms. Additionally, there are observations of chiral edge states. It's worth noting that the magnetization of the sample can be reversed by just applying a minor direct current. Despite the fact that the anomalous Hall resistance is not quantized and there's some dissipation, the measurements indicate that the system might display traits of an emerging Chern insulator.

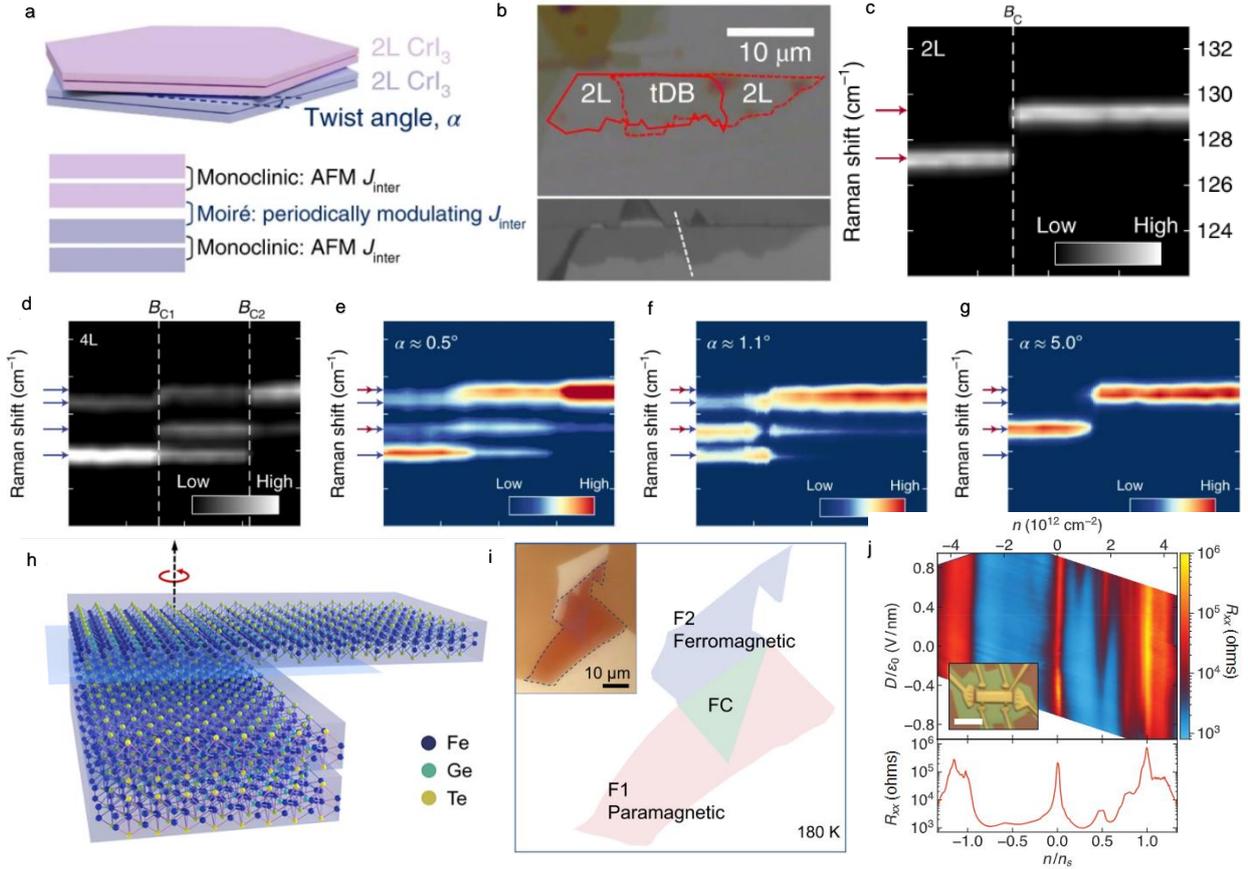

**Figure 8 | Moiré magnetic homostructure.** (a) Illustration depicting a twisted double bilayer (tDB) sample of chromium triiodide ($CrI_3$) composed of two layers (2L) of $CrI_3$ with a twist angle $\alpha$ between them. The interface between the two 2L $CrI_3$ layers gives rise to a moiré superlattice, resulting in a periodically modulated interlayer exchange coupling (Jinter). In contrast, individual 2L $CrI_3$ exhibits monoclinic stacking and corresponding antiferromagnetic (AFM) Jinter. (b) False-colored optical images showcasing a homostructure of tDB $CrI_3$. The top image highlights two 2L $CrI_3$ constituents outlined in red, while the bottom image displays the original large size 2L $CrI_3$. The tearing boundary is indicated by a white dashed line. (c-g) False-colored maps depicting the B⊥-dependent Raman spectra acquired for various samples. (c) Raman spectrum recorded on 2L $CrI_3$, (d) Raman spectrum captured on 4L $CrI_3$, (e) tDB $CrI_3$ homostructures with $\alpha = 0.5°$, (f) $\alpha = 1.1°$, and (g) $\alpha = 5.0°$. The measurements were performed in the crossed linear polarization channel at a temperature of 10 K. (a-g) Reprint with permission from ref.[163] Copyright 2021 Springer Nature. (h) Schematic representation of a homostructure composed of $Fe_3GeTe_2$ (FGT) layers. The blue plane represents the interface between two constituent FGT flakes. (i) Schematic representation of the magnetic phases observed in different regions of the sample at a temperature of 180 K, slightly above the Curie temperature (TC) of region F1. The inset shows an optical image of the sample with a scale bar of 10 μm. (h,i) Reprint with permission from ref[164] Copyright 2023 American Physical Society. (j) Top panel: The longitudinal resistance (Rxx) of the near-magic-angle TBG device is plotted as a function of carrier density (n) and perpendicular displacement field (D), which are controlled by the top- and back-gate voltages, at 2.1 K. The carrier density (n) is mapped relative to the superlattice density (ns), corresponding to four electrons per moiré unit cell, shown on the bottom axis. Inset: Optical micrograph of the completed device displaying the top-gated Hall bar region, electrical contacts 4, areas where the TBG has been etched (green), and regions that have not been etched (brown). Scale bar, 5 mm. Bottom panel: Line cut of Rxx

with respect to n taken at $D/\epsilon_0 = -0.22$ V/nm, revealing resistance peaks at full filling of the superlattice and additional peaks likely indicating the emergence of correlated states at intermediate fillings. Reprint with permission from ref.[166] Copyright 2019 American Association for the Advancement of Science.

### 3.1.6 Oxides

After observing great potential in other twisted layers, a novel central research arena has been established. Moiré patterns in oxides follow a similar rule and typically emerge from overlaying two slightly different periodicities. The emergence of moiré superposition has been reported in mechanically exfoliated vdW -bonded twisted perovskite oxides, specifically $Bi_2Sr_2CaCu_2O_{8+\delta}$. With a twist angle of approximately 45° (Figure 9a) at the temperature of 90 K, near the superconducting critical temperature in its bulk form, a robust and completely gapped topological phase is created and characterized by disrupted time-reversal symmetry and the presence of chiral modes.[167] It provides a promising opportunity for achieving a long-desired realization of topological superconductors at high temperatures. On the other hand, the electronic moiré patterns have been also reported in the structures of correlated transition metal oxides, which are usually not bonded by vdW interactions but it is well known that there are strong couplings and intricate interactions between charge, spin, bond, lattice, and orbital degrees of freedom, which offers a complex phase diagram to the material system and gives the possibility of achieving moiré patterns through strain engineering.[168] Zeng *et al.* first observed two distinct periodic domains from $La_{0.67}Sr_{0.33}MnO_3$ (LSMO) thin films on $LaAlO_3$ (LAO) substrates due to the interface coupling effects. One domain originated from the crystal terraces of the substrate, while the other emerged from the rhombohedral twin domains caused by the strain modulation within the LSMO film itself. Using scanning near-field optical microscopy (SNOM), they observed that when these two periodic potential fields overlapped at a small angle, their combined effect of periodic strain led to the formation of microscopic moiré profiles. This was attributed to the simultaneous presence of the two incommensurate patterns of strain modulation and their interactions (Figure 9b). Furthermore, the periodicity of the moiré patterns was significantly larger than that of each individual strain domain. In addition, there is a nonlinear dependence between conductivity and strain in this LSMO system, thus by slightly adjusting the period or curvature of one of the strain potentials, a substantial change in the moiré pattern can be achieved, transitioning from striped patterns to curved patterns (Figure 9c).[168]

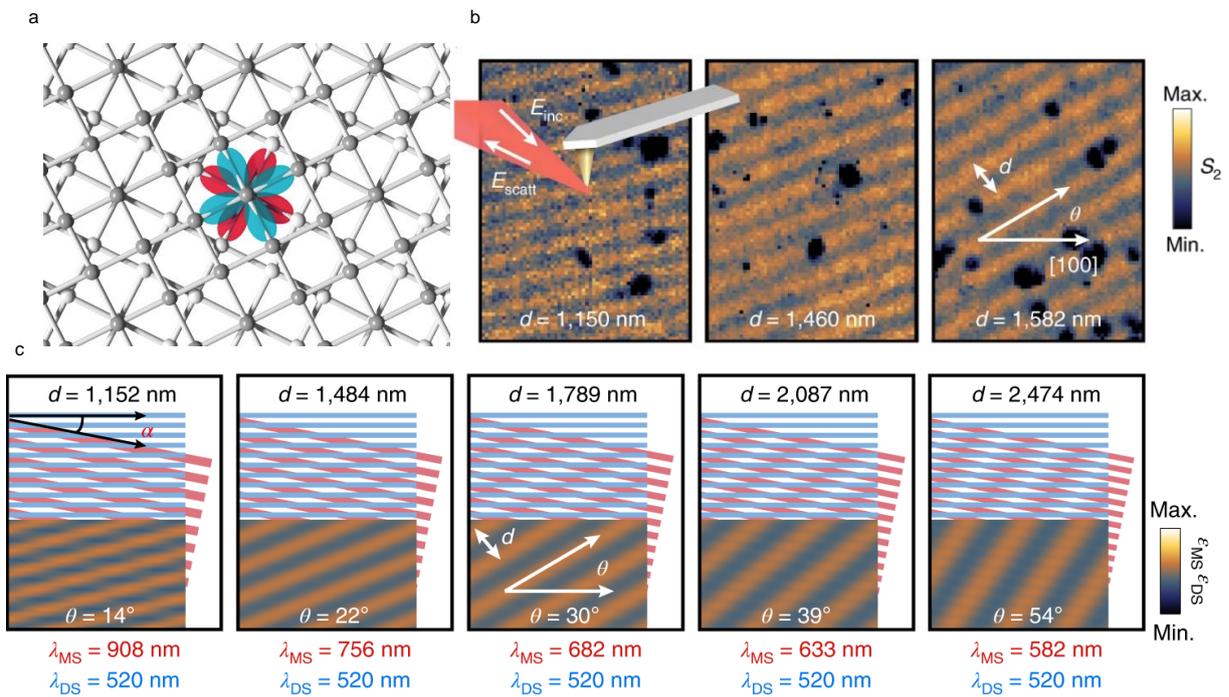

**Figure 9 | Moiré superlattice in twisted oxides.** (a) Schematic view of vdW -bonded two copper–oxygen square lattices with twist angle of 45°, indicating the emergence of moiré superstrucutre. Adapted with permission from ref.[167] Copyright 2021 Springer Nature. (b) Schematic demonstration of how the spatial coexistence of two distinct domains yields the combined moiré patterns comprising regions of local commensuration and incommensuration. (c) s-SNOM imaged electronic moiré patterns in representative LSMO/LAO films, demonstrating striped (left) or curved (right) patterns. (d) Different moiré patterns generated by superimposing two periodic striped motifs with the variation of one periodic domain, while another one and the twist angle between them are fixed. (b-d) Adapted with permission from ref.[168] Copyright 2020 Springer Nature.

The experimental findings align well with theoretical simulations, confirming that the period of the moiré pattern varies significantly by monotonically adjusting the periodicity of only one domain, and the angles between the orientation of moiré fringes and substrate direction will be changed accordingly, as depicted in Figure 9d. A variety of moiré textures were also observed in twisted bilayer oxide stacks consisting of $SrTiO_3$ nanomembranes with different crystallographic orientations.[169] These studies demonstrate the extension of moiré engineering beyond the conventional vdW stacking structures and highlight the potential of moiré modulation in other non-layered materials, which is not the primary focus of this review but it's worthy of further investigation.

### 3.2 Hetero-Structures

#### 3.2.1  Graphene/hBN

Due to the distinct electronic properties in instinct graphene, the advent of theoretically predicted magic twist angle in twisted heterostructures first sparked great interest in moiré superlattice that arises between graphene and other materials, by varying the mis-orientation angle between two materials. Of particular interest is the insulator layer of hBN with the same atomic structure but opposite electric performance. hBN has been recognized as a perfect substrate for graphene to achieve better quality devices with the highest mobility.[147] Compared to the most common substrate of silicon dioxide, hBN couples weakly to the graphene layer, exhibiting strongly reduced disorder and suppressed charge inhomogeneity.[122, 147, 170-171] It is devoid of the problem of high roughness and impurity-induced charge traps in $SiO_2$,[172-174] which significantly eliminates the limitation of device performance and makes the Dirac point physics accessible.[170] This makes it especially important to better understand the interlayer coupling between two materials. Due to the isomorphic structure of crystal lattice, they can share lots of similar properties except the character of bandgap,[175] a 1.8% mismatch between lattice constants[176] and the relative rotational angle between two layers lead to a moiré superlattice pattern in a hexagonal shape.[122, 170-171] Similar to the other materials, the moiré period can be tuned and it is directly related to the rotational angle between two lattices.[122, 170] Figure 10a-d shows STM topographs of graphene on hBN with different twist orientations, which resolved their corresponding onsite atomic lattices and presented periodic modulations that enable the observation of different moiré patterns. The difference in moiré wavelength and orientation results from the different twist angles between graphene and the underlying hBN lattice.[122] Different patterns can be also observed from the different locations within the same flakes owning to the different crystallographic orientations of a single hBN.[170] Although the lattice structure of hBN is not visible in the images, it can be studied by examining the Fourier transforms of the moiré patterns (Figure 10e-h), which also determines the relative angle of hBN with respect to the graphene, as illustrated in Figure 10e. The Fourier transform analysis of STM images reveals two separate sets of peaks, consisting of six outer spots and six inner spots which represent moiré pattern and atomic lattice of graphene, respectively. In different images, the positions of these points exhibit a rotational difference with respect to each other.[170] The relative orientation between graphene and the underlying hBN can be calculated based on the relationship between their reciprocal vectors ($\mathbf{k}_{BN} = \mathbf{k}_{graphene} - \mathbf{k}_{moiré}$ in Figure 10e).[122] The degenerated Dirac points in pristine graphene are protected by the equivalent A and B triangular sublattices in a honeycomb structure and lead to the semi-metallic behaviour of graphene.[124, 177] However, the generated moiré superlattice from graphene/hBN gives rise to periodic different local potentials that arise from the stacking of boron and nitrogen atoms on A and B sublattices, respectively, which breaks the symmetry between A and B sublattices (Figure 10i) and induces a large local bandgap (Figure 10j) on graphene band structure,[178-181] which potentially transforming graphene from a semimetal into a semiconductor. The theoretical calculation predicted that the electronic performance of graphene can be changed by external periodic potentials[182] because additional Dirac points can be formed from the moiré superlattice, the experimental effort reported a few years later proved the emergence of new Dirac points (Figure 10k) and suggests that the energy of these points will change as a function of the size of moiré supercell.[171] The moiré mini-Brillouin zone bands are further confirmed in measured transport results from aligned devices with longitudinal and Hall resistance.[123, 183] As indicated in Figure 10l, in addition to the typical standard peak at graphene's main charge neutrality point, two more additional peaks occur symmetrically on the two sides of the neutrality point (inset). The Hall resistance reverse sign at all of these positions because the Fermi energy passes through the band edge, those two additional reversals are attributed to the new neutrality points from the moiré minibands.[123, 183-184] There are many other modifications of graphene's electric properties such as Hofstadter states[123, 185] by introducing this moiré potential, which have been intensively studied and reported, however, most of them assumed that there are no structural changes in graphene when it is stacked with hBN. Interestingly, a study investigated the strain distribution in the graphene layer when it is on the top of hBN, suggesting that the vdW interaction between two materials still plays a non-negligible, albeit weak, role on the materials.[55, 179] Two different moiré superlattice patterns are schematically displayed in Figure 10m, incommensurate and commensurate structures with large and small twisting angles, their corresponding Young's modulus distributions were measured and extracted in Figure 10n along the cross-section of the conductive AFM (atomic force microscopy) image of materials.[55] In the incommensurate stacking, the signal displays close to a sine wave shape (left panel, Figure 10n) and follows the characteristic hexagonal structure, which is consistent with the prediction.[179] In contrast, when the twist angle is small with a longer moiré period, the almost constant Yong's modulus was separated by an abrupt high modulus with a narrow region of around 2nm (right panel, Figure 10n).[55] In addition, the Raman spectra of graphene on hBN also indicate distinct differences in FWHM between different structures (Figure 10o), which increases gradually with the decreased rotation angle.[186] This sensitive performance can be explained by the presence of a strain distribution in the perfectly aligned graphene. Figure 10p illustrates the ratio of FWHM of Young's modulus peak to the size of the moiré cell as a function of the moiré period. The ratio decreases as the sample becomes better aligned, but it saturates when the moiré length is shorter than 10 nm because no reconstruction occurres in the graphene lattice. The ratio then suddenly drops to the minimum when the alignment is perfect, due to the elastic deformation of graphene caused by the moiré effects.[55] This transition from incommensurate to commensurate induces the concentrated strain in a narrow region because the graphene needs to be stretched slightly to reach an energetically favourable state to realize the commensurate stacking for vdW interactions with the bottom layer of hBN, while for a large twist with small moiré pattern, the vdW energy is insufficient and two lattices behave independently, thus there are no discernible regions where strain accumulates.[55, 187-188] Furthermore, two or more superlattices can coexist by vertically stacking multiple layers together, creating individual moiré patterns between each pair of graphene-hBN layers and allowing for a generation of a more complicated interface (Figure 10q). It introduces additional controlling for significantly modifying the structure of graphene and it has been demonstrated that the multiple-order moiré cell can be also manipulated.[189] This technique is not limited to graphene on hBN but has been widely explored in the study of other materials and vdW heterostructure devices in recent years.[52, 190-192]

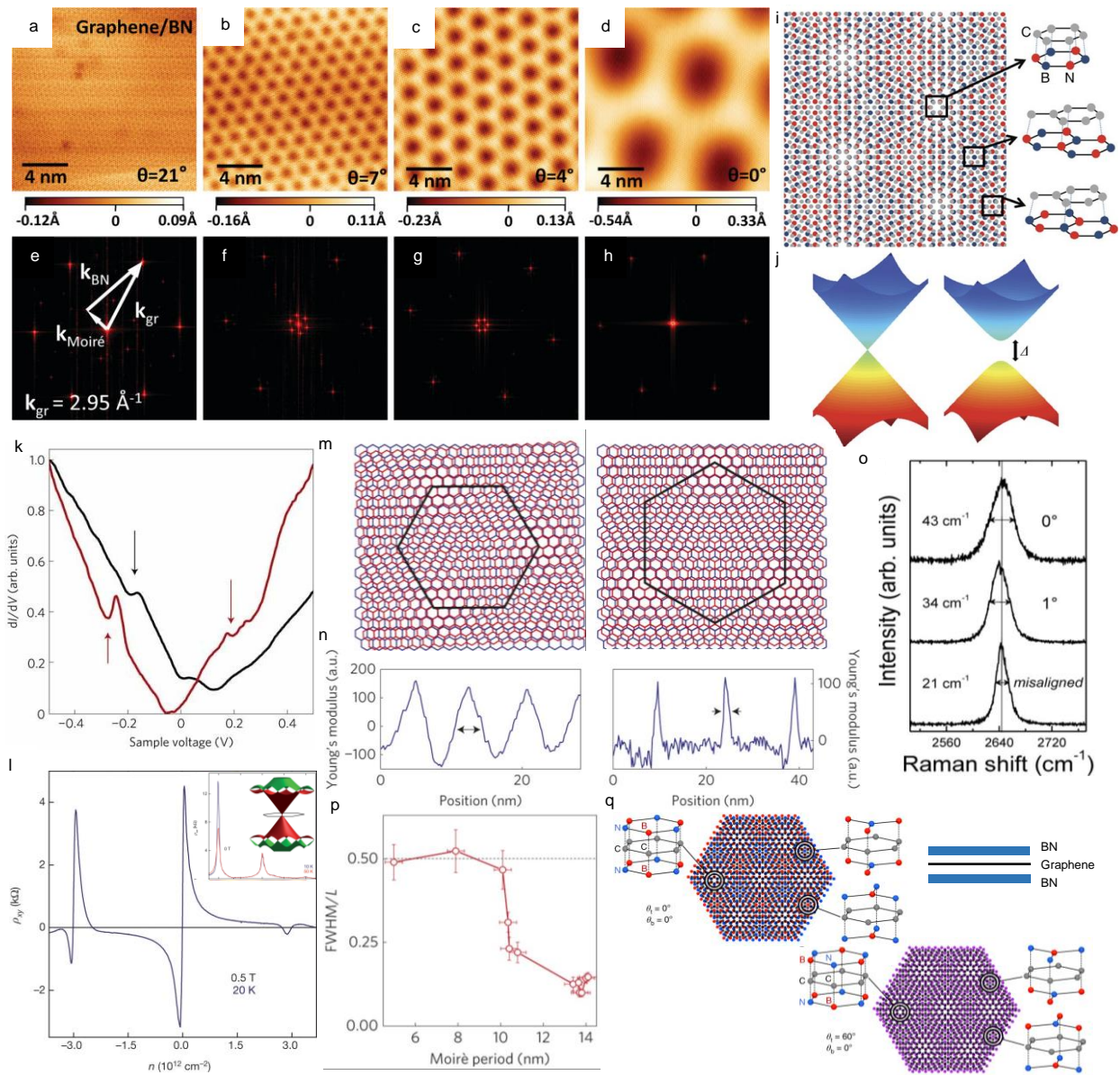

**Figure 10 | Moiré superlattice in graphene on hBN layer.** (a-d) Constant current STM images of graphene/BN structure with four different twist angles, showing that the moiré pattern changes depending on the rotational orientation between graphene and the BN lattices. (e-h) Fourier transforms of the topographs shown in (a-d). The outer spots show the graphene reciprocal lattice, the inner spots show the graphene/BN moiré pattern reciprocal lattice. (a-h) Adapted with permission from ref.[122] Copyright 2011 American Chemical Society. (i) Schematic of the moiré pattern in graphene on hBN with zero twist angle (carbon, grey; boron, blue; and nitrogen, red). The lattice alignments in different regions lead to different local sublattice symmetry breaking in graphene. (j) Energy spectrum of pristine graphene (left) and gapped graphene (right) when stacking with hBN layer. (i,j) Adapted with permission from ref.[181] Copyright 2014 Springer Nature. (k) Experimental dI/dV curves for two different moiré wavelengths, 9.0 nm (black) and 13.4 nm (red), in graphene/hBN heterostructure. The dips in the dI/dV curves are marked by arrows, revealing the different superlattice Dirac points from different moiré patterns. Adapted with permission from ref.[171] Copyright 2012 Springer Nature.

(l) The Hall resistivity, $\rho_{xy}$, changes sign and passes through zero at high electron and hole doping, revealing well-isolated secondary Dirac points. The inset shows the longitudinal resistivity, $\rho_{xx}$, as a function of electron and hole density. The band structure is plotted for the first and second superlattice Brillouin zone (shown in red and green, respectively). Adapted with permission from ref.[183] Copyright 2013 Springer Nature. (m) Schematic illustration of commensurate and incommensurate structure in graphene on hBN with a rotational angle of 0º and 3º. Black hexagons indicate the moiré patterns. (n) Cross-sections of Young's modulus distribution taken along the moiré patterns with the period of 8 nm and 14 nm. (m,n) Adapted with permission from ref.[55] Copyright 2014 Springer Nature. (o) Measured Raman spectra on graphene deposited on BN with a mismatch angle of 0° and 1° as compared to a misaligned structure. Adapted with permission from ref.[186] Copyright 2013 American Chemical Society. (p) Measured ratio between FWHM of the peak in Young's modulus distribution (as marked by arrows in n) and the period of the moiré structure L, as a function of the period of the moiré structure. Adapted with permission from ref.[55] Copyright 2014 Springer Nature. (q) Schematics of BN–graphene–BN heterostructures with all layers aligned at 0° (top) or 60° (bottom) rotational offset between the top and bottom BN layers, illustrating of the two individual moiré patterns and supermoiré patterns for three overlapping hexagonal lattices. Adapted with permission from ref.[189] Copyright 2019 Springer Nature.

### 3.2.2 TMD-TMD

Unlike the modulation on intralayer excitons from moiré homostructure, the new exciton state named interlayer excitons[193] normally arises from the interlayer coupling when the heterostructure is constructed. The type-II band alignment is typically formed in the TMD hetero-stack system,[194-195] which enables the rapid and efficient charge transfer[196-198] between two layers and provides a platform for the production of interlayer excitons at lower energies, with the electrons and holes being situated in separate layers (Figure 11a). The special pattern can be generated from the TMD stacking configuration with the atomic alignment, which induces a periodic potential landscape that displays varying separation distance and energy bandgap of heterobilayers[30, 36] based on the rotational angle and lattice mismatch between two TMD layers, allowing the effective modulation on the properties of exciton states. This in-plane periodic variation leads to the formation of moiré superlattice, and it can be divided into two categories according to the different stacking alignments. The heterobilayer with a twist angle (Figure 11b) near 0° refers to R-typed (rhombohedral stacking) moiré superlattice and near 60° is known as H-typed (hexagonal stacking). Three high-symmetry atomic sites can be identified in both typed stacking as depicted in Figure 11c for the 60° twist angle. The typical denotation of $H_h^h$, $H_h^M$, $H_h^x$ represent the specific atomic local overlapping in this position from vertically attached two layers. The superscript and subscript in the notation mean electron and hole layers respectively, h, x and M indicate the centre of the hexagonal lattice, chalcogen atom and metal site. Thus the produced moiré potentials are strongly position dependent, it can laterally and correspondingly modulate the topographic and electronic structures across the nanoscale supercell.[30] The different local separation distances at different sites enable the lateral modulation in the electric field produced across the unit cell (Figure 11d), and the global potential minima within the supercell can trap the interlayer excitons with a tunable depth, which is reported in the magnitude of around 100 to 200 meV.[24, 27, 30] Figure 11e schematically illustrates the calculated specific potential levels for both R- (top panel) and H-stacked (bottom panel) heterobilayers, which demonstrate the dependence on position and twist angle for the modulation of the energy landscape. The excepted moiré potential in R-stacking is much deeper than that in the H-stacking style because of the higher spatial variations with perfect alignment, which makes the excitons to be highly confined and the excitons in the 60° heterobilayer should be more flexible.[31] The misalignment in both rotational angle and lattice constant play a substantial role in the production of moiré supercell, these two factors can not only determine the construction of the superlattice and also decide the period size of the superlattice. The heterostructures with identical materials but different fabrication methods were reported for a clear comparison to better understand the formation mechanism of moiré superlattice (Figure 11f). For mechanically exfoliated samples, the heterostructure was made via dry transfer with a well-controlled twist angle, the moiré superlattice effects are typically produced with the characteristics of mutually twisted and superposition of crystalline lattices, and the size of supercell can be adjusted with the rotational angle between two monolayers (Figure 11f bottom), which reduces with an increase of the rotational misalignment[199]. In addition, the slightly incommensurate structure with two monolayers sharing the same chalcogen site has a broader tunable range in the moiré period than the incommensurate structure that has different lattice constants.[72, 200-201] However, in near-commensurate CVD(Chemical vapour deposition)-growth well-aligned heterostructures such as $MoSe_2$/$WSe_2$ heterobilayer, the lattice in each layer only distorts slightly because the crystalline axis is prone to align, leading to the formation of a completely commensurate structure, and the moiré crystal does not exist here (Figure 11f top).[199, 202] The moiré superlattice with deep potential levels and large period size can completely localize the excitons, while the smaller one with a larger twist angle can provide the opportunities for excitons to tunnel across the supercells. Therefore, moiré effects will be neglected for the heterostructures with rather large angles or very different lattice constants.

The emissions from some heterobilayers (for example, $MoSe_2$/$WS_2$) can also be hybridized when two TMD layers are closely aligned (Figure 11g). In both R- and H-stacking alignments, the Brillouin zones of two monolayers are appropriately well-overlapped in the momentum space, establishing a nearly direct bandgap for emissions. The conduction band edge of $MoSe_2$ and $WS_2$ are located at the similar energy position with a small difference,[195] while the hole band offset between two layers is large, which allows the interlayer electron hopping between states with the same spin and valley, enabling both the direct and indirect transitions to form the intra- and interlayer excitons and resulting in a hybridization effect between corresponding transitions that share the same hole state. Except for the trapping potentials, another important moiré effect that can modify the optical properties of the material is the spatially varied spin optical selection rules. The nanoscale patterning of distinct selection rules within the supercell (Figure 11h) indicates that the local regions are generally elliptically polarized in R-stacking heterobilayer, but the first two high-symmetry registries ($R_h^h$ and $R_h^x$) in the moiré pattern are oppositely circularly polarized. The wave packet falls into these two regions couples only to σ+ and σ− circularly polarized light propagating in the out-of-plane direction respectively, while the light coupling in the point of $R_h^M$ is not effectively allowed because of its perpendicular dipole with the 2D plane (Figure 11i).[24, 30] The H-stacked heterobilayers have different spin optical properties due to their bright triplet states.[203-204]

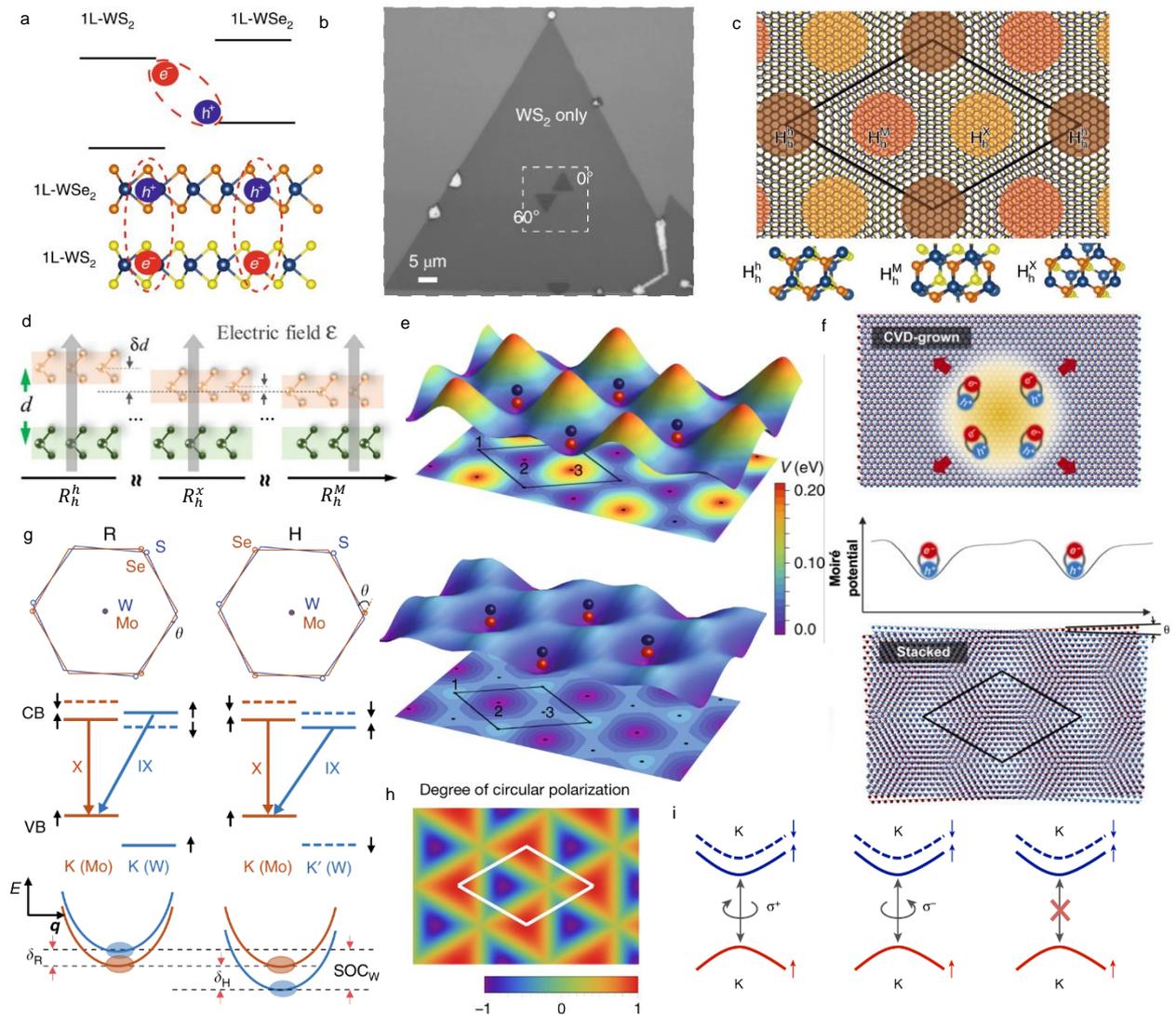

**Figure 11 | Moiré superlattice in TMD/TMD heterobilayers.** (a) Schematic of the type II band alignment in the WS$_2$/WSe$_2$ heterobilayers that shows interlayer exciton formation. Energy levels are indicated by solid black lines. Interlayer excitons are marked by dashed ovals. Navy, yellow and orange spheres represent W, S and Se atoms respectively. (b, c) Established moiré patterns of the WS$_2$/WSe$_2$ heterobilayers and their supercells (black lines) for a twist angle of 0° and 60° (c) Schematic of the moiré pattern of the WS2–WSe2 heterobilayers and

their supercells (black lines) for a twist angle of 60°. Brown, yellow and orange circles mark regions with high-symmetry stacking configurations, labelled as $H_h^h$, $H_h^M$, $H_h^X$, respectively. (a-c) Adapted with permission from ref.[31] Copyright 2020 Springer Nature. (d) Dependence of interlayer distance d on the atomic registries. Adapted with permission from ref.[30] Copyright 2017 American Association for the Advancement of Science. (e) Illustrations of the 2D K–K moiré potentials in both 3D graphs and 2D projections that can trap interlayer excitons (red and black spheres) in the local minima, for 0º (top) and 60º (bottom) twisted WS$_2$/WSe$_2$ heterostrucutres. 1, 2 and 3 indicate three different registries indicated in (b) and (c). The theoretical results show that the moiré pattern leads to a twist-angle-dependent modulation of the energy landscape. Adapted with permission from ref.[31] Copyright 2020 Springer Nature. (f) Schematics of excitons in a CVD-grown WSe$_2$/MoSe$_2$ bilayer with a commensurate structure, which does not form the moiré pattern (top). Trapped excitons (middle) by the moiré potential in a mechanically stacked heterobilayer (bottom). The moiré supercell size decreases with an increasing twist angle θ between the two layers. Adapted with permission from ref.[199] Copyright 2020 American Association for the Advancement of Science. (g) Top: Illustrations of a unit cell of R-stacking (left) and H-stacking (right) WS$_2$/MoSe$_2$ bilayers. Middle: The corresponding band alignments of WS$_2$ and MoSe$_2$, where X refers to the intralayer transition and IX labels the nearly resonant interlayer transition that shares the same hole state. Solid and dashed lines correspond to states of opposite spins. Bottom: The alignment between an intralayer MoSe$_2$ A exciton state (red) and the interlayer exciton state (blue) that it hybridizes with. Adapted with permission from ref.[47] Copyright 2020 Springer Nature. (h) The spatial map of the optical selection rules for K-valley excitons in a moiré superlattice. The high-symmetry points are circularly polarized with opposite signs and the regions in between are elliptically polarized. (i) K-valley excitons obey different optical selection rules depending on the atomic configuration $R_h^\mu$ within the moiré pattern. $R_h^\mu$ refers to R-type stacking with the μ site of the MoSe$_2$ layer aligning with the hexagon centre ($h$) of the WSe$_2$ layer. Exciton emission at $R_h^h$ ($R_h^X$) is left-circularly (right-circularly) polarized, but emission from the $R_h^M$ site is dipole-forbidden for normal incidence. (h,i) Adapted with permission from ref.[24] Copyright 2019 Springer Nature.

### 3.2.3 TMDs/graphene

After intensive interest arose from the t-BLG and twisted TMDs homo- or hetero-structure, the moiré pattern from the interface between graphene and TMD creates another additional path for studying moiré superlattice. The moiré pattern consists of TMD and graphene endows strong compatibility, which can be formed from various stacking configurations to give rise to a range of superpatterns. It is not limited by the stacking of monolayer TMD and graphene, the superlattice can also be generated from the heterostructures of monolayer TMD/nanolayer graphene, few-layer TMD/few-layer graphene, monolayer graphene/few-layer TMD and even monolayer TMD on graphite. The moiré superlattice formed from monolayer MoSe$_2$ and bilayer graphene (Figure 12a) is demonstrated via STM images[205]. In addition to the hexagonal atomic unit cell presented in bilayer graphene (Figure 12b), the heterostructure of 1L $MoSe_2$ and graphene displays a honeycomb atomic lattice (Figure 12c) with a unit cell of 3.3 Å, which is consistent with the atomic separation distance in basal Se and Mo planes of the TMD monolayer (Figure 12a left). The presence of this periodic superlattice can be attributed to the overlapping of $MoSe_2$ and graphene lattices. Specifically, because four graphene unit cells can fit three $MoSe_2$ unit cells, which leads to the formation of a quasi-commensurate pattern with a 9.87 Å × 9.87 Å moiré superstructure. In the larger-scale image of $MoSe_2$/bilayer graphene shown in Figure 12d, only one sublattice is visible due to the lower spatial resolution. While the atomic registries of superlattice were usually aligned accurately with the intrinsic lattice of graphene, as indicated by yellow arrows in images (Figure 12b and c), it also displayed a rotational angle as shown with the purple lines in Figure 12d. Besides, the periodic moiré superstructure can be observed if the TMD monolayer or a few layers were grown on HOPG (Highly Ordered Pyrolytic Graphite) substrate,[205-206] and this moiré pattern exhibits the strong dependent feature on the twist angle between materials. It reflects the commensurate alignment between TMD and graphene layer and affects the size of the moiré period, as well as tunes the quasiparticle bandgap because of the shift of the maximum valence band. The experimental data illustrated in Figure 12e and f provide strong evidence showing an almost monotonic decrease of quasiparticle bandgap and periodicity of moiré lattice with the increasing rotational misalignment, which are both reduced considerably when the relative angle between layers is changing from 0° and 30°. The similar variations here suggest a tight relationship between the moiré period and the optoelectronic properties observed from the samples.[206] The hetero-bilayers of $MoS_2$ and graphene with rotation angles of 0° and 30° are schematically illustrated in Figure 12e and f, which point out the different lattice mismatch between monolayer $MoS_2$ and graphene, showing a variation of the atomic registry at different angles. The formation mechanism of the moiré lattice in the $MoS_2$/graphene system is closely related to how the sulphur atoms of the $MoS_2$ layer are overlapped with the carbon atoms in the adjacent graphene layer. The effective number of coincident points can determine the periodicity size and this number is inversely proportional to the separation between the TMD and graphene layer. The maximum intensity in the moiré pattern occurred at the position where sulphur atoms coincide with carbon atoms or the distance between two atoms is the minimum. When the size of moiré supercell is decreased, the number of coincident positions per unit cell will be increased, resulting in an improvement of interlayer coupling, which reveals the connections between the quasiparticle bandgap and interlayer interaction in the heterostructure. Apart from the typical hexagonal TMDs, the representative distorted 1T structure of $ReSe_2$[207-208] produces a moiré pattern when it behaves as an overlayer to vertically stack onto the graphene, which differs from the normal twisted 2D heterostructures and has two distinct periods along different crystallographic orientations[209] (see $L_1$ and $L_2$ in Figure 12h-j). Both experimental observations (top panels of Figure 12h-j) from STM images and calculated geometrical analysis (bottom panels) of 1L $ReSe_2$/graphene consistently prove that the moiré patterns obtained from heterostructure are angle–dependent. Similarly, a one-dimensional striped super-pattern resulting from the moiré cell was observed across the interface between graphene and $ReS_2$ due to the stacking of triclinic lattice on the hexagonal lattice of graphene.[210] Although the moiré structures from TMD/graphene system have been massively reported and studied, the higher-indexed patterns with complex and rich textures can be arisen by adjusting the rotational angle, choosing appropriate lattice parameters and establishing the relationship between the potential modulations and electronic properties. A high-indexed moiré pattern was discovered from the heterostructure of $MoTe_2$/graphene by investigating surface topographies from STM image (Figure 12k), which is fabricated by growing a layer of *2H*-phase $MoTe_2$ on the top of high crystalline graphene by molecular beam epitaxy (MBE), with the twist angle is around 30°.[211] It can unexpectedly enhance the typically faint moiré potential modulations and allow direct and easier observation in microscopy investigations. The image covers three different periodic structures marked as (I), (II), and (III), which show a variety of patterns with different tunnelling conditions.

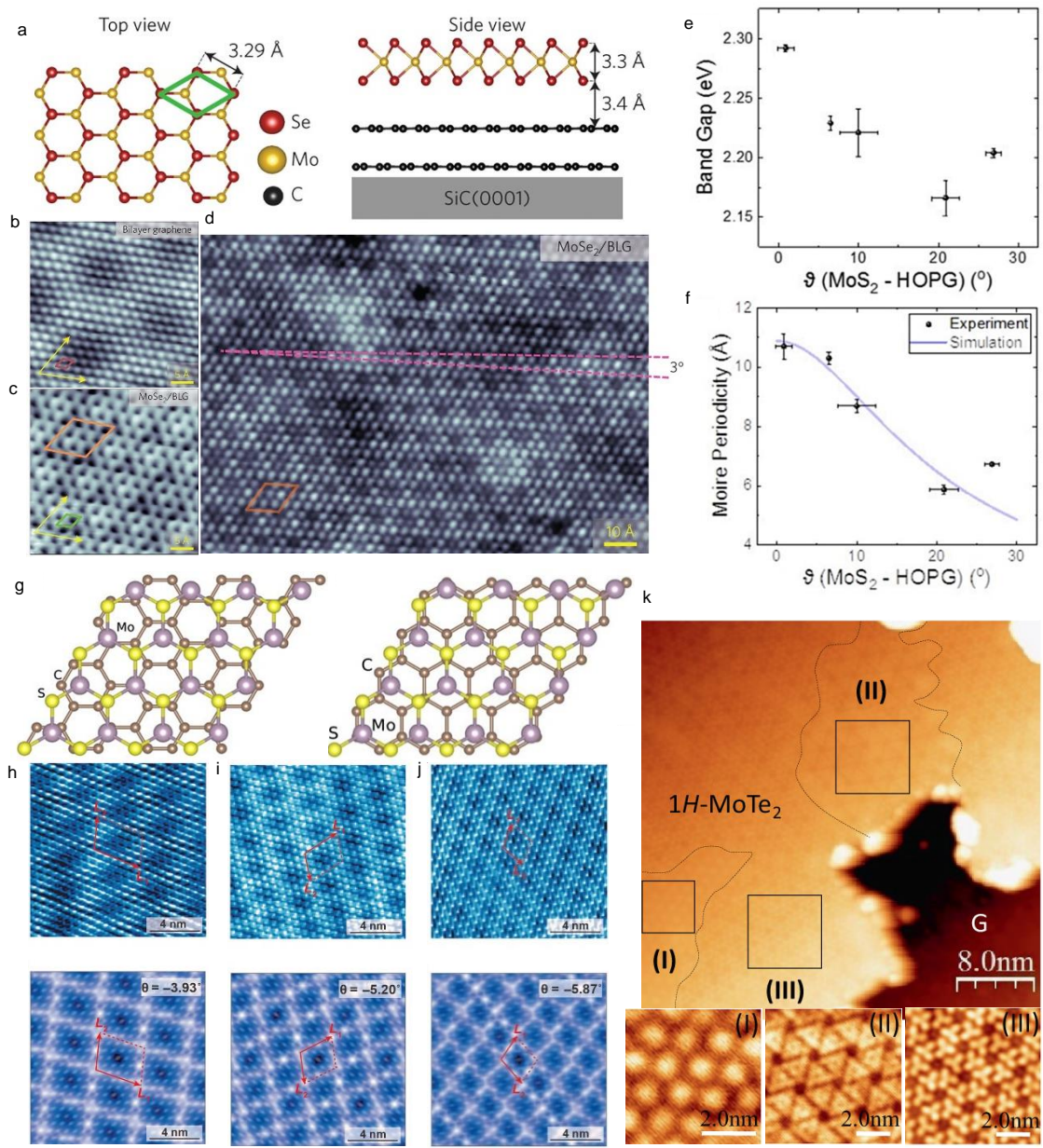

**Figure 12 | Moiré superlattice in TMD/graphene heterostructures.** (a) Top and side view sketches of monolayer MoSe₂ on the bilayer graphene. (b) High-resolution STM image of bilayer graphene. (c) High-resolution STM image of MoSe₂ on graphene. Unreconstructed unit cells are indicated in dark red and green, approximate moiré pattern unit cell for MoSe₂ on graphene is outlined in orange. (d) STM image (typical resolution) of monolayer MoSe₂ on bilayer graphene in a larger-scale. (a-d) Adapted with permission from ref.[205] Copyright 2014

Springer Nature. (e) Bandgap energies of 1L $MoS_2$/HOPG heterostructure as a function of the twist angle. (f) Moiré superlattice constant in 1L $MoS_2$/HOPG heterostructure as a function of the twist angle. (g) The geometry of two configurations for a $MoS_2$ overlayer with the underlying graphene substrate at a twist angle of $0°$ (left) and $30°$ (right). (e-g) Adapted with permission from ref.[206] Copyright 2019 Elsevier. (h-j) Top: Experimental observed moiré patterns by STM images in monolayer $ReSe_2$/graphene with different twist orientations. Bottom: Corresponding calculated moiré patterns obtained from the geometrical analysis. (h-j) Adapted with permission from ref.[209] Copyright 2019 American Association for the Advancement of Science. (k) STM image of 1L $MoTe_2$ on the graphene layer. The close-up images for the regions marked by black squares are indicated in (I) (II) and (III) below, revealing the observation of rich and complex moiré patterns on the local scan areas. Adapted with permission from ref.[211] Copyright 2022 Springer Nature.

### 3.2.4 TMDs/magnetic materials

Just as magnetic properties have been observed in homostructure, it is evident that heterostructures also possess the potential to exhibit magnetic behavior. In the case of twist-stacked heterostructures, the precise modulation of the angle between different layers allows for the formation of moiré superlattices. This control over the layer angle introduces the possibility of achieving profound magnetic properties within the material. By leveraging the tunability of the moiré potential, we can explore and exploit the magnetic characteristics of these layered structures, opening new avenues for investigating and harnessing their magnetic properties. Tong *et al.*[212] investigated the generation and manipulation of skyrmions in 2D magnets within vdW heterostructures, as illustrated in Figure 13a. By exploiting the moiré pattern (Figure 13b), skyrmions are formed through the manipulation of magnetic interaction between layers. The resulting skyrmions exhibit double degeneracy and are controlled through the application of strain and translation between layers. In the strong interlayer coupling regime, skyrmion lattices are observed with their vorticity and position modifiable via magnetic fields. In the weak coupling regime, controlled manipulation of metastable skyrmion formations within the ferromagnetic state becomes achievable through precise administration of current pulses. These findings demonstrate the potential of moiré skyrmions to serve as tunable carriers of information and topological structures that influence electron transport. Furthermore researchers have investigated the generation of valley splitting in monolayer TMDs through the lifting of valley degeneracy[45]. They demonstrated a correlation between the potency of the magnetic proximity impact, driven by charge transfer and Coulomb interaction between layers, and the degree of valley splitting in vdW heterostructures based on TMDs. Notably, achieving large valley splittings requires a type-III band alignment within these vdW heterostructures. First-principles calculations on TMD/NiY$_2$ heterostructures (where Y = Cl, Br, I) validate this correlation and foresee numerous heterostructures with substantial valley splittings as illustrated in Figure 13c and d. The study[213] demonstrates a significant advancement in manipulating valley splitting within vdW heterostructures. These structures combine a monolayer of MoTe$_2$ with layered MnSe$_2$ MnSe$_2$ exhibiting ferromagnetism at room temperature. The study unveils substantial enhancements in valley splitting within MoTe$_2$. The achieved maximum of 106 meV in different MoTe$_2$/MnSe$_2$ heterobilayer configurations (as depicted in Figure 13e, f) is similar to the impact of a magnetic field as potent as 530 T, characterized by effective Zeeman splitting. Furthermore, the investigation reveals that augmenting valley splitting is achievable through external vertical stress and biaxial compressive strain as illustrated in Figures 13g, h. For heightened operational efficacy at elevated temperatures, researchers applied a 2.3% biaxial tensile strain to the monolayer MnSe$_2$ magnetic substrate. This adaptation elevated the Curie temperature from its intrinsic 266 K to 353 K, while still upholding a significant valley splitting of 72 meV. Similarly, Tenzin *et al.*[214] pertains to the employment of the proximity effect via an EuS substrate. This strategy yielded an impressive valley exciton splitting of 16 meV/T in 1L WS$_2$, surpassing the effect of an externally imposed magnetic field by nearly two orders of magnitude. Notably, an intriguing observation emerged where a reversal in the valley splitting direction when compared to WSe$_2$ on an EuS substrate.

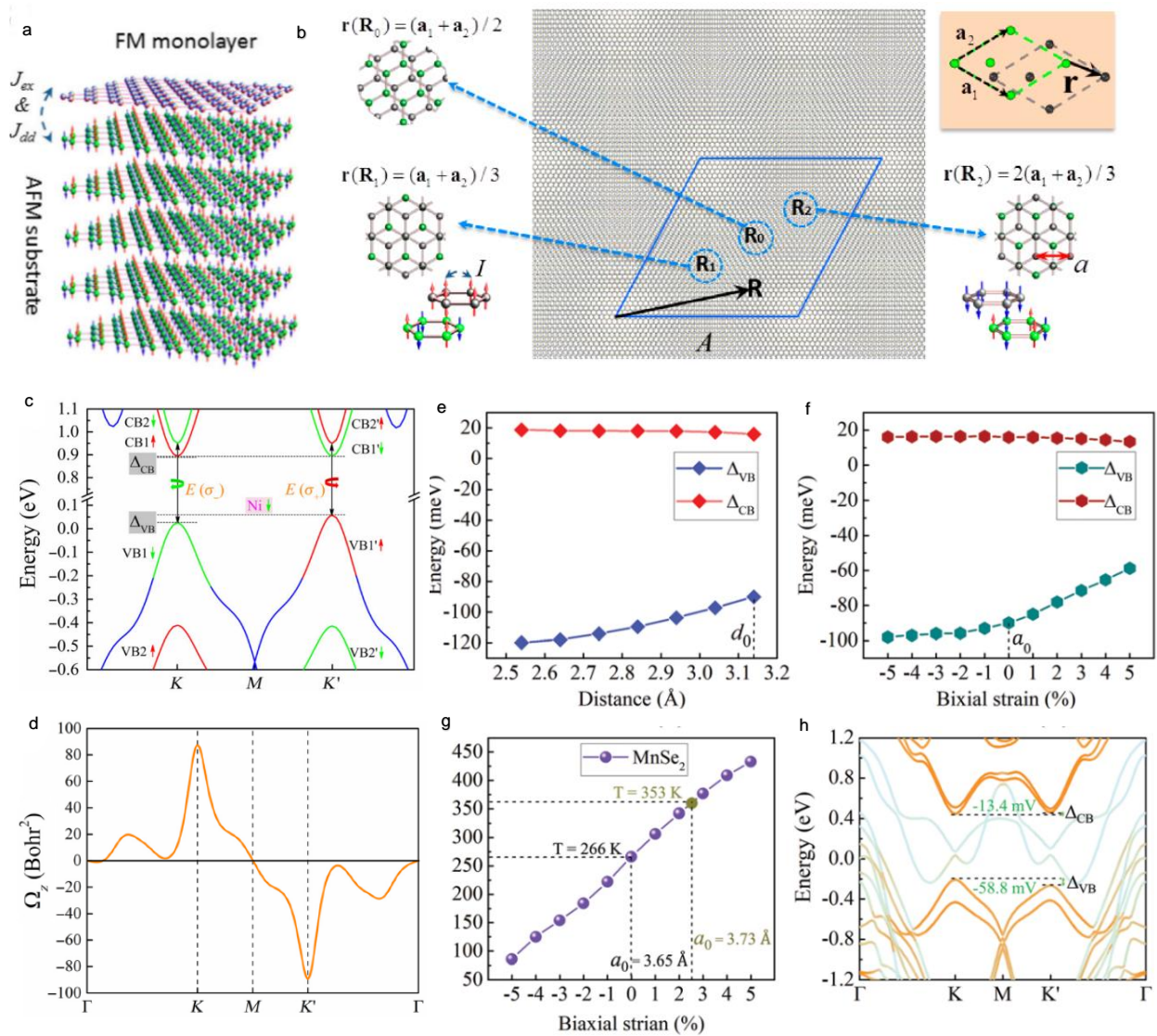

**Figure 13 | Moiré TMD/ magnetic heterostructures.** (a) An FM monolayer on a layered AFM substrate with lateral Néel order and perpendicular anisotropy. (b) Moiré pattern formation between the ferromagnetic (FM) (gray) and the top layer of the antiferromagnetic (AFM) substrate (green) resulting from lattice mismatch and/or twisting. Three local regions at R0, R1, and R2 are magnified to highlight their atomic registries, which closely resemble the registries of lattice-matched stacking with different interlayer translations. (a,b) Reprint with permission from ref.[212] Copyright 2018 American Chemical Society. (c) Schematic representation of moiré-induced modulation of interlayer magnetic coupling between a ferromagnetic (FM) monolayer and a layered antiferromagnetic (AFM) substrate, exhibiting lateral Néel order and perpendicular anisotropy. (d) Calculated Berry curvature of the monolayer WTe$_2$ in the heterobilayer at the high-symmetry k-points. (c,d) Reprint with permission from ref.[45] Copyright 2020 American Physical Society. (e) Variation of valley-splitting with interlayer distance, (f) effect of biaxial strain on valley-splitting, (g) changes in Curie temperatures with biaxial strain in monolayer MnSe$_2$, and (h) band structure of

MoTe$_2$/MnSe$_2$ under 5% tensile strain. (e-h) Reprint with permission from ref.[213] Copyright 2022 Royal Science of Chemistry.

### 3.2.5 Others

Following the preliminary success of moiré modulation on various kinds of material structures including t-BLG, trilayer graphene, TMD homobilayer, TMD heterobilayer and other vdWs heterostructures that may not be covered above, the study of moiré engineering has been extended to a wider range of material families, even in the system of 3D/3D oxides.[168] The periodic moiré superstructures were observed from the structure of Pd/IrTe$_2$ because of the interfacial lattice mismatch between Pd and IrTe$_2$.[215] The STM topography of the IrTe$_2$ surface at a low temperature shows a coexisting domain of hexagonal and stripe phases in IrTe$_2$ (Figure 14a), and the moiré patterns with different periodicity and orientation were observed in both regions The pattern from the Pb film on the hexagonal area gives rise to a regular moiré cell (Figure 14b), while the pattern from that on the stripe region exhibits a less regular structure (Figure 14c), owing to the extra strain produced in the hexagonal domain.[216] The induced moiré superlattice leads to a strong lateral modulation in electronic structures of Pb film with split sub-bands on its quantum well states, and the stripe charge ordering from IrTe$_2$ offers additional tunability and can form 1D electronic states in the ultrathin Pb films.[215] Beyond exploring the moiré effects by stacking atomic layers one on top of the other, starting from the heart of a sandwich structure is expected to open a new way for constructing more possible marvellous phenomena and creating more tenability to materials by decoupling the originally attached materials. The intercalation growth of a 2D oxide layer in between the metal substrate and graphene was investigated, which decoupled the graphene layer from the metal and enabled modifications of graphene locally on the electronic structure because of a periodic in-plane superstructure from the substrate.[217] The moiré superstructure of graphene on metal has been widely reported such as Ir,[218] Pd,[219] Ru,[220-221] Cu,[222] Co,[223] and Pt.[224-225] These moiré structures were created due to the superposition of graphene's honeycomb lattice and metal's hexagonal lattice[219] and most of them are evidenced by STM images. Figure 14d displays the superstructures of graphene on Pt(111), which displays multiple domains with different orientations. The largest size of the moiré supercell of graphene on Pt(111) was summarized as around 22 Å by comparing multiple superlattices with different rotation angles including the experimental data and theoretically predicted results.[225] At the same time, the moiré superlattice formed from the structure of 2D-FeO/Pt(111) was also extensively studied, which was formed due to the lattice mismatch between metal and 2D material as well as the varying adsorption sites, and the moiré period was measured around 26 Å.[226-229] The structure of graphene/FeO/Pt(111) was constructed after the growth of FeO from behind, inserting the oxide layer between the graphene and Pt(111) and STM studies confirmed the formation of a new interface in between.[217] as shown in Figure 14e. A long-range surface modulation was observed in addition to the perfect intrinsic honeycomb structure in graphene, which is the only superpattern and there are no different domains visible. The corresponding surface height profile along this moiré structure is plotted in Figure 14f, showing the periodic variations. Although the large-scale STM image (Figure 14g) did not resolve the superstructure, it shows the layer fashion of iron on the metal to indicate strong cohesive energy between iron and Pt as a driving force for the intercalation growth.[217] The new superlattice is different from the pattern observed from graphene/Pt(111), and the size of the unit cell was measured to be larger than any reported moiré pattern that formed of graphene on Pt(111).[225] It implies that the periodic surface modulation in the STM image of graphene/FeO/Pt(111) is from the underlying moiré structure generated in the bottom layers instead of the previous moiré pattern of graphene on Pt(111). It decouples the previous structure and the top layer of graphene will consistently conform to the modulation regardless of its orientation relative to the bottom structure. The properties of graphene were thereby tuned due to a periodic charge doping variation in graphene, which is caused by the fluctuation in the surface potential of the substrate and understood by studying the substrate work function.[225] Recently, based on the success of moiré engineering in 2D/2D systems, the moiré pattern in 2D/3D interface has been focused. The system of epitaxial aligned $MoS_2/Au(111)$ (Figure 14h) was studied by employing a combination of analytic convolution technique and a range of scanning transmission electron microscopy imaging methods[63] because the individual traditional technique such as STM or TEM is challenging for 2D/3D system with some limitations[230-231] that lead to discrepancies in measurement of moiré period. More detailed investigation regarding imaging techniques will be discussed and compared in the following sections of "Visualization". The uniform periodicity of the moiré cell demonstrates a weak binding between Au and underlying material because it lacks distortion, which is different from those cases of twisted vdW heterostructures with atomic lattice reconstructions.[46, 134] The calculated charge density of ground states in this $MoS_2/Au$ structure highlighted the impact of moiré periodicity, as the density difference at the interface shows the same moiré periodicity modulation (Figure 14i).[63] Similar to the previously mentioned graphene/metal system, the periodic superstructure occurs due to the varying adsorption sites of material atoms in relation to the substrate atoms, which leads to a consequent change in the material-substrate interactions.[232] The moiré-pattern of $MoS_2$ on Au (111) was also reported as a consequence of the formation of specific adsorption sites between lower S atoms and metal.[233] Figure 14j illustrates the three distinct positions in the moiré supercell: (A) the lower S atom is on the top of the Au atom, (B) the lower S atom is located at the hollow site of Au, (C) the lower S atom is still at the hollow site but the Mo atom is on the top of Au. However, those specific sites produced in TMDC–metal contact induce strain in TMD layers, which leads to the splitting of the Raman band and subsequently modulates the electronic structures of TMD materials.[233] The moiré-pattern was observed from MoS₂ on Au(111) but it cannot be seen from MoS₂ on Ploy Au or SiO₂, owing to the lack of varying adsorption sites (Figure 14k and l).[233]

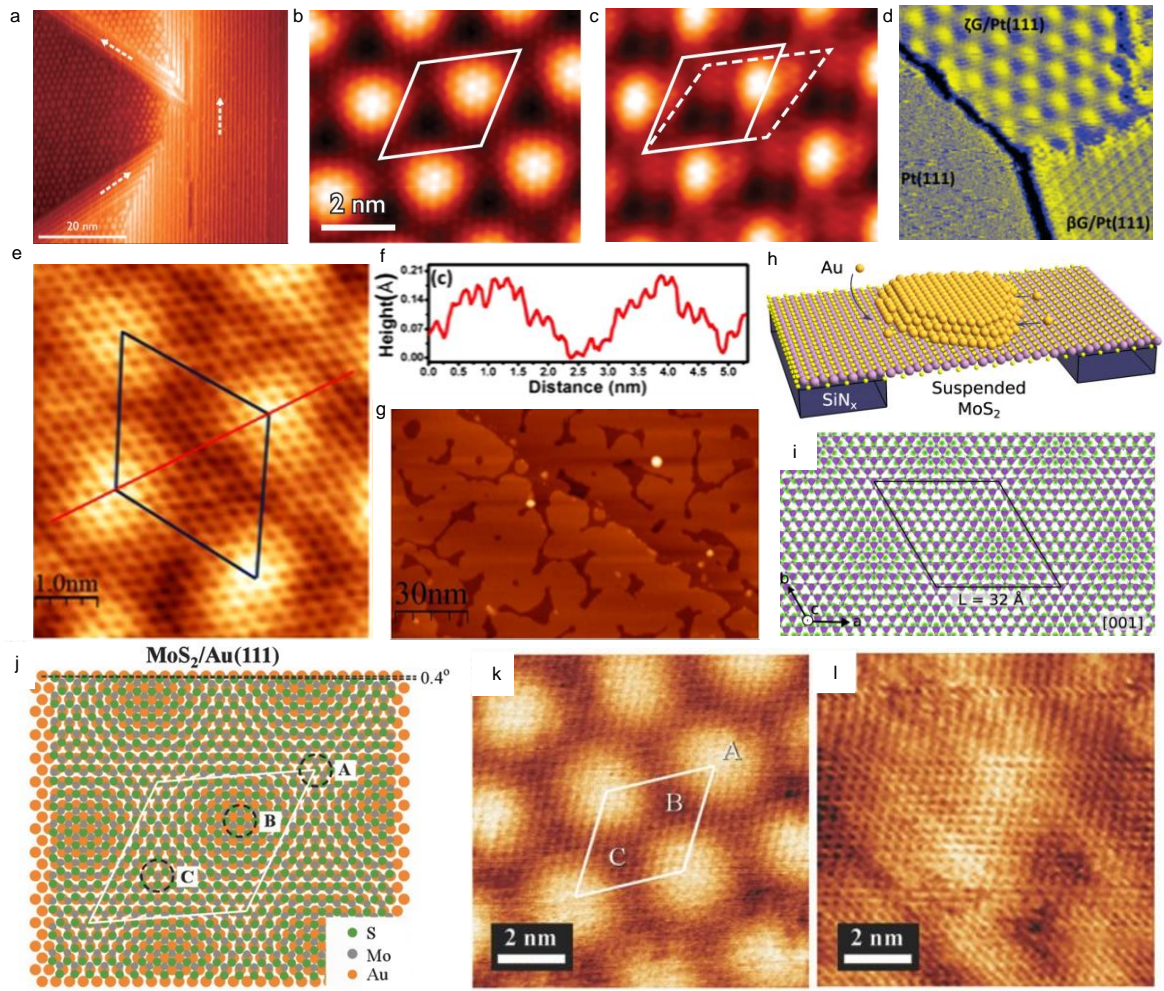

**Figure 14 | Moiré superlattice in other types of heterostructures.** (a) A typical cleaved IrTe2 surface shows hexagonal (H) and stripe (S) domains coexisting. (b, c) Moiré superstructures shown on the atomically resolved Pb islands on top of a cleaved IrTe₂ substrate with their unitcells marked with hexagonal (b) and stripe domains (c). (a-c) Adapted with permission from ref.[215] Copyright 2017 Springer Nature. (d) Constant current ($V$ = 100 mV, $I$ = 2 nA) STM image of monolayer graphene domains on Pt(111). Different graphene superstructures coexist together with clean Pt regions. Adapted with permission from ref.[225] Copyright 2011 American Chemical Society. (e) STM image of graphene on a 2D-FeO layer. The graphene honeycomb structure is resolved, while the superstructure (indicated by the black unit cell) is due to the underlying 2D-FeO moiré-unit cell. (f) Cross-sectional profile along a moiré structure. (g) STM image after intercalation of close to one monolayer of iron underneath graphene. (e-g) Adapted with permission from ref.[217] Copyright 2015 Springer Nature. (h) Schematic of epitaxially aligned Au deposited on suspended MoS₂ supported on a SiNₓ TEM grid. (i) Calculated charge density difference at the MoS₂/Au(111) interface, as viewed down the [001] axis, showing electronic modulation following the 32 Å moiré periodicity. Black line indicates the 32 Å crystallographic moiré unit cell. (h,i) Adapted with

permission from ref.[63] Copyright 2021 Springer Nature. (j) Top view of the moiré superstructure of MoS$_2$/Au(111) with a relative angle of 0.4° between lattices. (k, l) Magnified STM images of MoS$_2$ on Au(111) (k) and PolyAu (l), demonstrating the expected moiré pattern on Au(111) but not from the PolyAu. (j-l) Adapted with permission from ref.[233] Copyright 2017 John Wiley and Sons.

# 4. Fabrication and visualization of moiré superlattices

## 4.1 Fabrication

Various fabrication methods involving the relatively precise manipulation of layered materials have been proposed for creating moiré supercells. The lattice constant is an inherent, fixed parameter of the material, making lattice mismatch determination primarily relies on the appropriate choice of material layers and cannot be controlled during the fabrication process. In contrast, the rotation angle between two layers provides a much higher degree of controllability, emerging as a pivotal element in the fabrication of moiré superlattices. The fine control of the rotation angle serves as an effective means for manipulating the supercell. Mechanical exfoliation and controlled transfer approaches have proven to be surprisingly effective, capable of producing high-quality 2D superlattices. However, certain limitations to its scalability exist. It is necessary to consider alternative techniques such as one-step CVD growth commonly used in bilayer preparation. Easy and precise control with customized rotational misalignment is highly desired in this field.

### 4.1.1 Physical stacking methods

#### 4.1.1.1 Wet transfer stacking

Wet transfer stacking is a prevalent approach used in constructing twisted stacked structures. It enables the fabrication of a wide range of twist angles in a single straightforward step, typically achieved through the process of transferring one layer onto another[234], as illustrated in Figure 15a. However, achieving precise control over the twist angle presents a challenge. This uncertainty stems from the random stacking without an alignment system, but it's efficient for twist angle-dependent research. An example of this method is demonstrated by Puretzky *et al.* to produce twisted $MoSe_2$ bilayers showing a variety of stacking orientations[235]. The $MoSe_2$ monolayers were initially prepared on a substrate via CVD growth. This substrate was then split into two sections with one section coated with polymethyl methacrylate (PMMA). Etching of the coated substrate in a KOH (potassium hydroxide) solution yielded a floating PMMA/$MoSe_2$ film. Subsequently, the film was rinsed in deionised (DI) water to eliminate residual KOH. The $MoSe_2$ bilayer samples were prepared by stacking the cleaned film onto the other half of the substrate, covering the initially synthesised monolayers. Before the final step of annealing, the stacked sample underwent a soaking process in an acetone solution to eliminate PMMA. A batch of twisted $MoSe_2$ bilayers (Figure 15b) was produced through annealing at $300^{\circ}C$ with an environment of argon ($A_r$) and hydrogen ($H_2$) at atmospheric pressure for a duration of 2 hours. The same process has been also employed to create another similar structure with various twisted angles.[236] They used a similar method but coated both sections of the substrate with PMMA. They anticipated that incorporating PMMA into the lower substrate could potentially hinder the detachment and degradation of the monolayers during the assembly process in the water. However, the conventional acetone method posed challenges in completely removing the PMMA interlayer, which in turn affected the quality of the as-synthesised twisted samples. Because this traditional productive approach lacks both precise angle control and a clean interface, recent reports have introduced new approaches with the intervention of an alignment system. The fundamental and essential component of this type of system involves the use of an optical microscope[237]. The initial step remains the acquisition of a high-quality monolayer. Chen *et al.* proposed a novel methodology that involves femtosecond laser micromachining for controlling the twist angle between layers with good accuracy, around 0.1°.[238] The utilisation of femtosecond lasers in micromachining has become prevalent due to their superior machining precision compared to long-pulse lasers, which was used to cut the sample and produce a straight cutting edge for both monolayer parts (Figure 15c). These parallel cutting lines play a crucial role during the stacking process as a reference guide, enabling the fabrication of moiré superlattices in homostrucutres, similar to the "tear and stack" technique in the dry transfer approach. It also allows for a wide range of twist angles by adjusting the relative angle between two straight cutting edges. In addition, this methodology has been reported to yield twisted bilayer samples displaying predetermined patterns. The initial step involves the utilisation of a femtosecond laser to shape the graphene flakes into desired configurations, followed by the implementation of the cutting-rotation-stacking (CRS) technique.[238] Other comparable strategies with metal masks[141] as references or incorporating an additional step[239] to prevent interfacial contaminants have also been reported significantly enhance efficiency in moiré pattern fabrication. Moreover, Wang *et al.* demonstrated the fabrication of $WSe_2/WS_2$ heterostructures through a polymer-assisted wet transfer approach, similar to the first method explained with additional two procedures involved.[240] The high yield of the CVD-grown process allows for the generation of multiple bilayers with varied twist angles in a single stacking. The heterogeneous nature of the twisted heterostructures necessitated the separate growth of distinct materials on different substrates, utilising disparate precursors and synthesis conditions, which limits the application of above mentioned advanced techniques such as laser cutting. The determination of lattice orientation is necessary for precise controlling in heterobilayers.

### 4.1.1.2    Dry transfer stacking

The dry transfer method is a simple and powerful approach for fabricating both homo- and hetero-structures.[50] This method finds wide adoption in numerous studies, including the observation of superconductivity at magic angle,[37] extraordinary SHG behaviour,[50] orbital magnetism[166] and atomic reconstruction.[134] This is attributed to the cleaner interface devoid of polymer or chemical-etchant induced contamination or damage. The quality of the twisted layers partially depends on the quality of monolayers and the transfer procedures employed during stacking, which directly affect the interlayer coupling and the generation of moiré superlattices. The monolayers' excellence can be managed by controlling the critical parameters during CVD or exfoliation.[234] The most famous technique of "Tear-and-stack" has been widely used in the fabrication of homobilayers, which involves picking up a portion of a ultra-thin material, rotating it to a specific angle, and stacking it onto the remaining section of the same flake. To attain precise control over the twist angle at a sub-degree level, a high-resolution goniometer stage is employed, which is crucial in producing low-twist-angle heterostructures.[141, 241-242] The studies by Cao et *al.*[127] and Chung et *al.*[241] used this technique (Figure 15e-g) to explore t-BLG with rotational angles ranging from 0º to 2º. These studies unveiled the emergence of quantum states due to robust interlayer coupling enabled by the meticulous fabrication of high-quality superlattices. Xu *et al.* utilized the same technique recently to fabricate twisted bilayer $CrI_3$ as shown in Figure 15d.[242] Firstly, a section of the flake was lifted (black lines) from the substrate using a stamp made from polydimethylsiloxane/polycarbonate. Employing a high-precision rotation stage, the remaining segment (red lines) was then rotated by a certain angle on the substrate, and subsequently, the initial lifted part was attached at this angle, forming the desired pattern (Figure 15h). Similar to the previous case, this particular technique or a similar method[59] is constrained to the production of moiré homostructure. An alternative dry transfer method using CVD-grown samples to directly determine twist angle by observing the alignment of two TMD triangles (Figure 15i) has been reported.[236] Although some solutions were still involved in this process, the PMMA was not introduced between the upper and lower layers, which eliminates the interfacial residuals. This absence of external interference led to a strong coupling between the two layers. In contrast to the wet transfer technique, the dry transfer approach here relied solely on annealing to remove PMMA residual, eliminating the need for an acetone-soaking step. This largely reduces the detrimental effects of liquid-induced degradation and oxidation.[236] Furthermore, employing a glass slide and microscope, the stacking procedures of the dry transfer method exhibit enhanced controllability. Another powerful ultra-clean transfer technique requires the support of angle-resolved SHG that can determine the lattice orientation, thereby providing accurate guidance for stacking orientation. This has found extensive application across numerous studies,[34-35, 199, 243] particularly in the exploration of moiré patterns in TMD heterobilayers. The detailed stacking procedures are all dry, same as the method elucidated in viscoelastic stamping, where layers are stamped onto each other following a precise substrate rotation, achieving the desired twist angle for a 2D structure. As detailed by Castellanos-Gomez *et al*., the utilization of viscoelastic stamps holds the potential to achieve a fully dehydrated stacking procedure for 2D materials. In their study, a gel film was employed as a supportive layer for the top material in the heterostructure. This gel film was affixed to a glass slide to ensure structural stability. By applying pressure through the elastomer and aligning the upper and lower materials of the heterostructure via microscopic, the target substrate undergoes a stamping process.[244] The stamp displays elastic behavior over short durations and exhibits slow flow over longer periods.[245] Upon gradually lifting the stamp, vdW forces successfully extract the upper 2D material from the stamp due to the relatively weak interfacial force them, ultimately achieving the desired outcome in a dry manner.[244] This method has yielded rapid and effective results. In a separate investigation, $WSe_2$ flakes were initially exfoliated onto a PDMS substrate, following which a portion of the resulting $WSe_2$ monolayer was transferred onto a silicon wafer using stamping. The successful transfer is facilitated by a stronger interaction between substrate and $WSe_2$, contrasting with the interaction between PDMS and WSe$_2$. Subsequently, the remaining section of the WSe$_2$ monolayer was transferred onto the initial WSe$_2$ layer using the same stamping process, where the substrate is rotated to a specific orientation under an optical microscope to achieve the desired rotational misalignment. It was found that preheating the substrate before increased the transfer efficiency.[20]

A proper annealing process facilitates the achievement of efficient interlayer coupling in stacked bilayers, which is important for both wet and dry transfer methods. The elimination of adsorbates and polymer residues during annealing can further improve vdW bonding between two layers. A study conducted by Wang *et al.* highlighted the importance of the annealing process during the fabrication by analyzing AFM results of twisted WSe$_2$/WS$_2$.[240] It has been reported that the interlayer space between two layers will sufficiently affect the coupling between them.[246] The interlayer distance before and after the annealing for this heterobilayer was measured to be around 1.2 nm and 0.8 nm, respectively. This reduction suggests the establishment of effective vdW coupling between two previously uncoupled layers via an annealing process. In addition, The absence of annealing has been demonstrated to impact the characterization results of the sample, due to the resultant compromise in sample quality.[247]

On the other hand, Ma *et al.* have documented a technique that utilises capillary force to enable a clean stamping process for acquiring and transferring 2D materials.[248] Adhesion energy enhancement between PDMS and 2D materials is facilitated by a thin vapour layer, such as water, which plays an important role in the pickup process. Upon vapour evaporation, the adhesion energy decreases, allowing for the easy transfer of 2D materials onto the intended substrate. This approach exhibits significant potential for intra-chip material transportation. Tao *et al.* introduced an additional technique for peel-off procedures.[249] In this study, a thin film of polyvinyl alcohol (PVA), soluble in water, was employed for sample pickup. The film was subsequently removed through a floating dissolution process, which proved to be simple and effective. Other assisted stacking techniques, such as the thin film-assisted transfer (TAT) method[250] introduced by Zhang *et al*. have also been reported to provide better control over rotation angles. The integration of these innovative transfer techniques, combined with later advancements in precise angle control holds the potential to open up new prospects for producing high-quality t-TMDs over large surface areas.

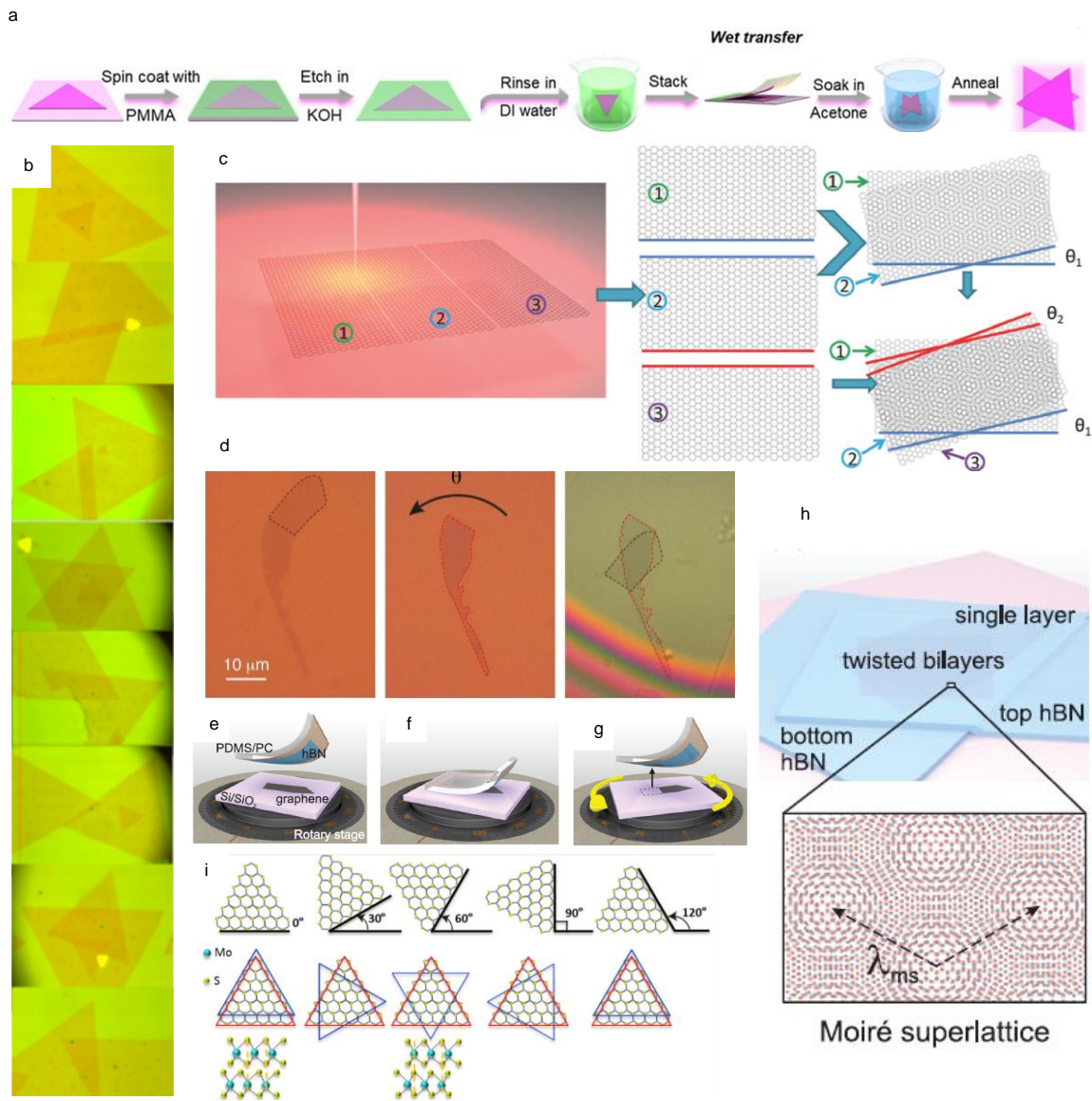

**Figure 15 | Physical stacking methods.** (a) Flowchart of wet transfer operations. Adapted with permission from ref.[234] Copyright 2021 John Wiley and Sons. (b) Optical images of bilayer MoSe$_2$ with different twist angles. Adapted with permission from ref.[247] Copyright 2016 American Chemical Society. (c) Schematic illustration of cutting-rotation-stacking process for creating twisted layers with controlled angle. Adapted with permission from ref.[238]

Copyright 2016 John Wiley and Sons. (d) Optical images to show the process of tear and stack technique for the fabrication of twisted bilayers. Scale bar is 10 µm. Adapted with permission from ref.[242] Copyright 2022 Springer Nature. (e-h) Schematics for assembling twisted bilayer graphene with hBN with well controlled twist angle between two layers. The resulted moiré pattern from tBLG is illustrated in (h). (e-h) Adapted with permission from ref.[241] Copyright 2018 American Physical Society. (i) Direct observation of twist angles in Bilayer MoS$_2$ samples, with the configurations of θ = 0, 30, 60, 90, and 120°. The 120° configuration is identical to 0°. Red and blue triangles indicate the bottom and top MoS$_2$ layers, respectively. All triangle flakes in the figure are S-terminated zigzag edges. Adapted with permission from ref.[236] Copyright 2014 American Chemical Society.

### 4.1.2   Chemical vapor deposition (CVD) growth

Chemical vapor deposition (CVD) is believed to be a promising technique in the creation of moiré superlattices within twisted stacking configurations. This is attributed to its ability to facilitate clean interface formation and scalability. Nonetheless, the precise manipulation of twist angles remains a technical challenge within the CVD approach. Typically, multilayer samples produced through CVD exhibit fixed twist angles of 0° or 60°, emphasizing the existing limitations in twist-angle control. Up to the present, the experimental work on the preparation of the twisted structure by the CVD method is relatively few, and the majority of directly grown twisted samples manifest as homo-bilayer structures. Notable examples include t-BLG and homostructures of twisted transition metal dichalcogenides (t-TMDs), both accomplished through the streamlined one-step growth process. In this section, we will introduce CVD methods used to fabricate t-BLG and t-TMD homo- and hetero-structures.

Optimizing the catalytic substrate and carefully managing the physical conditions during synthesis are vital in the fabrication of high-quality t-BLG using CVD. Cu has proven to be an influential catalytic substrate, empowering the yield and size of the twisted few-layer graphene. In 2014, the pyramid-like hexagonal bilayer and tri-layer graphene with a 0° or 30° rotation were successfully synthesized on the pretreated Cu foils, where the interlayer rotation is determined by the nucleation at the Cu step.[251] Indeed, twisted BLGs are energetically preferred on Cu surfaces with atomic steps, and the crystal plane of the Cu substrate also plays a crucial role in influencing the phenomenon of small-angle twisted rotation. For instance, t-BLG grown on a conventional Cu (111) surface predominantly encompasses 0° and 30° arising from the localized binding energy effects,[252] while the relatively lower stabilization energy of small-angle on atomic-stepped Cu (311) and Cu (110) surfaces enable t-BLG with smaller stacking angles in the range of 3° to 5°,[253] as shown in Figure 16a and b.

Moreover, adjusting the physical conditions for the growth, including temperature, flow rate and vacuum pressure, is another important way to increase the yield of small angle t-BLG (<30°). In 2016, Liu *et al*. achieved a successful generation of t-BLG domains with varying small twist angles, specifically, 4°, 13°, 21°, and 27°, by employing a low-pressure chemical vapor deposition (LPCVD) system.[254] These distinct graphene domains were prepared on copper foils, demonstrating the versatility of the low-pressure environment in facilitating controlled twist angle configurations. Hetero-site nucleation strategy was then developed to further improve the yield of t-BLG with a small twist angle, taking advantage of a gas-flow perturbation of $H_2$ and $CH_4$. This method leverages a sequential approach: subsequent nucleation of the second graphene layer is initiated after the initial graphene layer has been nucleated. This initiation occurs through a gas-flow perturbation that introduces a sudden increase of $H_2$ and $CH_4$ supply, which can amplify the presence of active hydrogen and carbon species, thereby promoting the nucleation and growth of the second graphene layer in a distinct local environment[255] (Figure 16c). As a result, a significant quantity of high-quality t-BLG with a diverse range of twist angles, including approximately 3°, 6°, 9°, 12°, 15°, 18°, 21°, 24°, 27°, and 30°, as shown in Figure 16d, can be reliably obtained.[256] Similarly, by adjusting the pressure ratio of CH$_4$ and H$_2$, for example, t-BLG with minor twist angles (4°, 7°, 14°) can be fabricated using the plasma-enhanced CVD approach.[199]

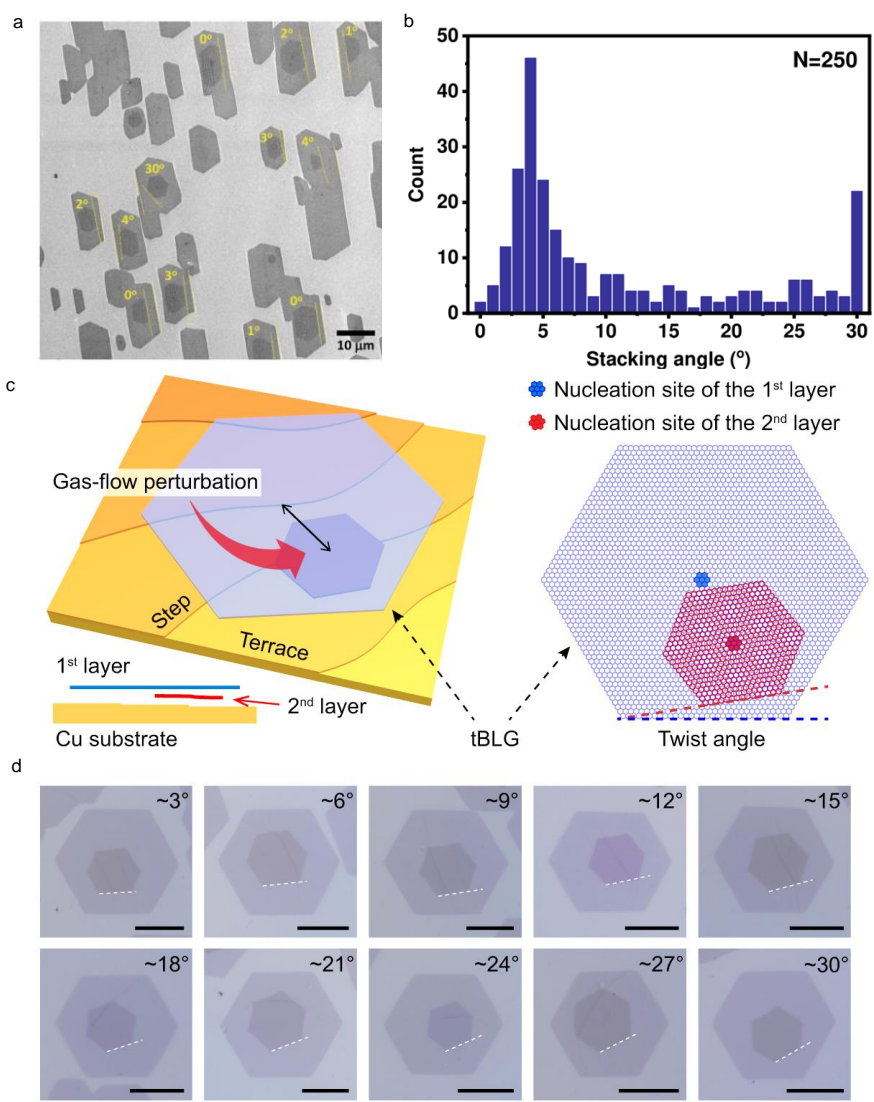

**Figure 16 | Chemical vapour deposition techniques for** t-BLG. (a) SEM image of BLG grown on Cu (301) and (110), where the stacking angle is determined by dashed lines, indicating BLGs have various twisted stacking order except the AB-stacking and 30°-twisted. (b) Statistic of twist angle distribution observed from BLG grown on Cu (301) and (110). (a,b) Adapted with permission from ref.[253] Copyright 2020 Springer Nature. (c) Schematic illustration of the hetero-site nucleation for the growth of tBLG on a Cu substrate, where the nucleation site for the second layer of graphene (depicted in red) differs from that of the first layer (depicted in blue). The side view provides a visual representation of how the second layer of graphene grows beneath the first layer. (d) Optical microscopy (OM) images showcasing as-grown tBLGs with various twist angles, including approximately 3°, 6°, 9°, 12°, 15°, 18°, 21°, 24°, 27°, and 30°. The scale bars in the images are equivalent to 10 μm. (c,d) Adapted with permission from ref.[256] Copyright 2021 Springer Nature.

The predominant configuration observed in the majority of CVD-grown t-TMD is the homojunction structure (t-MX$_2$). In fact, the fabrication process of twisted TMD emphasizes the growth conditions like temperature, reaction time, and carrier gas, with a lesser emphasis on catalytic substrates. Firstly, growth temperature can thermodynamically influence the stack configuration, so the higher growth temperature is commonly employed to enhance the yield of the sample featuring the small twist angles.[257] Therefore, in the work of fabrication of t-WS$_2$ bilayer with diverse twist angles including 0°, 13°, 30°, 41°, 60°, and 83° (Figure 17a-f), the growth temperature (>1100℃) primarily determine the nucleation and yield of bilayer t-WS$_2$.[258] The t-WSe$_2$ bilayer samples may occur between the deposition temperature range from 1100°C to 1120°C for AA/AB stacking.[259] Moreover, the reaction time or growth rate is also an important factor. For instance, in Figure 17g-l, the successful CVD-grown MoS$_2$ bilayers with twist angles of 0°, 15°, and 60° have been achieved by deliberately maintaining a low nucleation rate for more bilayer samples.[43] The synthesis of t-MoSe$_2$ relying on a longer reaction time require the support of the longer heating zone.[260] In fact, achieving a substantial yield of t-TMD with controlled and small twist angles is an ongoing challenge in this field. Statistic results suggest that the preference of twisted angle is dominated by the stacking energy densities, where the most prevalent configuration with twisted angles of 0°/60° exhibits the lowest values, followed by the 30° and 15° twisted arrangements.[261]

On the other hand, unlike the one-step growth method for the homojunction fabrication, CVD-grown hetero-structure commonly requires a two-step CVD growth method, where a monolayer TMD is firstly grown on a substrate, and the growth environment subsequently will be switched for the growth of another layer. While direct CVD has proven mature and reliable for synthesizing TMDs heterojunctions with clean interfacial stacking structures, the majority of resulting products primarily exhibit the 3R or 2H phases, like WS$_2$/WSe$_2$.[262] However, few works address the concept of twist angles within these heterojunctions. Recently, a MoSe$_2$/WSe$_2$ hetero-bilayer has been developed via three heating zone and a two-step growth method.[263] However, it's worth noting that the occurrence of samples with small twist angles remains relatively scarce when contrasted with the more prevalent $R$ or $H$-type configurations. On the other hand, by direct growth of MoS$_2$ on the exfoliated h-BN, an unconventional heterostructure with a rotation angle of less than 5° has been obtained.[264]

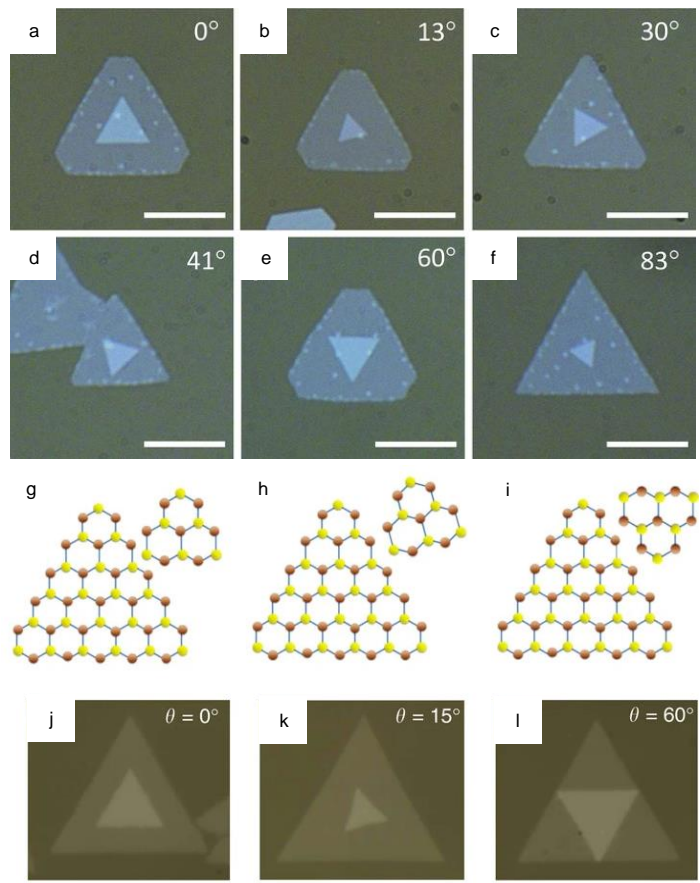

**Figure 17 | Chemical vapour deposition techniques for twisted TMD layers.** (a-f) Optical microscopic images of the twisted WS$_2$ bilayers with twist angles of 0°, 13°, 30°, 41°, 60°, and 83°, where the twist angles are determined by counter-clockwise rotation of the top triangle concerning the bottom triangle. The scale bars used in image equal to 10 µm. (a-f) Adapted with permission from ref.[258] Copyright 2015 John Wiley and Sons. (g-i) Schematics illustration of the atomic structure of a MoS$_2$ bilayers with twist angles of θ=0°, θ=15°, and θ=60° respectively. (j-l) Optical microscopic images of twisted MoS$_2$ bilayers corresponding to the stacking configurations shown in (g-i). The scale bars used in the image are 10 µm. (g-l) Adapted with permission from ref.[43] Copyright 2018 Springer Nature.

### 4.1.3 Other

Except for the previous physical stacking and CVD growth methods, creating the twist angle can be realized by the post-growth process. Common methods involve using SPM like AFM and STM to rotate and fold the layers via nano-manipulation. Due to the single atomic plane structure existing in h-BN and graphene, the interfacial superlubricity effect allows the rotation to form a twisted hetero-structure with external force applied.[265-266] As shown in Figure 18a, a tunable t-BLG has been assembled where the in-situ twist angle between the upper and lower graphene layers is adjusted by rotating the h-BN gear.[267] In addition, both AFM tip[268] and STM[269] tip have been reported to serve as tools to pick a portion of the film and fold it (Figure 18b), enabling good control of the twist angle[270] between homo-bilayer interface (Figure 18c) with a precision of approximately 0.1°. Chen *et al*. even demonstrated the repetitive folding and unfolding process of graphene (Figure 18d) through STM manipulation without inducing defects or causing any damage.[269] The technique of layer folding has been recently employed to massively produce controlled twisted $MoS_2$ bilayers with a wide range of angles[271] with TEM-confirmed high quality. Furthermore, recently an innovative method indicates the usage of an intercalation layer for twist angle generation.[272] As indicated by Figure 18e, organic molecules, such as 1,2-dichloroethane, are introduced between CVD-grown bilayer graphene to weaken the interlayer interaction, resulting in slide or rotation of the topmost graphene layer to fabricate the t-BLG with angle ranges of 0-10° and 20-30°.

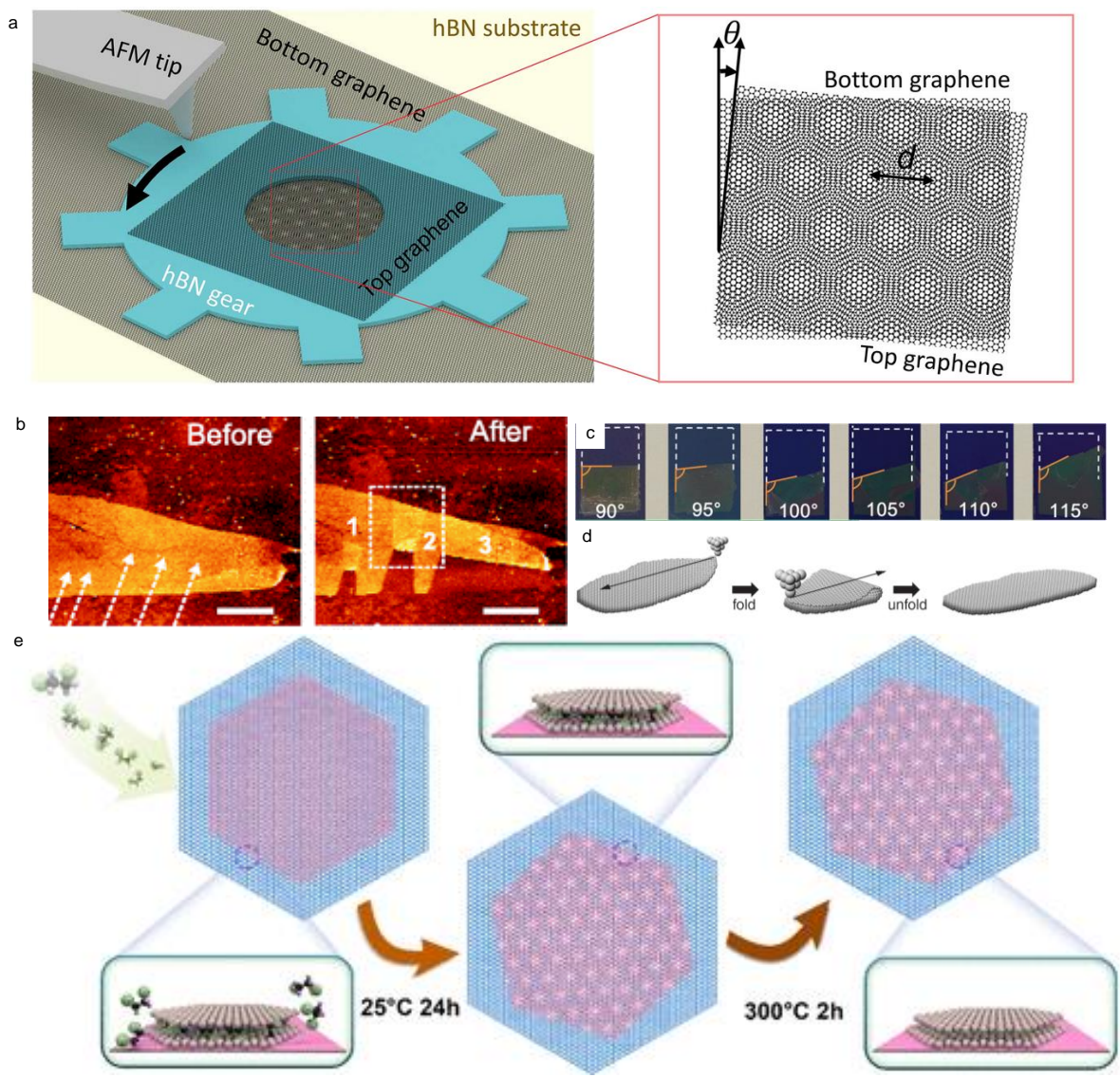

**Figure 18 | Self-folding technique for the fabrication of twisted structures.** (a) Schematic diagram of the configuration of an angle-tunable t-BLG device, which is constructed by two individual graphene monolayers separated by an h-BN gear on another h-BN substrate, resulting in the creation of moiré superlattice characterized by a period denoted as 'd,' emerging within the overlapping region situated at the center of the gear structure. Adapted with permission from ref.[267] Copyright 2022 Springer Nature. (b) AFM images of the monolayer graphene captured before and after parts of the film are picked and folded by the AFM tip in the contact mode. The scale bars in the image denote 4 μm. Adapted with permission from ref.[268] Copyright 2013 American Physical Society. (c) Optical images of folded graphene films with controlled stacking orientations. The fold lines are indicated in orange color. Adapted with permission from ref.[270] Copyright 2017 American Chemical Society. (d) Schematic illustration of folding and unfolding a graphene film along the direction of black arrows respectively. Adapted with permission from ref.[269] Copyright 2019 American Association for the Advancement of Science. (e) Schematic illustration of the intercalation process employed to prepare t-BLGs for the reconstruction of moiré superlattice: The 1,2-dichloroethane molecules (Cl, C, and H atoms are represented by green, black and violet balls respectively) intercalate into the interlayer spacing of CVD grown BLG (grey balls), promoting the rotation of the topmost layer (red) with the second layer (blue). The 1,2-dichloroethane molecules could be

subsequently removed after annealing at 300 °C for a duration of 2 hours. Reprinted with permission from ref.[272] Copyright 2023 American Chemical Society.

## 4.2 Visualization

Due to technical limitations imposed by the conventional optical microscopes, it is hard to achieve visualization of moiré super-lattices. In this section, we will present a summary of the common methods employed to determine artificial moiré super-lattices in the twisted homo/hetero structure, in terms of SPM and electron microscope (EM). SPM is a microscopy branch that generates surface images by physically scanning the specimen with a probe, providing impressive atomic resolution results. Currently, the mainstream imaging capability of SPM relies on STM system and AFM system with different functional modes. The electron microscope utilizes a beam of electrons as a source of illumination and provides significantly higher resolution than optical microscopes. In this section, we will introduce equipment and methods used for imaging the reconstructed superlattice formed by twisted stacking samples.

### 4.2.1   Scanning Probe Microscope (SPM)

Based on the resultant tunnelling current generated by the voltage driven electrons tunnelling between the conductive STM tip and the sample surface, STM can distinguish smaller than nano-scaled features and provide the scanning morphology images with atomic resolution. Therefore, the employment of STM can clearly image the morphology and probe the lattice change of the twisted vdW stacking structure.[55, 170] As shown in Figure 19a-b, the STM images not only exhibit the atomic resolution morphology of the sample surface, but they also show a clear periodic moiré pattern of the superlattice reconstructed by the twisted vdWs stacking configuration. Analysis of the 2D Fourier transformation (Figure 19c) implies that the stacking would cause the local changes in interatomic distances at different areas of the moiré patterns, where the lattice within the superlattice domain withstood the relatively largest strain than those within the domain walls highlighting the hexagon shape (Figure 19d).[55] Scanning tunnelling spectroscopy (STS) image obtained from the stacking sample (Figure 19e-g), probing the local DOS with respect to the energy of the electrons, reveals that the uneven strain distribution can result in local variation in the stacking energy.[273-274] This stacking energy landscape likewise exhibits a periodic moiré pattern, and the contrast is contributed by different energy governed by the stretched and less stretched lattice within the domain and domain wall areas.[274] Even though, the STM system can act as an effective tool for imaging reconstructed superlattices in the stacking structures, its operating conditions normally demand the high vacuum and low temperature, hindering its accessibility. Unlike STM, AFM system normally can work in ambient conditions. The AFM system can generate high resolution images by comparing the laser light reflected into the photodiode when the scanning probes interact with the sample surface and get deflected. Thus, researchers were successful in acquiring the morphologic image of the moiré structure and probing local variations via the AFM system (Figure 19h, i),[275] but it is more difficult to exhibit periodic superlattice patterns in larger scan sizes compared to STM system. However, the AFM system is also a multifunctional platform for imaging the variation in different properties, such as mechanical and electrical properties, distributed across the sample surface. The local variant properties induced by the stacking configuration enable the AFM system to visualize the reconstructed superlattice. In the non-contact mode, the moiré superlattice was observed in the phase channel which records the phase delay between oscillation of the tip and excitation of cantilever when the cantilever is distant from the sample (Figure 19j-m).[276-277] The pattern demonstrates the modulation of the vdW potential on the surface of sample, following AB and BA stacking domains as distinguished by the imaging contrast arisen from Debye force between interlayer permanent electric dipoles and the neutral tip.[276] Young's modulus mapping conducted on a twisted stacking sample (Figure 19n) reveals a hexagonal symmetry in the distribution of Young's modulus, displaying clear changes across each unit cell, where Young's modulus remain almost constant and low within the large areas of the domain but increased significantly on the walls.[55] This modification might be attributed to the fact that the stacking configuration would physically change the lattice constant, local stacking thickness and atom density, thus altering the local elasticity.

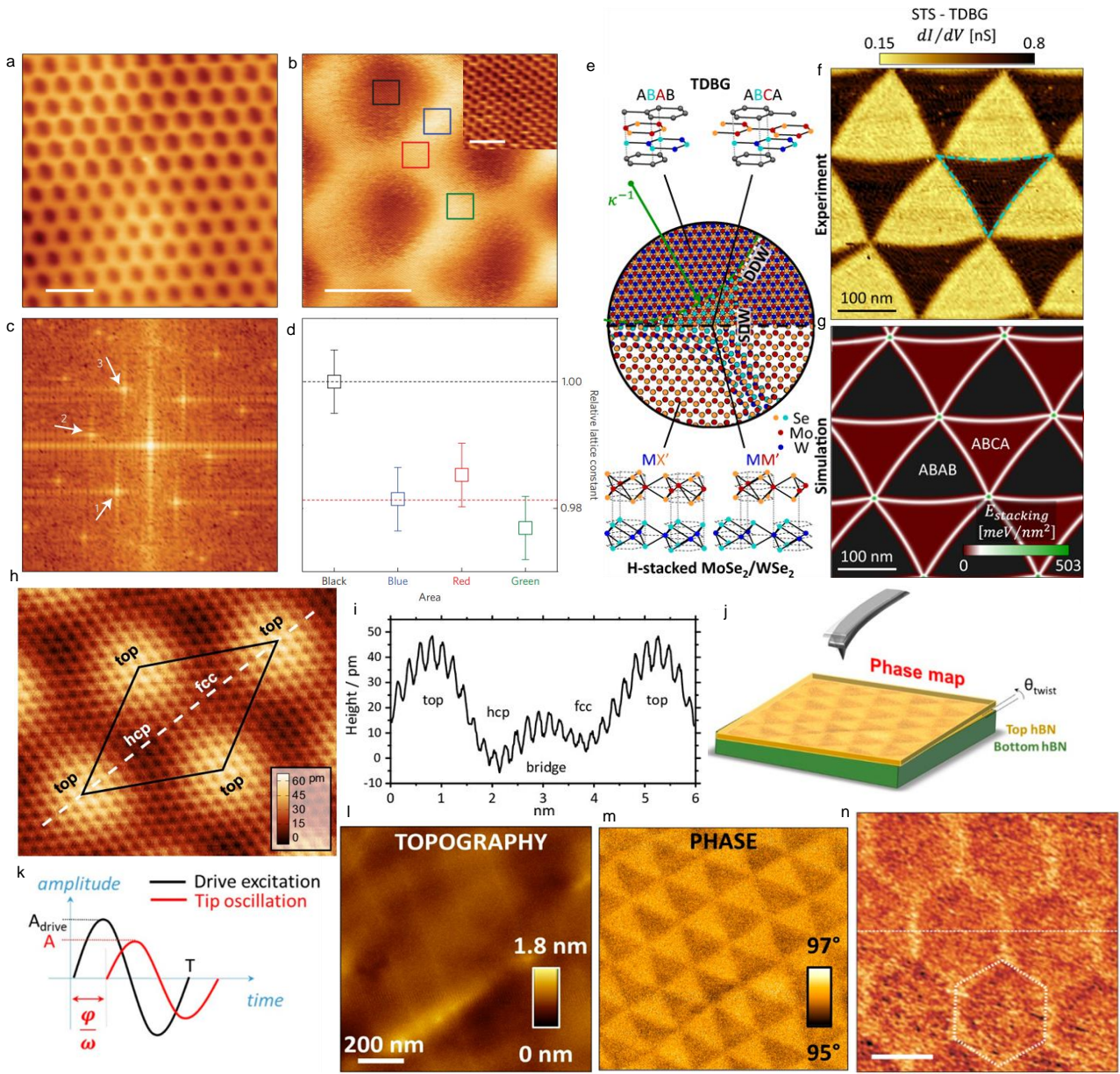

**Figure 19 | The employment of STM and AFM system for the spatial imaging of the reconstructed superlattice.** (a) STM image of one of our aligned samples. A moiré pattern is clearly visible. Scale bar, 30 nm. The sample bias is −0.1 V and the tunnelling current is 300 pA. (b) Zoom-in STM image for both the moiré pattern and the resolved atomic structure (inset).

Scale bar, 10 nm. The sample bias is −0.1 V and the tunnelling current is 800 pA. Coloured squares (3nm in size) indicate the fragments used for Fourier transformation to determine the interatomic distance. Inset: a blow up of the area marked by the black square, with the atomic structure clearly visible. Scale bar, 1 nm. (c) Example of the Fourier transform of the atomically resolved structure. In this case, as the starting image we used the 3nm × 3nm square image at the vertex of the hexagonal pattern (red square in (b)). The width of the panel is $19\text{nm}^{-1}$. (d) Relative lattice constants (with respect to those measured for the area marked by the black square in (b) for different areas within the moiré pattern (colours correspond to those in (b)), obtained from the positions of the first-order peaks in c and averaged over the three directions. The error bars are determined by the width of the peaks in the FFT of the atomically resolved structures in (c) and by the spread across the three directions as marked in (c). (a-d) Adapted with permission from ref.[55] Copyright 2014 Springer Nature. (e) Illustration of domain formations in a relaxed twisted bilayer structure. (f) Experimental STS map of TDBG revealing rhombohedral (ABCA—dark) and Bernal (ABAB—bright) domains with minimal external strain, where the rhombohedral phase bends inward (dashed turquoise line) revealing an energy imbalance between the two phases. (g) Full relaxation calculations for the stacking energy density based on the experimental case in (f). (e-g) Adapted with permission from ref.[274] Copyright 2021 Springer Nature. (h) AFM image showing both the moiré pattern and the atomic structure of the surface layer. (i) Height profile over the unit cell of moiré superlattice marked with a white dashed line in (h). (h,i) Adapted with permission from ref.[275] Copyright 2013 American Physical Society. (j) Schematic illustration of non-contact AFM mode for imaging the t-hBN sample. (k) Plot of the two main signals involved in AFM phase-imaging, i.e., drive excitation (black) and tip oscillation response (red). (l-m) Representative AFM topography and phase channel image of top hBN, showing a neglectable moiré superlattice pattern in the flat morphology but clear domain contrast corresponding to the local potential of surface attraction within the same region on the sample. (j-m) Adapted with permission from ref.[276] Copyright 2022 American Chemical Society. (n) Young's modulus distribution, measured in the PeakForce mode, for structures with a 14 nm moiré pattern. Adapted with permission from ref.[55] Copyright 2014 Springer Nature.

As the twisted stacking electrical properties, the AFM system equipped with electrical testing functions including PFM, EFM, KPFM and C-AFM are also able to probe the tip-sample interaction and realise the visualisation of the reconstructed superlattice when an electric field is applied to the stacked samples. When PFM measurements were conducted on twisted samples, the amplitude and phase images of the corresponding piezo response exhibited the ferroelectric nature with a triangular pattern and finite contrast between the adjacent triangle domains.[153, 278] The pronounced contrast at the domain wall between the neighboring domains is ascribed to the flexoelectric effect arising from the strain gradient at the domain boundary.[153, 278-279] As strain gradients are negligible inside the domain, the contrast stems from electric dipole moments induced by AB and BA stacking, leading to spontaneous opposite out-of-plane polarization in these domains (Figure 20a-f). EFM can be used to detect the local electric charges. In Figure 20g-i, despite no more pattern observed in the AFM morphologic image, the phase images of EFM for the same area exhibit ferroelectric-like domains arranged in triangular superlattices.[148] This observation is attributed to a substantial charge-density modulation induced by the interfacial elastic deformations in the twisted stacking configuration, resulting in the formation of out-of-plane dipoles belonging to opposite interfacial surfaces.[148] Similarly, KPFM can also detect the variations in the charge density and distribution by measuring the surface potential over the sample surface.[149, 280] Therefore, the potential map (Figure 20j) shows clear surface potential difference which distinguished domains induced by electric polarization density between AB/BA stacking regions. Moreover, the twisted stacking can also modulate the band gap of semiconductor and result in the regional conductivity. As a result of the C-AFM measurement (Figure 20k), the uniform contrast of discrete triangular domains in the superlattice suggested the constant but regional high and low conductivity caused by the smaller and lager bandgaps in BA and AB stacking respectively.[45-46]

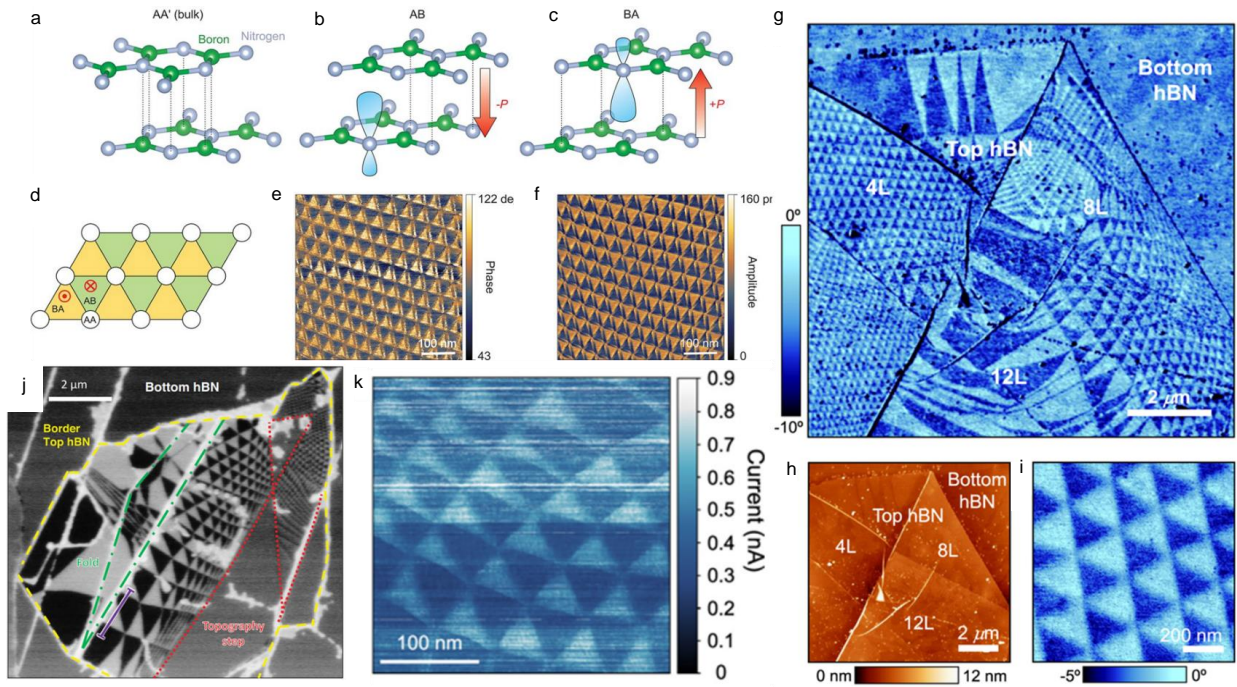

**Figure 20 | Imaging the moiré superlattice in the stacking sample via AFM system combing the applied electric field.** (a) Illustration of the atomic arrangement for AA′ stacking,

the bulk form of hBN. Nitrogen and boron atoms are shown in silver and green, respectively. (b, c) Illustration of the atomic arrangement for AB and BA stacking. The vertical alignment of nitrogen and boron atoms distorts the 2pz orbital of nitrogen (light blue), creating an out-of-plane electric dipole. (d) Illustration of a small-angle twisted bilayer BN after the atomic reconstruction. The reconstruction creates relatively large AB (green) and BA (yellow) domains, with small AA regions (white) and domain walls in between (black). The red circled dot and red circled X represent up and down polarization, respectively. (e, f) Vertical PFM phase (e) and amplitude (f) images of twisted bilayer BN displaying the AB and BA domain regions induced by opposite polarization distribution and the contrast at the domain wall originated from the flexoelectric effect. (a-f) Adapted with permission from ref.[153] Copyright 2021 American Association for the Advancement of Science. (g) Representative dc-EFM image (phase) of twisted hBN showing large areas with triangular potential modulation, where the changes in domains' shape and periodicity are due to small changes in θ caused by irregular strain and the wrinkles in the sample. (h) AFM topography image about the twisted stacking hBN crystal in (g) containing various layer conditions, where no reconstructed superlattice patterns are visible. (i) Zoom-in of a region in h with regular domain shape whose distinct contrast is induced by charge-involved polarization. (g-i) Adapted with permission from ref.[148] Copyright 2021 Springer Nature. (j) Surface potential mapping generated by KPFM showing oppositely polarized domains of AB/BA stacking (black and white), ranging in area between ~0.01 and 1 mm$^2$ and separated by sharp domain walls. Adapted with permission from ref.[149] Copyright 2022 Springer Nature. (k) Conductive AFM (C-AFM) image of small angle heterostructure showing alternating triangular domains with contrast arisen from different conductivity distribution. Adapted with permission from ref.[45] Copyright 2020 American Chemical Society.

Scanning near-field optical microscopy (SNOM) is a sort of SPM coupled with an excitation light source illuminating the sharp tip, creating a strong near field as indicated in Figure 21a. By utilizing various light sources and detecting tip-scattered light with changes of tip positions, SNOM enables the acquisition of continuous spectra and spatial mapping of optical characterizations such as tip-enhanced Raman spectroscopy (TERS) and nanoscale Fourier transform infrared spectroscopy (nano-FTIR). Similar as visualisation of the superlattice achieved by the AFM system cooperating with an applied electric field, the SNOM system can also capture the patterns of reconstructed superlattices. Considering the localization of lattice dynamics within the reconstructed superlattice of stacking samples, TERS can generate superlattice imaging based on the distribution of local vibrational states and the electronic structure. As displayed in Figure 21b and c, the commensurate triangular patterns are outlined by the stronger Raman responses at the AA sites and domain walls induced strain solitons (SP).[281] At the same time, this phonon vibration involved nature empowers the near-field scattering dynamics measurements of phonon-polariton. In Figure 21d, the clear contrast of the domain wall in the superlattice imaging is arisen from the local higher reflectivity of photon-polariton at domain walls and enhanced optical conductivity at solitons.[282-283] The presence of local lattice dynamics also allows nano-IR to obtain the spatial imaging of superlattice.[284] The near-field amplitude (S) and phase (φ) mapping show a distinct IR contrast and reveal elements including AB/BA regions, AA sites, and SP domain walls of the reconstructed superlattice. The nano-IR images have great consistency with the visualization of the moiré pattern governed by other methods such as PFM and EFM in Figure 21e. Due to the fact that the spatial variations of the optical resonance has high sensitivity towards light frequency, frequency modulation would result in a significant shift in near-field amplitude and phase contrast, depending on the stacking configuration, even reversing between AB/BA domains[284] (Figure 21f-k). By measuring the photoresponse and combining the electrical current readout, the photocurrent map features a clear periodicity of distorted triangular patterns with noticeable contrast induced by the local photoexcitation of carriers,[285] as illustrated in Figure 21l.

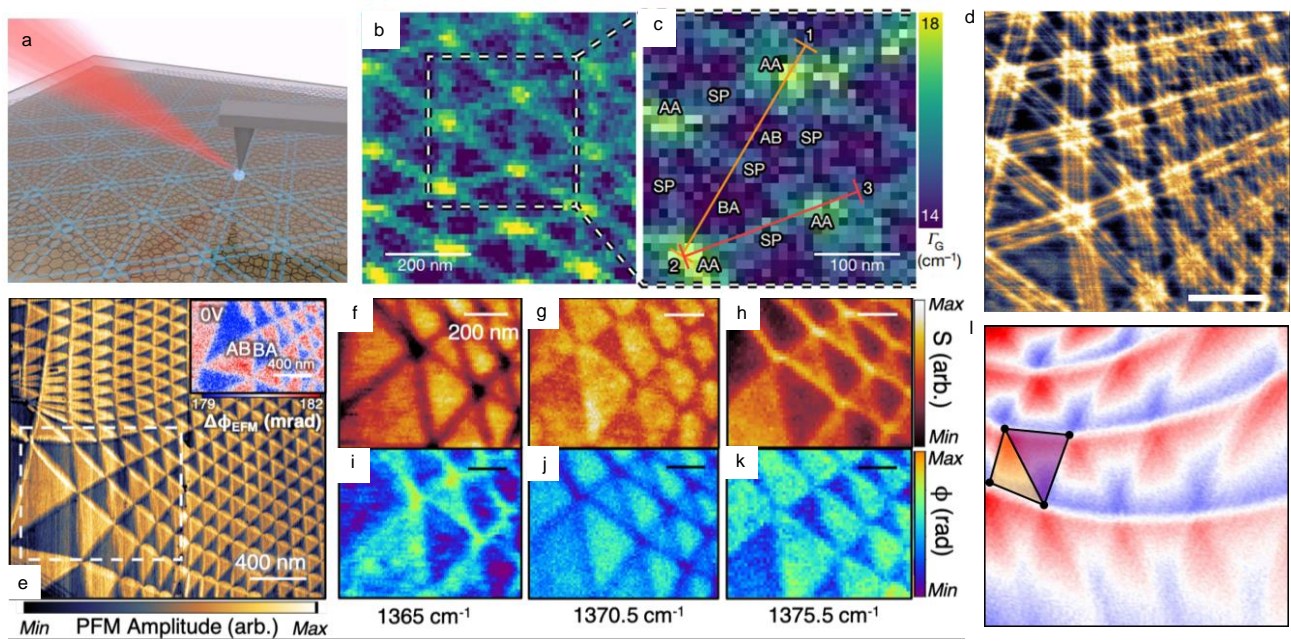

**Figure 21 | The spatial imaging of the reconstructed superlattice obtained by using near field scanning optical microscope (SNOM).** (a) Schematic illustration of infrared light beams incident and scattered by an atomic force microscope tip in the SNOM system, where the near-field response is detected through the scattered light in the far-field. Adapted with permission from ref.[282] Copyright 2020 Springer Nature. (b) Crystallographic hyperspectral image for the periodic pattern from the t-BLG based on the G'-band nano-Raman intensity. (c) Zoom-in image of the boxed region in (b) with the assignment of domain elements in the moiré pattern. (b,c) Adapted with permission from ref.[281] Copyright 2021 Springer Nature. (d) Near-field images of the normalized near-field scattering amplitude s(ω) in false color maps revealing the domain walls in TBG for the polariton, where the contrast seen of the domain walls is associated with the reflectivity of the polariton. Scale bar: 500 nm. Adapted with permission from ref.[282] Copyright 2020 Springer Nature. (e) Large-area PFM image and EFM image (inset) of the stacking sample. (f-h) Superlattice pattern images of the near-field amplitude taken at three selected frequencies for scanned region highlighted by dash line in (e). (i-k) Images of the near-field phase taken at three selected frequencies for the same region. (e-k) Adapted with permission from ref.[284] Copyright 2021 Springer Nature. (l) Reconstructed superlattice patterns with contrast domains (yellow/purple shaded triangles) *via* the photocurrent map of twisted stacking graphene, measured at excitation energy E = 117 meV and carrier density n ~$1 \times 10^{12}$ cm$^{-2}$, where the map is normalised by the maximum measured VPV, Colour code for blue and red is −1 and +1 respectively. Adapted with permission from ref.[285] Copyright 2021 Springer Nature.

### 4.2.2   Electron microscope

Transmission electron microscopy (TEM) forms images with a beam of electrons transmitted through a specimen in the bright field (BF) and dark field (DF) modes, which enables significantly higher resolution imaging compared to conventional optical microscopes. Therefore, TEM can map the stacking order to probe the generation of the commensurate domains through reconstruction. During the measurement, the specimen is tilted at a designated angle and an aperture is utilized to filter a distinct Bragg peak on the diffraction plane. Owing to the antisymmetric displacement of the lattice period within AB and BA domains, there is a noticeable difference in intensity between these domains when the specimens are tilted relative to the direction of the incident electron beam. This differential intensity offers a spatial depiction of the stacking order.[134] Although the reconstructed superlattice can also be visualized as commensurate triangular patterns highlighted by the domain walls via the use of the BF-TEM, the clear contrast between the adjacent domains obtained in DF-TEM mode are more intuitive (Figure 22a, b).[45] This is because DF-TEM only selects scattered electrons from the sample appearing bright and excludes unscattered electrons from areas around the sample, allowing to enhance domain contrast caused by the modified interfacial conditions in the stacking configurations. Another widely used electron microscope is SEM. Although the typical SEM imaging process is based on the detection of backscattered primary electrons and is believed to ineffectively probe ultrathin film specimen, researchers successfully imaged stacking domains based on the developed electron channelling contrast imaging technique, named secondary electron imaging technique. The secondary electron emission occurs dye to the interaction between the incident primary electron beam's scattering corss-section and the escape depth of the secondary electrons.[139] As illustrated in Figure 22c, after placing the sample with tilt ($\theta$) and azimuthal angles ($\phi$), the secondary electron emission from different stacking orders would be detected and exhibit a signal contrast between different channelling conditions. It is worth noting that the observed AB and BA domains (Figure 5d) appear at particular polar and azimuthal angles, as the electrons move through the stacking sample with minimal scattering when the incident electron beam was aligned parallel to the open cavities or 'channels' in the atomic lattice[139]. Unlike the special sample preparation required for TEM measurement, this approach can be conducted on the sample laying on common substrates, which enables further collaboration with other aforementioned SPM imaging techniques and characterization tests.

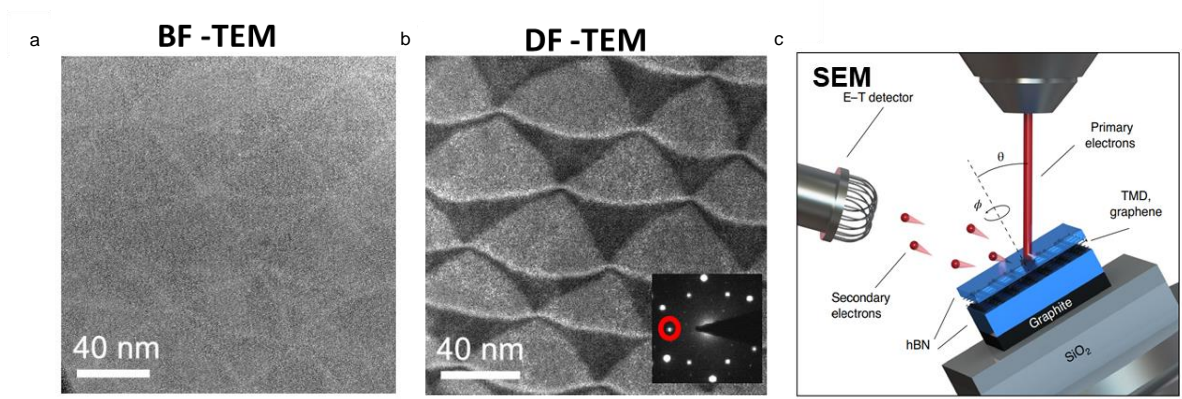

**Figure 22 | The acquisition of moiré pattern via electron microscope.** (a) BF-TEM image of the heterostructure showing triangular domains for the moiré pattern. (b) DF-TEM image of the same region on the heterostructure in (a) showing alternating triangular domains of different atomic arrangement with higher contrast. (a,b) Adapted with permission from ref.[45] Copyright 2020 American Chemical Society. (c) Schematic of operating principle: an accelerated electron beam is focused onto the device at $\phi$ and $\theta$. Channelling of the incident electrons and subsequent secondary electron generation depend on the local stacking order. Adapted with permission from ref.[139] Copyright 2021 Springer Nature.

## 5. Physical properties

### 5.1 Moiré conductivity

Although superconductivity has been discovered in certain compounds and alloy materials, the practical application of such superconducting materials is limited by the critical parameters (e.g., critical temperature $T_c$, critical magnetic field $H_c$, critical current density $J_c$, etc.) and fabrication techniques.[286] For example, conventional superconductors require extremely low temperatures to achieve superconductivity, leading to a high cost application. The emergence of 2D materials has spurred the investigation of 2D superconductors as a hotspot, with a goal of discovering room temperature superconductivity. It's worth noting that twisted TMDs heterostructures have not been proved to be superconductive despite other correlated phases having been discovered. Consequently, research into twisted carbon compounds dominates the current field of superconductivity.[287] Meanwhile, since the mechanism of the unconventional superconductivity in twisted graphene nanosheets cannot be elucidated through the conventional Bardeen–Cooper–Schrieffer (BCS) theory, which demonstrates the weak electron-phono interaction in conventional superconductors,[288] the mechanism behind the unconventional superconductivity remains to be elucidated. Over the past few years, numerous studies have been conducted to investigate the physical parameters underlying the superconductivity phenomenon in 2D graphene. The research approaches include modifying sample thickness, creating diverse heterostructures, controlling carrier density through electric fields, chemical doping, applying pressure and more.[289-290]

### 5.1.1 Superconductivity at magic angles of bilayer graphene

Graphene stands out as a promising 2D candidate to study superconductivity. Firstly, as the pioneering exfoliated 2D material, extensive research on graphene has been conducted, leading to the development of large-scale graphene thin films.[255, 291] Moreover, t-BLG is considered to be one of the most straightforward heterostructures to study quantum phenomena in superconductivity, which is fabricated by stacking two single-layer graphene sheets with a specific twist angle, illustrated in Figure 4a.[38] This arrangement forms a distinctive hexagonal moiré pattern in t-BLG, and such modulation in the superlattice induces the change of the band structure.[127] It has been proved that electrons in the AA-stacked regions are more likely to function as carriers rather than participating in bonding interactions, so the conductivity in AA-stacked regions is higher than that of the AB/BA-stacked regions due to the higher carrier density, as verified in Figure 23a.[135] Atomic-resolution STM images in Figure 23b-d depict t-BLG at various small angles, with bright spots corresponding to AA-stacked regions and dark regions representing AB/BA stacking.[67] Decreasing the twist angle results in a reduction in the relative area of AA-stacked regions, causing a lower conductivity owing to lower carrier density.[135] In 2018, Cao *et al.* discovered superconducting behaviour in t-BLG with a relatively high $T_c$ of 1.7 K at a small angle of about 1.1°, shedding light on the exploration of room temperature superconductivity.[37] When the twist rotation of t-BLG is tuned to 1.1º, the flat bands with zero Fermi energy occurs, showing an unconventional superconductivity phenomenon.[292-293] This angle of 1.1º is referred to as the first magic angle.[37] It has been reported that the Dirac-point velocity would vanish under a series of magic angles (Figure 23e), resulting in a very flat moiré band because of the strong interlayer coupling.[125] The existence of the flat band is the major contributing factor to the superconductivity. In order to observe the flat bands in t-BLG, Lisi and co-workers directly mapped the band dispersion by combining various imaging techniques, which is also calculated in Figure 23f (marked by red arrow).[293]

They also identified multiple hybridization gaps in the band dispersion (marked by black arrows in Figure 23g). The superconductivity would be suppressed by the hybridization gaps, which are caused by the lattice reconstruction when the twist angle is too low.[294] It's worth noting that the t-BLG would turn into a Mott-like insulator when the flat bands at magic angles are half filled, in spite of the metallic property of the monolayer graphene.[126] According to Figure 23h, the superconducting domes are adjacent to the insulating states,[37] so the domain wall should be broken by some external factors. An external magnetic/electric field or pressure would contribute to tailoring the band structure and lead to the transition of insulating states to superconducting states.[295] Up to now, plenty of theories are proposed to explain this unconventional superconductivity behaviour, including interlayer hybridization, electron-electron interaction and atomic reconstruction at low twist angles.[44, 126, 130, 296] Moreover, the modification of the carrier density plays a vital role in tuning the superconductivity. The carrier density would decrease with the reducing twist angles, causing a low conductivity in t-BLG.[135] Remarkably, carrier density in flat bands can be tuned electrically to the order of 10 meV, which is vital for achieving superconductivity.[37] In addition, t-BLG at small twist angles has revealed intriguing quantum phenomena such as the formation of van Hove singularities (VHSs)[297-298], enhancement in optical absorption[299] and chemical reactivity[298].

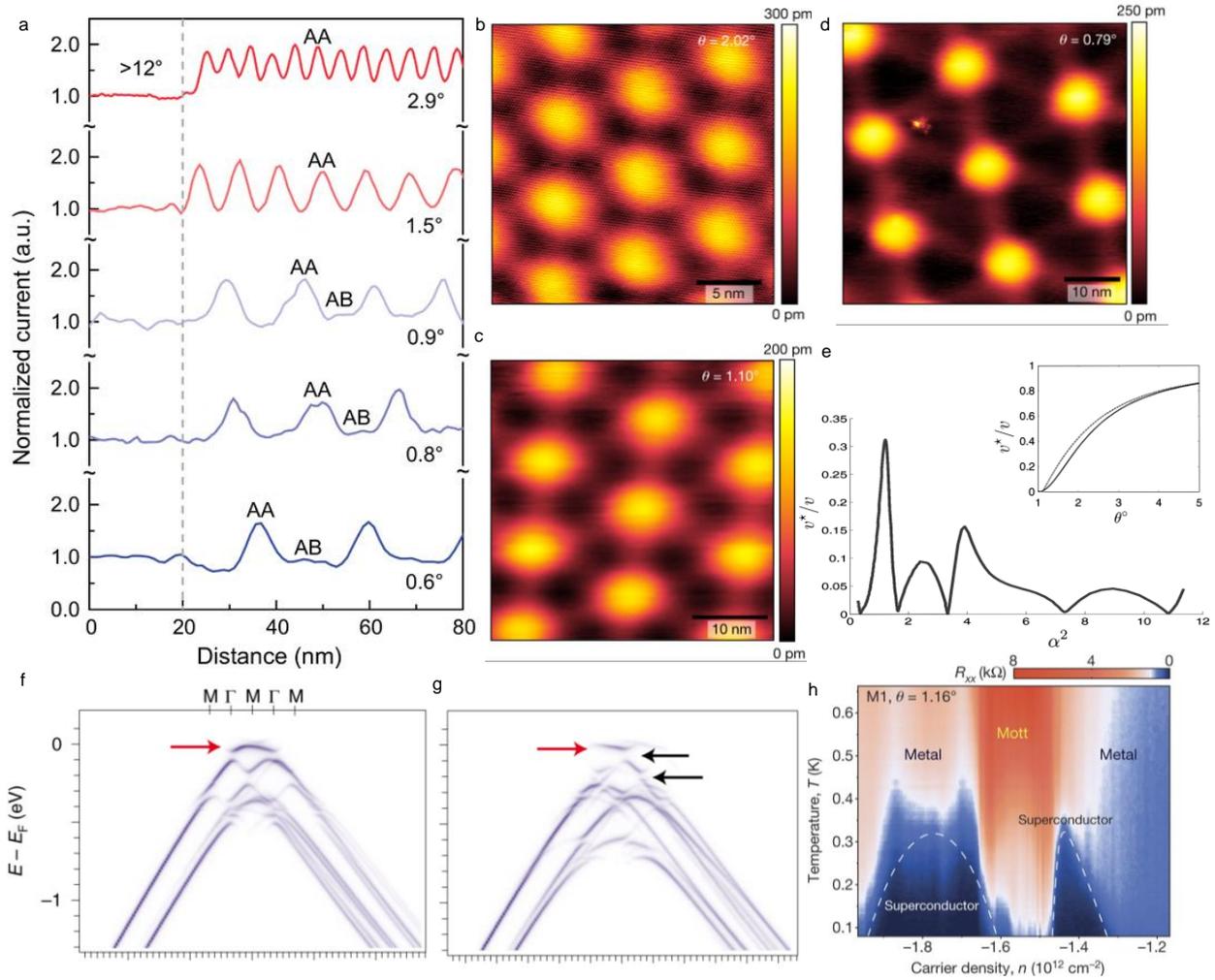

**Figure 23 | Superconductivity in** t-BLG. (a) Typical current profiles measured on TBG across two domains (one domain with a twist angle of >12° and the other domain with twist angles of 2.9°, 1.5°, 0.9°, 0.8°, and 0.6°, respectively). Adapted with permission from ref.[135] Copyright 2020 American Association for the Advancement of Science. (b-d) Atomic-resolution STM topographies on 2.02°, 1.10° and 0.79° t-BLG samples. Adapted with permission from ref.[67] Copyright 2019 Springer Nature. (e) Renormalized Dirac-point band velocity. (Inset) The renormalized velocity at larger twist angles. Adapted with permission from ref.[125] Copyright 2011 Proceedings of the National Academy of Sciences. (f, g) Calculations of the spectral weight distribution. The flat band and multiple hybridization gaps are marked by red and black arrows, respectively. Adapted with permission from ref.[293] Copyright 2020 Springer Nature. (h) Four-probe resistance $R_{xx}$, measured at densities versus temperature. Adapted with permission from ref.[37] Copyright 2018 Springer Nature.

### 5.1.2 Superconductivity in trilayer graphene

Magic-angle twisted trilayer graphene (MATTG) exhibits unconventional superconductivity attributed to a broken symmetry phase.[51, 300] Figure 24a shows the schematic for twist angle variation in TTG, where θ represents the twist angle, $t_1$ and $t_2$ are the moiré pattern lattice vectors. It has been proposed that the electronic structure and superconductivity of TTG are easier to modify than t-BLG. Park *et al.* demonstrated that the mirror symmetry phase of TTG would be broken by an electric displacement field, resulting in the hybridization of the flat bands and Dirac bands. Therefore, the band structure can be effectively controlled.[51] The flat band dispersion in TTG is sensitive to the moiré band filling factor $v$, as shown in Figure 24,[288] the flat band would be tuned to different locations due to the influence of different $v$. It has also been found that TTG would show the strongest superconductivity at the filling factor between -2 and -3 as well as $+2 + \boldsymbol{\delta}$, which indicates highly tunable electron interactions, resulting in adjustable superconductivity in TTG.[288] In Figure 24c, $R_{xx}$ versus T and $v$ plot displays superconducting regions near the hole-doped side ($v = -2$) and electron-doped side ($v = +2$), where $R_{xx}$ refers to the resistance of TTG.[51] Notably, stronger superconductivity is observed in the $+2 + \boldsymbol{\delta}$ and $-2 - \boldsymbol{\delta}$ regions, whereas weaker superconductivity is found at the $+2 - \boldsymbol{\delta}$ and $-2 + \boldsymbol{\delta}$ regions. Meanwhile, the superconducting domes can also be affected by temperature. Hence, many studies suggest that the superconductivity in TTG is dominated by electron interactions instead of weak electron-phonon interactions.[301] More comprehensive investigations of TTG are required to get a further understanding of the mechanism of this unconventional superconductivity. Ma *et al.* have calculated the band structures of TTG to provide further insight, explaining the unconventional superconductivity as depicted in Figure 24d-i. The band structures of TTG with twist angles of 5º (Figure 24d-e), 1.53º (Figure 24f-g) and 1.12º (Figure 24h-i) are calculated respectively. As can be seen in Figure 24d, the band structure of TTG can be made up of AB-stacked t-TLG and a monolayer graphene near the Dirac points. Figure 24f shows that the two bands near the Fermi level become narrow with the reducing twist angle, which is similar to the t-TLG. According to Figure 24h, the twist angle's reduction towards the magic angle results in nearly flat bands near the Fermi level. It's noteworthy that a hybridization gap arises as the twist angle reduces, as seen in Figure 24e, g and i. The opening gap would be larger with a smaller twist angle, which is unfavorable for superconductivity. The change of Chern numbers in Figure 24e, g and i also reflect the gap opening between the two bands near the Fermi level, which is a result of the broken symmetry phase with electron interaction in TTG.[302] Similarly, flat bands have also been investigated in twisted double bilayer graphene (TDBG). Nevertheless, some research demonstrated that the bands in TDBG can only be flattened by an external electric field, while others considered that the natural lattice relaxation is attributed to the flat bands in TDBG.[28] It emphasizes that further research is needed to elucidate the genuine mechanism behind flat band behavior in twisted graphene, as precise flat band modification can lead to correlated states like unconventional superconductivity, insulating states, and ferromagnetism. Moreover, the mechanism of unconventional superconductivity in twisted 2D materials remains to be extensively studied through a multitude of investigations.

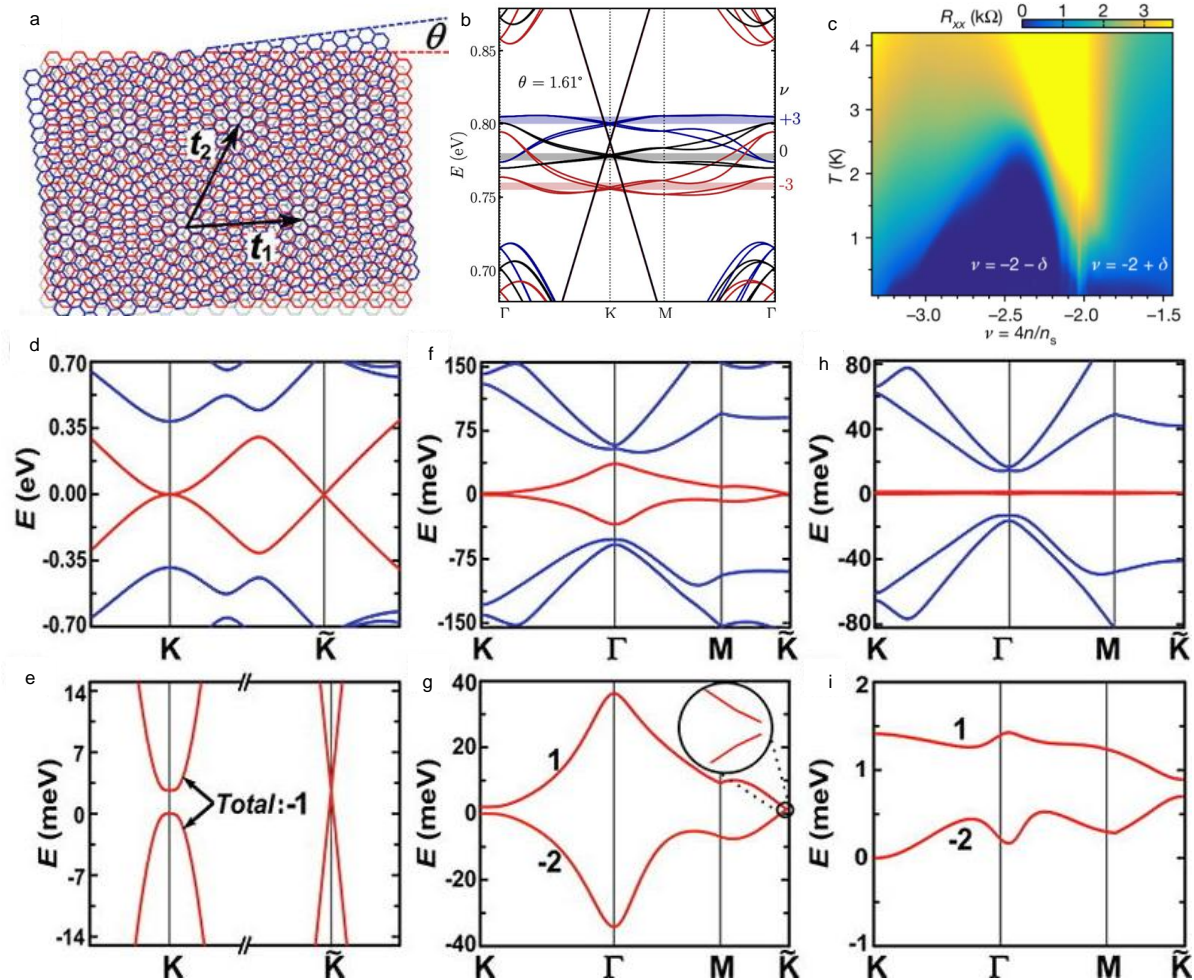

**Figure 24 | Superconductivity in twisted trilayer graphene**. (a) Lattice structure of the twisted trilayer graphene (TTG). θ is the twisting angle between the top monolayer (blue, n=3) and an AB-stacked bilayer(n=1,2). t1 and t2 are the moiré pattern lattice vectors. Adapted with permission from ref.[300] Copyright 2021 Elsevier. (b) Band structure of MATTG at θ = 1.61°. The flat band dispersion is strongly affected by the filling factor ν (blue: ν= +3, black: ν= 0, red: ν= −3), whereas the Dirac cone remains unaffected. Adapted with permission from ref.[288] Copyright 2022 Springer Nature. (c) Rxx versus T and ν showing the superconducting regions near ν= −2 and ν= +2. Adapted with permission from ref.[51] Copyright 2021 Springer Nature. (d, e) The band structure at = 5°. (f, g) The band structure at = 1.53°. (h, i) The band structure at = 1.12°. The parameters of the minimal model are: 0=2.464 eV, 1=400 meV, 3=4=Δ¨=0. The black numbers in (d), (f), (h) are the valley Chern numbers for two central bands (red lines). (d-i) Adapted with permission from ref.[300] Copyright 2021 Elsevier.

## 5.2 Moiré photons

### 5.2.1 Optical absorption

The most striking phenomenon regarding photons when the moiré supercell is involved in the heterostructure is that the regular moiré potential is notably stronger and more dominant than the kinetic energy in the strong-coupling regime within the mini-Brillouin zone. This results in a qualitative modification to the electronic band structure and thus generates multiple flat minibands for the corresponding photon absorption due to the intensive electron–hole correlation.[35] The optical absorption of 2D materials is normally directly measured by the reflection contrast spectroscopy. Figure 25a presents the reflection contrast spectrum of the $WS_2$/$WSe_2$ heterobilayer with subtracted background and a rotational angle of around 0.5°, which produces a moiré structure from the R-typed stacking in the system. As indicated in the top panel, the $WSe_2$ resonances are split into three distinct peaks with strong absorption instead of a single primary peak that is shown in the spectrum for the heterobilayer with a large twist angle (bottom panel), owning to the generated different moiré states from the minibands.[35] The emergence of these three peaks with moiré lattice is confirmed by more additional samples, which is present in all the aligned heterobilayers with rotational angles near 0° or 60°. However, their spectral weight decreases and eventually vanishes with increasing misalignments between the two layers, and the spectrum reverts to having only one peak. This is signature evidence of moiré superlattice and it exhibits a markedly different dependency on the gate voltage in the reflection contrast spectrum compared to the heterostructure that lacks the moiré pattern contribution. The gate-dependent optical transition in normal $WS_2/WSe_2$ heterostructure was reported that $WS_2$ and $WSe_2$ resonances demonstrate the opposite gate dependencies due to the type-II band alignment.[195, 197, 303] The electrostatically doped electrons and holes are mostly staying in the $WS_2$ and $WSe_2$ layers separately, which will not be able to significantly affect the performance of another layer. Therefore, resonances from $WS_2$ are sensitive on the electron-doping side, which experiences substantial modulation, but it only redshifts slightly when the heterostructure is doped with holes. Whereas, the three peaks of $WSe_2$ resonances from small twist heterobilayer exhibit particularly noteworthy gate-dependency under both electron and hole doping (Figure 25b) due to a large moiré period,[35] and their pronounced changes upon electrostatically induced electrons are depicted in Figure 25c. More intriguingly, the dependent feature on electron concentrations varies for different individual peaks. Peak I undergoes a significant blueshift and shifts its oscillator strength to a newly appeared peak at lower energy (I′) with increasing electron density, while peak III experiences a considerable blueshift with a decline in oscillator strength. Peak II, on the other hand, shows only a minor energy shift and remains mostly unchanged. Furthermore, the measured optical absorption from heterobilayers can validate the formation of hybrid states because of the large difference in intra- and interlayer transitions in terms of oscillator strength. The oscillator strength of interlayer emissions is typically weaker, with a difference of about two to three orders of magnitude compared to intralayer transition,[304-305] which makes it challenging to be observable in the spectrum of absorption or reflection contrast where the noise level is around 1% or even higher. The resonance peak from interlayer transition with appropriate spectral weight demonstrated in Figure 25d is direct evidence of the hybridization of two states because the hybrid states can gain the oscillator strength from the intralayer component, making it resolvable in the measurement.[31, 305] The hybrid resonances can be modulated with the varying moiré superlattice induced by different rotational angles on the energy and oscillator strength, which also affects the tunnelling and coupling between two layers.[47] When the twist angle is close to 0° or 60°, the two layers are almost perfectly aligned in momentum space and the lattice mismatch can be neglected, which allows a clear observation of hybrid resonance doublets as depicted in Figure 25e. The oscillator strengths and peak position of doublets continuously evolve when the two layers undergo relative twisting because the band minimum for interlayer emission is shifting away from the intralayer one by momentum mismatch. Meanwhile, the kinetic energy is enhanced with the increasing deviated angle from 0° and 60°, which leads to a strong blueshift of the doublet emissions and decreases the interlayer coupling because of the reduced tunnelling channels.[47] Moreover, the optical transitions observed in 1L/2L phosphorene heterostructures[161] explained above exhibit a pronounced dependence on the twist angle, supporting that the moiré potentials exert a substantial impact on the electronic band structures of the twisted 2D structures, for both the same or different stacked materials.

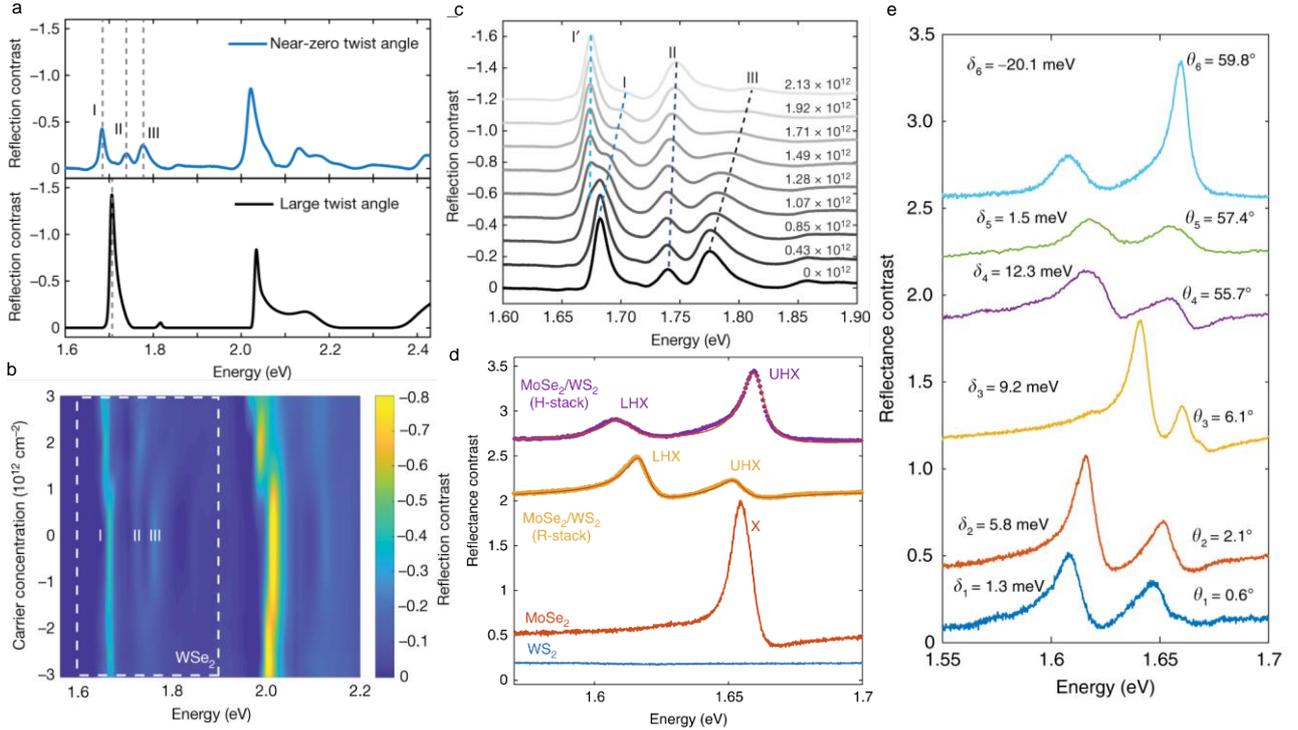

**Figure 25 | Optical absorption of twisted moiré superlattices.** (a) Reflection contrast spectrum of aligned WSe₂/WS₂ heterobilayer with $0.5 \pm 0.3°$ (top), compared to another sample with a large twist angle (bottom). (b) Gate dependent reflection contrast spectrum of nearly aligned WSe₂/WS₂ with both electron (positive carrier concentration) and hole (negative carrier concentration) doping. The white dashed box marks the photon-energy range close to that of the WSe₂ A exciton, where the three prominent moiré exciton states appear and are labelled as I, II and III. (c) Extracted reflection contrast spectra in the range of the WSe₂ A exciton on the electron-doping side. Upon doping, peak I shows a strong blueshift and transfers its oscillator strength to another emerging peak at lower energy (I′), peak II shows a strong blueshift with diminished oscillator strength, but peak II remains largely unchanged. (a-c) Adapted with permission from ref.[35] Copyright 2019 Springer Nature. (d) Reflectance contrast spectra for a monolayer WS₂ (blue), monolayer MoSe₂ (red), R-stacking MoSe₂/WS₂ bilayer (orange), and H-stacking MoSe₂/WS₂ bilayer (purple). (e) Reflectance contrast spectra of MoSe₂/WS₂ heterobilayers with different twist angles. (d,e) Adapted with permission from ref.[47] Copyright 2020 Springer Nature.

### 5.2.2  Second harmonic generation (SHG)

Second harmonic generation (SHG) is a nonlinear optical phenomenon in which photons interact with a material, resulting in the creation of new photons with twice the frequency and half the wavelength of the incident photons.[306-307] This process can occur in specific materials that lack inversion symmetry in their crystal lattice. It has been recognized as a powerful tool used to measure various properties in 2D materials,[308] such as lattice orientation,[309] twist angle,[50] inversion symmetry,[310-311] defects, [312-316] phase variation,[317] temperature variation,[307] piezoelectricity and ferroelectricity,[118, 318-320] layer number and thickness,[6, 48] strain,[306, 321-323] doping,[324-326] etc. This is because the efficiency of SHG depends on various factors such as the intensity of the incident light, the crystal structure and the phase matching conditions.[327-329] Phase matching refers to the condition in which the phase velocities of the incident and generated photons match, leading to constructive interference and maximizing SHG efficiency. For stacked 2D materials, SHG response is not only highly sensitive to lattice orientation, but twist angle variation has also been observed to tune SHG response in homo-structures,[50] folds[306, 330] and heterostructures.[331] Twist angle serves as a controlling knob for the SHG response. The SHG intensity ($I_s$) of stacked layers is the vector addition of SHG vectors from individual layers ($I_1$ and $I_2$) and is proportional to the combined SH intensity originating from the individual layers, given as[50] $I_s(\theta) \propto I_1 + I_2 + 2\sqrt{(I_1 I_2)}\cos3(\theta)$, where $\theta$ represents the angle of twist between the armchair directions of layers 1 and 2. According to the above equation, the resultant SHG from stacked structures becomes maximum for the bilayer with 0$^o$ twist angle (such as 3R stacking configuration of TMDs) and becomes zero for 60$^o$ twist angle (identical to 2H stacking configuration of TMDs). Figure 26a-c shows that the resultant SHG from stacked structures is the superposition of SHG coming from flake 1 and flake 2. Moreover, the phase difference from the twisted bilayer is also shown to be dependent on the twist angle both experimentally and theoretically.[50] The above superposition principle implies that the SHG response for 3R $MoS_2$, denoted as $SHG_{iL}$ can be expressed as $i^2(SHG_{1L})$, where $i$ is the layer number, and this has been verified by different studies. For instance, SHG response in samples with 2 layers, 3 layers and 4 layers of 3R-$MoS_2$ exhibits an increase of 4 times, 9 times and 16 times, respectively, compared to the SHG response of a single layer.[6] The same results are shown for different parallel and antiparallel stacking configurations of $MoS_2$.[310] Twisting is found to tune the resultant SHG response in both heterostructures[331] and folds.[306] Superposition principle is effectively

applied for multilayer folds of TMDs as well. For instance, stacking angle variation is used to

tune SHG response 1 to 9-times for trilayers folds.The folds were created using buckling of

$WS_2$, resulting in trilayer folding where the top and bottom layer of the fold is parallel to each

other, while the middle layers make an angle (called folding angle $\theta_f$) that is different for

different folds generated (Figure 26d). PSHG (polarization-resolved second harmonic

generation) shows an enhanced SHG response from the folded region ($\theta_f = 20^o$) as compared

to 1L $WS_2$ SHG (Figure 26e). A predicted 1 to 9 times SHG response for $0^o$ to $60^o$ variation

of folding angle is observed for the trilayer folded region of $WS_2$ according to superposition

principle. Experimental results for folding angles of $7^o$, $20^o$ and $30^o$ show ~1.8-, 2.4- and 3.2-

times enhancement, respectively, which agrees well with theoretical predictions (Figure 26f).

The superposition principle is independent of the excitation wavelength; therefore, similar enhancement is found for a variable range of wavelengths. The phase difference can also be calculated from the folding angle variation using superposition principle (Figure 26g).[306] This vector superposition rule is employed to find the resultant SHG in different stacking arrangement of multilayer system (Figure 26h).

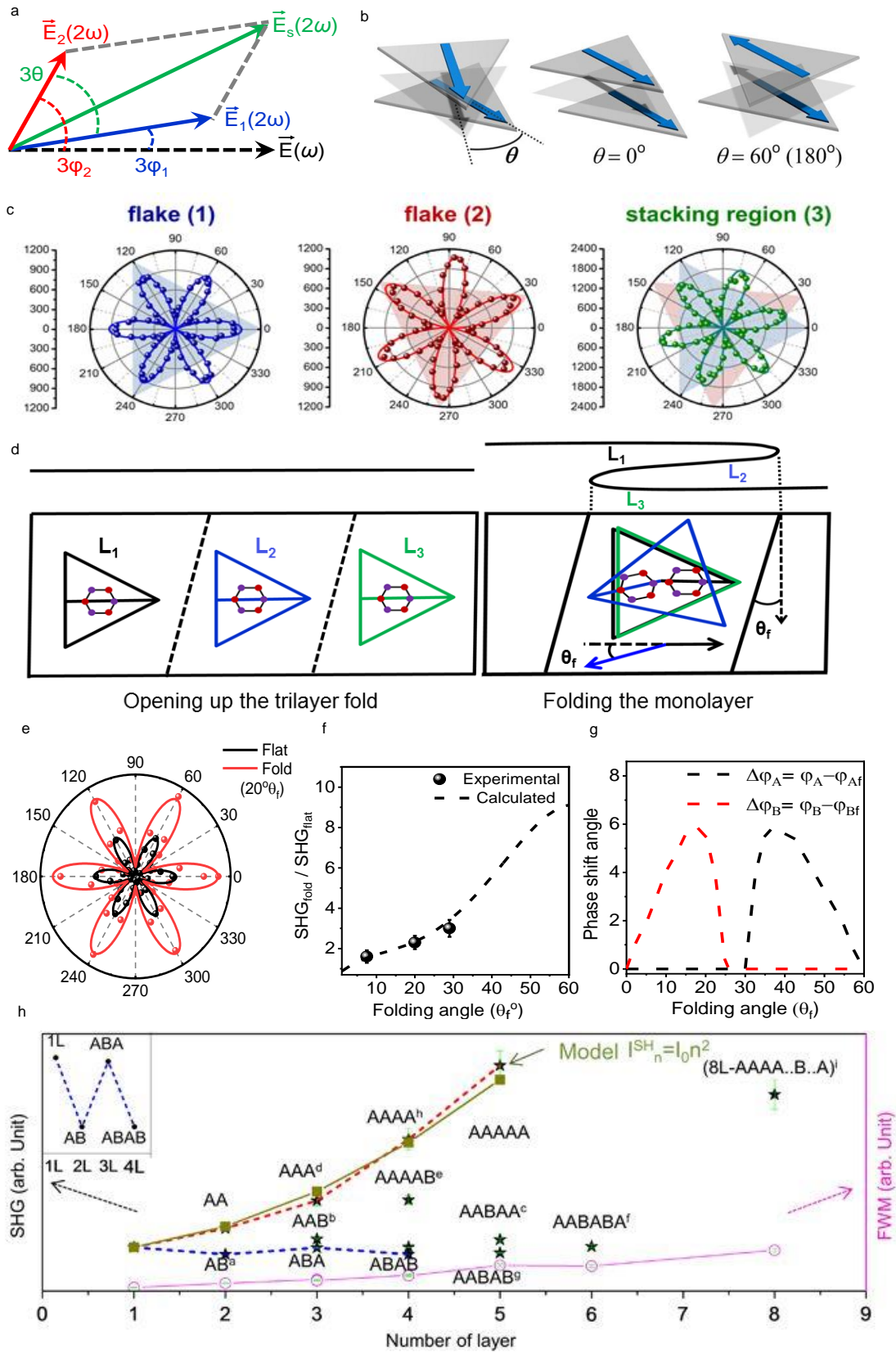

**Figure 26 | Stacking angle and strain determination via optical harmonic generation.** (a) The schematic showing the vector superposition of the SH electric fields $\vec{E}_1(2\omega)$ and $\vec{E}_2(2\omega)$. In the schematic, $\vec{E}(\omega)$ is the electric field of the fundamental light, $\vec{E}_1(2\omega)$ and $\vec{E}_2(2\omega)$ are the SH electric fields from the flake 1 and flake 2 respectively, $\vec{E}_s(2\omega)$ is the resulting SH electric field (as a result of vector superposition) from the stacking region. (b) Schematics for stacked bilayers with an arbitrary stacking angle θ, θ = 0° and θ = 60° (or θ = 180°). The resultant SHG from stacked bilayers is maximum at θ = 0° and minimum at θ = 60° due to vector superposition. (c) Polarization dependent SH plots as a function of azimuthal angle φ measured for monolayer flakes (1), (2), and stacking regions (3). (a-c) Reproduced with permission from ref.[50, 234] Copyright 2014 American Chemical Society. (d) A schematic illustration for the stacking of layers in a trilayer fold (*1L+1L+1L*). Upon folding, the top (layer 1) and bottom layer (layer 3) are parallel to each other, whereas armchair direction of mid layer (layer 2) makes an angle of $(180+\theta_f)$ with the armchair direction of top and bottom layer. The lines (black, blue and green) bisecting the triangles (black, blue and green) show the armchair direction] (e) Experimental investigation of polarization resolved SHG $I_\parallel(2\omega)$ intensity pattern for 1L flat and folded WS$_2$. Continuous lines are the fitted plots, whereas symbols are the experimental data points. (f) The folding angle dependence of SHG enhancement for fold, where enhancement= $\rho_A/\rho_{Af}$. Error bars represent the range of error in the measured values. (g) The folding angle dependence of SHG phase shift angle (degrees). (d-g) Reproduced with permission from ref.[306] Copyright 2020 American Chemical Society. (h) The graph shows the maximum intensity of the SHG for each layer with different stacking orientations. The inset represents the zoomed SHG intensity variation trend of the ABA…-stacking configuration. The olive color plot represents the results of the model to explain the SH intensity of the AA(A...)-stacking crystals. Reproduced with permission from ref.[310] Copyright 2018 Springer Nature.

Vector superposition is effectively applied in the case of heterostructures. For instance, 1L WSe$_2$ is stacked at various twisting angles (29.8º, 4.3º and 0.8º) on top of 1L MoS$_2$ to create 3 WSe$_2$/MoS$_2$ heterostructures as shown in Figure 27a-f. Twist angle variation and lattice mismatch between MoS$_2$ and WSe$_2$ give rise to the generation of moiré superlattices, which show a periodicity of 0.62 nm, 3.8 nm and 8.5 nm for 29.8º, 4.3º and 0.8º twisting angles, respectively. Due to a large variation in twist angle, very different SHG responses are observed for different heterostructures (Figure 27a-f). For HS1 = 29.8º, the maximum SHG is found along an axis that is midway between WSe$_2$ and MoS$_2$, not along either of them (Figure 27e). For HS2 = 4.3º and HS3 = 0.8º, the maximum SHG is slightly shifted (Figure 27d and f). According to the vector superposition principle, the resultant SHG of heterostructure HS2 and HS3 should be maximum along the maximum SHG direction of MoS$_2$ and WSe$_2$. Considerable phase shift is seen for HS1= 29.8º, which is in line with the vector superposition principle.[120] SHG phase difference and intensity response show the validity of vector superposition for the case of heterostructures as well. Twisting is seen as the controlling knob for quantum interference. For instance, an investigation with twisted bilayer WSe₂ shows a spectral dip triggered by quantum interference when SHG emitted photon energy is plotted against SHG exciton energy (Figure 27g). This quantum interference behavior is noted for twist angles between 21° and 45°, whereas no quantum interference is observed for 1° and 59° twist angles, indicating twist angle as a controlling tool that can be used to turn on and off quantum interference.[20]

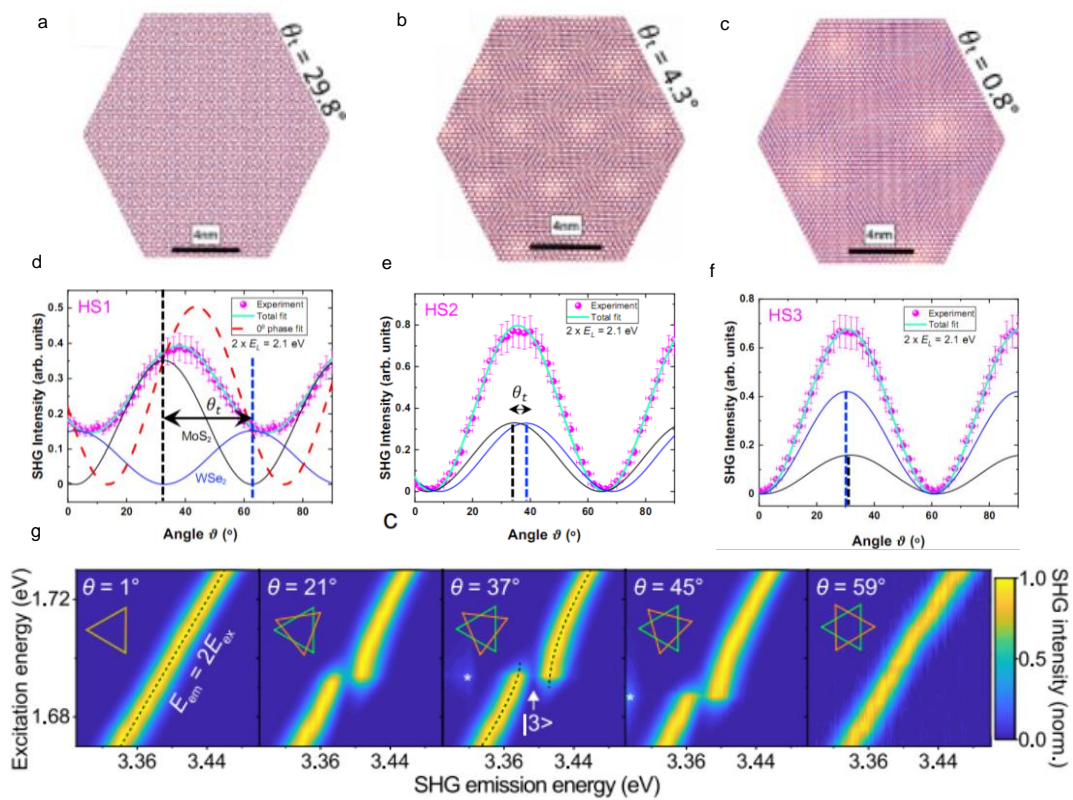

**Figure 27 | SHG investigation of twisted layers.** (a) Schematic representation of the moiré pattern of 1L-MoS₂/1L-WSe₂ heterostructures with twist angles θt of HS1=29.8°, (b) HS2=4.3°, and (c) HS3=0.8°. The extracted moiré period is 0.62, 3.84, and 8.45 nm, respectively. (d) Polarization angle dependent SHG and fitting, using excitation energy 2.1 eV, for HS1=29.8°, (e) HS2=4.3°, and (f) HS3=0.8° (a-f) Reprint with permission from ref.[331] Copyright 2022 American Physical Society. (g) Experimental twist-angle (θ) dependence of normalized SHG from bilayer WSe2 at 5 K, measured with 80 fs excitation pulses. The normalized SHG intensity is plotted as a function of emitted photon energy (horizontal axis, $E_{em}$) and central photon energy of the excitation laser (vertical axis, $E_{ex}$). The UPL feature is marked by asterisks. The SHG follows the canonical linear dependence on the excitation energy ($E_{em} = 2E_{ex}$, black dashed line in the left panel), as shown in the 1° and 59° twisted bilayers. Anti-crossing behavior (black dashed lines in the middle panel) becomes apparent when quantum interference

occurs, as shown in the 21°, 37°, and 45° twisted bilayers. Adapted with permission from ref.[20] Copyright 2021 Springer Nature.

## 5.3 Moiré excitons

### 5.3.1 Characteristics

Moiré exciton from TMD heterobilayers was first reported a few years ago by several groups with various representative distinct features. It arises from the moiré crystal and the formation of this superlattice pattern is closely related to the manipulation of twist angle engineering, and the moiré exciton exhibits a tight relationship to the induced potential depths. The period size is highly tunable, which varies ranging from a few to tens of nanometers depending on the mismatch of lattice constant and relative twist angle between layers.[30, 45, 200, 332] Both intralayer and interlayer excitons will undergo the modulation from moiré supercell owning to the periodic variation of band structures, endowing unique optical properties of excitons, which can be easily distinguished from typical exciton emissions. Similar to the absorption spectra presented in the above section, the in-plane periodic potential leads to an energy splitting of excitonic emissions, which can be observed in both R- and H-typed heterostructures.[23] Figure 28a presents the $MoSe_2$ PL emission from an hBN-encapsulated monolayer (top panel) and near 0° $MoSe_2/MoS_2$ heterostructure area (bottom panel) with both neutral and charged exciton peaks. The emergence of doublet peaks including an additional new energy peak at a higher energy for both exciton and trion is the signature of moiré excitons, in good agreement with theoretical predictions.[333] More heterostructure samples with different twist angles were further studied to reveal that such energy splitting is absent in heterobilayers with a large rotational angle and pure monolayer regions. The Bohr radius of excitons from monolayer TMD is reported on the order of one to a few nanometers[334] and the period of moiré superlattice in TMD heterobilayer decreases with increasing twist between the two layers relative to each other, therefore, the lateral size of moiré supercell can be much smaller with a large twist angle and may not be able to trap the excitons. As expected, a MoSe$_2$/MoS$_2$ heterostructure with an angle of approximately 19° is demonstrated without the appearance of peak splitting, because its moiré pattern size is considerably smaller than the spatial extent of the exciton wave function.[23] However, this moiré effect on the intralayer excitons is no longer available when the temperature exceeds 90 K even for the R-staking heterobilayer with a small twist angle, and the corresponding activation energy of intralayer exciton was estimated to be approximately 26 meV, which means the depth of moiré potential for that is around 26 meV. It is much smaller than the calculated deep potential level from R-typed heterostructure, because the intralayer excitons undergo much shallower moiré pattern induced effective potentials than interlayer excitons owing to the tiny modulation,[30] although their local band gaps both vary with the location, as depicted in Figure 28b. The detailed potential landscapes and optical selection rule for the moiré intra- and inter-layer excitons from R-type staked MoS$_2$/WSe$_2$ bilayer are illustrated in Figure 28c. It shows a significant difference in both depths of energy minima and spin-valley selection rules at high-symmetry points, which suggest opportunities for implementing dynamic controls to manage the quantum emitter arrays.[30]

The moiré potential for interlayer excitons can be very deep depending on the relative orientation between two monolayers and it may dominate over the kinetic energy of excitons. The moiré interlayer excitons in MoSe$_2$/WSe$_2$ heterobilayers with twist angles of around 1° and 2° are reported at low temperature of approximately 15 K, which were observed from PL spectra with multiple resonance peaks at lower energy side (Figure 28d).[24] The linewidth of interlayer excitons from the 1° heterobilayer is narrower than that from the 2° heterobilayer, which is in good agreement with the deeper potentials in the smaller-twist heterostructure. The energy splitting between resonances is almost a constant value and they were concluded stemming from the excitonic ground and excited states which are localised in the global potential minimum within the moiré lattice with a depth of around 100 meV. These interlayer emissions display alternating co- and cross-circularly polarized emission (Figure 28e), which originates from the distinctive spatial changes of the optical selection rules[24] and is consistent with the predictions claimed based on the considerations of rotational symmetry.[30] Subsequently, several moiré interlayer excitons with much narrower emission lines were observed (Figure 28f) from another 2° $MoSe_2$/$WSe_2$ heterobilayer under the excitation of much lower incident power, and each emission peak can be well fitted by the Lorentzian curves (inset). The PL signal from the same sample with a higher pumping power was also shown in Figure 28f for comparison, which demonstrates the transition from a broad PL spectrum to multiple sharp peaks, owing to the enhanced moiré modulation. The average full width at half maximum of observed discrete emissions is around 100 µeV, which is similar to the linewidths reported in quantum emitters found in 2D materials[335] including TMD single layer[336-337] and hBN.[338] Moreover, depending on the stacking configuration of two single layers, H- or R-stacking, the local minimum energy occurs at different locations, which enables modulation and control of circular polarization properties of the moiré trapped IXs (Figure 28g).[27]

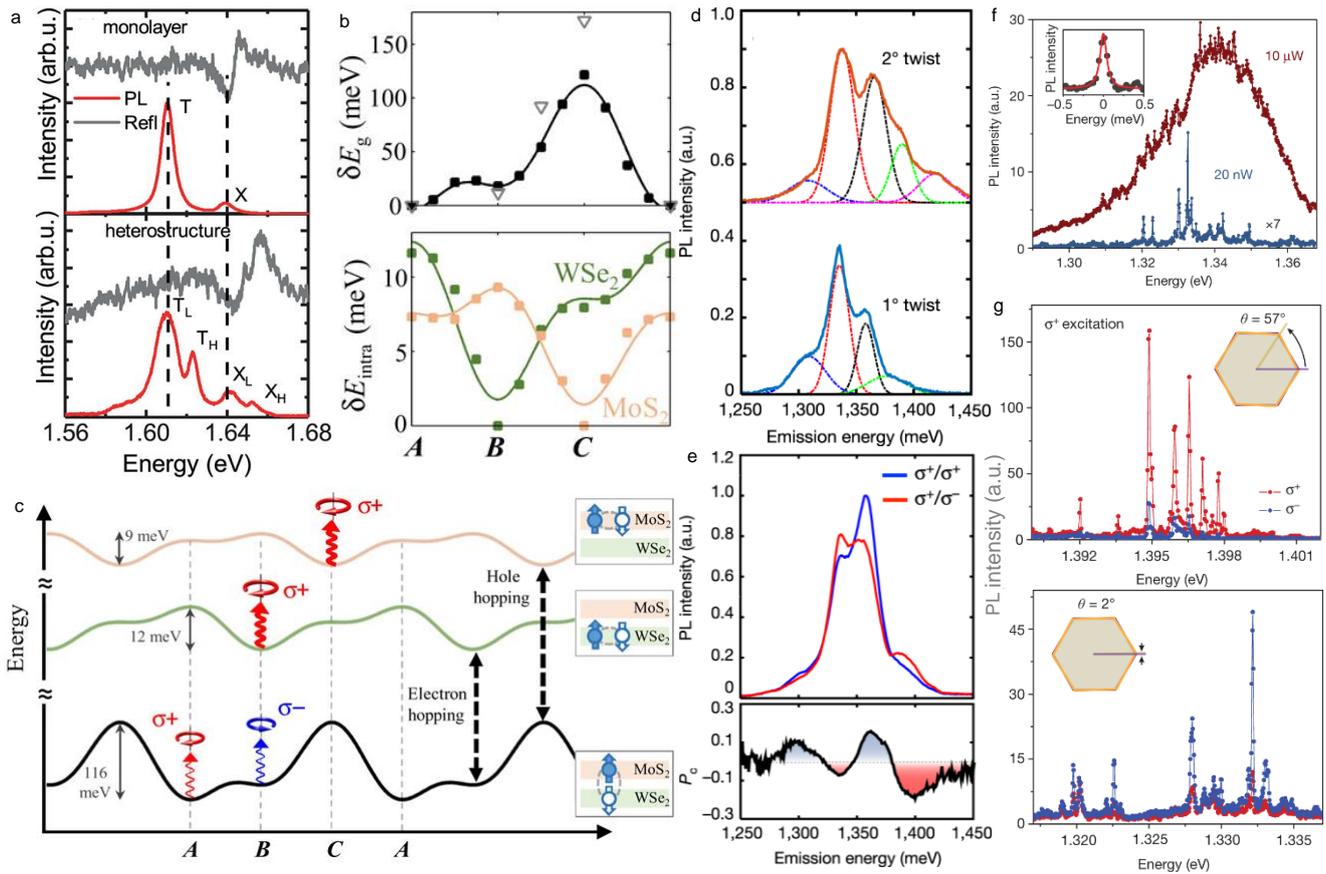

**Figure 28 | The Characteristic features of moiré excitons.** (a) PL and reflectivity spectra of MoSe$_2$ measured in monolayer (top) and MoSe$_2$/MoS$_2$ heterostructure (bottom). Intralayer exciton (X) and trion (T) both split into two peaks in the spectrum of twisted bilayers. Adapted with permission from ref.[23] Copyright 2018 American Chemical Society. (b) Top: Variation of the local bandgap E$g$ of the MoS$_2$/WSe$_2$ moiré excitons as a function of stacking registries. Bottom: Variation of the local intralayer bandgaps at different stacking registries. (c) Contrasted potential landscapes for both intra- and interlayer excitons in MoS$_2$/WSe$_2$ heterobilayer system, with the optical selection rule for the spin-up species shown at the energy minima. (b,c) Adapted with permission from ref.[30] Copyright 2017 American Association for the Advancement of Science. (d) PL spectra for MoSe$_2$/WSe$_2$ hetero-bilayers at twist angles of 2° (top) and 1° (bottom). The spectrum is fitted with five (2°) or four (1°) Gaussian functions. (e) Circularly polarized PL spectrum for σ+ excitation of the 1° MoSe$_2$/WSe$_2$ sample (top). The degree of circular polarization versus the emission wavelength is shown at the bottom, demonstrating multiple moiré interlayer excitons with alternating co-circularly and cross-circularly polarized emissions. (d,e) Adapted with permission from ref.[24] Copyright 2019 Springer Nature. (f) Comparison of interlayer exciton photoluminescence from a heterobilayer with a 2° twist angle at excitation powers of 10 μW (dark red) and 20 nW (blue; intensity scaled by 7×). Inset shows a Lorentzian fit to a representative photoluminescence peak, showing a linewidth of approximately 100 μeV under 20 nW excitation power. (g) Helicity-resolved PL spectra of trapped interlayer excitons of MoSe$_2$/WSe$_2$ heterobilayers with twist angles of 57° (top) and 2° (bottom). The insets illustrate the twist angles of the samples. The samples are excited with σ$^+$ polarized light at 1.72 eV. The σ$^+$ and σ$^-$ components of the photoluminescence are shown in red and blue, respectively. The PL from heterobilayers with twist angles of 57° is

co-circularly polarized, whereas that from the sample with a twist angle of 2° is cross-circularly polarized. (f, g) Adapted with permission from ref.[27] Copyright 2019 Springer Nature.

### 5.3.2    Many-body interactions

The moiré excitons are produced from the periodic moiré potentials trapped interlayer excitons,[339-340] which inherit the spatially separated electrons and holes, exhibiting a permanent dipole moment, therefore providing an opportunity for studying many-body interactions and facilitating the strong correlation in the system. The dipole-dipole repulsive interactions among IXs have been widely reported in recent years,[31, 341-342] which plays a crucial role in observing novel quantum phenomena such as superfluid, dipolar gases and dipolar crystals.[343-345] The effective long-range dipolar interactions can drive the system from the free-moving particles to a highly coherent aligned state, thus therefore to reach the exotic quantum phases. This dipolar interaction can be enhanced when the exciton concentration is increased. It is mainly reflected by the prominent induced blue shift of the average emission energy in the PL spectrum with an increasing exciton density,[201] as shown in Figure 29a. The peaks also broaden as a function of exciton density, especially when it exceeds a threshold excitation density. By recent experimental observations, it has been suggested that the states of charge-separated interlayer exciton in 2D heterobilayers are closely related to the incident density due to the highly dependent dipolar interactions.[27, 346-347] In a moiré supercell, the strong dipole-dipole interaction among IXs competes with the moiré potential generated from the superlattice to modulate and control the interlayer exciton dynamics. When the incident power is low, the exciton concentration will be reduced correspondingly, which will largely weaken the dipole-dipole repulsive interactions between IXs and bring the excitons to a localized state with sharp emissions. t-BLG with a magic angle exhibits flat minibands,[125] which leads to a strong electronic correlation and gives rise to diverse fascinating correlated phenomena like Mott insulators and superconductivity.[37, 126, 348] In contrast, the moiré coupling in TMD heterostructures can introduce flat minibands easier without the requirement of a specific angle owning to a large effective mass, which subsequently further reduced the kinetic energy in heterobilayers.[349-351] Figure 29b illustrates the moiré exciton properties of encapsulated MoSe2/WSe2 well-aligned hetero-bilayer (around 1°) with miniband structure under both low density and high density. It is found that the moiré exciton bands are completely flat for interlayer excitons (left panel), which is consistent with previous predictions[332, 352] and a series of straight bands here represent the localized excitonic transition paths of moiré exciton multiple resonance peaks. In this case, the interlayer exciton states can be completely trapped by the deep moiré potential landscapes and the repulsive interaction effects and kinetic energy of IXs can be ignored.[29] Moreover, the widely separated flat moiré minibands indicate that the supercell tunnelling ability of IX between neighbouring states is suppressed and negligibly weak.[332] However, when the density in the system is enhanced, there is a significant change in the band structure, which changes from flat to dispersed bands with smaller gaps, as displayed in the right panel of Figure 29b. The energy dispersion of exciton bands under high concentration indicates that interlayer excitons are no longer fully trapped in the superlattice and they can move through or even escape the potentials with sufficient kinetic energy arising from a strong dipole-dipole repulsion[29, 34]. The corresponding optical absorption spectra (Figure 29c) in both cases further evidenced the above-explained density dependent interactions among moiré excitons. The multiple sharp emissions are stemmed from the moiré states at low densities while only the peak from the lowest subband will appear at high densities because of less mixing of exciton centre-of-mass momenta.[352] Figure 29d schematically demonstrates the interplay between exciton–exciton interaction and moiré superlattice potential, providing a better understanding of many-body interactions in moiré crystal. The depth of effective potential wells will be changed with a variation of exciton concentration, and the localized states can become unstable at a high-density level due to the enhanced dipolar repulsive force, which can cause the excitons to be expelled from the trap and also significantly drive the hopping rate of excitons between the moiré cells.[29] In addition to the exciton density, the twist angle has been reported to be another crucial tool to modulate the dipole-dipole interactions and moiré potential levels.[31] The increased twist angle leads to a reduction in lateral confinement length, which can increase the kinetic energy of the localised excitons and the interactions among them will begin to manifest the effects.[352] The density-dependent curvatures of the lowest moiré minibands for both heterobilayers with 1° and 2° are depicted in Figure 29e. The completely flat minibands of the well-aligned sample at low density have been discussed above while the sample with the twist angle of 2° exhibits a noticeably curved exciton band with the same low level of concentration, although it shows the same density dependent performance. Also, the bands are more dispersed at a larger twist angle for each density and modification in the curvature is more pronounced when the density is changed, which demonstrates an enhanced interaction[29] and the interaction-induced delocalization will be easier. On the other hand, the modulation under the rotational angle is still related to the exciton density, because the twist angle highly influences the formation efficiency of the interlayer exciton,[21] which can therefore control the exciton density and accordingly affect the exciton-exciton interactions.

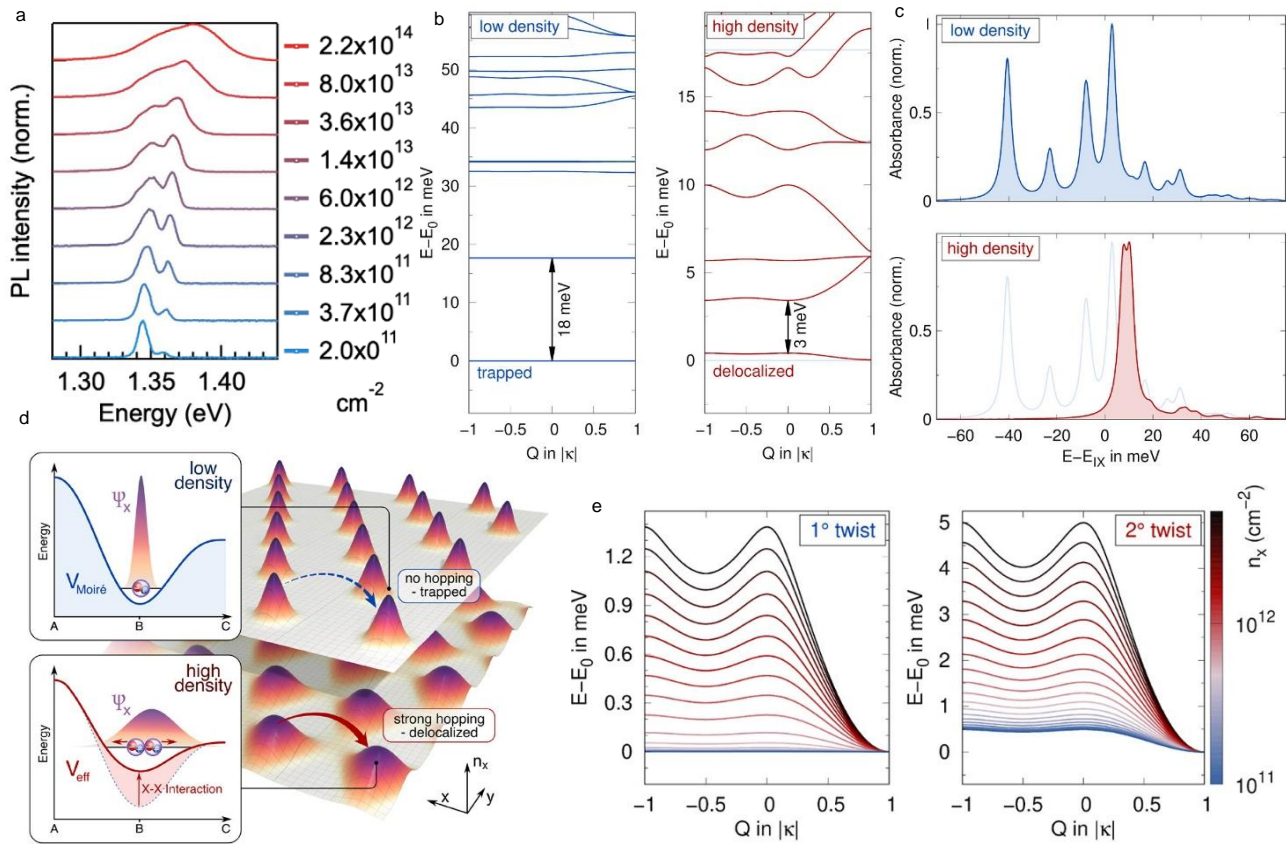

**Figure 29 | Many body interactions among moiré excitons.** (a) Density-dependent moiré exciton PL spectra from $MoSe_2/WSe_2$ heterobilayer with a twist angle of 2.6°. Reprint with permission from ref.[201] Copyright 2021 American Physical Society. (b) The miniband structures of interlayer excitons in hBN encapsulated $MoSe_2/WSe_2$ with 1° twist angle at cryogenic temperature (T = 4 K) with a low (left) and high (right) density. The widely separated flat bands of localized states at low density become continuous bands with smaller gaps reflecting delocalization at high density. (c) The corresponding absorbance spectra at both cases of low (top) and high (bottom) density. As a result of the interaction-induced delocalization under high density of interlayer exciton, the series of moiré resonances vanishes and a single dominant feature closely resembling the unperturbed exciton peak is found. (d) The moiré potentials trap excitons within their minima creating arrays of localized excitons. For large densities the interexcitonic repulsion gives rise to a decrease of the effective potential and a change of the exciton wave function with far-reaching consequences for optical properties and exciton transport. (e) Evolution of the lowest moiré miniband at different densities at T = 4 K and with twist angle of 1° (left) or 2° twist angle (right). Whereas at 1° the low density limit exhibits a completely flat band, the different confinement condition at 2° results in nonzero bandwidth and a stronger impact of interaction effects. (b-e) Adapted with permission from ref.[29] Copyright 2023 American Chemical Society.

### 5.3.3   Controlling carrier transport

Ultra-long transport of dipolar interlayer excitons has been reported not only because of their long-lived nature induced by spatially separated electrons and holes, but the strong and long-range dipolar interactions between them can also significantly modulate their motion and promote the IXs to travel farther away and sometimes across the whole sample.[31, 353-354] The techniques of the CCD image and APD system were employed to characterise the prorogation of interlayer excitons in terms of spatial and temporal distributions, respectively. Figure 30a-c show the spatial image of the focused laser spot and PL maps of IXs with different incident powers, indicating an expected prominent increase in the spatial profiles when the excitation density is increased due to the enhanced dipole-dipole repulsion.[354] Furthermore, when the excitation power is fixed, the spatial profiles of IXs still show an expansion after excitation as a function of time (Figure 30d-f). A linear increase of the exciton cloud in the PL image with time is expected if there is no spatial confinement and strong interaction effects.[354-355] The presence of dipole-dipole interactions can drive the excitons far away from the excitation spot and accelerate their propagation speed. However, when the expansion of IXs is close to the edge of the sample, which will be constricted by the finite size and the interlayer excitons will experience a strong repulsive interaction, thus resulting in nonlinearities of IXs in emission energy.[25] The moiré super-pattern induces the potential traps with different sizes and depths according to the lattice mismatch and angle misalignment, which can constrain interlayer excitons and impede their diffusion because of the reduced dipolar interaction and kinetic energy.[31, 199] The twist angle dependent modification on IX diffusion from the moiré cell is illustrated in Figure 30g-i with different fabricated heterobilayers of $WSe_2/MoSe_2$, which follows the rule of the interplay between dipolar interactions and moiré potential wells stated above. A longer diffusion length was observed from the CVD grown 0° heterobilayer because of the formation of a commensurate structure, and the moiré pattern does not exist in this structure. It means that the excitons are free and can move to the longest distance until limited by the physical lateral size. By contrast, no effective diffusion can be detected from the twisted sample created by mechanical exfoliation with a twist angle of around 1.1°, which was explained by the strong confinement resulting from the large moiré period (approximately 20 nm). The dipolar repulsive interaction was suppressed and excitons were trapped in the sufficiently deep potentials. When the rotational angle was further twisted to a large angle in another sample, the size of the moiré supercell decreased and the dipole-dipole interaction among IXs was activated, which thus enabled IX diffusion. This experimental observation of twist angle dependent moiré potentials is in good agreement with the theoretical analysis explained above.[352] However, the single layer spacer of hBN was reported can effectively weaken the moiré effect when it is inserted in between the TMD monolayers, attributed to the increased separation between electrons and holes, but the strong bound e-h pairs will be retained to host bright interlayer dipole transitions[356]. Therefore, an evident excitonic diffusion can be still observed from the sample with an hBN spacer even when the moiré pattern is present.[354] Figure 30j compared the diffusion lengths of excitons from both monolayers and heterostructures, indicating a much longer diffusion distance of interlayer excitons compared to the normal diffusion exhibited by intralayer excitons with randomly orientated in-plane dipoles. This anomalous diffusion is because of the contribution of strong dipolar interactions as discussed multiple times.[31] In addition, the heterobilayers with different stacking rotational angles convincingly demonstrated that the exciton diffusion can be highly modified and suggested the central role of moiré potentials. The migration speed of interlayer excitons at 60° is substantially faster than that at 0° due to the less and shallower diffusion barriers it has in the moiré pattern produced from 60° heterostructure.[31] The dipole-dipole repulsive interaction plays a dominant control role over the transport of interlayer excitons, but it can be simultaneously significantly re-manipulated by the moiré potentials.

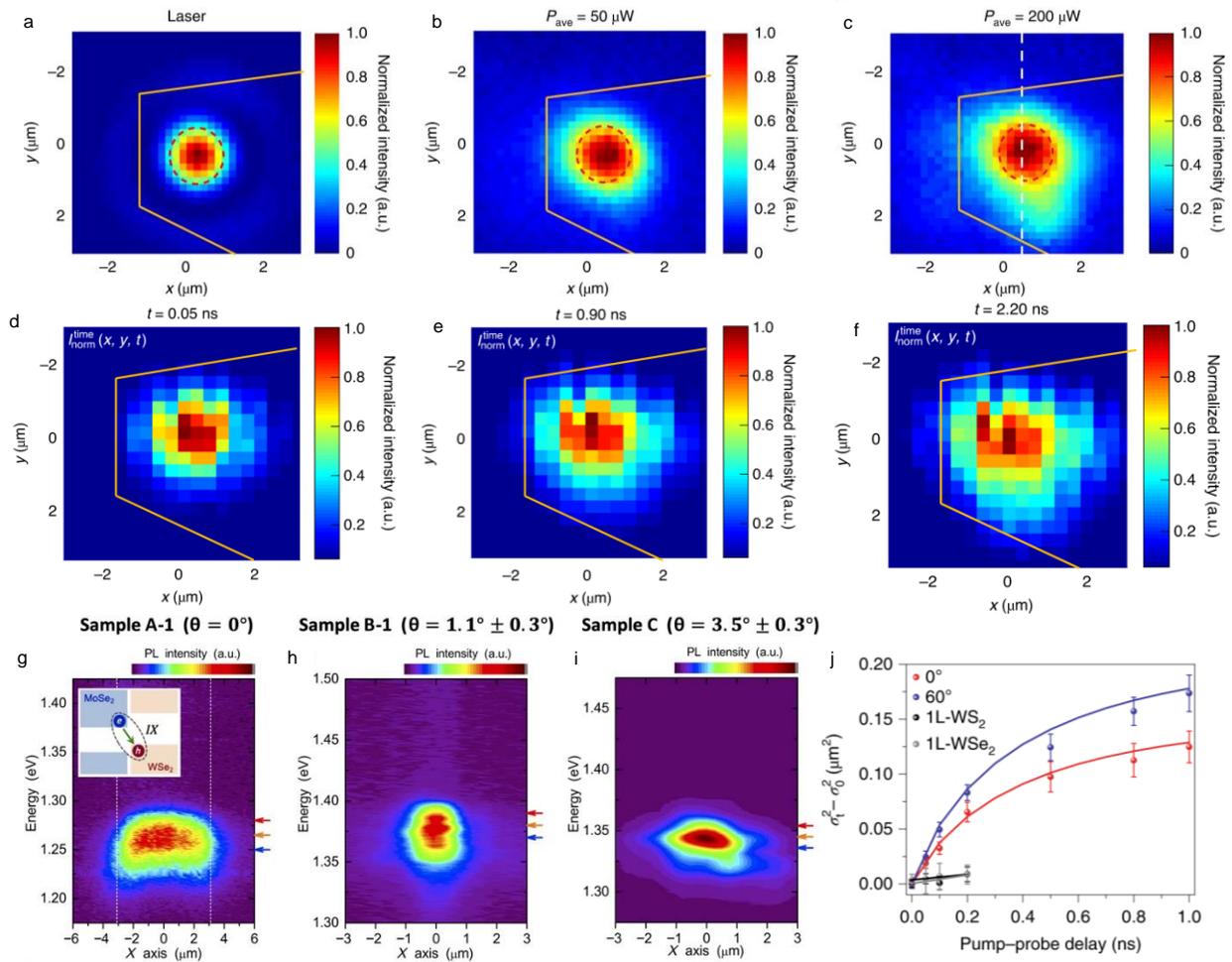

**Figure 30 | Engineering exciton diffusion through moiré potentials.** (a) CCD image of the excitation laser spot. (b, c) CCD images of the normalized IX PL intensity from a $WSe_2/hBN/MoSe_2$ heterostructure obtained under power of 50 μW (b) and 200 μW (c). (d-f) CCD images of the normalized IX PL intensity from the same $WSe_2/hBN/MoSe_2$ heterostructure acquired under the power of 200 μW at different times after the excitation, demonstrating the expansion of an exciton cloud driven by repulsive dipolar interaction. The yellow solid lines indicate the shape of the heterostructure. (a-f) Adapted with permission from ref.[354] Copyright 2022 Springer Nature. (g-i) 2D PL images of the IXs from a CVD-grown $WSe_2/MoSe_2$ sample A with zero twist (g), mechanical stacking $WSe_2/MoSe_2$ sample B with $1.1° \pm 0.3°$ twist angle (g) and mechanical stacking sample C with $3.5° \pm 0.3°$ twist angle (i). A much longer IX diffusion length in sample A-1 is clearly observed. (g-i) Adapted with permission from ref.[199] Copyright 2020 American Association for the Advancement of Science. (j) Twist-angle-dependent diffusion of IXs at room temperature, comparing with the diffusion measured from monolayers, indicating an anomalous diffusion of interlayer excitons due to the interplay between the moiré potentials and strong many-body interactions. Adapted with permission from ref.[31] Copyright 2020 Springer Nature.

### 5.3.4 Integration with microcavity

Given that the integration of intralayer exciton from TMD monolayer with microcavity not only gives rise to a giant enhancement in photoluminescence efficiency but also provides opportunities to observe exciton-polaritons, the coupling of the interlayer exciton to microcavity should induce some novel physical phenomena with more flexible controlling by generating strongly enhanced light-matter interactions. The moiré exciton provides a new degree of freedom, which may bring about more complicated interactions with tuneable resonate energy and oscillation strength. Based on the investigation of 2D TMD single layers, it is expected to observe two different regimes when the interlayer excitons are integrated into the resonant cavities. When IX is coupled with the cavity modes, a Purcell enhancement[357-360] will be generated, which is mainly reflected in the two characteristics of improved emission rate and shortened exciton decay and the system is considered entering a weak coupling regime. In contrast, the presence of exciton-polariton[361-362] is a signature of the regime with strong light-matter coupling, implying a significantly enhanced interaction between excitons and cavity photons. A variety of different cavities can be considered for integrating interlayer excitons, such as plasmonic nanocavity, photonic crystal cavity and Fabry–Perot cavity. A plasmonic gap cavity created with silver nanocube and gold mirror film (Figure 31a) was reported to integrate with $WS_2/MoS_2$ heterobilayer, demonstrating a giant enhancement in PL intensity (Figure 31b) because of the Purcell effect and increased absorption efficiency when the IXs interact with the cavity modes.[363] Due to the same effect, it shows a decrease in decay curves for interlayer excitons at both room and low temperatures, corresponding to a reduction of around 20% and 25%, respectively, compared with the off-cavity state (Figure 31c). The cavity also works as a tuneable source to modulate the photonic environment for exciton emissions, affecting the coupling strength of light-matter interactions. The photoluminescence from $MoSe_2/WSe_2$ heterobilayer in a designed cavity with translational degrees of freedom presents cavity-controlled dynamics based on the different cavity lengths (Figure 31d).[364] Figure 31e and f display the transmission and PL maps of the hetero-bilayer at a cryogenic temperature of around 4 K to monitor both absorption and scattering as well as the emission state within the cavity. The strongest absorption region coincides with the highest emission area at the coupling regime, and the detailed spectrum of the marked point in Figure 30f was extracted to illustrate the cavity-length-dependent IX performance. The lifetime of IX was shortened as a decreasing cavity length (Figure 31g), but it was accompanied by an enhancement in total PL emissions (Figure 31h), which indicates an induced Purcell effect inside the cavity and it can be tuned with an assigned cavity length. The strong cavity light-matter coupling can promote the formation of the new quasiparticle, which occurs at a superposition of exciton and photon and is named exciton-polariton. The formation of exciton-polariton and the effective engineering of cavity-matter coupling by varying the dielectric screening in vdW heterostructure were theoretically proven by a first principle study,[365] with a dielectric encapsulated material embedded in an optical cavity (Figure 31i). An experimental observation of exciton-polariton with moiré excitons was reported from a fabry-perot cavity (Figure 31j) integrated $MoSe_2$/$WS_2$ heterobilayer,[366] indicating that the moiré excitons strongly coupled to the light in the microcavity. It was observed from the anti-crossing behaviour of dispersive modes at moiré exciton resonances in angle resolved white-light reflectance spectra at 5 K (Figure 31k) and 70 K (Figure 31l). A more detailed study will be further explained in the moiré polariton section, and a series of other related devices by coupling the IX to the cavity (such as lasers) will also be detailed in the Applications.

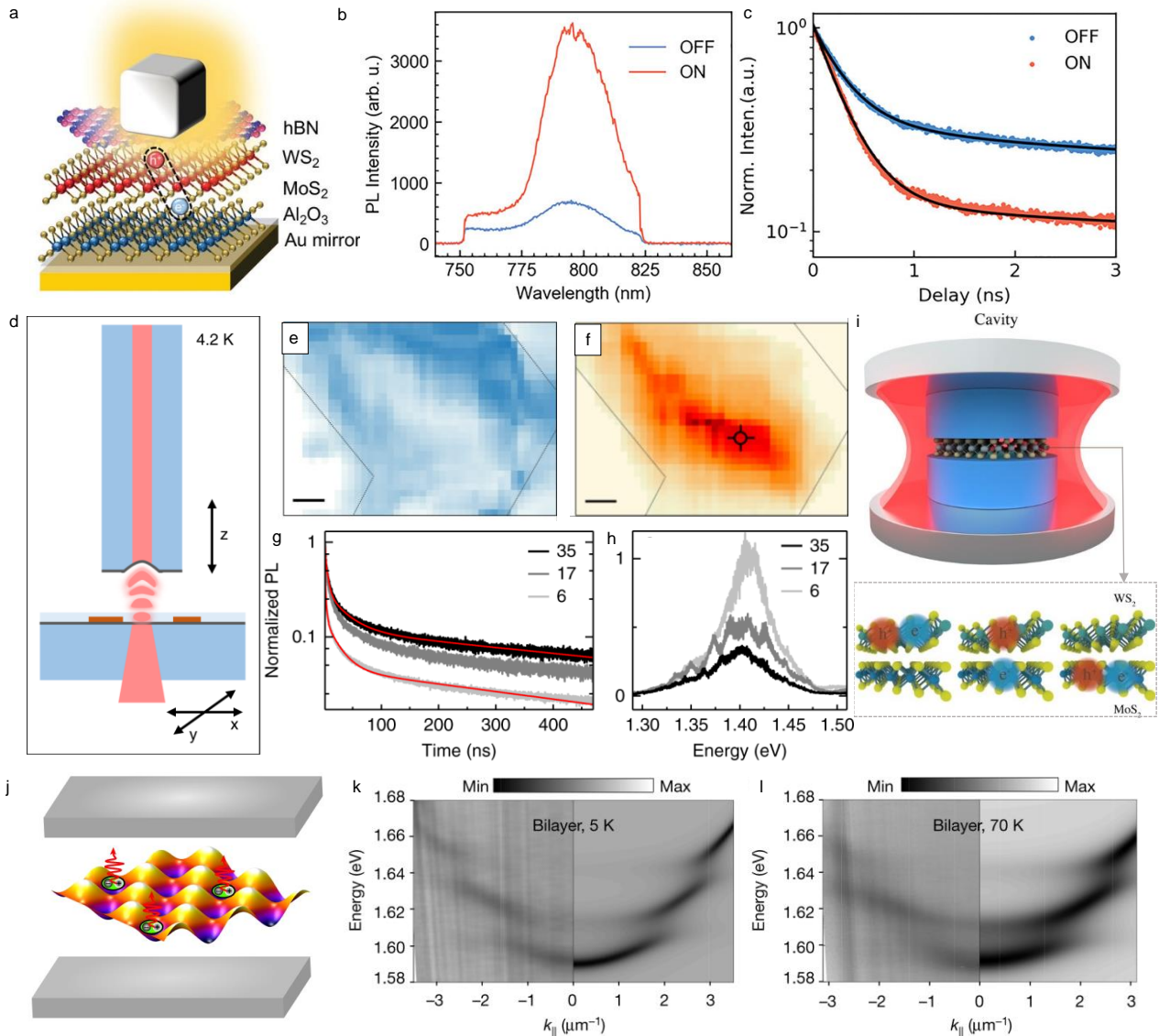

**Figure 31 | Integration of twisted heterobilayers with microcavity.** (a) Schematic illustration of integrated WS₂/MoS₂ heterostructure with a plasmonic gap cavity. The gap cavity consists of a 100 nm silver nanocube and an ultra-flat gold film. The gold film is coated with 3 nm thick Al₂O₃. A thin hBN flake is inserted as a space to tune the resonance wavelength to match the interlayer emission. (b) PL comparison between cavity coupling ON (red) and non-coupling OFF (blue) mode at 5 K. PL intensity increases about five times in the coupling region. (c) Time resolved PL of cavity coupled IX (red) and intrinsic IX (blue), showing the reduction in longed-lived interlayer exciton in both fast and slow decay channels. (a-c) Adapted with permission from ref.[363] Copyright 2021 John Wiley and Sons. (d) Cavity setup at 4.2 K to integrate a MoSe₂/WSe₂ heterostructure, the fiber-based micro-mirror forms the cavity together with a planar macro-mirror with CVD-grown heterobilayer on top. Independent translational degrees of freedom enable lateral sample displacement and cavity length detuning. (e) Transmission map recorded through the cavity with laser excitation at 635 nm (blue color corresponds to reduced transmission due to local variations in absorption and scattering). (f) Map of integrated PL intensity recorded simultaneously with the transmission map (dark red color represents the maximum intensity). The cross indicates the position on the flake used in the measurements illustrated in (g) and (h). The gray dashed lines indicate the boundaries of

the flake. (g) Measured interlayer exciton PL decay shown for three different cavity lengths of 35, 17, and 6 μm. The decay speeds up with the reduction of the cavity length. (h) PL spectra of interlayer exciton for the corresponding three selected cavity lengths. (d-h) Adapted with permission from ref.[364] Copyright 2019 Springer Nature. (i) Schematic representation of an encapsulated TMD heterobilayer in an optical resonator (cavity), inset shows the intralayer excitons in both monolayers and the formation of interlayer exciton. Adapted with permission from ref.[365] Copyright 2019 American Chemical Society. (j) Schematic of the moiré excitons confined in the superlattice and coupled with a planar cavity, forming a moiré polariton system. (k, l) Angle-resolved white-light reflection spectra, demonstrating strong coupling between moiré excitons and cavity photons at 5 K (k) and 70 K (l). (j-l) Adapted with permission from ref.[366] Copyright 2021 Springer Nature.

### 5.4 Moiré Phonons

Moiré Superlattices (MSL) have led to the emergence of multifarious phenomena, stirring the material science community to unexpected reality. After initially observed in graphene and its related heterostructures, these distinctive superlattice formations have gradually surfaced in few layer TMDs and their heterostructures where complicated molecular bonding and lattice arrangement play pivotal roles.[54] Such unique periodic patterns give rise to spectacular features and aspects that are reflected in numerous spectroscopy. For example, moiré excitons, band structures, topological transitions and moiré phonons appear to be strongly modulated by the periodic variation.[54, 367] Compared to atomically thin or monolayer TMDs, the moiré superlattices in multi-layer and bilayer heterostructures frequently exhibit diverse and distinct unit cells; hence the phonon properties witnessed in twisted structures have wide variation resulting in stark disparity in non-linear optical spectroscopy, particularly in Raman signatures.[367] Consequently, this gives rise to distinct moiré phonons stemming from these modulated moiré patterns. One of the earliest 2D system used to investigate such features includes twisted bilayer $MoS_2$ (tBLM) where remarkable variation was observed in phonon properties due to the stacking angle. Early studies in 2016 on tBLM delved into probing and analysing these features, leading to the detection of multiple vibrational signatures in Raman spectroscopy and are usually classified by low and high frequency regimes. For instance shearing and breathing modes appearing in low frequency Raman spectrum provided undeniable proof of robust interlayer interactions and presence of a moiré pattern.[368-369] Lin's team extensively explored moiré phonons in tBLM through comprehensive theoretical analysis and experimental observations.[368]

### 5.4.1   Low frequency Modes

Over the past few years, multiple other TMD systems have been scrutinized for moiré phonons through Raman spectroscopy, providing an extraordinary phonic response. This was largely facilitated by the significant development of high-density optical filters and attenuators, allowing even ultra-low frequency modes to be detected with ease, which are more responsive to the twist angle. Interlayer breathing and shearing modes offer several exquisite information, such as the layer number, stacking sequence, interlayer interaction and presence of moiré pattern.[54] Figure 32a shows the schematic illustration of twisted bilayer $WSe_2$ indicating how these ultra-low-frequency modes would be affected by the presence of a moiré pattern. Recent reports have also indicated that the interlayer breathing mode has a heightened sensitivity to the twist angle and can enable the determination of such periodic structures.[370] Furthermore, separate moiré phonon modes are expected to arise in twisted homo and hetero bilayers and offer unique sensitivity to the twist angle. Figure 32b-c represents the Raman spectra in the low-frequency range for twisted bilayer $WSe_2$.[371] The stacking angle was varied from $0^o$ (3R stacking) to $60^o$ (2H stacking) in the heterostructures and prominent peaks appeared below $50cm^{-1}$, which are expected to be shearing and breathing modes. In addition, moiré phonons also emerge in certain twisted samples but disappear when the twist angle is very small or large. The breathing modes manifest strong scattering with a stark change in position when the twist angle is modulated. Furthermore, the calculated moiré period matched coincidentally with the variation pattern of breathing mode with twist orientation as summarized in Figure 32b. Figure 32d plots the low-frequency Raman spectra of three distinct regions, α, β and γ identified by the same researchers in different areas of the bilayer twisted at 5, while Figure 32e represents a similar spectrum for two domains in the β region. As witnessed, there is an undeniable difference in the Raman spectrum from the three different regions while the spectra from two domains in β region are rather similar. In addition to the breathing modes, different moiré phonons are prominent in β and γ regions, especially the folded transverse and longitudinal acoustic ones. It was found missing in α region due to weak interlayer interaction.[371] Therefore low-wavenumber Raman provides an inevitable way to probe the moiré phonons and eventually MSL. Similar analysis has been extended to vraious 2D material systems. For instance, Quan *et al.*[121] reported on the low-frequency Raman spectra of twisted $MoS_2$ bilayer as shown in Figure 32f-i. In this study, the researchers correlated strain effects with the emergence of moiré phonons, alongside the formation of periodic moiré patterns. Additionally, within the low-frequency range, the Raman spectra captured interlayer shearing and breathing modes, defined by the relative parallel and perpendicular motion of the individual monolayers respectively,[372-373] as shown in Figure 32f. Furthermore, the researchers analysed the Raman spectra's evolution by monitoring the peak positions and linewidths relative to the twist angle's variations. Based on the three distinct regimes, rigid, transition and relaxed, the linewidth and peak position of the modes revealed systematic but drastic variation with the stacking angle, indicating substantial sensitivity to the moiré pattern. Figure 32j also represents low frequency Raman spectra of twisted $MoS_2$ heterolayers taken by Lin *et al.*,[368] using a 2.54 ev excitation laser but the twist angle was varied with a bigger step size. Similar layer breathing and shearing mode were observed with strong vibrational intensity at around 60°.

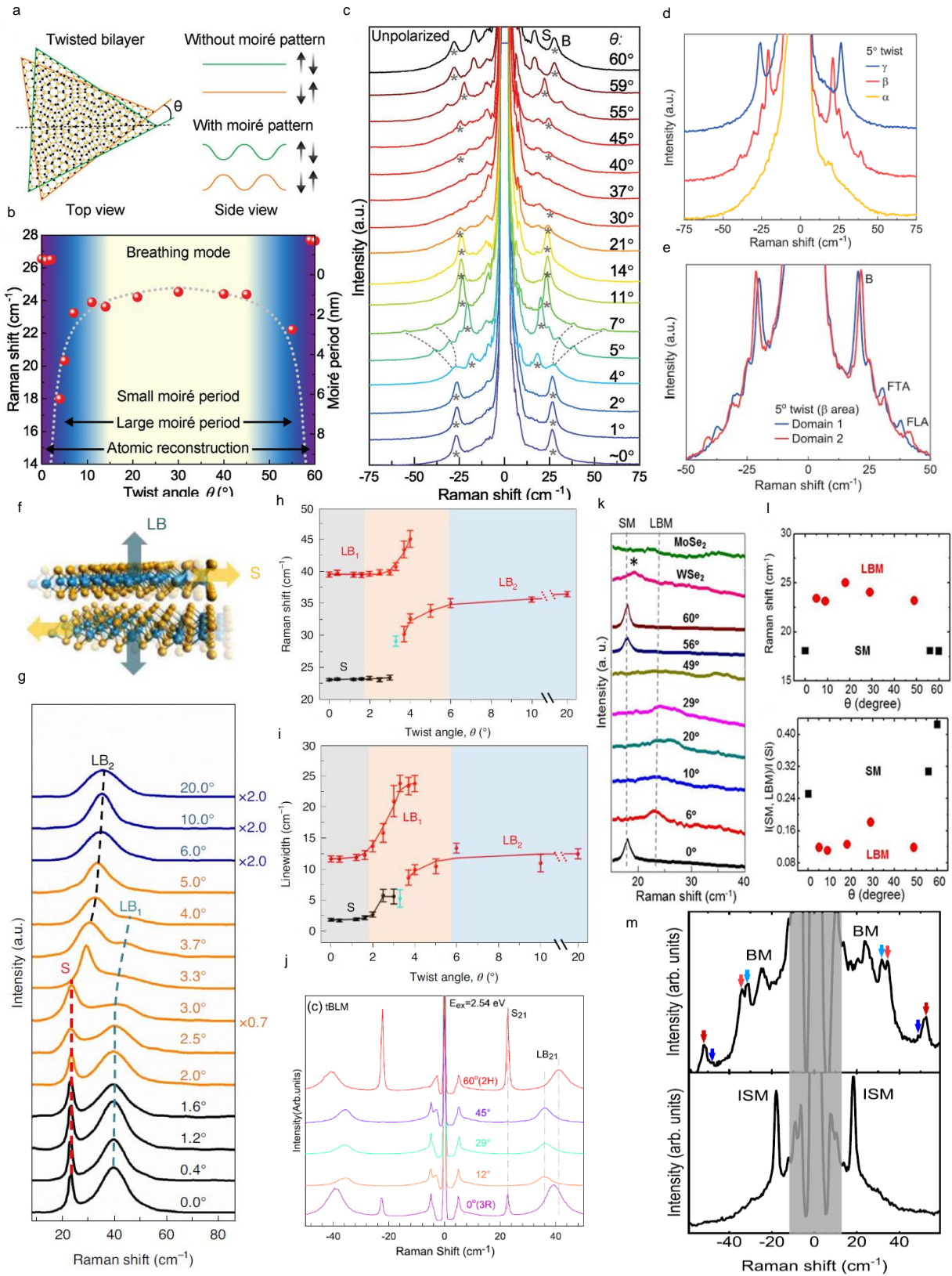

**Figure 32 | Moiré induced Shearing and Breathing Modes.** (a) Schematic illustration of WSe2 twisted bilayers separated by an angle, θ. Periodic patterns are formed due to moiré supperlattice effecting the low energy interlayer modes. (b) Low frequency Raman spectrum of bilayer WSe₂ as a function of stacking angle. Asterisk indicate the layer breathing (LB)

modes, appearing to be highlighted sensitive to the stacking angle. For comparison, shearing (S) and breathing mode in 2H natural bilayer (60) is also shown. (c) Evolution of breathing mode frequency with twist angle, distinguishing three regimes of interest: atomic reconstruction, small and large moiré period. Peak position remains nearly constant where period is small. (d) Typical low frequency raman spectra taken from different position from the same twisted $WSe_2$ homobilayers. (e) Representative Raman spectra from different domains within the same area, showing both moiré phonons and interlayer breathing mode. (a-e) Adapted with permission from ref[371]. Copyright 2021 John Wiley and Sons. (f) Molecular schematic showing the origin of layer breathing and shearing mode. (g) Low frequency Raman spectrum in twisted $MoS_2$ bilayers showing the two interlayer LB and one S mode. (h, i) Tend of central frequency and linewidth with twist angle respectively for the LB and modes. (f-i) Reprinted with permission from ref.[121] Copyright 2021 Springer Nature. (j) The Raman spectra of 3R- and 2H bilayer $MoS_2$ and other twisted bilayers, in the ultralow-frequency region. Excitation energy used is 2.54 eV. The vertical dashed lines are guides to eyes. Reprinted with permission from ref.[368] Copyright 2018 American Chemical Society. (k, l) Low frequency Raman spectrum in $MoSe_2$/$WSe_2$ twisted heterobilayer, showing the LB and S modes. Figures reprinted with permission from ref.[21] Copyright 2017 American Chemical Society. (m) Stokes and antistokes low frequency Raman spectrum in $MoSe_2$/$WSe_2$ twisted heterostructure, showing the LB and S modes. Reprinted with permission from ref.[374] Copyright 2021 IOP Publishing.

The intriguing phenomenon of moiré phonons has extended its influence to diverse material systems encompassing various TMD heterostructures. Nayak *et al.*[21] investigated the impact of coherently stacked angles within the $MoSe_2$/$WSe_2$ bilayer system, as summarized in Figure 32k-l. High symmetry stacking configuration in near zero and 60º twisted heterobilayers led to shear modes at low frequency, whereas mismatched atomic misalignment in intermediate twisted samples gave rise to layer breathing modes.[21, 375] As expected the low wavenumber modes depicted extraordinary sensitivity to twist angle and also provides optical evidence for the existence of moiré patterns. Figure 32m provides room temperature stoke and anti-stoke low frequency Raman spectra of $MoSe_2$/$WSe_2$ heterobilayers probed by Phillepe *et al.*.[376] Moiré phonons marked by red arrows are found to have low intensity; interlayer shearing and breathing modes in contrast have much stronger signatures indicating the occurrence of atomic reconstruction.

### 5.4.2 High frequency Modes

The emergence of moiré patterns also affects the more common or high-frequency vibration modes that can be easily probed by a regular Raman spectrometer. This generally includes the in-plane $E^I_{2g}$ mode and out-of plane $A_{2g}$ mode, represented schematically in Figure 33a for a typical $MoS_2$ sample. $E^I_{2g}$ mode is generated by the opposite motions of two sulphur atoms in relation to the Mo atom within the 2D plane, while the $A_{1g}$ mode is a result of the out-of-plane relative vibrations of the sulphur atoms.[370] Figure 33b shows the high frequency Raman spectrum of $MoS_2$ homobilayers twisted at different angles with small step sizes. Notably, the two prominent intralayer modes in the high frequency range exhibited a significant sensitivity to the twist angle, particularly the in-plane mode. The $E^I_{2g}$ mode, which arises from the in-plane vibration of sulphur atoms, portrayed strong evidence of atomic reconstruction by evolving into a doublet between 21° and 61° twist.[121] This occurrence is attributed to the redistribution of local strain induced by the moiré pattern, resulting in a transitional state that distorts the hexagonal unit cell and disrupts its original three-fold rotational symmetry, thus causing mode splitting.[377-378] The researchers further utilized tip-based piezo mapping to validate the existence of the MSL structure. Figure 33c shows similar backtracking of the phonons' peak position. High frequency $E_{2g}$ mode clearly reveals three reconstruction regimes as highlighted by coloured boxes in Figure 33c. In the relaxed region, strain is nearly imperceptible except for the domain walls, which comprise of small fraction of the total sample area. In addition, strain is negligible in the rigid regime. Therefore, these regimes had a negligible effect on the far field Raman measurement. In contrast, the local strain arising from the atomic reconstruction leads to a doublet of the in-plane mode to $E^+_{2g}$ and $E^-_{2g}$ in the transition regime. Figure 33d shows a similar high-frequency Raman spectrum for twisted bilayer $WSe_2$ over a wide range highlighting all the peaks including the moiré phonons and their shift with twist angle. Peaks in high-frequency regimes undergo similar modulation with the stacking angle.[371] The cleanest and most pristine interface for analysing moiré phonons and patterns is achieved through dry exfoliation and vDW technique; however, advanced CVD technique and finely articulated transfer methods can also be utilized to achieve such nanoengineering marvels at the atomic level. Epitaxial growth process allowed the preparation of large-scale $MoS_2$ homobilayers and tri-layer structures.[367] Liao $et$ $al.$[59] recorded the the captivating dependence of the out-of-plane mode ($A_{1g}$) on the stacking angle in samples grown by CVD as shown in Figure 33e-g. The angle-dependent behavior of the $A_{1g}$ peak was influenced by both interlayer vdW interactions and long-range Coulombic interactions. The strongest interlayer coupling was found to occur at near-zero twist angles, as indicated by the softening of the $A_{1g}$ phonon with increasing twist angle. Moiré phonons were also manifested in these CVD-grown samples, originating from the off-centre phonons of the monolayer aligned with lattice vectors of the moiré reciprocal space.[368] The intensity of the Raman peaks exhibited a sequential change with the twist angle, attributed to dominant electron-phonon coupling.[59] However, in the case of twisted $MoS_2$ bilayers grown by CVD, Liu $et$ $al.$[43] observed slightly different behaviour. While no significant intensity variation in the Raman modes was captured except for the $A_{1g}$ mode, they detected interesting blue- and red-shifts for the $E^1_{2g}$ and $A_{1g}$ modes, respectively.[43] This twist dependence of the Raman modes provided a platform to investigate the strength of interlayer mechanical coupling. The separation between the $E^1_{2g}$ and $A_{1g}$ peak was utlized to characterize the effective interlayer coupling strength, where greater separation meant stronger coupling.

High frequency Raman spectrum from $MoS_2$ homobilayer is shown in Figure 33h and j for several twist angles. They were investigated by Lin $et$ $al.$ in CVD-grown samples.[368] The folded transverse acoustic (FTA) and longitudinal acoustic (FLA) phonons displayed progressive but quadratic shifts in the peak positions when the stacking angle was varied. Interestingly, these modes exhibited mirror symmetric behaviour when the twist angle was elevated beyond 30° (Figure 33i and j respectively). Furthermore, the twist-dependant FLA mode positions could be found between double segments of the longitudinal acoustic (LA) phonon dispersion projected toward the G–M and G–K directions in atomically thin MoS$_2$. A similar distribution was observed for the FLA mode positions along the G–M and G–K branches due to the sensitivity of the basic vectors of the moiré reciprocal lattices to the angle of interlayer orientation (Figure 33i). In the case of tBLM, six moiré phonons persisted in every phonon branch due to the interplay between the basic vectors of the moiré reciprocal lattice and the stacking angle. These phonons could degenerate into a single folded[379] mode akin to the FLA and FTA modes, regardless of interlayer interaction.[368] Figure 33j and k illustrate the presence of moiré phonons in tBLM analogous to other phonon branches,[380] in addition to the LA phonon branch. Notably, the characteristic of non-folded $E^1_{2g}$ and $A_{1g}$ modes manifested as strong peaks at 385 and 405 cm$^{-1}$; furthermore, the second-order Raman modes of A1(M)–TA(M) were detected in tBLM. The Raman active E``(Γ) mode, absent in the monolayer due to the backscattering configuration, was also observed. This mode is associated with the degenerate LO and TO phonons at the zone centre.[367-368] Effect of stacking angle on higher-order Raman modes has also been carried out in twisted heterobilayers of sulphides and selenides. Figure 33l summarizes the emergence of in-plane and out-of-plane modes of MoS$_2$ and WS$_2$ as a function of twist in the vdW heterostructure.[375] $E^1_{2g}$ experienced a softening of up to 2 cm$^{-1}$ for WS$_2$ and 0.6 cm$^{-1}$ for MoS$_2$ in the heterostructure when the twist angle was approximately 0°, 60° or 120°. In contrast, the $A_{1g}$ only underwent modulation for the 30° and 90° twisted samples, softening by 1.5 cm$^{-1}$ and stiffening by 0.7 cm$^{-1}$ respectively. Researchers claimed that the carrier doping and charge transfer was the main reason behind the drastic effect; these also affected the band alignment that directly triggers the dynamics of A$_{1g}$ mode.[381] Furthermore, the researchers argued that the space-related repulsion between sulphur atoms led to the peak separation of the in-plane mode which was utilized to determine the interlayer mechanical coupling strength. Rahman *et al*.[382] have also demonstrated intriguing moiré-based phonon dynamics in twisted heterobilayers of $WS_2/WSe_2$, fabricated mechanically from bulk crystals. Stacking angle revealed the extraordinary movement of phonon peaks registered by the lattice and heterostrain arising from MSL. In addition, twist angle also influenced the first-order linear coefficients and FWHM of high frequency Raman modes. Furthermore, the researchers also concluded that critical twist angle can lead to a large enhancement in Raman vibrational intensity which has been demonstrated for the first time in TMD heterostructures.[382] Moiré phonons exhibit a high degree of complexity, challenging to attribute specific peaks to phonons in bilayer TMDs.  . This complexity is further compounded in the case of folded modes, which are highly influenced by moiré superlattices. In addition, investigating moiré phonons becomes even more intricate in thin-layered TMDs (with more than two layers) and their heterostructures due to the intricate nature of their superlattice unit cells and computational models. However, a few experimental studies have emerged in this area, contributing to a deeper understanding of moiré phonons. In all the cases, Raman spectra remain an indispensable tool for identifying changes and modes associated with the moiré superlattice.

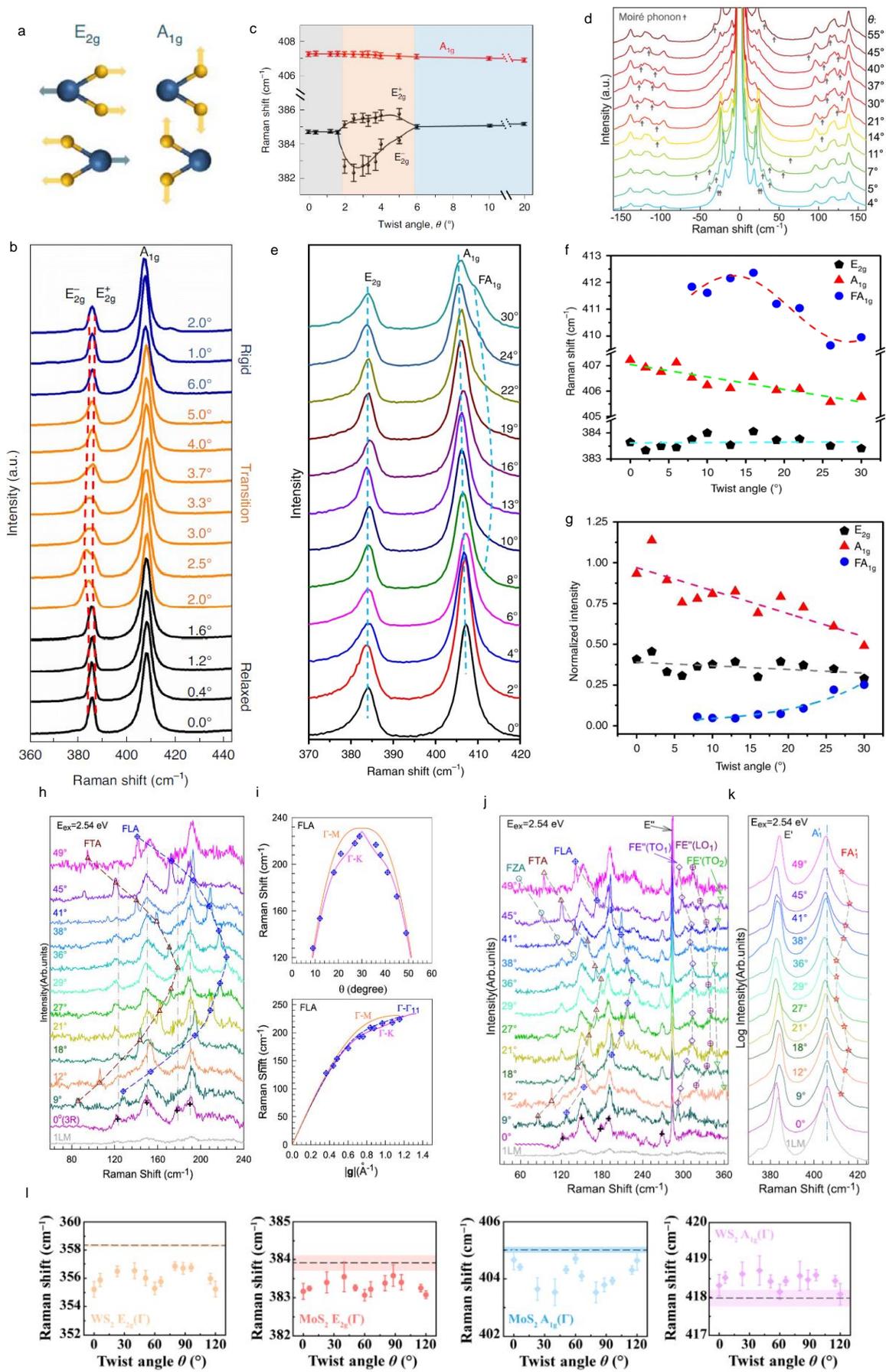

**Figure 33 | Twist Correlated High frequency Raman signatures**. (a) Schematic depiction of high frequency intralayer $E_{2g}$ mode and $A_{1g}$ mode. $E_{2g}$ mode corresponds to in-plane relative motion between molybdenum and sulphur atoms; and the $A_{1g}$ mode corresponds to out-of-plane sulphur atom vibrations. (b) Measured Raman spectrum of various $MoS_2$ twisted bilayers, showing the two $E^+_{2g}$ ($E^-_{2g}$) mode and $A_{1g}$ mode. (c) Trend of central frequencies of intralayer modes as a function of stacking angle. $A_{1g}$ mode appears to be insensitive to the twist angle while the in-plane mode splits into two peaks with increasing twist angle. (a-c) Reprinted with permission from ref.[121] Copyright 2021 Springer Nature. (d) Raman spectra of twisted bilayer $WSe_2$ spanning over a broad frequency range. Arrows indicate the moiré phonon peaks. Reproduced with permission from ref.[371] Copyright 2021 John Wiley and Sons. (e) High frequency Raman spectra of multiple twisted $MoS_2$ homobilayers with controlled twist angle. Each spectrum is calibrated and normalized by the position and intensity of silicon peak 520.7cm$^{-1}$. (f, g) Intensity and Peak position evolution of $E_{2g}$, $A_{1g}$ and $FA_{1g}$ mode as function of twist angle. Dashed lines are linear ($E_{2g}$, $A_{2g}$) ad sinusoidal ($FA_{1g}$) fitting. (e-g) Reproduced with permission from ref.[59] Copyright 2020 Springer Nature. (h, i) Raman spectrum of twisted bilayer $MoS_2$ in region of 50-365 cm$^{-1}$ and 370-425 cm$^{-1}$ . Various shapes and colour symbols represent the different phonon branches and their evolution with twist. For comparison, Raman spectra of 1L $MoS_2$ and 3R bilayer $MoS_2$ (0 degree) are also shown. (j, k) The Raman spectra of 3R- and 2H bilayer $MoS_2$s and other twisted samples in the ultralow-frequency region, using $E_{ex} = 2.54$ eV. The vertical dashed lines are guides to eyes. (h-k) Adapted with permission from ref.[368] Copyright 2018 American Chemical Society. (l) Summary of $E_{2g}$ and $A_{1g}$ peak of $WS_2$ and $MoS_2$ in twisted $WS_2/MoS_2$ heterostructures. Twist appears to be a sensitive factor in the phonon peak position. Adapted with permission from ref.[375] Copyright 2021 Springer nature.

### 5.5 Moiré polariton

The realm of polaritons has witnessed significant exploration in recent years, encompassing a diverse range of phenomena and materials. While surface plasmon polaritons, arising from electron oscillations in metals, have been extensively studied, there is a rich diversity of other polaritons yet to be discovered. These include polaritons originating from atomic vibrations in polar insulators, excitons in semiconductors and spin resonances in ferromagnetic or antiferromagnetic materials.[383-386] Spanning a vast electromagnetic spectrum from microwave to ultraviolet frequencies, this broad array of polaritons presents intriguing possibilities for foundational investigations and potential applications in photonics and optoelectronics. Notably, the wealth of polaritonic modes is further enriched by the abundant variety of 2D materials and the inherent anisotropy of the materials themselves. Tunable graphene polaritons,[387] chiral polaritons,[388] anisotropic and hyperbolic polaritons,[389] long-lived hyperbolic phonon polaritons with fascinating characteristics such as slow light and exciton-polaritons with remarkable binding energies and anisotropic properties are among the notable examples that highlight the diverse flavours of polaritons available for exploration. In the realm of nanophotonics and optoelectronics, polaritons have emerged as hybrid excitations that allow for accurate manipulation of light on the nanoscale. Within this context, moiré superlattices have garnered significant interest as a platform for polariton engineering and the advancement of photonic quantum technologies. These artificial structures, formed by layering vdW materials with slightly misaligned crystal lattices, give rise to a novel class of polaritons known as moiré polaritons.[383] The intricate interplay between the twisted atomic arrangements within moiré superlattices results in the emergence of unique electronic and optical properties. Exploring the characteristics and behaviour of moiré polaritons provides a promising avenue for gaining new insights into light-matter interactions and unlocks unprecedented possibilities for the design and control of polaritonic devices at the nanoscale.

### 5.5.1   Exciton polaritons

Exciton-polaritons are a specific type of polariton formed through the strong interaction between photons and excitons in semiconductors. These unique quasiparticles inherit the strong interparticle interaction from their excitonic component, which is facilitated by Coulomb interaction. However, they also acquire the characteristics of photons, such as a significantly smaller mass (approximately 105 times lighter than free electrons) and long propagation distances. Furthermore, exciton-polaritons retain properties of the host material's excitons, including spin and valley polarization.[390] These intriguing properties can be probed through the photons that escape from the cavities, as the photons carry all the necessary information due to conservation laws. Moiré exciton-polaritons are generated by integrating moiré superlattices with resonantly hybridized moiré excitons with a microcavity. They offer a promising avenue for exploring polaritons with quantum nonlinearity and cooperative effects. These polaritons possess appreciable oscillator strengths and built-in electric dipoles, making them particularly intriguing. The initial study[366] on moiré exciton polaritons highlighted the potential of moiré superlattices as a pathway to achieve significant nonlinearities in polariton systems. Through the integration of $MoSe_2$-$WS_2$ heterolayers into a microcavity as illustrated in Figure 31j, the authors successfully established collaborative connection between moiré-lattice excitons and microcavity photons at temperatures achievable with liquid helium (Figure 34a,b). The resulting moiré polaritons exhibited strong nonlinearity, which was attributed to exciton blockade, diminished energy alteration, and reduced excitation-induced dephasing, indicating the quantum confined characteristics of moiré excitons. This moiré polariton setup merges robust nonlinearity with precise control of matter excitations through cavity designing and long-range light coherence. Consequently, it provides an ideal platform for exploring collective phenomena in customizable arrays of quantum emitters. In another recent investigation,[391] the interaction between cavity photons and intralayer moiré excitons is explored within an angularly displaced AA-stacked $MoSe_2$/$WSe_2$ heterostructure integrated into a Fabry–Perot cavity as illustrated in Figure 34c. The results emphasize the considerable influence of the twist angle on various aspects including the coupling between exciton and light, energy of polaritons and the number of polariton branches (Figure 34d, e). Notably, a twist angle of 3° induces remarkable exciton hopping between moiré supercells, resulting in a nearly parabolic energy dispersion and an amplified Rabi splitting effect. Furthermore, the energies of moiré excitons exhibit a clear dependence on the twist angle (Figure 34f), with a 14 meV shift observed over a range of angles from 0.5° to 3°. The study uncovers a spectrum of bright excitons that rely on the twist angle, with their number diminishing as the angle increases. The research[392] investigated the interaction between excitons residing between layers and photons within a cavity in a TMD hetero bilayer. The study demonstrated the formation of moiré polaritons with distinct optical and electric dipole properties. These moiré polaritons displayed phenomena related to topological transport, such as the spin/valley Hall and polarization Hall effects, which could be electrically controlled through interlayer bias. The moiré pattern generated unique exciton dispersions and pseudospin splitting, enabling bulk topological transport with larger energy scales. The intensive coupling between IX and the cavity photon (Figure 34g, h) was crucial for moiré polariton formation, and the optical dipole of IX could be enhanced through interlayer distance reduction. The findings have implications for potential applications in TMD heterobilayers and cavity photon systems. The tunability of moiré excitons within heterostructures having type I band alignment or pronounced interlayer and intralayer exciton intermingling can be transcribed to moiré polaritons. A significant enhancement in the interaction strength of dipolar polaritons was observed by Emre *et al*.[393] by modifying the gate voltage, thereby increasing the size of the dipoles. To achieve this, they utilize coupled quantum well structures (Figure 34i, j) within a microcavity, where coherent electron tunnelling generates excitonic dipoles. The modifications in interaction strength are quantified by measuring changes in reflected light intensity when driving polaritons with a resonant laser. The observed 6.5-fold increase in the ratio of interaction strength to linewidth suggests the potential of dipolar polaritons to enable the polariton blockade effect. A similar study[394] demonstrates the possibility of amplifying local interactions and non-linear effects by manipulating the dipole moment of exciton-polaritons within semiconductor microcavities with broad quantum wells under the influence of electrical bias. Through the application of an electrical field, we achieve a twofold increase in exciton-exciton interactions (Figure 34k) and more effective transitioning to polariton states with lower energy. This unique feature allows for the tuning of coupling strength between excitons and photons, exciton and photon detuning, and the nonlinearity of the polariton modes through interlayer hybridization. The integration of moiré superlattice into cavity has introduced a novel category of polariton system, offering opportunities for controlling electronic phases and exploring topological transport phenomena through cavity interactions.

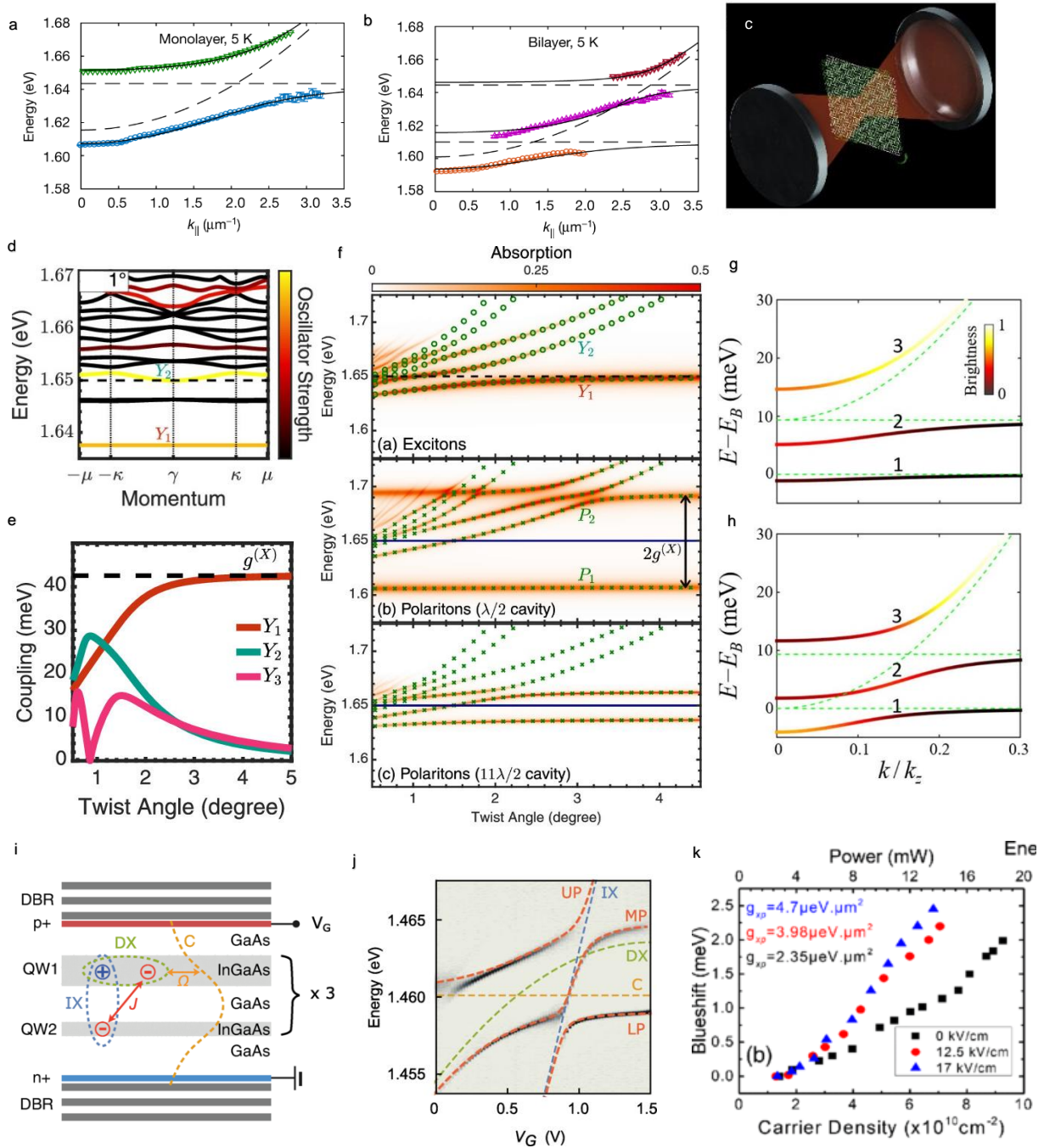

**Figure 34 | Generation and control of exciton Polariton in moiré superlattices (MSL).** (a, b) Comparison of polariton energies versus in-plane wavenumber obtained from the angle-resolved white-light reflection spectra for monolayer excitons and cavity photons (a) and ML excitons and cavity photons (b). (a,b) Reprint with permission from ref.[366] Copyright 2021 Springer Nature. (c) Illustration of the Fabry-Perot cavity with a hetero-bilayer positioned at the center. (d) Moiré exciton minibands at a 1° twist angle, depicting the intralayer exciton. The oscillator strength is color-coded, with the black dashed line representing the MoSe₂-based

exciton energy without moiré effects. (e) Variation of coupling strength for the three lowest energy moiré excitons with respect to the twist angle. The black dashed line represents the radiative coupling of the intralayer exciton without moiré effects. (f) Absorption and twist-angle dependence of intralayer moiré excitons without a cavity. The exciton energy is indicated by green circles. Additionally, twist-angle dependence and absorption of moiré polaritons (represented by green crosses) in the presence of a cavity. (c-f) Reprint with permission from ref.[391] Copyright 2022 American Chemical Society. (g, h) The AB-polariton dispersions for two different IX-cavity energy detuning. The coupling strengths are $g_A = 3$ meV and $g_B = 3$ meV. The green dashed curves represent the cavity mode and $A_\sigma$/ $B_\sigma$ IX dispersions without coupling. The color scale indicates the brightness (photon fraction) of the polariton. (g,h) Reprint with permission from ref.[392] Copyright 2020 Elsevier. (i) Illustration of the sample structure comprising three pairs of $In_{0.04}Ga_{0.96}As$ coupled quantum wells (QWs) positioned at three distinct antinodes within a $5\lambda/2$ cavity formed between GaAs/AlAs distributed Bragg reflectors (DBRs). (j) Reflection spectrum under zero in-plane momentum excitation as a function of applied gate voltage VG. The energy variations of the three coupled states—upper polariton (UP), middle polariton (MP), and lower polariton (LP)—with VG are depicted as red dashed lines. (i,j) Reprint with permission from ref.[393] Copyright 2018 American Physical Society. (k) Polariton energy blueshift at three distinct electric field values. Reprint with permission from ref.[394] Copyright 2018 American Physical Society.

### 5.5.2   Phonon polaritons

Phonon polaritons, captivating quasi-particles resulting from the coupling of phonons to electromagnetic radiation, have attracted substantial interest in condensed matter physics and nanophotonics due to their extraordinary properties and potential for driving advanced technological advancements. The integration of phonon polaritons with moiré patterns unveils a vast landscape of opportunities for both fundamental scientific exploration and practical applications. This interplay unlocks a rich tapestry of effects, allowing precise manipulation and tailoring of the dispersion properties of these quasi-particles. By harnessing the moiré - induced effects, researchers gain unprecedented control over essential characteristics such as wavelength, propagation direction, and group velocity of phonon polaritons, enabling the engineering of light and sound wave behavior at the nanoscale. This integration holds great promise across various scientific pursuits and practical applications, offering insights into the interaction between light and lattice vibrations, as well as advancements in nanoscale light propagation, ultramicroscopy, coherent thermal emission, and the development of advanced nanophotonic devices with enhanced functionalities and control over light-matter interactions.

The study[395] focused on manipulating the photonic dispersion in vdW bilayer samples of α-phase molybdenum trioxide (α-MoO$_3$) by controlling the twist angle as illustrated in Figure 35a. The research demonstrated the ability to achieve tunable topological transitions (Figure 35b) and flat polariton bands, facilitating low-loss and diffraction less polariton propagation with high resolution. The optical images and the near field distribution measurements illustrated in Figure 35d-g shows the hyperbolic to elliptic transition based on the twist orientation. These transitions, occurring at the photonic magic twist angle, were governed by polariton hybridization and a topological parameter. The flattening of the bilayer dispersion at these transitions enabled efficient, low-loss tunable polariton channeling, and exhibited exceptional resolution in diffraction-free propagation. Similarly, another research[396] presents a technique devoid of lithography for modifying the hyperbolic nature of dual layers of vdW α-MoO$_3$ crystals via twist-dependent interactions among phonon polaritons. (Figure 35h-k). By adjusting the twist angle between the two layers, the polariton isofrequency contours can be modified, offering control over the polaritonic characteristics. This twist-induced tailoring of hyperbolic phonon polaritons[396-397] provides opportunities for controlling nanoscale light flow in stacked vdW crystals. Moreover, it demonstrates that the engineered hyperbolic phonon polaritons in the double-layer system retain the advantages of effective light trapping and minimal propagation loss.

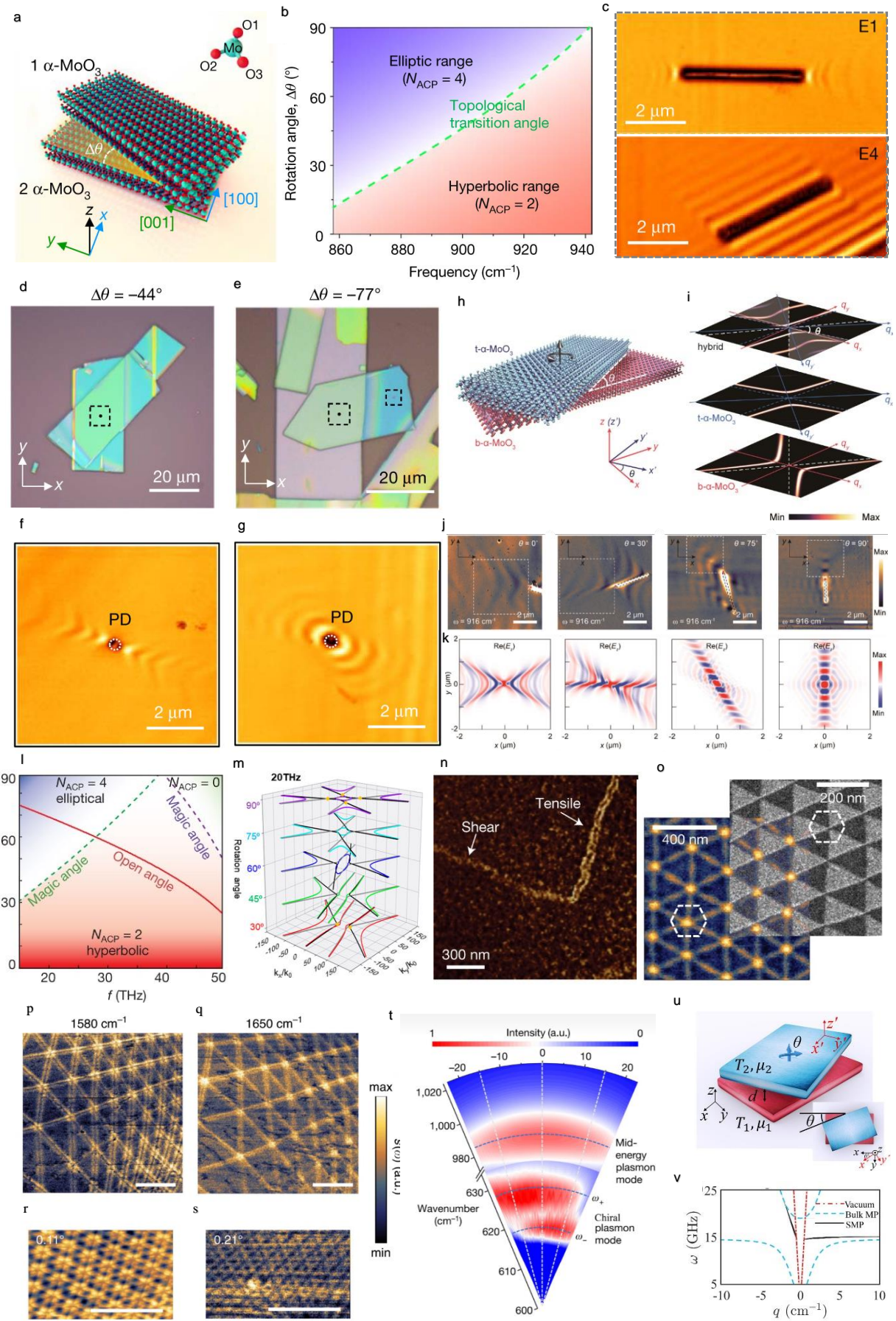

**Figure 35 | Moiré Phonon polaritons and other types of polaritons in moiré superlattices.**
(a) Schematic illustration of s-SNOM nanoimaging of PhPs. When exposed to infrared light, the metallic tip initiates the generation of polaritons (PhPs) within the sample. These polaritons subsequently disperse into the surrounding free space and can be collected for analysis. (b) Topological mapping of PhP dispersion concerning twist angle and frequency. The figure displays the relationship between the twist angle ($\Delta\theta$) and frequency, highlighting the distinct dispersion regimes. The dashed green line represents the topological transition angle, which serves as the boundary between the hyperbolic dispersion regime (red gradient) and the elliptical dispersion regime (blue gradient). (c) Near-field images (s3) showcasing the sample. The electric field E1 is aligned parallel to the [100] crystal axis of the bottom layer, while the electric field E4 is aligned parallel to the [001] crystal axis of the top layer. (d, e) Optical images of twisted bilayer α-MoO3 samples with twist angles $\Delta\theta$ = -44° and $\Delta\theta$ = -77°. The dashed squares indicate the specific regions measured in (f), (g), while the black dots represent the point defect. (f, g) Near-field distribution (s3) around a point defect (PD) indicated by white dashed circles, as measured experimentally. (a-g) Reprint with permission from ref.[395] Copyright 2020 Springer Nature. (h) Schematic representation of the twisted double-layer α-MoO3 laminates. (i) Calculated isofrequency contours illustrating the phonon polaritons in the α-MoO3 laminates. Bottom: Isolated bottom layer; Middle: Isolated top layer; Top: Twisted double-layer. The dashed lines denote the asymptotes of the in-plane hyperbolic dispersions in each isolated α-MoO3 layer. (j) Experimental verification of phonon polaritons in twisted double-layer α-MoO3 using real-space near-field optical images. (k) Calculated real parts of the z-components of the electric field distributions, Re (Ez). (h-k) Reprint with permission from ref.[396] Copyright 2020 American Chemical Society. (l) Theoretical behavior of plasmon polaritons in twisted bilayer metasurfaces as a function of rotation angle. (m) Dispersion curves of moiré hyperbolic metasurface (HMTS) as a function of the twist angle at 20. The solid lines in red, green, blue, cyan, and purple represent the dispersion of moiré HMTSs with rotation angles of 30°, 45°, 60°, 75°, and 90°, respectively. The gray (black) solid lines in each plot depict the dispersion of the first (second) individual HMTSs, while the yellow dots indicate the anticrossing points of dispersion between the two individual HMTSs. (l,m) Reprint with permission from ref.[398] Copyright 2020 American Chemical Society. (n) Near-field optical image showcasing a sharply bent L-shape domain wall. On one segment, a single-bright-line feature is observed, while on the other segment rotated by 90°, a double-bright-line feature is present, indicating distinct infrared responses. Reprint with permission from ref.[399] Copyright 2016 Springer Nature. (o) (Left) Nano-light photonic crystal formed by the soliton lattice visualized, where the enhanced local optical conductivity at solitons results in contrast. (Right) Dark-field TEM image of a t-BLG sample revealing contrast between AB and BA triangular domains. The dashed hexagons represent the unit cells of the crystals. The intensity is measured in arbitrary units (a.u.). Reprint with permission from ref.[283] Copyright 2018 American Association for the Advancement of Science. (p, q) Near-field images of the normalized amplitude showcasing moiré patterns in TBG captured at the same position but with varying excitation frequencies. The twisted angle between the layers is approximately 0.05°. (r, s) Near-field images of the normalized amplitude s(ω) illustrating different twisted angles of 0.11° and 0.21° at an excitation frequency of ω = 1560 cm$^{-1}$. Scale bars represent 500 nm. (p-s) Reprint with permission from ref.[282] Copyright 2020 Springer Nature. (t) Extinction mapping of t-BLG spiral nanoribbons, showing the dependence of chiral berry plasmons on wavenumber and polarization detection angle. The nanoribbons are excited by a linearly polarized broad-band light source, with the incident polarization direction perpendicular to the long ribbons of the spiral array (0°). The blue and white dashed lines indicate the intensity of various plasmon peaks and extinction spectra observed at different polarization detection angles. Reprint with permission from ref.[400] Copyright 2022 Springer Nature. (u) Schematic illustration of the experimental setup showcasing radiative heat transfer between two ferromagnetic insulators (FMIs) separated by a vacuum gap of distance d. The bottom and top slabs are characterized by temperatures T1 and T2, respectively. The y (y') axis corresponds to the direction of the saturation magnetization in the bottom (top) FMI. A magnetic field is applied along the direction of the respective saturation magnetization in each slab. θ is the twist angle. (v) The dispersion relation of a nonreciprocal surface magnon-polariton (SMP) with a single interface between vacuum and FMI. (u,v) Reprint with permission from ref.[401] Copyright 2021 American Chemical Society.

To address the need for a non-destructive diagnostic tool to visualize nanometre-scale properties of moiré superlattices in vdW heterobilayer devices, a scanning probe microscopy technique utilizing infrared light was developed.[282] This technique enabled direct imaging of moiré patterns in t-BLG encapsulated by hBN (Figure 35p-s). By mapping the scattering dynamics involving phonon polaritons initiated within the capping layers, the researchers successfully visualized the buried moiré superstructures, providing valuable insights into the induced electronic and optical phenomena. Additionally, the origin of double-line features observed in the images was investigated, revealing the underlying mechanism of effective phase change in the reflectance of phonon polaritons at domain walls. This non-destructive approach offers a versatile and in situ method for studying moiré engineered heterostructures, facilitating the observation of changes in local conductivity and advancing our understanding of these complex systems.

### 5.5.3  Other polaritons

In addition to exciton and phonon polaritons, significant research efforts are directed toward the exploration of other types of polaritons, including plasmon polaritons and magnon polaritons. Plasmon polaritons arise from the collective oscillations of electrons in metals, while magnon polaritons emerge from the coupling between photons and spin waves in magnetic materials. The study[398] focuses on the exploration of moiré hyperbolic plasmons in pairs of hyperbolic metasurfaces (HMTSs) and their potential applications in metasurface optics. By rotating evanescently coupled HMTSs, the researchers revealed a diverse range of dispersion engineering, observed topological transitions at specific magic angles (Figure 35l), noted the presence of broadband field canalization, and identified plasmon spin-Hall effects. They demonstrate that specific rotation angles induce parallel dispersion curves as illustrated in Figure 35m, facilitating wide-range field concentration with minimal damping. The chemical potential of graphene is shown to offer tunability in the band dispersion of moiré stacks. Additionally, the study highlights the possibility of substantial strong spin-orbit interactions and plasmonic spin-Hall effects, controlled by the rotation angle. The research suggests the versatility of different HMTS configurations and examines the adjustability of the response by varying the coupling distance. In the presence of any displacement or misalignment between the two layers, a soliton-like domain wall (Figure 35n) emerges near the transition region where two distinct lowest-energy stacking orders coexist but remain spatially separated.[399] This phenomenon has led to the discovery of quantum valley Hall edge states that are safeguarded by topological properties in the field of electronics.

Furthermore, it has been observed that graphene plasmons experience strong reflection when encountering these domain walls. Plasmon polaritons in t-BLG have been investigated[283] as a means of controlling nano-light. Infrared nano-imaging revealed that small twist angles induce atomic reconstruction in TBG, resulting in a natural plasmon photonic crystal (Figure 35o) that can propagate nano-light. This finding offers a promising avenue for manipulating nano-light by harnessing the quantum characteristics of graphene and other vdW materials, without the requirement for complex nanofabrication. The nano-light photonic crystal exhibits unique features rooted in topological electronic phenomena taking place at solitons, setting it apart from conventional photonic crystals. Its periodicity and band structure can be continuously adjusted through electrostatic and/or nanomechanical means, providing versatility for device engineering. The investigation of plasmon modes in highly ordered macroscopic tBLG structures reveals intriguing findings.[400] Chiral plasmons, arising from the uncompensated

Berry flux induced by optical pumping (Figure 35t), exhibit a dependence on pumping intensity, electron fillings, and distinctive resonance splitting. Furthermore, a slow plasmonic mode is observed, originating from interband transitions within relaxed AB-stacked domains. This mode holds significant potential for facilitating strong interactions between light and matter within the mid-wave infrared spectral range. These discoveries shed light on the distinctive electromagnetic properties exhibited by small-angle tBLG, establishing it as a remarkable platform for advancements in the field of quantum optics. The investigation on near-field radiative heat transfer between two ferromagnetic insulator (FMI) slabs as illustrated in Figure 35u, under twisted magnetic fields reveals significant findings.[401] Notably, large damping conditions in FMIs result in a notable near-field thermal switch ratio due to twisting, whereas small damping conditions result in a nonmonotonic control of heat transfer. These effects are primarily influenced by nonreciprocal surface magnon-polaritons. Moreover, when ultrasmall damping conditions are present, the near-field radiative heat transfer experiences a substantial enhancement through twist-nonresonant surface magnon-polaritons (Figure 35v). Importantly, these twist-induced effects extend beyond FMIs and can be applied to anisotropic slabs that exhibit a disruption of time-reversal symmetry, presenting an opportunity for effective thermal regulation at the nanoscale.

## 5.6 Moiré magnons

### 5.6.1   Generation of magnons

The generation of magnons, which are quanta of magnonic energy representing collective oscillations of electron spins in magnetic materials, can be accomplished by means of various techniques. One method includes periodic modulation of the magnetic system.[402-403] This modulation can be accomplished through patterned structures[404-405] or modulated magnetic properties.[403] Another method involves controlling and manipulating spin waves using external stimuli, such as spin texture,[406] electric fields,[407-408] and magnetic fields. These stimuli allow for active control of magnons, including their excitation, propagation, and detection, and enable tuning and modulation of magnonic properties. Additionally, specific phenomena and effects in magnonic systems, such as magnon Bose-Einstein condensation,[409] spin pumping,[410-411] spin-orbit coupling,[412-413] spin Hall effect,[414] and Dzyaloshinskii-Moriya interaction (DMI),[415-416] can be utilized to generate magnons and control their behaviour. A new and universal approach is presented in ref[417], aiming to enable the Bose-Einstein condensation (BEC) of quasiparticles. By utilizing magnons as the model system for Bose particles, the key aspect lies in introducing a disequilibrium between magnons and the phonon bath (Figure 36a). Through the process of heating the system and subsequently rapidly reducing the temperature of the phonons, a surplus of incoherent magnons is generated. This leads to a spectral redistribution of the magnons, triggering the Bose-Einstein condensation. Experimental verification showcases the successful creation of a magnon Bose-Einstein condensate in an individual magnetic nanostructure, achieved by applying low-power current pulses instead of high-power microwave pumping. In the realm of spintronics, STNOs[414] (Spin-Torque Nano-Oscillators) with a point contact on an extended free layer exhibit the emission and propagation of spin-torque-driven spin waves outward from the nanocontact (Figure 36b,c). Experimental observations have directly confirmed the emission of spin waves from STNOs with nanoscale electrical contacts, which can propagate over several micrometres. This emission of propagating spin waves forms the basis for synchronizing arrays of STNOs, enabling their application in microwave technologies. Consequently, STNOs hold great promise as wave emitters for future magnonic devices. In the field of topological magnonics, layered structures, heterostructures, and moiré structures hold promise for generating magnons. Topological effects in condensed matter exhibit robustness against perturbations and offer the potential for advanced information processing with reduced dissipation. Experimental studies in this area have focused on two main categories: the detection of the magnon Hall effect and the observation of topological magnon bands. In insulating collinear ferromagnets, such as $Lu_2V_2O_7$ (Figure 36d) and pyrochlore compounds, transverse magnon currents have been observed by driving a temperature gradient.[418] The magnon Hall effect[418-419] has been detected in these materials through thermal Hall effect measurements or direct temperature difference observations. The Dzyaloshinskii-Moriya interaction plays a pivotal role in creating Berry curvature within magnon bands, resulting in the lateral deviation of the magnon current. In materials characterized by strong spin-orbit coupling, like specific antiferromagnetic substances displaying non-collinear arrangements, experiments involving thermal Hall effects and inelastic neutron scattering have identified magnon bands with distinct topological properties in nearly 2D structures such as Kagome or honeycomb materials. The work discussed in ref.[420] explores the analysis of band topology within crystalline substances, centering on the revelation of topological spin excitations (referred to as magnons) within a three-dimensional antiferromagnetic material. Through the application of inelastic neutron scattering (depicted in Figure 36e,f), the scientists visualize two magnon bands intersecting at points known as Dirac points, which remain protected by the symmetry of spin rotation. This finding represents a new addition to experimentally confirmed topological band structures. The study highlights the potential of neutron scattering in exploring band topology in magnon systems and emphasizes the distinct observables provided by these experiments.

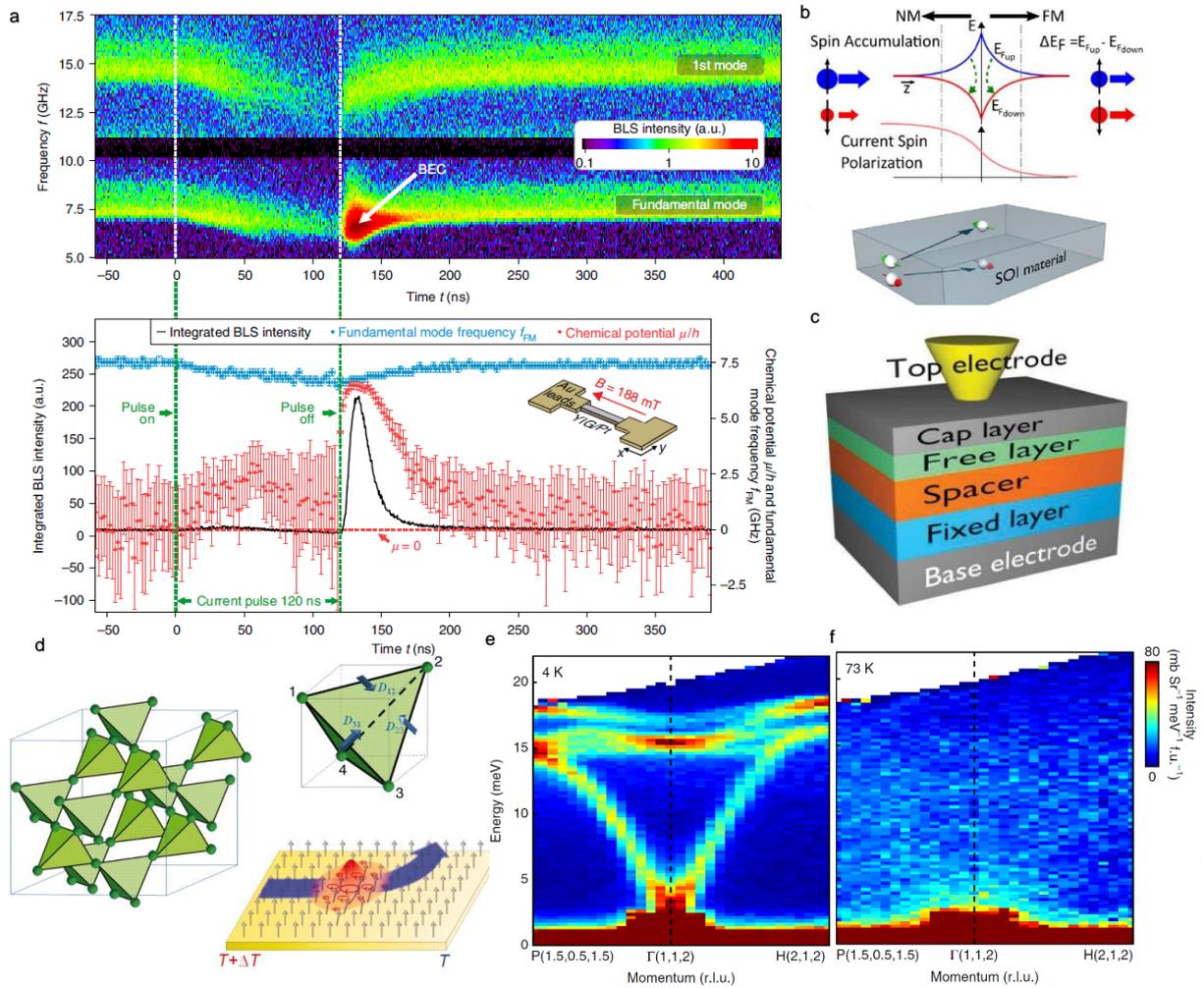

**Figure 36 | Generation methods of magnons.** (a) top panel: Brillouin light scattering (BLS) spectroscopy measurements for B∥x, revealing the density of magnons (colour-coded, log scale). The pulse start and end are indicated by vertical dashed lines. Upon pulse termination, a prominent magnon signal emerges at the lower edge of the magnon band. Bottom panel: Normalized BLS intensity plotted over time (black line), along with the frequency of the fundamental mode (blue dots) and measured chemical potential (red dots) integrated from 4.95 GHz to 8.1 GHz.) Reprint with permission from ref.[417] Copyright 2020 Springer Nature. (b) Top Panel: Schematic illustrating the fundamental mechanism behind the emergence of spin accumulation zones in both the Ferromagnetic (FM) and Nonmagnetic (NM) layers resulting from the transport of a charge current through an FM/NM bilayer. Bottom Panel: Schematic depiction of the Spin Hall Effect (SHE). Initially, an unpolarized charge current propagates through a material with a significant Spin-Orbit Interaction (SOI), thereby generating a transverse spin current. (c) Schematic representation of common device architectures for STNOs emphasizing the configuration of magnetic and nonmagnetic layers, as well as the electrode arrangements of a point contact. (b,c) Reprint with permission from ref.[414] Copyright 2016 IEEE. (d) The crystal structure of $Lu_2V_2O_7$ and the magnon Hall effect. Reprint with permission from ref.[418] Copyright 2010 American Association for the Advancement of Science. (e, f) Inelastic neutron scattering data showcasing representative measurements obtained using an incident neutron energy Ei = 28 meV, captured within the ordered and paramagnetic states. The data presented here correspond to the vicinity of a magnetic ordering wavevector Q = (1,

### 5.6.2 Moiré magnons in magnetic 2D materials and heterostructures

Moiré magnons originating from the interplay of spin waves in twisted vdW magnets, have emerged as a fascinating area of research with promising implications for spin-based devices and applications. The study presented in[421] focuses on moiré magnons in twisted bilayer magnets exhibiting collinear order. The authors reveal that the twist angle between the layers significantly impacts the magnon dispersion (Figure 37a), leading to the appearance of new types of magnon bands arising from the moiré superlattice (Figure 37b-e). Studies emphasize that the valley moiré bands display striking planarity across an extensive scope of twisting angles. However, the topological Chern numbers of the bottommost flat bands undergo notable fluctuations in response to the twist angle. Furthermore, the lowermost topological flat bands within bilayer antiferromagnets facilitate intricate transverse thermal spin conveyance. In ref[422], researchers investigate the observation of spin-wave moiré edge and cavity modes in twisted magnetic lattices (Figure 37f,g). By precisely tuning the twist angle between the layers, they demonstrate the formation of moiré patterns and the emergence of localized spin-wave modes at the lattice edges as illustrated through the results in Figure 37h and i. This discovery provides valuable insights into the interaction of magnons with twisted structures, paving the way for novel spin-wave-based applications. The study discussed in[423] investigates the formation of a magic-angle magnonic nanocavity in a magnetic moiré superlattice. By engineering the twist angle, researchers observe the confinement and manipulation of magnons within the nanocavity (Figure 37j-m). This discovery holds great potential for designing compact and efficient magnonic devices, as well as for exploring fundamental aspects of magnon propagation in moiré superlattices. Another intriguing aspect of moiré magnons is the creation and characteristics of magnons associated with stacking domain walls in twisted vdW magnets.

The study by the researchers in[424] explores this phenomenon and demonstrates the emergence of spin-wave moiré patterns (Figure 37n-q). These patterns confine and localize magnons at the domain walls, giving rise to distinct magnon modes with unique dispersion relations. Such localized magnons present exciting opportunities for controlling and manipulating spin waves in vdW heterostructure.

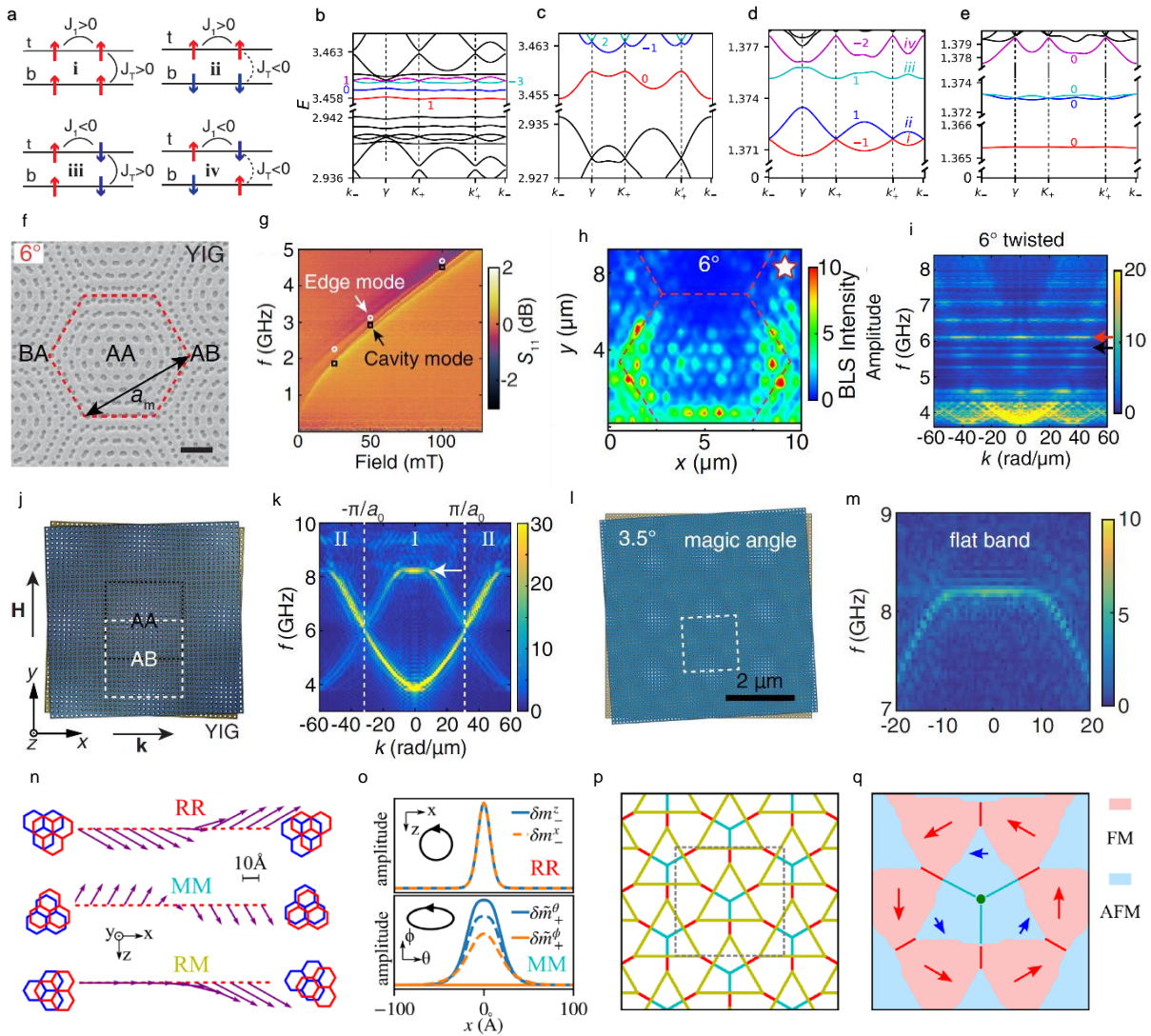

**Figure 37 | Moiré magnons in magnetic 2D materials and heterostructures.** (a) schematic Illustration of four distinct cases of twisted bilayer magnets, categorized based on the signs of J1 and JT. (b-e) The moiré bands of valley magnons corresponding to the four cases illustrated in (a). The band Chern numbers are indicated on the top of each band. It should be noted that the band Chern number may not be well-defined in cases where two bands intersect. In panels (b) and (c), the DMI value is set to 0.05, while in panels (d) and (e), it is set to 0.35. The twist angle is fixed at θ = 1.0°. (a-e) Reprint with permission from ref.[421] Copyright 2020 American Physical Society. (f) SEM image of a moiré magnonic lattice based on YIG grown on a GGG substrate with a twist angle of 6°. The red dashed line indicates a moiré unit cell with a

commensurate AA region at its centre and incommensurate AB (BA) region at its edge. Moiré lattice constant $a_m$ is marked by the black arrow. Scale bar, 2 μm. (g) Spin wave reflection spectra S11 obtained using all-electrical spin-wave spectroscopy. The field-dependent frequencies of the edge mode and cavity mode, detected by micro-focused Brillouin light scattering (μBLS), are represented by white circles and black open squares, respectively. (h) 2D intensity maps of spin waves measured μBLS at twist angles of 6°. The excitation frequency is 3.10 GHz, and the applied magnetic field is 50 mT. Red dashed lines serve as visual guides for moiré unit cells. The white star indicates the profile of the edge mode near the optimal twist angle, often referred to as the "magic angle". (i) Simulation-derived magnonic band structure of the moiré magnonic lattice at twist angle of 6°. The red arrow highlights the presence of the edge mode at a frequency of 6.10 GHz. (f-i) Reprint with permission from ref.[422] Copyright 2023 American Physical Society. (j) Schematic representation of a magnetic moiré superlattice based on yttrium iron garnet (YIG) with a twist angle of $\theta = 3.5°$. The white dashed square indicates the moiré unit cell with AB stacking region at its center, while the black dotted square represents the AA stacking region. A magnetic field of 50 mT is applied along the y direction. (k) Spin wave dispersion obtained from micro-magnetic simulations, considering an interlayer exchange coupling of A12 = 11 μJ/m². The white dashed lines correspond to the boundaries of the Brillouin zone (BZ), defining the first BZ ("I") and second BZ ("II"). (l) Schematic illustrations of twisted bilayer magnonic crystals with a twist angle of 3.5° and a fixed interlayer exchange coupling of 6 μJ/m². The white dashed square highlights a moiré unit cell with dimensions of 1.6 μm × 1.64 μm. (m) Spin wave dispersion simulated near the flat-band region at the center of the moiré unit cell for magnetic moiré superlattices with a twist angle of 3.5°. (j-m) Reprint with permission from ref.[423] Copyright 2022 American Physical Society. (n) Top view illustrating three types of stacking domain walls, indicated by purple arrows, with the stacking configurations of the left and right domains depicted on the corresponding sides. (o) Bounded magnons observed within the RR (top) and MM (bottom) domain walls. In the bottom panel, solid lines represent kz = 0.0 1/Å, while dashed lines indicate kz = 0.05 1/Å. Insets depict the motion of the magnons. (p) Sketch of the magnon network in twisted bilayer CrI₃ with a twist angle of 0.1°. (q) Enlarged view of the region highlighted by the gray rectangle in (p), illustrating the stacking and magnetic domain patterns. The red and blue arrows represent stacking vectors for rhombohedral and monoclinic stackings, respectively. Red and cyan lines represent the RR and MM stacking domain walls, respectively. (n-q) Reprint with permission from ref.[424] Copyright 2020 American Physical Society.

## 6. Device integration and applications

In recent years, the integration of twisted vdW materials into practical devices has shown promising prospects due to the unique properties arising from moiré superlattices. These materials have expanded the horizons of quantum science, sensor technology, and photonics, leading to the conception of new classes of devices with unprecedented functionalities.

## 6.1 Quantum science and information processing

Twisted vdW materials have opened up avenues for advancing a host of quantum technologies, including quantum communication, simulation and sensing. The wide variety of physical and mechanical properties they displayed, the atomically thin nature, and the ease of integration empower the development of 2D material-based quantum devices with unprecedented characteristics. In this part, we provide an overview of the possibilities that 2D materials have brought to the quantum information community. In particular, we highlight the development and state-of-the-art quantum emitters hosted by 2D materials. We outline the recent research on quantum simulation using 2D materials. We discuss the challenges ahead and the technical hurdles that need to be overcome to advance the associated science and technology in truly practical regimes.

### 6.1.1   Generation of quantum light sources

Non-classical light sources constitute the cornerstone for a wide range of leading quantum technologies. In particular, the single-photon sources play a critical role in discrete-variable quantum information processing where information is encoded on variables with discrete eigenvalues.[425] Despite tremendous progress in the advancement of high-performance single-photon sources, the generation of truly on-demand single photons remains onerous. An optimal on-demand single photon emitter (SPE) emits precisely one photon at one time into a specific given spatiotemporal mode with maximal efficiency, and all emitted photons are indistinguishable. This means, when two such photons are combined on a beamsplitter, they yield complete interference. This definition of the ideal SPE pinpoints the metrological parameters commonly used to measure the performance of single photon sources: 1) purity that quantifies the multiphoton component of the emitted state relative to the single photon component, independently of losses. It is often characterized by the second-order intensity correlation function $g^{(2)}(0)$, which is measured by a Hanbury Brown and Twiss experiment. 2) Indistinguishability that refers to the degree of modal similarity among photons, including spectral, temporal and polarization characteristics. The indistinguishability can be measured through Hong-Ou-Mandel (HOM) interference[426] where two light wavepackets are combined on a beamsplitter and detected at each output. Ideal single photons exhibit anti-bunching, causing two photons to be either both transmitted or both reflected, leading to full destructive interference. This effect is observed in well-defined single spatial and spectral-temporal modes (e.g., Fourier-transform-limited Gaussian modes). The correlation measurement reveals the interference fringe of the two wavepackets, yielding the HOM visibility $V_{HOM}$, with $V_{HOM} = 1$ indicating perfect indistinguishability. The visibility $V_{HOM}$ can be expressed in terms of the two-detector coincidence counts $C_\parallel$ and $C_\perp$ in the presence and absence of quantum interference, respectively. $C_\perp$ can be evaluated by introducing asymmetries between the incident photon wavepackets, such as delaying or changing the polarization of one photon. Therefore, $V_{HOM}$ is given by the equation of $(C_\perp - C_\parallel)/C_\perp$. Indistinguishability is necessary for achieving high-fidelity photon-photon interactions, a primary component for photonic quantum computing. 3) Brightness that quantifies the efficiency of the single photon source. A plethora of definitions of photon source brightness are in use among various communities. However, a platform-agnostic definition involves the probability of having a single photon present in a single-mode fiber during a specific clock cycle, irrespective of whether a heralding detection is received or not.[427]

Single photon sources can be categorized into two groups: first, isolated single quantum systems capable of emitting just one photon at a time. Examples include atomic systems that marked the initial generation of single photons,[428] and solid-state SPEs based on atom-like emitters,[429] such as fluorescent atomic defects and quantum dots. The second category comprises heralded single photon sources, which rely on nonlinear processes such as spontaneous parametric down-conversion (SPDC) and spontaneous four-wave-mixing (FWM). Each approach has its distinct advantages and drawbacks. While SPDC holds the promise of high purity and indistinguishability, the advantages are offset by its low source brightness due to its inherent probabilistic nature. Various multiplexing schemes, including spatial, temporal, and hybrid multiplexing, have been developed to enhance the source brightness.[427] The state-of-the-art SPDC single photon source relies on temporal multiplexing facilitated by a programmable hybrid delay line comprised of a short fiber, a Pockels Cell switch and a single-pass ring cavity[430] (Figure 38a). The results demonstrated photon indistinguishability of 91%, photon brightness of 0.05 (10k counts/s) with $g^{(2)}(0)=0.007$, or brightness of 0.67 (130k counts/s) at the expense of a reduced $g^{(2)}(0)=0.27$. In contrast, solid-state SPEs promise deterministic sources and significantly higher photon generation rates. However, emitter inhomogeneity and additional charge noise considerably degrade photon indistinguishability. In addition, photon extraction faces limited efficiencies due to the refractive index contrast between the host material and the collection fiber (or waveguide). The foremost performance of current solid-state SPEs relies on an InAs/GaAs quantum dot.[431] The authors demonstrated the production of single photons with 95.4% indistinguishability, a brightness of 0.34 (equivalent to a generation rate of $1.6\times10^7$ counts/s) and $g^{(2)}(0)=0.025$. The single photons were exploited in a Boson sampling experiment involving 20 pure single photons in a $60 \times 60$ interferometer (Figure 38b). To this end, persistent efforts have been made to advance existing technologies on one hand, and to explore new paradigms on the other. Recently, a variety of 2D layered materials have showcased their capability to host SPEs.

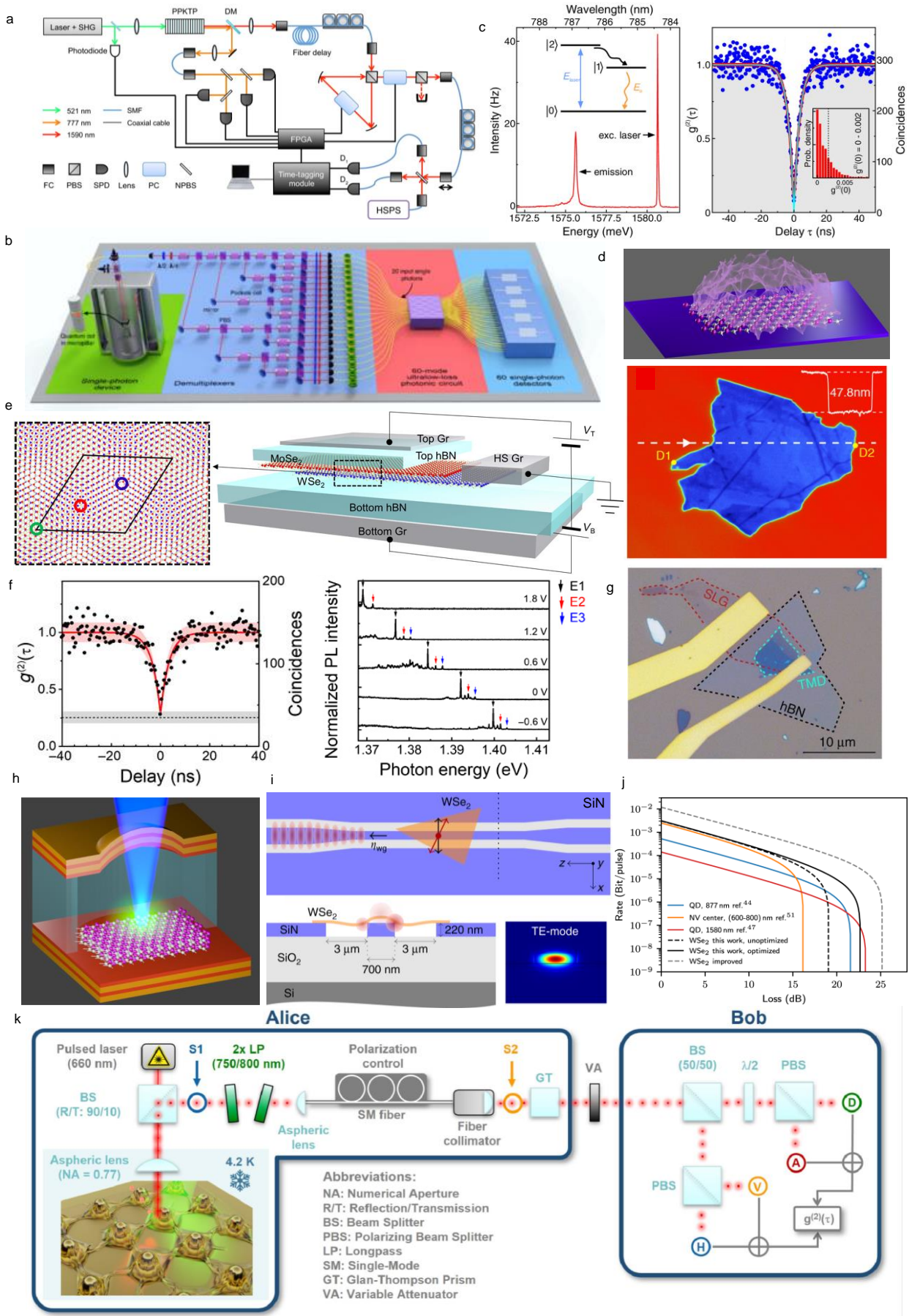

**Figure 38 | Single photon sources from various platforms.** (a) Experimental scheme of temporal-multiplexed single photon sources. Single photons pairs are generated from spontaneous parametric down conversion by pumping a nonlinear periodically poled potassium titanyl phosphate (PPKTP) crystal. Triggered by the detection of idler photons, a field-programmable gate array module controls the transmission of a Pockels Cell which in turn tailors the delay of the signal photon and therefore realizing the photon multiplexing in time domain. Reprinted with permission from ref.[430] Copyright 2019 American Association for the Advancement of Science. (b) Boson sampling machine. Single photons generated from a microcavity coupled InAs/GaAs quantum dot are multiplexed into 20 spatial modes. The 20 input photons are injected into a 3D integrated, 60-mode photonic circuit comprised of beam splitters and phase rotations. The circuit output are detected using 60 superconducting nanowire single-photon detectors to sample the photon distributions. Reprinted with permission from ref.[431] Copyright 2019 American Physical Society. (c) Left: the fluorescence spectrum of the low-energy bright exciton emission. Inset: schematic of resonant excitation of the blue-shifted exciton (BS-X) and emission via the ground-state exciton. Right: The second correlation and coincidence counting, revealing a $g^{(2)}(0) = 0.002$. Reprinted with permission from ref.[432] Copyright 2016 Optica. (d) Top: 2D hBN quantum emitter. The hBN flakes are exfoliated from the bulk crystal and transferred by dry contact to Si/SiO$_2$ substrates (280 nm thermally grown). Bottom: phase-shift interferometry image of the hBN crystal. Reprinted with permission from ref.[10] Copyright 2018 American Chemical Society. (e) Left: moiré superlattice is formed with a twist angle of ~60°, represented as a black solid line. Right: layouts of the heterostructures. Graphite layers are used as contacts for top, bottom, and heterostructure, and hBN layers are used for both dielectric layers and encapsulation of heterobilayer. (f) Left: Black dots represent the experimental second-order correlation at 760 nm, showing clear antibunching effect, while the red curve is the fitting revealing a $g^{(2)}(0) = $ ~0.28. The gray shaded area shows the error interval of the $g^{(2)}$. Right: dependence of moiré IXs on the external electric field. (e,f) Adapted from permission from ref.[32] Copyright 2020 American Association for the Advancement of Science. (g) Optical microscope image of the TMD-based LED structure. The light-emitting diode is comprised of a single-layer graphene (SLG), thin hBN and TMD mono- and bi-layers in between. Adapted from permission from ref.[433] Copyright 2016 Springer Nature. (h) Multilayer hBN flake hosting the quantum emitter is coupled into a confocal microcavity to facilitate the Purcell enhancement. The hemisphere spatially confines the cavity mode to the location of the emitter. Adapted from permission from ref.[434] 2019 American Chemical Society. (i) Top: A WSe$_2$ flake is integrated on a 220nm thick single mode SiN waveguide. The orientation of the dipole moment of the quantum emitters (red arrow) is random with respect to the polarization of the waveguide mode (black arrow). Bottom: cross-section of the sample. The waveguide is sandwiched by two air trenches so that it is separated from the bulk SiN. The air trenches are 3 um wide whilst the waveguide is 700 nm wide. Adapted from permission from ref.[435] Copyright 2019 Springer Nature. (j) BB84 quantum key distribution protocol using single photon sources generated from a WSe$_2$ monolayer kept at cryotemperature (~4.2 K). (k) Positive key rates vs the channel loss, equivalent to the transmission distance. The performance of the setup is benchmarked against other platforms based on solid-state single photon emitters. It is evidenced that the present 2D system is capable of delivering a positive key over the longest distance. (j,k) Reprinted with permission from ref.[436], Copyright 2023 Springer Nature.

The most widely studied 2D hosts of quantum emitters are TMDs and hBN. TMDs hosts for

SPEs are found in mono-and multi-layered WSe$_2$,[336-337, 437-438] WS$_2$,[439] MoSe$_2$,[379, 440] MoS$_2$[441-

[443] and MoTe$_2$[444-445] where quantum defects are attributed to localized, weakly bound excitons. However, the exact confinement mechanism remains inconclusive. TMDs SPEs hold the potential for high-purity single photon emissions, typically achieving purity levels of <0.1 or less and photon lifetimes ranging from a few ns to several tens of ns, as revealed by time-resolved photoluminescence (PL) measurements. In particular, the monolayer WSe$_2$ has demonstrated high-purity single photon sources with $g^{(2)}(0) \sim 0.022$ at cryogenic temperatures (T = 4K) and photon generation rates exceeding 3MHz at saturation under nonresonant excitation from a highly spectrally and spatially isolated emitter.[432] Under resonant excitation, it was observed that blue-shifted excitons, upon decaying to the ground state, produced single photons with record-high purity $g^{(2)}(0) = 0.002$ (Figure 38c). The origin and properties of these blue-shifted excitons are yet to be explored. The wide-bandgap hBN constitutes an alternative host for SPEs (Figure 38d). The photon emission in hBN is attributed to deep-level point defects within the crystal lattice, introducing energy states within the electronic bandgap. Techniques have been developed to reliably create and activate these defects, including plasma etching followed by high-temperature annealing,[10, 338, 446] electron-beam irradiation[447] and irradiation with ultrafast laser pulses or focused ion beams.[448] The hBN emitters hold promise for achieving high brightness, on the orders of millions of counts per second, for single photon generation at room temperature, rendering them a particularly attractive SPE candidate in extreme operational environments, such as space instruments exposed to extensive radiation like gamma-rays, energetic protons and electrons where energy consumption is a critical concern. Experimental investigations show that the performance of the hBN-based SPEs remains intact under radiation levels common for satellite altitudes up to geostationary orbit.[449] Single photons with a purity of $g^{(2)}(0) = 0.033$[10] and a lifetime as short as 294 ps have been reported. The emission wavelength and linewidth of these SPEs can be tuned by strain[450] and electric field.[451] Reversible Stark-shift

tuning of the emission wavelength can reach up to 43 meV/(V/nm), which is four times larger than the emission linewidth at room temperature. Nevertheless, the diversity of zero phonon lines (ZPL), ranging from emitter to emitter and covering the entire visible spectrum[447] down to ultraviolet,[452-454] suggests that the precise identification of defect classes and underlying mechanisms remain inconclusive and merit further investigations.

Twisted vdW materials have recently gained prominence as another fascinating platform for SPEs. Being a category of 2D materials, these vdW materials are characterized by their single layer lattices held together by intralayer covalent bonds and weak interlayer bonds.[455] The tunability of their electronic and optical properties, coupled with the ability to integrate them into nano-scale devices, makes them ideal candidates for quantum emission applications. Theoretical simulations indicate that the moiré pattern naturally creates superstructures of nano-spot confinements for long-lived interlayer excitons, suggesting their potential to host arrays of quantum emitters.[30] The quantum nature of the moiré heterostructures was first experimentally demonstrated in a $MoSe_2/WSe_2$ heterobilayer with $H$-stacked confirmation (~60°).[32] Figure 38e illustrates the superlattice composed of 1L $MoSe_2$ and 1L $WSe_2$ encapsulated by hBN layers. Photon antibunching was observed, indicated by a second-order correlation of ~0.28 (Figure 38f). Time-resolved PL measurement revealed a photon lifetime of 12.1 ns under the excitation at 1.63 eV with an excitation power of 4 uW on average. In contrast to SPEs in monolayers, where only very small permanent dipoles may exist in the vertical direction, significantly larger permanent dipoles of interlayer excitons were observed and validated. These vertical dipoles can be exploited to Stark-tune the emission energy position of moiré IXs by up to 40 meV, a record significantly larger than what has been reported for other solid-state emitters.

Other vdW heterostructures (non-moiré) have also been proposed as sources of quantum emitters. Trapping and dipole interactions between interlayer excitons have been vastly reported recently.[25, 456-458] TMD SPEs can be incorporated into heterostructure devices to enable all-electrical generation of single photon emissions, a desirable feature for system integration[433] (Figure 38g). The electrically driven TMD-based LED consists of mono- or bi-layered TMD positioned between a thin layer of hBN and a single-layer graphene, forming a vertical stacking of layered materials. The achieved second-order correlation is approximately 0.29, without correction for background emission or detector dark noise. Implementing spectral filtering to suppress the background emission can enhance the photon purity. Photon emission driven by the quantum tunneling effect was demonstrated in a vdW heterostructure comprised of gold, hBN and graphene.[459] Interestingly, the electronic properties of the device are decoupled from its optical properties, where the former can be tailored by a combination of stacked 2D atomic layers, while the latter is managed using an external photonic setup. An intriguing observation was made regarding the potential enhancement of the photon emission rate per tunneling electron within an Au/hBN/graphene tunnel junction by leveraging the anisotropy of hBN.[460] Theoretical modeling was proposed to simulate the effect. The underpinning rationale is that by modifying the tunneling barrier itself, the efficiency of inelastic tunneling rates can be heightened. Electrically controllable quantum emission was demonstrated in hBN-graphene heterostructures,[451] attributed to electrically-induced changes in the charge states of hBN defects. At 4K, the $g^{(2)}$ measured was 0.48 using a bias voltage of -10V.

The 2D SPE hosts offer distinctive features compared to other emitters. Unlike conventional solid-state SPEs such as the color centers in diamonds and quantum dots, 2D monolayer materials promise near-unity extraction efficiency, as the emitters are not embedded in high refractive index materials and remain unaffected by Fresnel or total internal reflection.[461] Furthermore, the atomically thin nature of these materials enables reliable stacking, unlocking a variety of new properties unavailable in monolayers. Complex artificial heterostructures[27, 30] can also be constructed with radically different chemical and physical properties, including electrical, optical, and magnetic characteristics. The variety of available vdW materials and their distinct optical and electronic properties, coupled with the extensive library of the vdW material-based SPEs that have been established, implies a potentially revolutionary paradigm for photon generations and quantum photonic devices. 2D materials are characterized by one or a few atomic layers, inherently allowing quantum confinement in one dimension. Their atomically thin composition, combined with flexibility in fabrication processes, facilitates integration with various optical cavities and waveguides, leading to the Purcell enhancement of quantum light emission due to the reduction in mode volumes. The implications of the Purcell enhancement are three-fold. First, the cavity, functioning as a low-pass filter,[462] can narrow the spectrum, improving single-photon purity by suppressing off-resonant noise. Second, the Purcell effect mitigates spectral-diffusion-induced inhomogeneous broadening, paving the way for generating indistinguishable photons. Third, the brightness of the emitters can be enhanced, partly due to the shortened photon lifetime and enhanced collection efficiency provided by the cavities. The Purcell effect can be exploited to enhance the photon count rate by more than 3 times compared to bare material emissions.[434] Progress in this regard has been made across a wide range of 2D materials (Figure 38h, i). For example, quantum emitters hosted by hBN have been integrated with a confocal open microcavity,[434] metallo-dielectric antennas[463] and $Si_3N_4$ photonic crystal cavities.[464] Photon emitters in monolayer TMDs have been successfully coupled to plasmonic nanocavities.[360, 465] Cavities and waveguides can also be fabricated directly from the 2D material itself, facilitating monolithic integration,[454, 466] open microcavities,[467] photonic chips[435] and waveguides.[468-470]

Despite the rapid progress in this research area, the significant disparities in optical characteristics, like the ZPL position in the spectrum or excitonic lifetime, across emitters within the same or different flakes underscore the uncertainty regarding the precise nature of the defects. This underscores the need for more comprehensive foundational investigations. Such efforts are crucial for achieving the generation of indistinguishable single photons, a task that remains challenging and currently lacks conclusive outcomes. The generation of indistinguishable single photons holds immense significance for quantum computation[471] and numerous quantum communication applications beyond point-to-point quantum key distributions.[472-475] For a wider and more thorough look at the 2D-material-host SPEs, we refer the reader to some excellent reviews.[455, 476-478]

### 6.1.2    Quantum communications and computations

A new frontier of research in photonic quantum technologies is the development of fully integrated platforms that incorporate components such as quantum emitters, quantum circuits, quantum memory, detectors and controls.[479-482] This integrated system constitutes a versatile and powerful toolkit for quantum information processing. The platform can function as quantum circuits that empower certain quantum algorithms such as Boson sampling,[483-485] and has the potential to evolve into a universal fault-tolerant quantum computer.[486-487] It can also serve as the relay node in the memoryless quantum repeaters,[488-489] thus facilitating the establishment of quantum networks. The challenge lies in identifying a material platform that encapsulate all these building blocks. VdW materials, particularly those presenting quantum-dot-like single-photon emitters, have emerged as promising platforms for the development of scalable quantum light source on integrated circuits. They hold the potential to enable various electro-optical devices that offer precise control, switching and manipulation of photons, along with efficient photon detections. However, despite the extensive demonstrations so far, the coupling efficiency of 2D materials onto integrated photonic devices remains quite limited.[490] Current approaches for transferring 2D materials onto photonic waveguides report coupling efficiencies below 10%.[435, 470] The major technical challenge resides in achieving mode matching between the 2D SPE and the cavity mode. Deterministic positioning with high lateral accuracy of the emitters is essential to achieve high-efficiency coupling to the small mode volume of the cavities. Nonetheless, ongoing research and development in this field are paving the way toward more efficient and scalable quantum photonic circuits. vdW materials have found application in quantum communications, driven by recent advances in their fabrication and SPEs.[436] Notably, a $WSe_2$ single photon source has been utilized to demonstrate the BB84 quantum key distribution (QKD) protocol (Figure 38j,k). The emitted single photons have a wavelength of 807 nm and have purity of $g^{(2)}(0) = 0.034$, which has been improved by spectral filtering.

Performance of the setup is benchmarked against previous QKD results using other solid-sate quantum emitters, such as quantum dots and NV centres in diamond. The current protocol delivers positive key rates, around $10^{-9}$ bis/pulse, at a transmission distance corresponding to 25 dB loss, which outperforms other solid-state systems.[491-492] The development of single photon emission at telecom wavelengths could be beneficial for future quantum communications using vdW materials. Similar to the development pathway of other solid-state SPEs, future perspectives include generating entangled photon pairs from spontaneous parametric down conversion. The capability of entanglement generation would unlock a range of quantum information protocols using vdW materials, including quantum teleportation, quantum repeater, distributed quantum computing and sensing. Additionally, the realization of hybrid spin-photon entanglement realized in quantum dots[493-495] and NV centers[496-498] presents another possibility for developing complex quantum networks.

### 6.1.3  Quantum simulations

Quantum simulators are specialized devices that utilize one quantum system to model or simulate another. They offer a novel approach of studying and understanding the complexities of quantum phenomena. Twisted vdW materials, particularly moiré superlattices, are currently recognized as effective platforms for these quantum simulations due to their diverse and tunable electronic properties.

The tunability of these materials primarily depends on the twist angle applied, which can be manipulated to simulate different quantum phenomena. This unique characteristic potentially provides invaluable insights into unresolved issues within the field of quantum physics. Furthermore, these twisted materials can exploit quantum interference to suppress effective kinetic energy scales. This suppression enables the transition of the system into new states at low energies, a feature that holds considerable significance for quantum simulations. Moreover, these twisted vdW materials offer an unprecedented level of control in quantum simulations. Parameters such as the angle of twist, stacking configuration, choice of substrate, and utilization of gating techniques can be meticulously managed to provide clean and efficient quantum systems. The potential scope of applications for these materials is vast. predictions indicate the existence of over a thousand unique 2D vdW materials, which suggest millions of potential configurations for twisted bilayer systems[69]. Consequently, these materials present substantial opportunities for advancements in fundamental studies, materials science, and quantum technologies. Twisted vdW materials are regarded as a versatile toolbox for realizing a wide range of quantum model systems (Figure 39). They can manifest exotic and elusive phases of matter, potentially contributing significantly to the material sciences and quantum technologies. By manipulating these materials, researchers can create "quantum matter on-demand".[86] Overall, twisted vdW heterostructures are viewed as an efficient, reliable, and scalable platform for quantum simulations. They enable a seamless realization and control of multiple interacting quantum models within a solid-state framework. Therefore, these materials are poised to revolutionize our ability to study complex quantum systems.

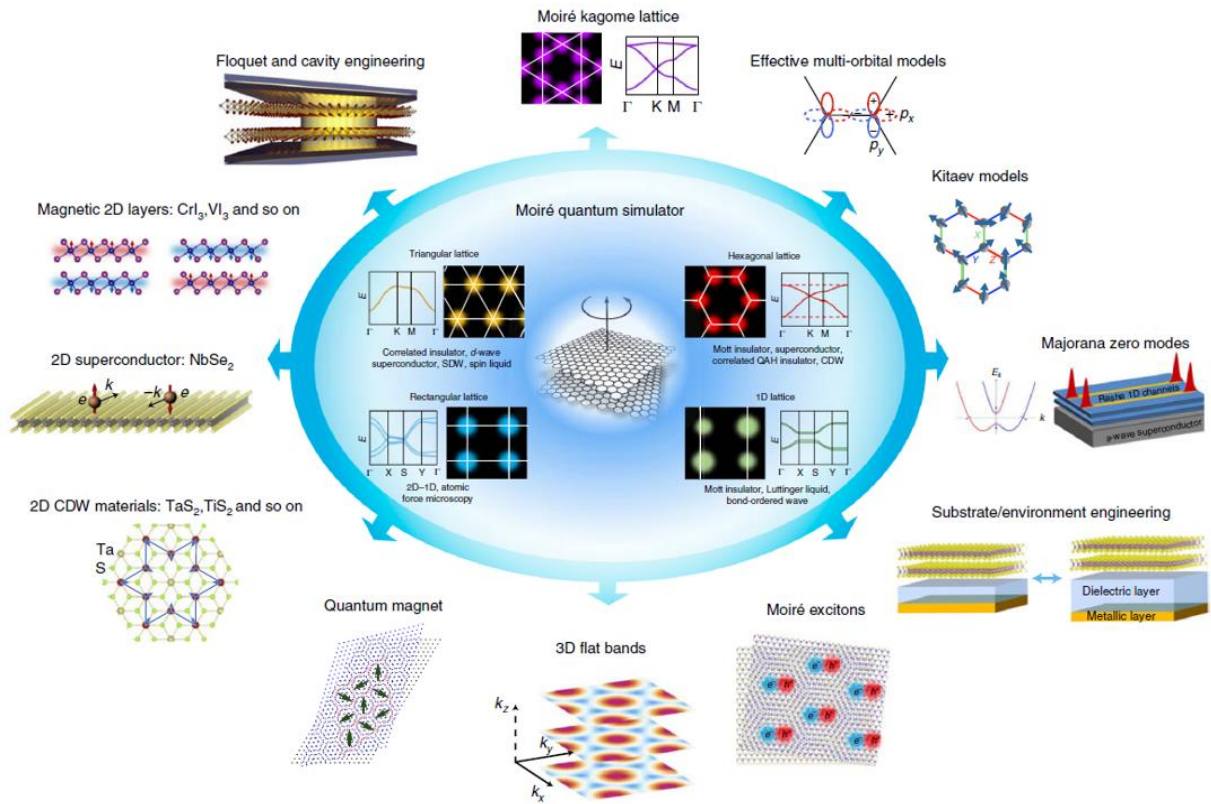

**Figure 39 | Moiré quantum simulator.** There are several efficient low-energy Hamiltonians produced by twisting and stacking sheets of vdW heterostructures. The inner circle of the picture includes some of these realizations that have been explored in the literature along with the matching lattice, band structure, and probable realisable phases of matter. The dashed lines for the red band structure show the extra flat bands that are present in twisted bilayer $MoS_2$ compared to TBG. In the outer circle, a perspective on potential future advancements for moiré quantum simulators is sketched.[86] Copyright 2021, Springer Nature.

Recent research further reinforces the potential of twisted vdW materials as quantum simulators. They have shown versatility in modeling diverse quantum systems exhibiting exotic phases of matter. The tunable properties of moiré superlattices, a specific type of twisted vdW material, have been harnessed to simulate various quantum phenomena. Scientists are now exploring these materials to understand elusive phases of matter like the spin-liquid phase, and systems possessing topological properties that hold promise for quantum technologies. Driven by the controllability of moiré superlattices, these advancements herald a promising era for quantum simulations, empowering scientists to explore uncharted questions in quantum physics.

We envisage that elucidating the underlying physics of vdW material-based quantum emitters will be a fundamental breakthrough. Further investigations will incorporate these SPEs into integrated photonics and quantum information protocols that are under development, to enhance performance, scalability and application scope. This in turn would advance a broad range of quantum communications, computations and sensing technologies.

### 6.2 Sensors

Twisted vdW materials find applications in a variety of sensor technologies due to their high sensitivity to external stimuli and strong interaction with light and other particles.

### 6.2.1　Bio sensors

VdW materials have been explored widely for their potential in nanophotonic biosensing applications. Due to their biocompatibility and sensitivity to environmental changes, these materials can be used in biological sensors to detect biomolecules and monitor biological processes. These materials possess unique properties such as low dimensionality, which can be harnessed for enhancing various parameters in biosensors. Specifically, their capability to confine tightly polaritonic waves has proven beneficial for boosting sensitivity and efficiency.[499] Materials like 2D graphene are prime examples of VdW materials that exhibit these traits. The diminished dimensionality of such materials significantly amplifies the confinement of plasmonic fields, providing a notable edge over traditional nanophotonic devices relying on surface plasmon resonance within metallic films. Ultrathin, self-supporting nanomembranes have considerable potential in flexible optoelectronic applications, including wearable light-based devices, implantable biosensors,[500] and photo-diagnostic tools (Figure 40a). Twisted VdW materials, which form when layers of these substances are stacked with a slight rotational offset, represents a captivating advancement in this domain. These materials open up fresh avenues for both fundamental scientific exploration and the potential utilization within materials science and quantum information technologies.[86] The vdW assembly of several functional layers enables the creation of innovative multi-purpose optical systems. An instance of this is the development of a multi-functional contact lens that can simultaneously capture images and monitor temperature and lactate levels,[501] opening up new possibilities for wearable optoelectronic applications (Figure 40b). The formation of vdW heterostructures (VDWHs) is another approach taken in this field. It involves combining different 2D materials to offset the drawbacks of individual materials. For example, the inertness and zero-band gap of pristine graphene can be counterbalanced by the less environmentally stable TMDs.[502] Recently, a platform built upon a 2D electronic material enclosed within a planar cavity comprised of ultrathin polar vdW crystals has been proposed. The setup uses hBN layers that are nanometer thick, showing promise in achieving the ultrastrong coupling regime for single-electron cyclotron resonance in graphene bilayers.[503] Tightly-confined polaritons on a dielectric background and subwavelength polariton resonators have the potential for robust light-matter interaction, making them viable for sensitive sensing and surface-enhanced infrared absorption spectroscopy. They're typically occur in the infrared range, where many biomolecule fingerprints exist (Figure 40c). Specifically, the resonance of graphene plasmons allows for label-free, ultra-sensitive detection of proteins and other molecules.[504] A concern with these resonators, however, is their low quality (Q) factor due to the inherent Ohmic losses from free electron scattering. In contrast, polar crystals' phonon polaritons, like those in hBN and α-MoO$_3$, avoid Ohmic loss and have high Q factors up to

200. This feature enhances their potential for high-sensitivity sensing applications. In summary, twisted vdW materials are demonstrating significant promise for the advancement of nanophotonic biosensors, with their unique properties and the possibility of creating heterostructures to overcome individual material limitations.

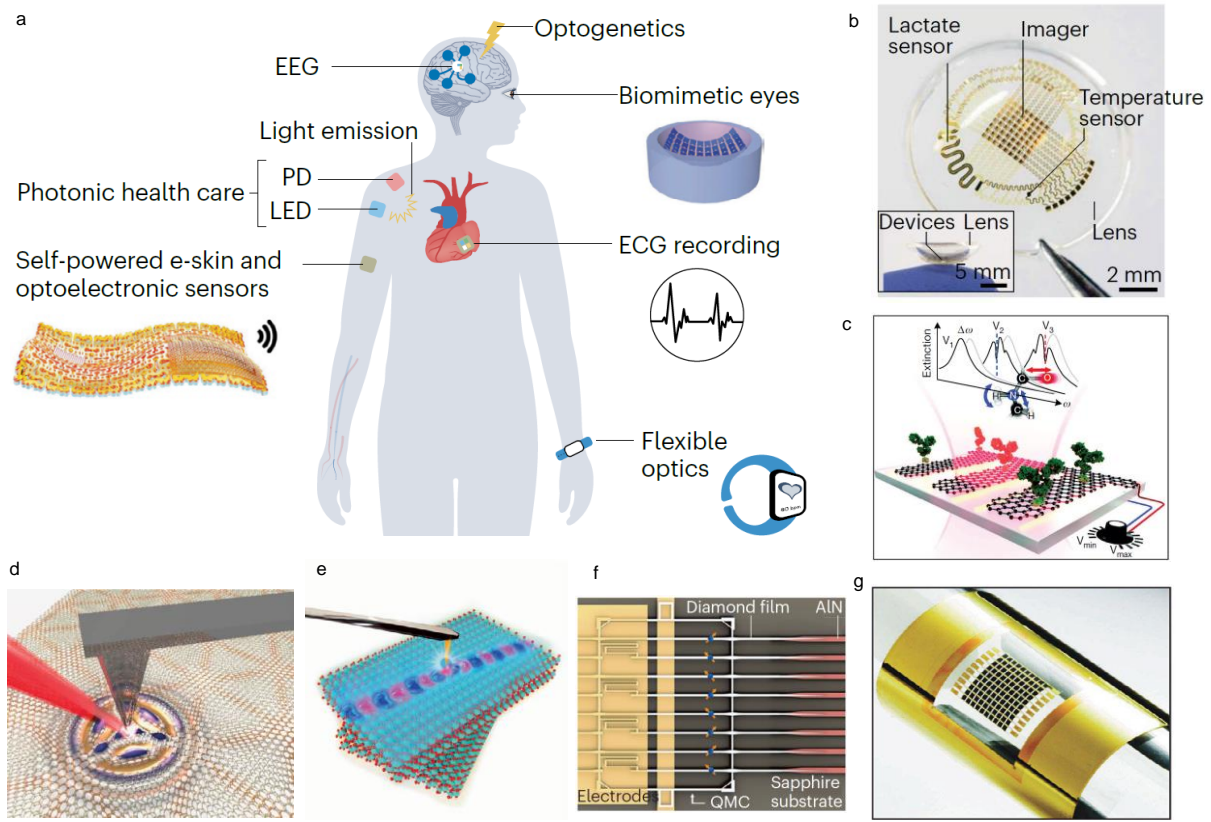

**Figure 40 | Applications of 3D vdW films and moiré superlattices (MSL).** (a) Wearable, implantable, and biocompatible optoelectronic applications are made possible by 2D and 3D freestanding films, including the creation of a biomimetic eye using Si films and an e-skin that runs on its own power using GaN films. (b) Multifunctional lens detector and sensor. Reprinted with permission from ref.[501] Copyright 2019 Springer Nature. (c) Electric adjustable infrared biosensors by localized plasmon resonance in graphene nanoribbons. Reprinted with permission from ref.[504] Copyright 2015 American Association for the Advancement of Science. (d) Diagram of a photonic crystal generated by a network of solitons in the t-BLG for near-field nano-imaging. Reprinted with permission from ref.[283] Copyright 2018 American Association for the Advancement of Science. (e) Twisted bilayer α-MoO₃ enables the manipulation of canalized polariton propagation. Reprinted with permission from ref.[395] Copyright 2020 Springer Nature. (f) Chip with a hybrid diamond quantum emitter and optics. Adapted with permission from ref.[505] Copyright 2020 Springer Nature. (g) GaAs solar cell array that is stretchable atop a polyethylene terephthalate substrate. Reprinted with permission from ref.[506] Copyright 2010 Springer Nature.

### 6.2.2 Gas sensors

VdW materials are showing substantial potential for use in gas sensors, owing to their unique properties including high surface areas, tunable bandgaps, and various crystal structures.[507] They have especially drawn interest for their utilization in vdW junctions, which demonstrated significant promise in electronic devices including gas sensors. Particularly, TMDs such as $WS_2$, $MoS_2$, $MoTe_2$ and more, are gaining momentum in sensing applications, specifically in the realm of room-temperature gas sensors. This interest stems from their captivating electronic, optical, and mechanical properties.[507] In t-BLG with small rotational angles, a moiré superlattice emerges, exhibiting a periodicity that closely matches the graphene plasmon wavelength. This gives rise to a sequence of reflections of graphene plasmons within periodic soliton domain walls, effectively forming a photonic crystal for graphene plasmons (Figure 40d).[283] These photonic crystal plasmons can be observed through scattering-type SNOM. The resonance of graphene plasmons facilitates label-free, highly sensitive detection of inorganic molecules[508] and the identification of gases.[509] The impressive combination of a high surface-to-volume ratio and the flat nature of VdW layered materials positions them as an exceptional platform for gas sensing applications. This is exemplified in field effect transistor gas sensors that rely on mechanically exfoliated VdW materials. These sensors offer an appropriate framework to explore the Langmuir absorption model[510]. Furthermore, gas sensors employing these materials frequently operate by detecting alterations in resistance or conductance upon exposure to gas molecules. Some sensors also utilize Schottky junctions composed of 2D materials and metal electrodes, or heterostructures such as graphene/$MoS_2$ and BP/$MoS_2$ VdW hetero-junctions.[511] In the context of twisted VdW materials, these have been identified as an exciting development with considerable potential for fundamental scientific research and potential applications, although the specific use of these twisted materials for gas sensors has not been explicitly reported in the literature.

However, considering their potential in other sensing applications, it is plausible that these twisted VdW materials might find relevance in gas sensing applications in the future as the field progresses. In general, although there's a substantial body of research on the use of VdW materials and their junctions in gas sensing, there remains ample space for additional exploration, particularly in the context of twisted vdW materials, essentially to fully harness their potential within this domain.

### 6.2.3  Chemical sensors

Twisted vdW materials are an exciting development in the field of material science and have great potential for chemical sensor applications. They provide a highly tunable platform for various applications, including "twistronics".[86, 512] Similarly, they can be employed in chemical sensors, offering high sensitivity and selectivity for the detection of various chemical species. Significant efforts have been put into enhancing the performance of vdW materials in sensor applications by tuning their properties. For example, doping vdW materials has been studied as a method to enhance strain sensors' sensitivity.[513] In the context of chemical sensors, vdW heterostructures (VDWHs) formed by combining different 2D materials have shown significant promise. These structures can effectively address the limitations inherent in individual materials, such as the inertness and and lack of a bandgap in pristine graphene by capitalizing on the advantageous characteristics of different materials.[502] Recent advancements in chemical sensors utilizing VDWHs of 2D materials have illuminated novel understandings of sensing mechanisms and offered pathways for future research directions, including challenges and opportunities in the pursuit of developing next-generation chemical sensors with potential environmental science impacts.[502] Despite this, the exact potential of twisted vdW materials specifically in the context of chemical sensors has not been reported in the literature. However, considering their versatility and high tunability, along with the potential demonstrated by non-twisted VdW heterostructures, it can be speculated that they would have valuable applications in chemical sensing. Continued research in this area is needed to more fully understand and harness the potential of twisted vdW materials in chemical sensor applications.

### 6.3 Other applications

Aside from the aforementioned applications, twisted vdW materials can find utility in numerous other technologies, fields (Table 1) and even beyond.

#### 6.3.1   Ultrafast modulators

Twisted vdW materials' electronic and optical properties make them attractive for ultrafast modulators. These are essential components in optoelectronic devices, where they control the intensity, phase, polarization, or propagation direction of light. The materials' tunability can enable modulators that operate at unprecedented speed and efficiency. The twist angle between layers in twisted vdW structures offers a tuning knob that can drastically alter the electronic properties of these materials. For instance, it can transform graphene from a semimetal to a Mott insulator or a superconductor,[145] indicating the promising capability of these materials in the development of tunable electronic and optoelectronic devices. One key property that could be useful for ultrafast modulators is the saturable absorption characteristic of 2D vdW materials. This property, which refers to a material's decreasing absorption coefficient with increasing incident light intensity, has led to a series of proposed and confirmed applications in ultrafast photonics, encompassing roles as high-threshold, broad-spectrum, and rapidly responsive saturable absorbers (SAs).[514] Saturable absorbers are essential components in mode-locked lasers, a crucial technology in ultrafast optics. In addition, the control of light propagation at deeply subwavelength scales is crucial for integrated photonics. VdW stacked structures offer an ideal platform for these highly localized light-matter interactions, known as vdW polaritons. The orthorhombic α-MoO₃, polar vdW crystals with pronounced in-plane anisotropy, is particularly suitable for supporting hyperbolic polaritons, quasiparticles exhibiting highly anisotropic permittivity. The polariton dispersion in α-MoO₃'s twisted homobilayers can be manipulated by adjusting the twist angle, causing a transition from a hyperbolic to an elliptical configuration. This occurs due to the conservation of topological invariant, specifically the count of anti-crossing points in the dispersion curves of the two α-MoO₃ layers. At a specific twist orientation, the polariton dispersion changes from closed to open, forming a flattened dispersion profile that facilitates the propagation of polaritons without diffraction (Figure 40e).[395] This twist strategy can be extended to other types of quasiparticles for the purpose of engineering dispersion. For example, by altering the twist angle between nanostructured graphene bilayers with hyperbolic dispersion, moiré plasmons characterized by field confinement and minimal damping can be achieved. Furthermore, 2D vdW materials have been used in mixed-dimensional vdW heterostructures,[515] which have shown rapid advancements in wearable, stretchable, implantable, and portable electronic and optoelectronic devices.[516] This suggests the potential for similar advancements in ultrafast modulators, given the importance of these properties in such applications. While the results provide indirect evidence supporting the potential of twisted vdW materials in ultrafast modulators, more research is needed to fully realize and validate this potential.

### 6.3.2   LED devices

In the field of optoelectronics, light-emitting diode (LED) devices based on twisted vdW materials could offer a new class of light emitters. Such devices may leverage the materials' quantum mechanical properties and unique band structures to emit light at different wavelengths, potentially creating highly efficient, tunable LEDs. The twist configuration is a crucial factor in the material's electronic properties, which can alter significantly based on the rotation degree.[69, 266] This tunability is advantageous for designing LED devices, as the emission properties of LEDs depend largely on the electronic properties of the material used. Creating vdW heterostructures by vertically stacking 2D materials has proven to be an effective strategy to harness and synergize the optoelectronic properties inherent in various types of 2D materials.[320] Since LEDs are essentially optoelectronic devices that convert electrical energy into light, this approach could be beneficial in developing LEDs with improved or unique characteristics. The ability to create new architectures through vdW integration, such as coupling diamond films with optical cavities and waveguides, has shown considerable promise. For instance, diamond films with color centres that can be optically addressed were transferred onto aluminum nitride waveguides (Figure 40f) in a hybrid integrated photonic chip.[505] This innovative heterostructure combines the superior linear and nonlinear waveguiding properties of aluminum nitride with the controlled transitions of single-photon emitters within diamond films. Germanium membranes, on the other hand, have been utilized to create high-quality infrared photodetectors[517] and flexible/stretchable LEDs,[518] expanding the versatility and potential applications of these materials. Additionally, the discovery of ferroelectricity in some 2D vdW materials could bring essential functionalities to LED devices.[519] Ferroelectric materials possess a spontaneous inherent electric polarization that can be reversed by an external electric field, which might be useful for creating LEDs with tunable emission properties.

### 6.3.3 Lasers

The moiré superlattice in vdW materials greatly impacts the electronic and optical properties of materials.[154] It offers novel routes to develop lasers based on twisted vdW materials.[520]

The efficient lasing behavior has been observed in TMD heterostructures with the integration of cavities.[521-522] By controlling the stacking configuration and twist angle, we can tune the emission wavelength of these lasers, leading to compact, efficient, and tunable laser sources. The materials' intrinsic properties also support low-threshold lasing, a sought-after feature in the design of energy-efficient devices (Figure 40g).[506] The twist angle-dependent optical properties of these materials could be particularly useful for lasers, where control over light emission is crucial. Furthermore, these materials exhibit ultrafast optical responses and significant light-matter interactions, properties that are particularly desirable for laser applications.[514] It is important to note that more research is needed in this area to fully understand and exploit their potential for this application.

### 6.3.4   Photodetectors

The convergence of moiré superlattices and graphene technology is paving the way for innovative photodetectors. Graphene, renowned for its exceptional carrier mobility and rapid photo response, holds immense potential for high-speed photo sensing. However, the limited optical absorption capacity (approximately 2.3%) of 1L graphene restrains its practicality in high-responsivity, highly selective photodetectors. In a recent breakthrough,[254] Yin *et al*. devised a novel approach by introducing a twist angle of 13° in trilayer graphene (TBG) photodetectors. By incorporating a plasmonic structure within the device, they achieved a remarkable enhancement in photocurrent, reaching up to 80 times amplification for specific incident radiation wavelengths. The TBG with various twist angles was synthesized using CVD on copper foil and subsequently transferred onto a SiO2/Si substrate (as illustrated in Figure 41a). The fabrication process involved creating photodetectors using two different configurations: one utilizing bilayer graphene twisted at 7° and the other employing a twist angle of 13°. This innovative combination of t-BLG and plasmonic enhancements lays the groundwork for advanced photodetection technologies with substantially improved performance characteristics. Zhenjun *et al.*[239] present a groundbreaking study demonstrating the utilization of t-BLG with van Hove singularities (VHSs) to develop high-performance photodetectors with enhanced light-matter interaction. The scientists establish a simple technique to generate extensive domain t-BLG with precisely controlled twist angles, ensuring minimal contamination from interfacial polymers. This meticulous process of transferring two graphene monolayers in a layer-by-layer manner encourages robust interactions between π-bond electrons across layers. This interaction leads to the formation of small energy gaps in the electronic structure and the emergence of van Hove singularities in the DOS. This distinctive alteration in the band structure significantly enhances the intensity of the Raman G-band (~20 times), serving as an indicator of the high quality of the tBLG. The resultant photodetectors based on tBLG demonstrate a discernibly amplified generation of photocurrent, reaching levels approximately 6 times higher than their monolayer counterparts. Notably, the photocurrent enhancement mainly occurs at the tBLG/metal junction due to charge-transfer doping effects. This research highlights the potential of VHS-rich tBLG for advanced photodetector design, harnessing strong light-matter interaction to enhance performance. (Figure 41b illustrates the device structure.) Based on Wei *et al.*'s[523] research, innovative Metal-Graphene-Metal (MGM) architectures (Figure 41c) are showcased using a conventional Kretschmann configuration as illustrated in Figure 41d, which integrates gold films with various graphene samples (Figure 41e). Notably, the photoresponse behaviour exhibits a proportional increase with augmented graphene thickness, accompanied by a distinct sensitivity to the twist angle between the graphene layers. A particularly compelling outcome emerges from a 10° t-BLG photodetector, demonstrating a remarkable sevenfold amplification in photovoltage at the optimal incident angle. This enhancement is characterized by intriguing polarization-dependent and antisymmetric properties, the elucidation of which involves a synergistic interplay of Photovoltaic Effect (PVE), DOS theory, and Surface Plasmon Resonance (SPR). Deng *et al.*[524] have achieved a notable breakthrough in mid-infrared photodetection, showcasing strong photoresponsivity in t-BLG devices (Figure 41f) with a specific twist angle of 1.81°, as opposed to larger angles. This achievement was attributed to the creation of moiré bandgaps in the electron and hole branches, achieved through fine lattice structures at minor twist angles. Notably, the researchers employed this approach instead of manipulating the DOS gap in the combined moiré Dirac band. With a gate bias of 43.5 V (Figure 41g) and the Fermi level aligned at the centre of the superlattice gap, the extrinsic responsivity (Rex) reached 26 mA/W at a distance of 12 µm under 1200 nm radiation.

Researchers have achieved breakthroughs in atomically thin electronic devices using vdW heterostructures, stacking diverse 2D materials. Zixing *et al.*[525] successfully grew β-$In_2Se_3$/$MoS_2$ heterostructures with an aligned lattice through a two-step CVD method despite significant lattice mismatch. Strong interlayer charge transfer was observed, evidenced by photoluminescence quenching. Electronic devices based on these heterostructures exhibited a high on/off ratio, rectifying behaviour, and exceptional photosensing properties. Under a 450 nm laser and 1V bias, responsivity reached 4.47 A/W with a detectivity of $1.07 \times 10^9$ Jones. The inclusion of β-$In_2Se_3$ enabled near-infrared detection and fast response at room temperature, showing potential for high-performance broadband photodetectors. Wei *et al.*[526] investigated photoeffects in vdW stacked BP junctions. Crystal orientation and stacking morphology strongly influence interlayer carrier transport, revealing type-I and II band alignments. Dual vertical and lateral transport modes are observed, emphasizing the role of crystal orientation for orientation-based nanodevice design. Xiong *et al.*[527] employed stacked vdW layers for an ultracompact polarimeter integrated with fibres (Figure 41h). The device detects linear and circular polarizations through photo-galvanic effects, enabling real-time, self-calibrated polarization analysis and imaging. This approach holds potential for miniaturized optical systems and multifunctional fibre-integrated devices, revolutionizing polarization-related applications. Choi *et al.*[108] investigated the role of lattice matching in vertical heterostructures (HS). Twisted p-type $WSe_2$ and n-type $MoSe_2$ layers were fabricated with twist angles. Well-matched lattice led to a distinct p-n junction and efficient diode behaviour. Twisting introduced a potential barrier due to lattice mismatch. Illumination enabled tunnelling of excited carriers beyond the barrier, yielding current flow. Spectral responsivity revealed higher sensitivity in untwisted HS under IR light. Twisting weakened coupling confirmed through photoluminescence and Raman studies, highlighting crystal orientation's importance for optoelectronic devices. In a parallel discovery[528], Srivastava *et al.* unveiled new observations. The previously identified rectification phenomenon is now replaced by a symmetric nonlinear voltage-current curve. Additionally, the researchers showcased resonant tunnelling diodes featuring intricate BP-MS structures, highlighting a distinct characteristic: the negative differential resistance phenomenon. These findings hold promise for advancing photoreactor developments.

**Table 1.** Summary of some typical twisted structure applications.

| Twist system | Twist angle (°) | Parameters details | Applications | Ref. |
|---|---|---|---|---|
| Graphene | 0.6 | Photoresponsivity of ~80 mV/W at $432\times10^3$ nm | Photodetector | [529] |
| | 1.81 | Photoresponsivity of 26 mA/W at 1200 nm | | [524] |
| | 10 | Photoresponsivity of 2.2 V/W at 633 nm | | [530] |
| | 10 | Photoresponsivity of 2.5 mA/W 633 nm | | [239] |
| | 12 | Photoresponsivity of 1.6 V/W at 532 nm | | [530] |
| | 10 | Photoresponsivity of 9 mA/W at 633 nm | | [531] |
| | 13 | Photoresponsivity of 1.0 mA/W at 532 nm | | [254] |
| Graphene/hBN | - | Photoresponsivity of 0.3A/W at 660 nm | | [532] |
| $MoSe_2/WSe_2$ | 15 | Photoresponsivity of 2 A/W at 365 nm | | [108] |
| BP | 90 | Photoresponsivity of 4.6 mA/W at 532 nm | | [533] |
| $WS_2$ | 14 | overpotential of 60 mV at 10 mA cm$^{-2}$ | Energy storage | [534] |
| $MoS_2$ | 7.3 | overpotential of -153 mV at -10 mV cm$^{-2}$ | | [535] |
| Graphene | 1.1 | superconductivity transition at ~1.7 K | Superconductivity | [37] |
| | 0.93 | superconductivity transition at ~0.3 K | | [536] |
| | 1.56 | superconductivity transition at ~2.1 K | | [537] |
| | >1.1 | superconductivity transition at ~3 K | | [39] |
| Graphene | 10 | ~3 times more photo absorption | Photo absorption | [531] |
| Perovskite | 30 | 28.5% enhancement of light harvesting | | [538] |
| $MoSe_2/WS_2$ | 2 | Tunable PL peaks around ~0.1 eV | Luminescence | [539] |
| $WSe_2$-$MoSe_2$ | 0 | 100-fold intensity enhancement of emission | | [521] |
| Photonic crystal | - | lasing threshold of ~6 kW cm$^{-2}$ | | [540] |
| Graphene | 17.5 | 5-Time enhancement in PL | | [541] |

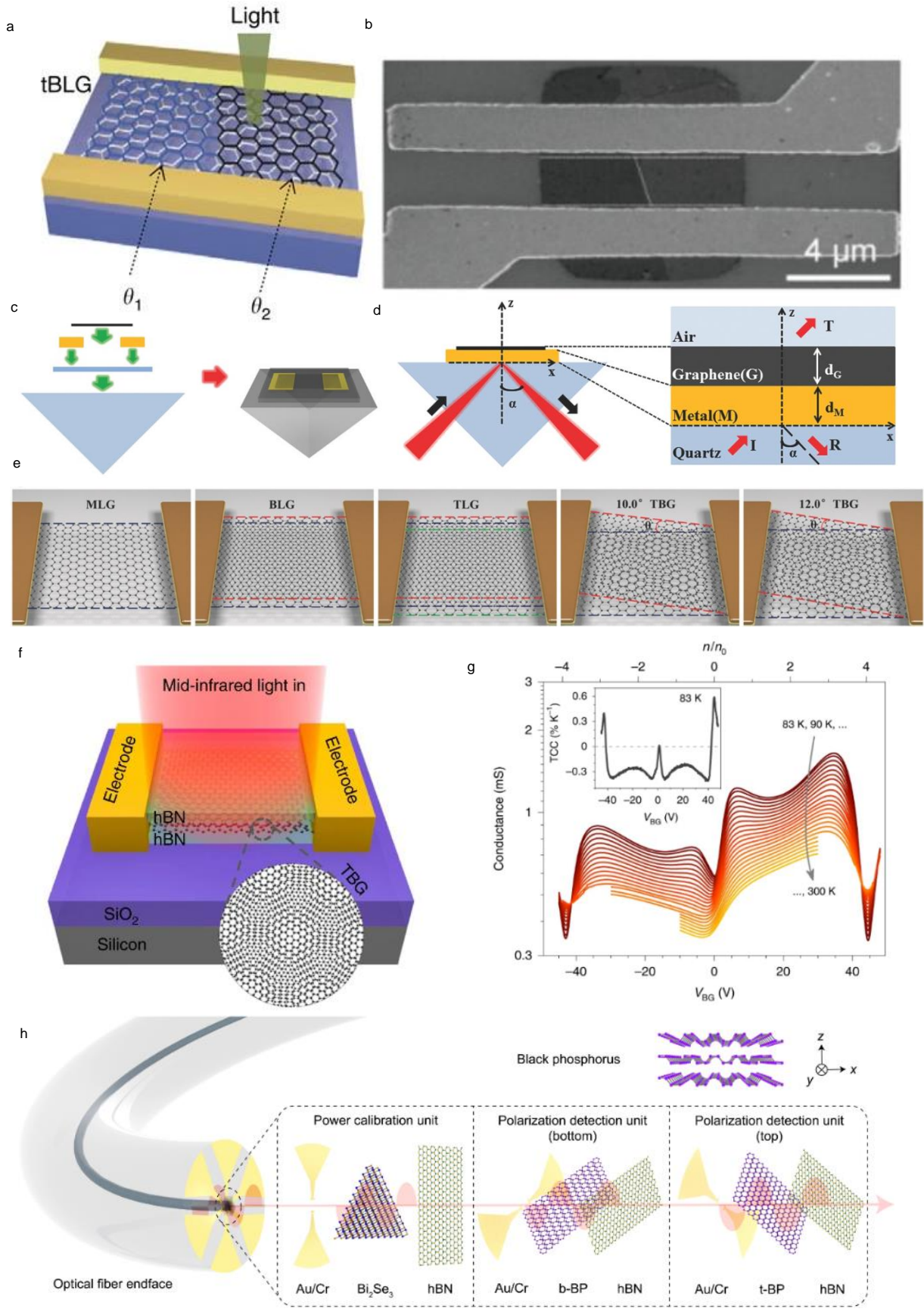

**Figure 41 | Moiré superlattice based photodetectors.** (a) Schematic illustration of a tBLG photodetection device. The channel comprises of two adjacent tBLG domains with different twist angles of θ1 and θ2, respectively. Reprint with permission from ref.[254] Copyright 2016 Springer Nature. (b) SEM image of the t-BLG photodetector featuring an approximate twist angle of 10°. The bilayer channel is situated to the left of the dashed line, with the monolayer region located on the right side. Reprint with permission from ref.[239] Copyright 2016 American Chemical Society. (c) Illustration of the Metal-Graphene-Metal (MGM) configuration. The layers, from top to bottom, consist of graphene, gold electrodes, a quartz plate, and a rectangular prism. (d) A cross-sectional view of the Kretschmann model. (e) Schematic representations of distinct samples including Monolayer Graphene (MLG), Bilayer Graphene (BLG), Trilayer Graphene (TLG), 10.0° t-BLG, and 12.0° TBG. Here, θ represents the interlayer twist angle. (c-e) Reprint with permission from ref.[523] Copyright 2016 John Wiley and Sons. (f) The schematic of the hBN-encapsulated TBG device and the photocurrent measurement scheme. (g) Temperature-dependent conductance of the 1.81° twisted bilayer graphene (TBG) device is depicted from 83 K to 300 K, revealing thermal excitation behaviour at around ±43.5 V VBG values, indicative of superlattice-induced bandgap formations. (f,g) Reprint with permission from ref.[524] Copyright 2020 Springer Nature. (h) Schematic of the fibre-integrated polarimeter design, featuring stacked functional units on an optical fibre end face. BP's puckered honeycomb lattice with armchair (x) and zigzag (y) crystal orientations is shown. 'z' denotes layer stacking direction. Reprint with permission from ref.[527] Copyright 2022 American Association for the Advancement of Science.

# 7. Perspectives

Twisted vdW quantum materials represent a rapidly growing area of research due to their promising properties and potential applications. This type of structure utilizes twisting and vdW weakly bonded atomic layers to expand the family of materials and create a platform for designing new quantum materials.[16, 542] Such twisted 2D vdW materials have exhibited great versatility in studying exotic quantum phenomena, particularly with anisotropic layered materials, providing a higher degree of tunability.[512] This capability and controllable band and emission properties also support the moiré superlattices for exploring a plethora of astonishing discoveries, which has already yielded the immense potential of scientific advancements in a variety of related fields, including unconventional superconductivity[37] and insulator phases,[126] moiré quantum emitters,[32] multi-color photon emission[543] and variations in phonon modes.[544] These breakthroughs open up multiple new avenues for exploration and will pave the way for more future advancements if the scope is further expanded, such as exploring the similar study of the emerging new family of materials.

A recent interesting example of novel vdW materials is the layered 2D superconducting material 2H-NbSe$_2$, which is employed to construct a vdW heterostructure, enabling the induction of a proximity-coupled superconducting order in the interfacing materials.[545] Moreover, twisted vdW quantum materials are considered promising platforms for exploring the intriguing nature of the quantum spin-liquid state, demonstrating the versatility of magnetic vdW heterostructures. Beyond the fundamental science, twisted vdW provides entirely new ways of manipulating and controlling quantum materials, creating an exciting opportunity to realize "quantum matter on demand".[86] However, realizing the engineering of materials with exactly desired quantum properties is still challenging, which will require further investigation and the discovery of the correct methodologies. In addition, there has been a particular emphasis on investigating the magnetic properties of twisted structures. One notable area of research is the examination of the Magnons we explained above, which are quasiparticles representing collective excitations of electron spin structures in crystal lattices. In the context of twisted ferromagnetic bilayers (tFBL), collinear magnetic order has been observed at critical angles known as magnon magic angles. Similar to t-BLG, t-FBL exhibits flat moiré minibands that possess Dzyaloshinskii-Moriya interactions (DMI) at these magnon magic angles.[546] Moreover, in twisted bilayer CrI$_3$, twisted domain walls have been found to harbor more stable 1D magnons compared to magnetic domain walls, and small twist angles offer greater stability for 1D magnon networks within domain walls.[424] These properties position twisted bilayer magnets as an perfect platform for exploring the magnonic counterparts of moiré electrons.[421] Consequently, the twist angle becomes a crucial parameter for controlling magnon conductivity, and it is expected to manipulate not only optical and electric properties but also magnetic conductivities within these structures, which has unveiled a new horizon of research possibilities.[546] The interplay between twist angles and various properties has ignited significant interest among researchers, leading to a surge of studies that aim at further exploring and understanding the phenomena. However, there are still some challenges that hinder the full realization of its value and role in practical applications.

One significant hurdle is the absence of rapid and precise synthesis methods. Although scientists and researchers have developed certain techniques[234] for stacking 2D materials at fixed twist angles, the field still requires more accurate and efficient methods for precisely controlling the twist alignment. An automated stacking fabrication technique is required and expected to be an ideal approach to achieve the rapid synthesis of desired structures with a customised angle. The future of 2D crystals and heterostructures prepared by top-down exfoliation from layered bulk crystals remains promising, with a focus on larger, higher quality films, by improving control over interfacial interactions and optimizing crystal growth processes at a large scale. Deterministic assembly of vdW heterostructures by mechanical stacking faces two major challenges: reducing residues between stacked layers and avoiding variability during manual dry transfer.[547] Implementation of the cleaner transfer process in a controlled ultrahigh-vacuum environment and the use of autonomous robotic approaches to material stacking could address these (Figure 42). Some current bottom-up synthesis techniques have demonstrated their capability to create vertical stacks, lateral heterostructures and superlattices,[548] but are limited by materials compatibility and high growth temperatures. Materials compatibility plays an important role in bottom-up synthesis, as it determines the feasibility of integrating different materials to form desired structures. The variations in lattice constants, crystal structures, and interfacial phases among different materials sometimes pose challenges in achieving compatibility. The advancements in CVD and other synthesis methods hold promise for addressing this issue. Potential approaches include the use of gaseous precursors and the creation of heterostructures of metallic and semiconducting components for integrated circuits.[549-550] In addition, the development of low-temperature growth methods is an active area of research.[551-552] It aims to reduce thermal decomposition and induced interdiffusion of the constituent materials, achieving improved material quality and enhancing the structural integrity of the synthesized structures. Beyond these, high-order vdW superlattices hold the potential to exhibit intriguing physics phenomena. The construction of quasi-3D superlattices may overcome limitations observed in purely 2D systems and allow for novel conditions, including the exciting prospect of Bose-Einstein condensation at room temperature.[553] While mechanically stacking vdW layers for superlattices presents a challenge, alternative methods such as assisted CVD growth, intercalation, and assembly and roll-up techniques offer promising solutions.[554-555] The progress made in both top-down and bottom-up approaches, along with the development of hybrid methods that integrate mechanical stacking, photolithographic or laser patterning, as well as vertical or lateral growth, holds great potential for the creation of heterostructures of remarkable intricacy and utility. The ideal multifaceted approach enables the integration of diverse materials and the precise control of their arrangement, resulting in structures with tailored properties. A crucial direction for upcoming research involves the predictive design of novel materials with predetermined attributes by vdW integration of diverse 2D and 3D materials featuring customized layer characteristics. This design-driven approach offers opportunities to engineer materials with desired functionalities and performance. The implications of such advancements span across numerous domains, encompassing energy-efficient electronics, optoelectronics, and emerging fields like quantum information processing.[556-557] Moiré photonics and optoelectronics represent a swiftly evolving field with a myriad of phenomena discovered in less than half a decade, including moiré excitons, polaritons, and strong mid-infrared photoresponses, among others. The unveiling of these phenomena has opened up a vast array of possibilities for both scientific exploration and technological innovations. The recent rapid progress suggests that we have barely scratched the surface of what is achievable in the realm of moiré photonics and optoelectronics. An essential future direction involves the improvement of scanning probe techniques, which play a crucial role in the characterization and understanding of moiré systems. Current methods, based on far-field techniques, often yield complex and convoluted results due to the involvement of thousands of moiré unit cells. To overcome this challenge and gain deeper insights, it is imperative to develop techniques capable of probing individual supercells with ultra-high resolution. By achieving this level of precision (Figure 42), it becomes possible to resolve existing mysteries and shed light on unresolved questions, thereby propelling further advancements in the field.[46, 558] Moreover, the techniques that can enable the simultaneous imaging of supercells, assessment of the moiré potential, and characterization of correlated states through optical spectroscopy measurements hold significant value in advancing the understanding of moiré systems. Several promising techniques,[559-562] including near-field scanning optical microscopy, STM tip-induced luminescence and cathodoluminescence (CL) image could show potential in this regard. Some studies have explored and calculated the status of trapped moiré excitons in the moiré potential,[243] but the techniques that specifically target the determination of moiré exciton positions are highly desired although it is more challenging, which will address uncertainties and validate assumptions made during the interpretation of early experimental results.

Another important avenue is manipulating moiré photonic and optoelectronic properties through various external factors.[30, 563] For instance, the application of an electric field has the capacity to modify the overall moiré potential minima due to distinct local static dipole moments. Similarly, strain can reshape the moiré potential landscape and impact the optoelectronic response, while light can highly influence spin-spin interactions, offering the potential for all-optical control of emergent phenomena. On the other hand, the lattice reconstruction has been found in vdW TMD heterostructures recently[45], but the specific effects of this on the more complex moiré exciton performance have not been fully explored. The improvement of the moiré exciton emission rate is also another crucial direction that requires further exploration although it poses certain challenges. Typically, increasing the excitation power can lead to a higher emission intensity of excitons, but this approach also raises the possibility of helping the localized excitons to escape the trapping caused by moiré potentials. Thus the optimization of the emission rate in the moiré system has to be considered through incorporation with some external means, such as reducing the dielectric effect by adjusting the surrounding environment[34] or integrating with plasmonic[564] or metasurface[362] microcavities. Furthermore, moiré optoelectronics holds great promise for on-chip applications, such as the development of good-performance moiré lasers, programmable quantum light sources and quantum simulations of the Bose-Hubbard model. Notable recent achievements include the realization of ultra-low-threshold broadband excitonic lasing and the creation of spatially ordered quantum emitters through enhanced Coulomb interactions in moiré superlattices. Moiré quantum light sources exhibit high degree of programmability, with polarization configurations and wavelengths controlled by various factors, providing a foundation for exploring coherent quantum phenomena and developing quantum technologies.

In conclusion, twisted vdW quantum materials have garnered significant attention for their great potential in fundamental science and applications. Their outstanding performance makes them promising for future scientific and technological advancements. The continuing further exploration in this field, including the expansion of material families, the development of more precise construction techniques, advanced characterization methods, and improved performance, holds great promise for transformative applications in energy efficiency, optoelectronics, and quantum information processing. Integrating moiré superlattices with other optical systems could lead to a plethora of applications in photonics and optoelectronics, like quantum nonlinear optics, laser technologies, imaging arrays, modulation and switching devices, as well as polarization devices, further brightening the future of the field. Moiré superlattices are expected to continue to be an exhilarating platform for more electronic and photonic quantum breakthroughs in the coming decade.

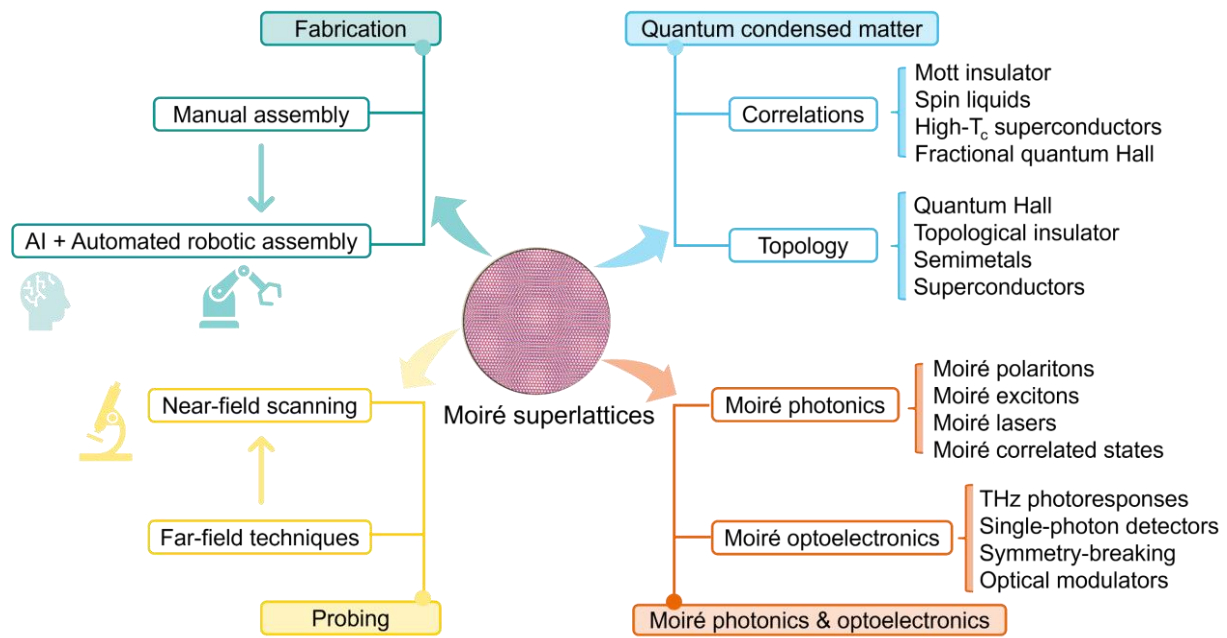

**Figure 42 | Perspectives for twisted vdW quantum materials.**

**Author information**

**Corresponding Author**

Yuerui Lu - *School of Engineering, College of Engineering and Computer Science, the Australian National University, Canberra, ACT, 2601, Australia; Australian Research Council Centre of Excellence for Quantum Computation and Communication Technology, the Australian National University, Canberra, ACT, 2601 Australia;* Email: yuerui.lu@anu.edu.au

**Authors**

Xueqian Sun - *School of Engineering, College of Engineering and Computer Science, the Australian National University, Canberra, ACT, 2601, Australia*

Manuka Suriyage - *School of Engineering, College of Engineering and Computer Science, the Australian National University, Canberra, ACT, 2601, Australia*

Ahmed Khan - *Department of Industrial and Manufacturing Engineering, University of Engineering and Technology Lahore (Rachna College Campus), Gujranwala, Pakistan*

Mingyuan Gao - *School of Engineering, College of Engineering and Computer Science, the Australian National University, Canberra, ACT, 2601, Australia; College of Engineering and Technology, Southwest University, Chongqing 400716, China*

Jie Zhao - *Department of Quantum Science & Technology, Research School of Physics, The Australian National University, Canberra ACT 2601 Australia; Australian Research Council Centre of Excellence for Quantum Computation and Communication Technology, the Australian National University, Canberra, ACT, 2601 Australia*

Boqing Liu - *School of Engineering, College of Engineering and Computer Science, the Australian National University, Canberra, ACT, 2601, Australia*

Mehedi Hasan - *School of Engineering, College of Engineering and Computer Science, the Australian National University, Canberra, ACT, 2601, Australia*

Sharidya Rahman - *Department of Materials Science and Engineering, Monash University, Clayton, Victoria, 3800 Australia; ARC Centre of Excellence in Exciton Science, Monash University, Clayton, Victoria, 3800 Australia*

Ruosi Chen - *School of Engineering, College of Engineering and Computer Science, the Australian National University, Canberra, ACT, 2601, Australia*

Ping Koy Lam - *Department of Quantum Science & Technology, Research School of Physics, The Australian National University, Canberra ACT 2601 Australia; Institute of Materials Research and Engineering (IMRE), Agency for Science, Technology and Research (A\*STAR), 2 Fusionopolis Way, 08-03 Innovis 138634, Singapore; Australian Research Council Centre of Excellence for Quantum Computation and Communication Technology, the Australian National University, Canberra, ACT, 2601 Australia*

## Author contributions

Y. L. conceived and supervised this study; X. S. writing-original draft, writing-review & editing; M. S. writing-original draft; A. K. writing-original draft; M. G. writing-original draft; J. Z. writing-original draft; B. L. writing-original draft; M. H. writing-original draft; S. R. writing-original draft; R. C. writing-original draft; P. L. writing-review & editing and Y. L. conceptualization, writing-review & editing; All authors contributed to the preparation of the manuscript.

## Notes

The authors declare no competing interests.

## Biographies

Xueqian Sun received her B.E degree in Material and Mechanical Engineering from the Australia National University in 2017 with first-class honors, and now she is a PhD candidate under the supervision of Professor Yuerui (Larry) Lu in the School of Engineering, ANU. Her research interests include the nano-engineering of exciton science in novel 2D semiconducting materials, with an emphasis on optical properties and device fabrication, and fundamental theories in light−matter interactions.

Manuka Suriyage earned his B.Sc. in Mechanical Engineering with a specialization in Mechatronics from the University of Moratuwa, Sri Lanka, in 2020. During his undergraduate studies, he delved into the intricacies of Microelectromechanical Systems (MEMS) and Nanoelectromechanical Systems (NEMS) technology, conducting research under the guidance of Prof. Ranjith Amarasinghe. Now he is a PhD candidate under the supervision of Professor Yuerui (Larry) Lu in the School of Engineering, ANU. His central research focuses on the compelling fields of 2D materials, moiré superlattices, and photonics, concentrating on investigating phonon polaritons and exciton-polaritons.

Ahmed Raza Khan is working as Assistant Professor in Industrial and Manufacturing Engineering Department, University of Engineering and Technology Lahore (Rachna Campus Gujranwala). Ahmed did his PhD and research assistantship under the supervision of Professor Yuerui Lu in Research School of Engineering, Australian National University. His research interests include linear and nonlinear optics, strain-engineering of nano-materials and non-conventional machining processes.

Mingyuan Gao is an Associate Professor in the College of Engineering and Technology at Southwest University, China, and a Postdoctoral Researcher in the College of Engineering and Computer Science at the Australian National University. He earned his B.E. and M.E.

from Huazhong University of Science and Technology, China, in 2008 and 2010, and his Ph.D. from Southwest Jiaotong University in 2015. He was a Senior Mechanical Engineer at Siemens AG from 2011 to 2018. His research interests include flexible electronics, energy harvesting, wearable devices, and nonlinear dynamics. He won the 2010 GE Foundation Tech Award-First Place Award.

Jie Zhao received her Ph.D. degree from the Australian National University working on continuous-variable quantum information. After a short period working at ANU as a postdoctoral fellow, she moved to Canada and joined Xanadu Quantum Technologies. There she worked on developing hardware prototypes for a universal photonic quantum computer. In 2020, she moved to the US and started her journey as a postdoctoral researcher at the Joint Quantum Institute, NIST & University of Mayland. Her research in the US focused on quantum applications based on warm atomic vapor. She is now a research fellow at the Australian National University. Her research interests include photonic quantum computation and networking, and quantum information.

Boqing Liu received his B.E degree in Material and Mechanical Engineering from the Australia National University, and now he is a PhD candidate under the supervision of Professor Yuerui Lu in the School of Engineering, ANU. His research interests include the strain-engineering induced linear and nonlinear optics studies of 2D materials.

Mehedi Hasan is a Ph.D. candidate under the supervision of Professor Yuerui (Larry) Lu in the School of Engineering, ANU. His research interests include the nanoengineering of exciton science in 2D organic semiconductors.

Sharidya Rahman received his BEng Degree from the Faculty of Engineering, University of Malaya, Malaysia in 2017 and Doctorate (PhD) from the School of Engineering, Australian National University (ANU), Australia. Currently he is a Research Fellow at Material Science and Engineering Department at Monash University within ARC Centre of Excellence in Exciton Science. His research interests focus on novel and emerging nanomaterials including perovskite materials, Thin Films, Colloidal Quantum Dots, 2D materials and their Heterostructures for future applications in multifarious optoelectronics and nano-devices.

Ruo-Si Chen is a PhD student in School of Engineering, Australian National University. Her research interests include 2D material-based electronic devices, 2D interfaces engineering combined with their applications in bio-emulation and artificial synapse and neuron field.

Ping Koy Lam completed his degree with a double major in Mathematics and Physics from the University of Auckland in 1990. He worked as a process engineer for Sony (Audio Electronics) and Hewlett-Packard (Semiconductor LED) for 3 years prior to his post-graduate studies at the ANU where he obtained a Masters in theoretical physics, and a PhD in experimental physics. He was awarded the Australian Institute of Physics Bragg Medal and the ANU Crawford Prize for his PhD in 1999. He was an Alexander von Humboldt fellow at the Erlangen-Nürnberg Universität in 2000 and a CNRS visiting professor at Paris University in 2007. He was awarded the 2003 British Council Eureka Prize for inspiring science (Quantum Teleportation) and the 2006 UNSW Eureka Prize for innovative research (Quantum Cryptography). Ping Koy was the chief scientist and co-founder of QuintessenceLabs Pty. Ltd, a spin-off company of his group that commercialises quantum communication technology. His research interests include quantum optics, optical metrology, nonlinear optics and quantum information. Within the Centre, he manages the ANU node and is the quantum communication work package leader. He is also A*STAR chief quantum scientist at Institute of Materials Research and Engineering (IMRE). He has published more than 200 scientific articles with more than 35 papers appearing in Physical Review Letters, Science and the Nature suite journals. His research expertise is in the field of quantum optics and metrology.

Yuerui (Larry) Lu received his Ph.D. degree from Cornell University, the school of Electrical and Computer Engineering, in 2012. He holds a B.S. degree from department of Applied Physics at University of Science and Technology of China. In 2013, he joined the Australian National University as a research fellow and lecturer under the Future Engineering Research Leadership Fellowship. He is now a full professor at the ANU. He is a chief investigator and program manager at the Australian Research Council Centre of Excellence for Quantum Computation and Communication Technology (CQC2T). His research interests include 2D quantum materials and optoelectronic devices, MEMS sensors and actuators, biomedical and energy devices, etc. He has published 2 books (as editor), 7 book chapters, and more than 100 papers in high-impact journals, including Nature, Science, Nature Physics, Nature Communications, etc. He is serving as a reviewer for many top-tier journals, including Nature, Science, Nature Materials, Nature Photonics, Nature Communications, etc. He is serving as an associate editor for the nature publishing group journal Scientific Reports.

## Acknowledgments

The authors would like to acknowledge funding support from the ANU PhD student scholarship, Australian Research Council (grant No.: DP220102219, DP180103238, LE200100032), ARC Centre of Excellence in Quantum Computation and Communication Technology (project number CE170100012) and National Heart Foundation (ARIES ID: 35852).

# References

1.      Zhang, L.; Hasan, M. M.; Tang, Y.; Khan, A. R.; Yan, H.; Yildirim, T.; Sun, X.; Zhang, J.; Zhu, J.; Zhang, Y., 2D organic single crystals: Synthesis, novel physics, high-performance optoelectronic devices and integration. *Materials Today* **2021**, *50*, 442-475.

2.      Zhu, Y.; Sun, X.; Tang, Y.; Fu, L.; Lu, Y., Two-dimensional materials for light emitting applications: Achievement, challenge and future perspectives. *Nano Research* **2021**, *14*, 1912-1936.

3.      Li, X. L.; Han, W. P.; Wu, J. B.; Qiao, X. F.; Zhang, J.; Tan, P. H., Layer-number dependent optical properties of 2D materials and their application for thickness determination. *Advanced Functional Materials* **2017**, *27* (19), 1604468.

4.      Pandey, S. K.; Das, R.; Mahadevan, P., Layer-dependent electronic structure changes in transition metal dichalcogenides: the microscopic origin. *ACS omega* **2020**, *5* (25), 15169-15176.

5.      Ruppert, C.; Aslan, B.; Heinz, T. F., Optical properties and band gap of single-and few-layer MoTe2 crystals. *Nano letters* **2014**, *14* (11), 6231-6236.

6.      Zhao, M.; Ye, Z.; Suzuki, R.; Ye, Y.; Zhu, H.; Xiao, J.; Wang, Y.; Iwasa, Y.; Zhang, X., Atomically phase-matched second-harmonic generation in a 2D crystal. *Light: Science & Applications* **2016**, *5* (8), e16131-e16131.

7.      Pei, J.; Yang, J.; Xu, R.; Zeng, Y. H.; Myint, Y. W.; Zhang, S.; Zheng, J. C.; Qin, Q.; Wang, X.; Jiang, W., Exciton and trion dynamics in bilayer MoS2. *Small* **2015**, *11* (48), 6384-6390.

8.      Neupane, G. P.; Ma, W.; Yildirim, T.; Tang, Y.; Zhang, L.; Lu, Y., 2D organic semiconductors, the future of green nanotechnology. *Nano Materials Science* **2019**, *1* (4), 246-259.

9.      Vogl, T.; Doherty, M. W.; Buchler, B. C.; Lu, Y.; Lam, P. K., Atomic localization of quantum emitters in multilayer hexagonal boron nitride. *Nanoscale* **2019**, *11* (30), 14362-14371.

10.     Vogl, T.; Campbell, G.; Buchler, B. C.; Lu, Y.; Lam, P. K., Fabrication and deterministic transfer of high-quality quantum emitters in hexagonal boron nitride. *ACS Photonics* **2018**, *5* (6), 2305-2312.

11.     Vogl, T.; Lu, Y.; Lam, P. K., Room temperature single photon source using fiber-integrated hexagonal boron nitride. *Journal of Physics D: Applied Physics* **2017**, *50* (29), 295101.

12.     Tebyetekerwa, M.; Zhang, J.; Liang, K.; Duong, T.; Neupane, G. P.; Zhang, L.; Liu, B.; Truong, T. N.; Basnet, R.; Qiao, X., Quantifying Quasi-Fermi Level Splitting and Mapping Its Heterogeneity in Atomically Thin Transition Metal Dichalcogenides. *Advanced Materials* **2019**, *31* (25), 1900522.

13.     Blundo, E.; Di Giorgio, C.; Pettinari, G.; Yildirim, T.; Felici, M.; Lu, Y.; Bobba, F.; Polimeni, A., Engineered creation of periodic giant, nonuniform strains in MoS2 monolayers. *Advanced Materials Interfaces* **2020**, *7* (17), 2000621.

14.     Calman, E.; Fowler-Gerace, L.; Choksy, D.; Butov, L.; Nikonov, D.; Young, I.; Hu, S.; Mishchenko, A.; Geim, A., Indirect excitons and trions in MoSe2/WSe2 van der Waals heterostructures. *Nano letters* **2020**, *20* (3), 1869-1875.

15.     Miller, B.; Steinhoff, A.; Pano, B.; Klein, J.; Jahnke, F.; Holleitner, A.; Wurstbauer, U., Long-lived direct and indirect interlayer excitons in van der Waals heterostructures. *Nano letters* **2017**, *17* (9), 5229-5237.

16.     Novoselov, K. S.; Mishchenko, A.; Carvalho, o. A.; Castro Neto, A., 2D materials and van der Waals heterostructures. *Science* **2016**, *353* (6298), aac9439.

17.     Jiang, Y.; Chen, S.; Zheng, W.; Zheng, B.; Pan, A., Interlayer exciton formation, relaxation, and transport in TMD van der Waals heterostructures. *Light: Science & Applications* **2021**, *10* (1), 72.

18.     Ohta, T.; Robinson, J. T.; Feibelman, P. J.; Bostwick, A.; Rotenberg, E.; Beechem, T. E., Evidence for interlayer coupling and moiré periodic potentials in twisted bilayer graphene. *Physical Review Letters* **2012**, *109* (18), 186807.

19.     Liao, M.; Wu, Z.-W.; Du, L.; Zhang, T.; Wei, Z.; Zhu, J.; Yu, H.; Tang, J.; Gu, L.; Xing, Y., Twist angle-dependent conductivities across MoS2/graphene heterojunctions. *Nature communications* **2018**, *9* (1), 1-6.

20.     Lin, K.-Q.; Faria Junior, P. E.; Bauer, J. M.; Peng, B.; Monserrat, B.; Gmitra, M.; Fabian, J.; Bange, S.; Lupton, J. M., Twist-angle engineering of excitonic quantum interference and optical nonlinearities in stacked 2D semiconductors. *Nature communications* **2021**, *12* (1), 1553.

21.     Nayak, P. K.; Horbatenko, Y.; Ahn, S.; Kim, G.; Lee, J.-U.; Ma, K. Y.; Jang, A.-R.; Lim, H.; Kim, D.; Ryu, S., Probing evolution of twist-angle-dependent interlayer excitons in MoSe2/WSe2 van der Waals heterostructures. *ACS nano* **2017**, *11* (4), 4041-4050.

22.     Wang, Z.-J.; Dong, J.; Cui, Y.; Eres, G.; Timpe, O.; Fu, Q.; Ding, F.; Schlögl, R.; Willinger, M.-G., Stacking sequence and interlayer coupling in few-layer graphene revealed by in situ imaging. *Nature Communications* **2016**, *7* (1), 13256.

23.     Zhang, N.; Surrente, A.; Baranowski, M.; Maude, D. K.; Gant, P.; Castellanos-Gomez, A.; Plochocka, P., Moiré intralayer excitons in a MoSe2/MoS2 heterostructure. *Nano letters* **2018**, *18* (12), 7651-7657.

24.     Tran, K.; Moody, G.; Wu, F.; Lu, X.; Choi, J.; Kim, K.; Rai, A.; Sanchez, D. A.; Quan, J.; Singh, A., Evidence for moiré excitons in van der Waals heterostructures. *Nature* **2019**, *567* (7746), 71-75.

25.     Li, W.; Lu, X.; Dubey, S.; Devenica, L.; Srivastava, A., Dipolar interactions between localized interlayer excitons in van der Waals heterostructures. *Nature Materials* **2020**, *19* (6), 624-629.

26.     Alden, J. S.; Tsen, A. W.; Huang, P. Y.; Hovden, R.; Brown, L.; Park, J.; Muller, D. A.; McEuen, P. L., Strain solitons and topological defects in bilayer graphene. *Proceedings of the National Academy of Sciences* **2013**, *110* (28), 11256-11260.

27.     Seyler, K. L.; Rivera, P.; Yu, H.; Wilson, N. P.; Ray, E. L.; Mandrus, D. G.; Yan, J.; Yao, W.; Xu, X., Signatures of moiré-trapped valley excitons in MoSe2/WSe2 heterobilayers. *Nature* **2019**, *567* (7746), 66-70.

28.     Haddadi, F.; Wu, Q.; Kruchkov, A. J.; Yazyev, O. V., Moiré Flat Bands in Twisted Double Bilayer Graphene. *Nano Letters* **2020**, *20* (4), 2410-2415.

29.     Brem, S.; Malic, E., Bosonic Delocalization of Dipolar Moiré Excitons. *Nano Letters* **2023**.

30.     Yu, H.; Liu, G.-B.; Tang, J.; Xu, X.; Yao, W., Moiré excitons: From programmable quantum emitter arrays to spin-orbit–coupled artificial lattices. *Science advances* **2017**, *3* (11), e1701696.

31.     Yuan, L.; Zheng, B.; Kunstmann, J.; Brumme, T.; Kuc, A. B.; Ma, C.; Deng, S.; Blach, D.; Pan, A.; Huang, L., Twist-angle-dependent interlayer exciton diffusion in WS2–WSe2 heterobilayers. *Nature materials* **2020**, *19* (6), 617-623.

32.     Baek, H.; Brotons-Gisbert, M.; Koong, Z. X.; Campbell, A.; Rambach, M.; Watanabe, K.; Taniguchi, T.; Gerardot, B. D., Highly energy-tunable quantum light from moiré-trapped excitons. *Science advances* **2020**, *6* (37), eaba8526.

33.     Kremser, M.; Brotons-Gisbert, M.; Knörzer, J.; Gückelhorn, J.; Meyer, M.; Barbone, M.; Stier, A. V.; Gerardot, B. D.; Müller, K.; Finley, J. J., Discrete interactions between a few interlayer excitons trapped at a MoSe2–WSe2 heterointerface. *npj 2D Materials and Applications* **2020**, *4* (1), 8.

34.     Sun, X.; Zhu, Y.; Qin, H.; Liu, B.; Tang, Y.; Lü, T.; Rahman, S.; Yildirim, T.; Lu, Y., Enhanced interactions of interlayer excitons in free-standing heterobilayers. *Nature* **2022**, *610* (7932), 478-484.

35.     Jin, C.; Regan, E. C.; Yan, A.; Iqbal Bakti Utama, M.; Wang, D.; Zhao, S.; Qin, Y.; Yang, S.; Zheng, Z.; Shi, S., Observation of moiré excitons in WSe2/WS2 heterostructure superlattices. *Nature* **2019**, *567* (7746), 76-80.

36.     Zhang, C.; Chuu, C.-P.; Ren, X.; Li, M.-Y.; Li, L.-J.; Jin, C.; Chou, M.-Y.; Shih, C.-K., Interlayer couplings, Moiré patterns, and 2D electronic superlattices in MoS2/WSe2 hetero-bilayers. *Science advances* **2017**, *3* (1), e1601459.

37.     Cao, Y.; Fatemi, V.; Fang, S.; Watanabe, K.; Taniguchi, T.; Kaxiras, E.; Jarillo-Herrero, P., Unconventional superconductivity in magic-angle graphene superlattices. *Nature* **2018**, *556* (7699), 43-50.

38.     Balents, L.; Dean, C. R.; Efetov, D. K.; Young, A. F., Superconductivity and strong correlations in moiré flat bands. *Nature Physics* **2020**, *16* (7), 725-733.

39.     Yankowitz, M.; Chen, S.; Polshyn, H.; Zhang, Y.; Watanabe, K.; Taniguchi, T.; Graf, D.; Young, A. F.; Dean, C. R., Tuning superconductivity in twisted bilayer graphene. *Science* **2019**, *363* (6431), 1059-1064.

40.     Huang, S.; Kim, K.; Efimkin, D. K.; Lovorn, T.; Taniguchi, T.; Watanabe, K.; MacDonald, A. H.; Tutuc, E.; LeRoy, B. J., Topologically protected helical states in minimally twisted bilayer graphene. *Physical review letters* **2018**, *121* (3), 037702.

41.     Vaezi, A.; Liang, Y.; Ngai, D. H.; Yang, L.; Kim, E.-A., Topological edge states at a tilt boundary in gated multilayer graphene. *Physical Review X* **2013**, *3* (2), 021018.

42.     Shimazaki, Y.; Schwartz, I.; Watanabe, K.; Taniguchi, T.; Kroner, M.; Imamoğlu, A., Strongly correlated electrons and hybrid excitons in a moiré heterostructure. *Nature* **2020**, *580* (7804), 472-477.

43.     Liu, K.; Zhang, L.; Cao, T.; Jin, C.; Qiu, D.; Zhou, Q.; Zettl, A.; Yang, P.; Louie, S. G.; Wang, F., Evolution of interlayer coupling in twisted molybdenum disulfide bilayers. *Nature Communications* **2014**, *5* (1), 4966.

44.     Morell, E. S.; Vargas, P.; Chico, L.; Brey, L., Charge redistribution and interlayer coupling in twisted bilayer graphene under electric fields. *Physical Review B* **2011**, *84* (19), 195421.

45.     Rosenberger, M. R.; Chuang, H.-J.; Phillips, M.; Oleshko, V. P.; McCreary, K. M.; Sivaram, S. V.; Hellberg, C. S.; Jonker, B. T., Twist angle-dependent atomic reconstruction and moiré patterns in transition metal dichalcogenide heterostructures. *ACS nano* **2020**, *14* (4), 4550-4558.

46.     Weston, A.; Zou, Y.; Enaldiev, V.; Summerfield, A.; Clark, N.; Zólyomi, V.; Graham, A.; Yelgel, C.; Magorrian, S.; Zhou, M., Atomic reconstruction in twisted bilayers of transition metal dichalcogenides. *Nature Nanotechnology* **2020**, *15* (7), 592-597.

47.     Zhang, L.; Zhang, Z.; Wu, F.; Wang, D.; Gogna, R.; Hou, S.; Watanabe, K.; Taniguchi, T.; Kulkarni, K.; Kuo, T., Twist-angle dependence of moiré excitons in WS2/MoSe2 heterobilayers. *Nature communications* **2020**, *11* (1), 5888.

48.     Säynätjoki, A.; Karvonen, L.; Rostami, H.; Autere, A.; Mehravar, S.; Lombardo, A.; Norwood, R. A.; Hasan, T.; Peyghambarian, N.; Lipsanen, H., Ultra-strong nonlinear optical processes and trigonal warping in MoS2 layers. *Nature communications* **2017**, *8* (1), 893.

49.     Yu, H.; Yang, Z.; Du, L.; Zhang, J.; Shi, J.; Chen, W.; Chen, P.; Liao, M.; Zhao, J.; Meng, J., Precisely Aligned Monolayer MoS2 epitaxially grown on h-BN basal plane. *Small* **2017**, *13* (7), 1603005.

50.     Hsu, W.-T.; Zhao, Z.-A.; Li, L.-J.; Chen, C.-H.; Chiu, M.-H.; Chang, P.-S.; Chou, Y.-C.; Chang, W.-H., Second harmonic generation from artificially stacked transition metal dichalcogenide twisted bilayers. *ACS nano* **2014**, *8* (3), 2951-2958.

51.     Park, J. M.; Cao, Y.; Watanabe, K.; Taniguchi, T.; Jarillo-Herrero, P., Tunable strongly coupled superconductivity in magic-angle twisted trilayer graphene. *Nature* **2021**, *590* (7845), 249-255.

52.     Andrei, E. Y.; Efetov, D. K.; Jarillo-Herrero, P.; MacDonald, A. H.; Mak, K. F.; Senthil, T.; Tutuc, E.; Yazdani, A.; Young, A. F., The marvels of moiré materials. *Nature Reviews Materials* **2021**, *6* (3), 201-206.

53.     Du, L.; Molas, M. R.; Huang, Z.; Zhang, G.; Wang, F.; Sun, Z., Moiré photonics and optoelectronics. *Science* **2023**, *379* (6639), eadg0014.

54.     He, F.; Zhou, Y.; Ye, Z.; Cho, S.-H.; Jeong, J.; Meng, X.; Wang, Y., Moiré patterns in 2D materials: A review. *ACS nano* **2021**, *15* (4), 5944-5958.

55.     Woods, C.; Britnell, L.; Eckmann, A.; Ma, R.; Lu, J.; Guo, H.; Lin, X.; Yu, G.; Cao, Y.; Gorbachev, R. V., Commensurate–incommensurate transition in graphene on hexagonal boron nitride. *Nature physics* **2014**, *10* (6), 451-456.

56.     Saveljev, V.; Kim, J.; Son, J.-Y.; Kim, Y.; Heo, G., Static moiré patterns in moving grids. *Scientific Reports* **2020**, *10* (1), 14414.

57.     Fu, Q.; Wang, P.; Huang, C.; Kartashov, Y. V.; Torner, L.; Konotop, V. V.; Ye, F., Optical soliton formation controlled by angle twisting in photonic moiré lattices. *Nature Photonics* **2020**, *14* (11), 663-668.

58.     Amidror, I., *The Theory of the Moiré Phenomenon: Volume I: Periodic Layers*. Springer Science & Business Media: 2009; Vol. 38.

59.     Liao, M.; Wei, Z.; Du, L.; Wang, Q.; Tang, J.; Yu, H.; Wu, F.; Zhao, J.; Xu, X.; Han, B., Precise control of the interlayer twist angle in large scale MoS2 homostructures. *Nature communications* **2020**, *11* (1), 2153.

60.     Marcellina, E.; Liu, X.; Hu, Z.; Fieramosca, A.; Huang, Y.; Du, W.; Liu, S.; Zhao, J.; Watanabe, K.; Taniguchi, T., Evidence for moiré trions in twisted MoSe2 homobilayers. *Nano Letters* **2021**, *21* (10), 4461-4468.

61.     Wang, L.; Zihlmann, S.; Liu, M.-H.; Makk, P.; Watanabe, K.; Taniguchi, T.; Baumgartner, A.; Schönenberger, C., New generation of moiré superlattices in doubly aligned hBN/graphene/hBN heterostructures. *Nano letters* **2019,** *19* (4), 2371-2376.

62.     Jin, C.; Tao, Z.; Li, T.; Xu, Y.; Tang, Y.; Zhu, J.; Liu, S.; Watanabe, K.; Taniguchi, T.; Hone, J. C., Stripe phases in WSe2/WS2 moiré superlattices. *Nature Materials* **2021,** *20* (7), 940-944.

63.     Reidy, K.; Varnavides, G.; Thomsen, J. D.; Kumar, A.; Pham, T.; Blackburn, A. M.; Anikeeva, P.; Narang, P.; LeBeau, J. M.; Ross, F. M., Direct imaging and electronic structure modulation of moiré superlattices at the 2D/3D interface. *Nature communications* **2021,** *12* (1), 1290.

64.     Yi, C.-H.; Park, H. C.; Park, M. J., Strong interlayer coupling and stable topological flat bands in twisted bilayer photonic Moiré superlattices. *Light: Science & Applications* **2022,** *11* (1), 289.

65.     Song, T.; Sun, Q.-C.; Anderson, E.; Wang, C.; Qian, J.; Taniguchi, T.; Watanabe, K.; McGuire, M. A.; Stöhr, R.; Xiao, D., Direct visualization of magnetic domains and moiré magnetism in twisted 2D magnets. *Science* **2021,** *374* (6571), 1140-1144.

66.     Wang, X.; Xiao, C.; Park, H.; Zhu, J.; Wang, C.; Taniguchi, T.; Watanabe, K.; Yan, J.; Xiao, D.; Gamelin, D. R., Light-induced ferromagnetism in moiré superlattices. *Nature* **2022,** *604* (7906), 468-473.

67.     Kerelsky, A.; McGilly, L. J.; Kennes, D. M.; Xian, L.; Yankowitz, M.; Chen, S.; Watanabe, K.; Taniguchi, T.; Hone, J.; Dean, C., Maximized electron interactions at the magic angle in twisted bilayer graphene. *Nature* **2019,** *572* (7767), 95-100.

68.     Frisenda, R.; Navarro-Moratalla, E.; Gant, P.; De Lara, D. P.; Jarillo-Herrero, P.; Gorbachev, R. V.; Castellanos-Gomez, A., Recent progress in the assembly of nanodevices and van der Waals heterostructures by deterministic placement of 2D materials. *Chemical Society Reviews* **2018,** *47* (1), 53-68.

69.     Carr, S.; Fang, S.; Kaxiras, E., Electronic-structure methods for twisted moiré layers. *Nature Reviews Materials* **2020,** *5* (10), 748-763.

70.     Morimoto, Y.; Fujigaki, M. In *Theory and application of sampling moire method*, Recent Advances in Mechanics: Selected Papers from the Symposium on Recent Advances in Mechanics, Academy of Athens, Athens, Greece, 17-19 September, 2009, Organised by the Pericles S. Theocaris Foundation in Honour of PS Theocaris, on the Tenth Anniversary of His Death, Springer: 2011; pp 227-248.

71.     Kotova, L. V.; Rakhlin, M. V.; Galimov, A. I.; Eliseyev, I. A.; Borodin, B. R.; Platonov, A. V.; Kirilenko, D. A.; Poshakinskiy, A. V.; Shubina, T. V., MoS 2 flake as a van der Waals homostructure: luminescence properties and optical anisotropy. *Nanoscale* **2021,** *13* (41), 17566-17575.

72.     He, J.; Hummer, K.; Franchini, C., Stacking effects on the electronic and optical properties of bilayer transition metal dichalcogenides MoS 2, MoSe 2, WS 2, and WSe 2. *Physical Review B* **2014,** *89* (7), 075409.

73.     Kang, P.; Zhang, W.; Michaud-Rioux, V.; Wang, X.; Yun, J.; Guo, H., Twistronics in tensile strained bilayer black phosphorus. *Nanoscale* **2020,** *12* (24), 12909-12916.

74.     Amidror, I.; Hersch, R. D., The role of Fourier theory and of modulation in the prediction of visible moiré effects. *Journal of Modern Optics* **2009,** *56* (9), 1103-1118.

75.     Amidror, I.; Hersch, R. D., Mathematical moiré models and their limitations. *Journal of Modern Optics* **2010,** *57* (1), 23-36.

76.     Hermann, K., Periodic overlayers and moiré patterns: theoretical studies of geometric properties. *Journal of Physics: Condensed Matter* **2012,** *24* (31), 314210.

77.     Kim, K.; DaSilva, A.; Huang, S.; Fallahazad, B.; Larentis, S.; Taniguchi, T.; Watanabe, K.; LeRoy, B. J.; MacDonald, A. H.; Tutuc, E., Tunable moiré bands and strong correlations in small-twist-angle bilayer graphene. *Proceedings of the National Academy of Sciences* **2017,** *114* (13), 3364-3369.

78.     Wang, J.; Zheng, Y.; Millis, A. J.; Cano, J., Chiral approximation to twisted bilayer graphene: Exact intravalley inversion symmetry, nodal structure, and implications for higher magic angles. *Physical Review Research* **2021,** *3* (2), 023155.

79.     Tang, K.; Qi, W., Moiré-pattern-tuned electronic structures of van der Waals heterostructures. *Advanced Functional Materials* **2020,** *30* (32), 2002672.

80.	Törmä, P.; Peotta, S.; Bernevig, B. A., Superconductivity, superfluidity and quantum geometry in twisted multilayer systems. *Nature Reviews Physics* **2022,** *4* (8), 528-542.

81.	Abbas, G.; Li, Y.; Wang, H.; Zhang, W. X.; Wang, C.; Zhang, H., Recent advances in twisted structures of flatland materials and crafting moiré superlattices. *Advanced Functional Materials* **2020,** *30* (36), 2000878.

82.	Chatterjee, S.; Das, S.; Gupta, G.; Watanabe, K.; Taniguchi, T.; Majumdar, K., Probing biexciton in monolayer WS2 through controlled many-body interaction. *2D Materials* **2021,** *9* (1), 015023.

83.	Zhao, Y.-X.; Zhou, X.-F.; Zhang, Y.; He, L., Oscillations of the spacing between van Hove singularities induced by sub-Ångstrom fluctuations of interlayer spacing in graphene superlattices. *Physical Review Letters* **2021,** *127* (26), 266801.

84.	Bansil, A.; Lin, H.; Das, T., Colloquium: Topological band theory. *Reviews of Modern Physics* **2016,** *88* (2), 021004.

85.	Sinha, S.; Adak, P. C.; Chakraborty, A.; Das, K.; Debnath, K.; Sangani, L. V.; Watanabe, K.; Taniguchi, T.; Waghmare, U. V.; Agarwal, A., Berry curvature dipole senses topological transition in a moiré superlattice. *Nature Physics* **2022,** *18* (7), 765-770.

86.	Kennes, D. M.; Claassen, M.; Xian, L.; Georges, A.; Millis, A. J.; Hone, J.; Dean, C. R.; Basov, D.; Pasupathy, A. N.; Rubio, A., Moiré heterostructures as a condensed-matter quantum simulator. *Nature Physics* **2021,** *17* (2), 155-163.

87.	Xu, Y.; Liu, S.; Rhodes, D. A.; Watanabe, K.; Taniguchi, T.; Hone, J.; Elser, V.; Mak, K. F.; Shan, J., Correlated insulating states at fractional fillings of moiré superlattices. *Nature* **2020,** *587* (7833), 214-218.

88.	Wilson, N. P.; Yao, W.; Shan, J.; Xu, X., Excitons and emergent quantum phenomena in stacked 2D semiconductors. *Nature* **2021,** *599* (7885), 383-392.

89.	Xie, Y.; Lian, B.; Jäck, B.; Liu, X.; Chiu, C.-L.; Watanabe, K.; Taniguchi, T.; Bernevig, B. A.; Yazdani, A., Spectroscopic signatures of many-body correlations in magic-angle twisted bilayer graphene. *Nature* **2019,** *572* (7767), 101-105.

90.	Classen, L.; Honerkamp, C.; Scherer, M. M., Competing phases of interacting electrons on triangular lattices in moiré heterostructures. *Physical Review B* **2019,** *99* (19), 195120.

91.	Tang, H.; Carr, S.; Kaxiras, E., Geometric origins of topological insulation in twisted layered semiconductors. *Physical Review B* **2021,** *104* (15), 155415.

92.	Angeli, M.; Schleder, G. R.; Kaxiras, E., Twistronics of Janus transition metal dichalcogenide bilayers. *Physical Review B* **2022,** *106* (23), 235159.

93.	Zhang, Q.; Yu, J.; Ebert, P.; Zhang, C.; Pan, C.-R.; Chou, M.-Y.; Shih, C.-K.; Zeng, C.; Yuan, S., Tuning band gap and work function modulations in monolayer hBN/Cu (111) heterostructures with Moiré patterns. *ACS nano* **2018,** *12* (9), 9355-9362.

94.	Xie, L.; Wang, L.; Zhao, W.; Liu, S.; Huang, W.; Zhao, Q., WS2 moiré superlattices derived from mechanical flexibility for hydrogen evolution reaction. *Nature communications* **2021,** *12* (1), 5070.

95.	Linnera, J.; Sansone, G.; Maschio, L.; Karttunen, A. J., Thermoelectric properties of p-type Cu2O, CuO, and NiO from hybrid density functional theory. *The Journal of Physical Chemistry C* **2018,** *122* (27), 15180-15189.

96.	Dos Santos, J. L.; Peres, N.; Neto, A. C., Continuum model of the twisted graphene bilayer. *Physical review B* **2012,** *86* (15), 155449.

97.	Mele, E. J., Commensuration and interlayer coherence in twisted bilayer graphene. *Physical Review B* **2010,** *81* (16), 161405.

98.	Shallcross, S.; Sharma, S.; Kandelaki, E.; Pankratov, O., Electronic structure of turbostratic graphene. *Physical Review B* **2010,** *81* (16), 165105.

99.	Jones, R. O., Density functional theory: Its origins, rise to prominence, and future. *Reviews of modern physics* **2015,** *87* (3), 897.

100.	Larson, D. T.; Carr, S.; Tritsaris, G. A.; Kaxiras, E., Effects of lithium intercalation in twisted bilayer graphene. *Physical Review B* **2020,** *101* (7), 075407.

101.	Minkin, A. S.; Lebedeva, I. V.; Popov, A. M.; Knizhnik, A. A., Atomic-scale defects restricting structural superlubricity: Ab initio study on the example of the twisted graphene bilayer. *Physical Review B* **2021,** *104* (7), 075444.

102.    Gharekhanlou, B.; Khorasani, S., An overview of tight-binding method for two-dimensional carbon structures. *Graphene Prop. Synth. Appl* **2011**, 1-37.

103.    Krowne, C. M., Tight-binding formulation of electronic band structure of 2D hexagonal materials. In *Advances in imaging and electron physics*, Elsevier: 2019; Vol. 210, pp 23-45.

104.    Nam, N. N.; Koshino, M., Lattice relaxation and energy band modulation in twisted bilayer graphene. *Physical Review B* **2017**, *96* (7), 075311.

105.    Solís-Fernández, P.; Ago, H., Machine learning determination of the twist angle of bilayer graphene by Raman spectroscopy: implications for van der Waals heterostructures. *ACS Applied Nano Materials* **2022**, *5* (1), 1356-1366.

106.    Choi, J.; Zhang, H.; Choi, J. H., Modulating optoelectronic properties of two-dimensional transition metal dichalcogenide semiconductors by photoinduced charge transfer. *ACS nano* **2016**, *10* (1), 1671-1680.

107.    Du, L.; Yu, H.; Liao, M.; Wang, S.; Xie, L.; Lu, X.; Zhu, J.; Li, N.; Shen, C.; Chen, P., Modulating PL and electronic structures of MoS2/graphene heterostructures via interlayer twisting angle. *Applied Physics Letters* **2017**, *111* (26).

108.    Choi, W.; Akhtar, I.; Rehman, M. A.; Kim, M.; Kang, D.; Jung, J.; Myung, Y.; Kim, J.; Cheong, H.; Seo, Y., Twist-angle-dependent optoelectronics in a few-layer transition-metal dichalcogenide heterostructure. *ACS applied materials & interfaces* **2018**, *11* (2), 2470-2478.

109.    Lopatin, S.; Aljarb, A.; Roddatis, V.; Meyer, T.; Wan, Y.; Fu, J.-H.; Hedhili, M.; Han, Y.; Li, L.-J.; Tung, V., Aberration-corrected STEM imaging of 2D materials: Artifacts and practical applications of threefold astigmatism. *Science Advances* **2020**, *6* (37), eabb8431.

110.    Chang, Y.-Y.; Han, H. N.; Kim, M., Analyzing the microstructure and related properties of 2D materials by transmission electron microscopy. *Applied Microscopy* **2019**, *49*, 1-7.

111.    Sousa, A. A.; Leapman, R. D., Development and application of STEM for the biological sciences. *Ultramicroscopy* **2012**, *123*, 38-49.

112.    Guo, X.; Zou, C.-l.; Schuck, C.; Jung, H.; Cheng, R.; Tang, H. X., Parametric down-conversion photon-pair source on a nanophotonic chip. *Light: Science & Applications* **2017**, *6* (5), e16249-e16249.

113.    Shibata, N.; Seki, T.; Sánchez-Santolino, G.; Findlay, S. D.; Kohno, Y.; Matsumoto, T.; Ishikawa, R.; Ikuhara, Y., Electric field imaging of single atoms. *Nature communications* **2017**, *8* (1), 15631.

114.    Dumcenco, D.; Ovchinnikov, D.; Marinov, K.; Lazic, P.; Gibertini, M.; Marzari, N.; Sanchez, O. L.; Kung, Y.-C.; Krasnozhon, D.; Chen, M.-W., Large-area epitaxial monolayer MoS2. *ACS nano* **2015**, *9* (4), 4611-4620.

115.    Yamashita, S.; Kikkawa, J.; Yanagisawa, K.; Nagai, T.; Ishizuka, K.; Kimoto, K., Atomic number dependence of Z contrast in scanning transmission electron microscopy. *Scientific reports* **2018**, *8* (1), 12325.

116.    Hovden, R.; Liu, P.; Schnitzer, N.; Tsen, A. W.; Liu, Y.; Lu, W.; Sun, Y.; Kourkoutis, L. F., Thickness and stacking sequence determination of exfoliated dichalcogenides (1T-TaS2, 2H-MoS2) using scanning transmission electron microscopy. *Microscopy and Microanalysis* **2018**, *24* (4), 387-395.

117.    Shi, J.; Yu, P.; Liu, F.; He, P.; Wang, R.; Qin, L.; Zhou, J.; Li, X.; Zhou, J.; Sui, X., 3R MoS2 with broken inversion symmetry: a promising ultrathin nonlinear optical device. *Advanced Materials* **2017**, *29* (30), 1701486.

118.    Wu, W.; Wang, L.; Li, Y.; Zhang, F.; Lin, L.; Niu, S.; Chenet, D.; Zhang, X.; Hao, Y.; Heinz, T. F., Piezoelectricity of single-atomic-layer MoS2 for energy conversion and piezotronics. *Nature* **2014**, *514* (7523), 470-474.

119.    Wang, H.; Qian, X., Giant optical second harmonic generation in two-dimensional multiferroics. *Nano letters* **2017**, *17* (8), 5027-5034.

120.    Moutinho, M. V.; Venezuela, P.; Pimenta, M. A., Raman Spectroscopy of Twisted Bilayer Graphene. *C* **2021**, *7* (1), 10.

121.    Quan, J.; Linhart, L.; Lin, M.-L.; Lee, D.; Zhu, J.; Wang, C.-Y.; Hsu, W.-T.; Choi, J.; Embley, J.; Young, C.; Taniguchi, T.; Watanabe, K.; Shih, C.-K.; Lai, K.; MacDonald, A. H.; Tan, P.-H.; Libisch, F.; Li, X., Phonon renormalization in reconstructed MoS2 moiré superlattices. *Nature Materials* **2021**, *20* (8), 1100-1105.

122.     Decker, R.; Wang, Y.; Brar, V. W.; Regan, W.; Tsai, H.-Z.; Wu, Q.; Gannett, W.; Zettl, A.; Crommie, M. F., Local electronic properties of graphene on a BN substrate via scanning tunneling microscopy. *Nano letters* **2011,** *11* (6), 2291-2295.

123.     Dean, C. R.; Wang, L.; Maher, P.; Forsythe, C.; Ghahari, F.; Gao, Y.; Katoch, J.; Ishigami, M.; Moon, P.; Koshino, M., Hofstadter's butterfly and the fractal quantum Hall effect in moiré superlattices. *Nature* **2013,** *497* (7451), 598-602.

124.     Hunt, B.; Sanchez-Yamagishi, J. D.; Young, A. F.; Yankowitz, M.; LeRoy, B. J.; Watanabe, K.; Taniguchi, T.; Moon, P.; Koshino, M.; Jarillo-Herrero, P., Massive Dirac fermions and Hofstadter butterfly in a van der Waals heterostructure. *Science* **2013,** *340* (6139), 1427-1430.

125.     Bistritzer, R.; MacDonald, A. H., Moiré bands in twisted double-layer graphene. *Proceedings of the National Academy of Sciences* **2011,** *108* (30), 12233-12237.

126.     Cao, Y.; Fatemi, V.; Demir, A.; Fang, S.; Tomarken, S. L.; Luo, J. Y.; Sanchez-Yamagishi, J. D.; Watanabe, K.; Taniguchi, T.; Kaxiras, E., Correlated insulator behaviour at half-filling in magic-angle graphene superlattices. *Nature* **2018,** *556* (7699), 80-84.

127.     Cao, Y.; Luo, J.; Fatemi, V.; Fang, S.; Sanchez-Yamagishi, J.; Watanabe, K.; Taniguchi, T.; Kaxiras, E.; Jarillo-Herrero, P., Superlattice-induced insulating states and valley-protected orbits in twisted bilayer graphene. *Physical review letters* **2016,** *117* (11), 116804.

128.     Yu, Y.; Zhang, K.; Parks, H.; Babar, M.; Carr, S.; Craig, I. M.; Van Winkle, M.; Lyssenko, A.; Taniguchi, T.; Watanabe, K., Tunable angle-dependent electrochemistry at twisted bilayer graphene with moiré flat bands. *Nature Chemistry* **2022,** *14* (3), 267-273.

129.     Popov, A. M.; Lebedeva, I. V.; Knizhnik, A. A.; Lozovik, Y. E.; Potapkin, B. V., Commensurate-incommensurate phase transition in bilayer graphene. *Physical Review B* **2011,** *84* (4), 045404.

130.     Dos Santos, J. L.; Peres, N.; Neto, A. C., Graphene bilayer with a twist: Electronic structure. *Physical review letters* **2007,** *99* (25), 256802.

131.     Brihuega, I.; Mallet, P.; González-Herrero, H.; De Laissardière, G. T.; Ugeda, M.; Magaud, L.; Gómez-Rodríguez, J.; Ynduráin, F.; Veuillen, J.-Y., Unraveling the intrinsic and robust nature of van Hove singularities in twisted bilayer graphene by scanning tunneling microscopy and theoretical analysis. *Physical review letters* **2012,** *109* (19), 196802.

132.     Luican, A.; Li, G.; Reina, A.; Kong, J.; Nair, R.; Novoselov, K. S.; Geim, A. K.; Andrei, E., Single-layer behavior and its breakdown in twisted graphene layers. *Physical review letters* **2011,** *106* (12), 126802.

133.     Liu, J.; Dai, X., Orbital magnetic states in moiré graphene systems. *Nature Reviews Physics* **2021,** *3* (5), 367-382.

134.     Yoo, H.; Engelke, R.; Carr, S.; Fang, S.; Zhang, K.; Cazeaux, P.; Sung, S. H.; Hovden, R.; Tsen, A. W.; Taniguchi, T., Atomic and electronic reconstruction at the van der Waals interface in twisted bilayer graphene. *Nature materials* **2019,** *18* (5), 448-453.

135.     Zhang, S.; Song, A.; Chen, L.; Jiang, C.; Chen, C.; Gao, L.; Hou, Y.; Liu, L.; Ma, T.; Wang, H., Abnormal conductivity in low-angle twisted bilayer graphene. *Science Advances* **2020,** *6* (47), eabc5555.

136.     Yuk, J. M.; Jeong, H. Y.; Kim, N. Y.; Park, H. J.; Kim, G.; Shin, H. S.; Ruoff, R. S.; Lee, J. Y.; Lee, Z., Superstructural defects and superlattice domains in stacked graphene. *Carbon* **2014,** *80*, 755-761.

137.     Plochocka, P., Excitons in a twisted world. *Nature Nanotechnology* **2020,** *15* (9), 727-729.

138.     Villafañe, V.; Kremser, M.; Hübner, R.; Petrić, M. M.; Wilson, N. P.; Stier, A. V.; Müller, K.; Florian, M.; Steinhoff, A.; Finley, J. J., Twist-Dependent Intra-and Interlayer Excitons in Moiré MoSe 2 Homobilayers. *Physical Review Letters* **2023,** *130* (2), 026901.

139.     Andersen, T. I.; Scuri, G.; Sushko, A.; De Greve, K.; Sung, J.; Zhou, Y.; Wild, D. S.; Gelly, R. J.; Heo, H.; Bérubé, D., Excitons in a reconstructed moiré potential in twisted WSe2/WSe2 homobilayers. *Nature Materials* **2021,** *20* (4), 480-487.

140.     Wu, B.; Zheng, H.; Li, S.; Ding, J.; He, J.; Zeng, Y.; Chen, K.; Liu, Z.; Chen, S.; Pan, A., Evidence for moiré intralayer excitons in twisted WSe2/WSe2 homobilayer superlattices. *Light: Science & Applications* **2022,** *11* (1), 166.

141.    Kim, K.; Yankowitz, M.; Fallahazad, B.; Kang, S.; Movva, H. C.; Huang, S.; Larentis, S.; Corbet, C. M.; Taniguchi, T.; Watanabe, K., van der Waals heterostructures with high accuracy rotational alignment. *Nano letters* **2016**, *16* (3), 1989-1995.

142.    McGilly, L. J.; Kerelsky, A.; Finney, N. R.; Shapovalov, K.; Shih, E.-M.; Ghiotto, A.; Zeng, Y.; Moore, S. L.; Wu, W.; Bai, Y., Visualization of moiré superlattices. *Nature Nanotechnology* **2020**, *15* (7), 580-584.

143.    Carr, S.; Massatt, D.; Torrisi, S. B.; Cazeaux, P.; Luskin, M.; Kaxiras, E., Relaxation and domain formation in incommensurate two-dimensional heterostructures. *Physical Review B* **2018**, *98* (22), 224102.

144.    Yu, H.; Yao, W., Luminescence anomaly of dipolar valley excitons in homobilayer semiconductor moiré superlattices. *Physical Review X* **2021**, *11* (2), 021042.

145.    Merkl, P.; Mooshammer, F.; Brem, S.; Girnghuber, A.; Lin, K.-Q.; Weigl, L.; Liebich, M.; Yong, C.-K.; Gillen, R.; Maultzsch, J., Twist-tailoring Coulomb correlations in van der Waals homobilayers. *Nature communications* **2020**, *11* (1), 2167.

146.    Kretinin, A.; Cao, Y.; Tu, J.; Yu, G.; Jalil, R.; Novoselov, K.; Haigh, S.; Gholinia, A.; Mishchenko, A.; Lozada, M., Electronic properties of graphene encapsulated with different two-dimensional atomic crystals. *Nano letters* **2014**, *14* (6), 3270-3276.

147.    Dean, C. R.; Young, A. F.; Meric, I.; Lee, C.; Wang, L.; Sorgenfrei, S.; Watanabe, K.; Taniguchi, T.; Kim, P.; Shepard, K. L., Boron nitride substrates for high-quality graphene electronics. *Nature nanotechnology* **2010**, *5* (10), 722-726.

148.    Woods, C.; Ares, P.; Nevison-Andrews, H.; Holwill, M.; Fabregas, R.; Guinea, F.; Geim, A.; Novoselov, K.; Walet, N.; Fumagalli, L., Charge-polarized interfacial superlattices in marginally twisted hexagonal boron nitride. *Nature communications* **2021**, *12* (1), 347.

149.    Vizner Stern, M.; Waschitz, Y.; Cao, W.; Nevo, I.; Watanabe, K.; Taniguchi, T.; Sela, E.; Urbakh, M.; Hod, O.; Ben Shalom, M., Interfacial ferroelectricity by van der Waals sliding. *Science* **2021**, *372* (6549), 1462-1466.

150.    Gilbert, S. M.; Pham, T.; Dogan, M.; Oh, S.; Shevitski, B.; Schumm, G.; Liu, S.; Ercius, P.; Aloni, S.; Cohen, M. L., Alternative stacking sequences in hexagonal boron nitride. *2D Materials* **2019**, *6* (2), 021006.

151.    Ribeiro, R.; Peres, N., Stability of boron nitride bilayers: Ground-state energies, interlayer distances, and tight-binding description. *Physical Review B* **2011**, *83* (23), 235312.

152.    Pease, R., Crystal structure of boron nitride. *Nature* **1950**, *165* (4201), 722-723.

153.    Yasuda, K.; Wang, X.; Watanabe, K.; Taniguchi, T.; Jarillo-Herrero, P., Stacking-engineered ferroelectricity in bilayer boron nitride. *Science* **2021**, *372* (6549), 1458-1462.

154.    Zhao, P.; Xiao, C.; Yao, W., Universal superlattice potential for 2D materials from twisted interface inside h-BN substrate. *npj 2D Materials and Applications* **2021**, *5* (1), 38.

155.    Li, L.; Wu, M., Binary compound bilayer and multilayer with vertical polarizations: two-dimensional ferroelectrics, multiferroics, and nanogenerators. *ACS nano* **2017**, *11* (6), 6382-6388.

156.    Hu, Q.; Zhan, Z.; Cui, H.; Zhang, Y.; Jin, F.; Zhao, X.; Zhang, M.; Wang, Z.; Zhang, Q.; Watanabe, K.; Taniguchi, T.; Cao, X.; Liu, W.-M.; Wu, F.; Yuan, S.; Xu, Y., Observation of Rydberg moiré excitons. *Science* **2023**, *380* (6652), 1367-1372.

157.    Li, L.; Yu, Y.; Ye, G. J.; Ge, Q.; Ou, X.; Wu, H.; Feng, D.; Chen, X. H.; Zhang, Y., Black phosphorus field-effect transistors. *Nature nanotechnology* **2014**, *9* (5), 372-377.

158.    Rodin, A.; Carvalho, A.; Neto, A. C., Strain-induced gap modification in black phosphorus. *Physical review letters* **2014**, *112* (17), 176801.

159.    Qiao, J.; Kong, X.; Hu, Z.-X.; Yang, F.; Ji, W., High-mobility transport anisotropy and linear dichroism in few-layer black phosphorus. *Nature communications* **2014**, *5* (1), 4475.

160.    Li, L.; Kim, J.; Jin, C.; Ye, G. J.; Qiu, D. Y.; Da Jornada, F. H.; Shi, Z.; Chen, L.; Zhang, Z.; Yang, F., Direct observation of the layer-dependent electronic structure in phosphorene. *Nature nanotechnology* **2017**, *12* (1), 21-25.

161.    Zhao, S.; Wang, E.; Üzer, E. A.; Guo, S.; Qi, R.; Tan, J.; Watanabe, K.; Taniguchi, T.; Nilges, T.; Gao, P., Anisotropic moiré optical transitions in twisted monolayer/bilayer phosphorene heterostructures. *Nature communications* **2021**, *12* (1), 3947.

162.     Kang, P.; Zhang, W.-T.; Michaud-Rioux, V.; Kong, X.-H.; Hu, C.; Yu, G.-H.; Guo, H., Moiré impurities in twisted bilayer black phosphorus: effects on the carrier mobility. *Physical Review B* **2017,** *96* (19), 195406.

163.     Xie, H.; Luo, X.; Ye, G.; Ye, Z.; Ge, H.; Sung, S. H.; Rennich, E.; Yan, S.; Fu, Y.; Tian, S., Twist engineering of the two-dimensional magnetism in double bilayer chromium triiodide homostructures. *Nature Physics* **2022,** *18* (1), 30-36.

164.     Liang, S.; Liang, J.; Kotsakidis, J. C.; Arachchige, H. S.; Mandrus, D.; Friedman, A. L.; Gong, C., New coercivities and Curie temperatures emerged in van der Waals homostructures of Fe 3 GeTe 2. *Physical Review Materials* **2023,** *7* (6), L061001.

165.     Chen, W.; Sun, Z.; Wang, Z.; Gu, L.; Xu, X.; Wu, S.; Gao, C., Direct observation of van der Waals stacking–dependent interlayer magnetism. *Science* **2019,** *366* (6468), 983-987.

166.     Sharpe, A. L.; Fox, E. J.; Barnard, A. W.; Finney, J.; Watanabe, K.; Taniguchi, T.; Kastner, M. A.; Goldhaber-Gordon, D., Emergent ferromagnetism near three-quarters filling in twisted bilayer graphene. *Science* **2019,** *365* (6453), 605-608.

167.     Can, O.; Tummuru, T.; Day, R. P.; Elfimov, I.; Damascelli, A.; Franz, M., High-temperature topological superconductivity in twisted double-layer copper oxides. *Nature Physics* **2021,** *17* (4), 519-524.

168.     Chen, X.; Fan, X.; Li, L.; Zhang, N.; Niu, Z.; Guo, T.; Xu, S.; Xu, H.; Wang, D.; Zhang, H., Moiré engineering of electronic phenomena in correlated oxides. *Nature Physics* **2020,** *16* (6), 631-635.

169.     Shen, J.; Dong, Z.; Qi, M.; Zhang, Y.; Zhu, C.; Wu, Z.; Li, D., Observation of Moiré Patterns in Twisted Stacks of Bilayer Perovskite Oxide Nanomembranes with Various Lattice Symmetries. *ACS Applied Materials & Interfaces* **2022,** *14* (44), 50386-50392.

170.     Xue, J.; Sanchez-Yamagishi, J.; Bulmash, D.; Jacquod, P.; Deshpande, A.; Watanabe, K.; Taniguchi, T.; Jarillo-Herrero, P.; LeRoy, B. J., Scanning tunnelling microscopy and spectroscopy of ultra-flat graphene on hexagonal boron nitride. *Nature materials* **2011,** *10* (4), 282-285.

171.     Yankowitz, M.; Xue, J.; Cormode, D.; Sanchez-Yamagishi, J. D.; Watanabe, K.; Taniguchi, T.; Jarillo-Herrero, P.; Jacquod, P.; LeRoy, B. J., Emergence of superlattice Dirac points in graphene on hexagonal boron nitride. *Nature physics* **2012,** *8* (5), 382-386.

172.     Martin, J.; Akerman, N.; Ulbricht, G.; Lohmann, T.; Smet, J. v.; Von Klitzing, K.; Yacoby, A., Observation of electron–hole puddles in graphene using a scanning single-electron transistor. *Nature physics* **2008,** *4* (2), 144-148.

173.     Chen, J.-H.; Jang, C.; Adam, S.; Fuhrer, M.; Williams, E. D.; Ishigami, M., Charged-impurity scattering in graphene. *Nature physics* **2008,** *4* (5), 377-381.

174.     Rossi, E.; Sarma, S. D., Ground state of graphene in the presence of random charged impurities. *Physical review letters* **2008,** *101* (16), 166803.

175.     Watanabe, K.; Taniguchi, T.; Kanda, H., Direct-bandgap properties and evidence for ultraviolet lasing of hexagonal boron nitride single crystal. *Nature materials* **2004,** *3* (6), 404-409.

176.     Liu, L.; Feng, Y.; Shen, Z., Structural and electronic properties of h-BN. *Physical Review B* **2003,** *68* (10), 104102.

177.     Wallace, P. R., The band theory of graphite. *Physical review* **1947,** *71* (9), 622.

178.     Giovannetti, G.; Khomyakov, P. A.; Brocks, G.; Kelly, P. J.; Van Den Brink, J., Substrate-induced band gap in graphene on hexagonal boron nitride: Ab initio density functional calculations. *Physical Review B* **2007,** *76* (7), 073103.

179.     Sachs, B.; Wehling, T.; Katsnelson, M.; Lichtenstein, A., Adhesion and electronic structure of graphene on hexagonal boron nitride substrates. *Physical Review B* **2011,** *84* (19), 195414.

180.     Kindermann, M.; Uchoa, B.; Miller, D. L., Zero-energy modes and gate-tunable gap in graphene on hexagonal boron nitride. *Physical Review B* **2012,** *86* (11), 115415.

181.     Chen, Z.-G.; Shi, Z.; Yang, W.; Lu, X.; Lai, Y.; Yan, H.; Wang, F.; Zhang, G.; Li, Z., Observation of an intrinsic bandgap and Landau level renormalization in graphene/boron-nitride heterostructures. *Nature communications* **2014,** *5* (1), 4461.

182.     Park, C.-H.; Yang, L.; Son, Y.-W.; Cohen, M. L.; Louie, S. G., New generation of massless Dirac fermions in graphene under external periodic potentials. *Physical review letters* **2008,** *101* (12), 126804.

183.    Ponomarenko, L.; Gorbachev, R.; Yu, G.; Elias, D.; Jalil, R.; Patel, A.; Mishchenko, A.; Mayorov, A.; Woods, C.; Wallbank, J., Cloning of Dirac fermions in graphene superlattices. *Nature* **2013**, *497* (7451), 594-597.

184.    Wallbank, J.; Patel, A.; Mucha-Kruczyński, M.; Geim, A.; Fal'Ko, V., Generic miniband structure of graphene on a hexagonal substrate. *Physical Review B* **2013**, *87* (24), 245408.

185.    Yu, G.; Gorbachev, R.; Tu, J.; Kretinin, A.; Cao, Y.; Jalil, R.; Withers, F.; Ponomarenko, L.; Piot, B.; Potemski, M., Hierarchy of Hofstadter states and replica quantum Hall ferromagnetism in graphene superlattices. *Nature physics* **2014**, *10* (7), 525-529.

186.    Eckmann, A.; Park, J.; Yang, H.; Elias, D.; Mayorov, A. S.; Yu, G.; Jalil, R.; Novoselov, K. S.; Gorbachev, R. V.; Lazzeri, M., Raman fingerprint of aligned graphene/h-BN superlattices. *Nano letters* **2013**, *13* (11), 5242-5246.

187.    Bak, P., Commensurate phases, incommensurate phases and the devil's staircase. *Reports on Progress in Physics* **1982**, *45* (6), 587.

188.    Pokrovsky, V.; Talapov, A., Ground state, spectrum, and phase diagram of two-dimensional incommensurate crystals. *Physical Review Letters* **1979**, *42* (1), 65.

189.    Finney, N. R.; Yankowitz, M.; Muraleetharan, L.; Watanabe, K.; Taniguchi, T.; Dean, C. R.; Hone, J., Tunable crystal symmetry in graphene–boron nitride heterostructures with coexisting moiré superlattices. *Nature nanotechnology* **2019**, *14* (11), 1029-1034.

190.    Xie, S.; Faeth, B. D.; Tang, Y.; Li, L.; Gerber, E.; Parzyck, C. T.; Chowdhury, D.; Zhang, Y.-H.; Jozwiak, C.; Bostwick, A., Strong interlayer interactions in bilayer and trilayer moiré superlattices. *Science advances* **2022**, *8* (12), eabk1911.

191.    Zheng, H.; Wu, B.; Li, S.; Ding, J.; He, J.; Liu, Z.; Wang, C.-T.; Wang, J.-T.; Pan, A.; Liu, Y., Localization-enhanced moiré exciton in twisted transition metal dichalcogenide heterotrilayer superlattices. *Light: Science & Applications* **2023**, *12* (1), 117.

192.    Zheng, H.; Wu, B.; Li, S.; He, J.; Chen, K.-Q.; Liu, Z.; Pan, A.; Yanping, L., Moiré Enhanced Potentials in Twisted Transition Metal Dichalcogenide Trilayers Homostructures. **2022**.

193.    Tebyetekerwa, M.; Zhang, J.; Saji, S. E.; Wibowo, A. A.; Rahman, S.; Truong, T. N.; Lu, Y.; Yin, Z.; Macdonald, D.; Nguyen, H. T., Twist-driven wide freedom of interlayer exciton emission in MoS2/WS2 heterobilayers. *Cell Reports Physical Science* **2021**, *2* (8).

194.    Zheng, B.; Ma, C.; Li, D.; Lan, J.; Zhang, Z.; Sun, X.; Zheng, W.; Yang, T.; Zhu, C.; Ouyang, G., Band alignment engineering in two-dimensional lateral heterostructures. *Journal of the American Chemical Society* **2018**, *140* (36), 11193-11197.

195.    Kang, J.; Tongay, S.; Zhou, J.; Li, J.; Wu, J., Band offsets and heterostructures of two-dimensional semiconductors. *Applied Physics Letters* **2013**, *102* (1), 012111.

196.    Ceballos, F.; Bellus, M. Z.; Chiu, H.-Y.; Zhao, H., Ultrafast charge separation and indirect exciton formation in a MoS2–MoSe2 van der Waals heterostructure. *ACS nano* **2014**, *8* (12), 12717-12724.

197.    Hong, X.; Kim, J.; Shi, S.-F.; Zhang, Y.; Jin, C.; Sun, Y.; Tongay, S.; Wu, J.; Zhang, Y.; Wang, F., Ultrafast charge transfer in atomically thin MoS2/WS2 heterostructures. *Nature nanotechnology* **2014**, *9* (9), 682-686.

198.    Li, Y.; Cui, Q.; Ceballos, F.; Lane, S. D.; Qi, Z.; Zhao, H., Ultrafast interlayer electron transfer in incommensurate transition metal dichalcogenide homobilayers. *Nano Letters* **2017**, *17* (11), 6661-6666.

199.    Choi, J.; Hsu, W.-T.; Lu, L.-S.; Sun, L.; Cheng, H.-Y.; Lee, M.-H.; Quan, J.; Tran, K.; Wang, C.-Y.; Staab, M., Moiré potential impedes interlayer exciton diffusion in van der Waals heterostructures. *Science advances* **2020**, *6* (39), eaba8866.

200.    Ding, Y.; Wang, Y.; Ni, J.; Shi, L.; Shi, S.; Tang, W., First principles study of structural, vibrational and electronic properties of graphene-like MX2 (M= Mo, Nb, W, Ta; X= S, Se, Te) monolayers. *Physica B: Condensed Matter* **2011**, *406* (11), 2254-2260.

201.    Wang, J.; Shi, Q.; Shih, E.-M.; Zhou, L.; Wu, W.; Bai, Y.; Rhodes, D.; Barmak, K.; Hone, J.; Dean, C. R., Diffusivity reveals three distinct phases of interlayer excitons in MoSe 2/WSe 2 heterobilayers. *Physical Review Letters* **2021**, *126* (10), 106804.

202.    Enaldiev, V.; Zolyomi, V.; Yelgel, C.; Magorrian, S.; Fal'Ko, V., Stacking domains and dislocation networks in marginally twisted bilayers of transition metal dichalcogenides. *Physical Review Letters* **2020**, *124* (20), 206101.

203.     Yu, H.; Liu, G.-B.; Yao, W., Brightened spin-triplet interlayer excitons and optical selection rules in van der Waals heterobilayers. *2D Materials* **2018**, *5* (3), 035021.

204.     Zhang, L.; Gogna, R.; Burg, G. W.; Horng, J.; Paik, E.; Chou, Y.-H.; Kim, K.; Tutuc, E.; Deng, H., Highly valley-polarized singlet and triplet interlayer excitons in van der Waals heterostructure. *Physical Review B* **2019**, *100* (4), 041402.

205.     Ugeda, M. M.; Bradley, A. J.; Shi, S.-F.; Da Jornada, F. H.; Zhang, Y.; Qiu, D. Y.; Ruan, W.; Mo, S.-K.; Hussain, Z.; Shen, Z.-X., Giant bandgap renormalization and excitonic effects in a monolayer transition metal dichalcogenide semiconductor. *Nature materials* **2014**, *13* (12), 1091-1095.

206.     Trainer, D. J.; Putilov, A. V.; Wang, B.; Lane, C.; Saari, T.; Chang, T.-R.; Jeng, H.-T.; Lin, H.; Xi, X.; Nieminen, J., Moiré superlattices and 2D electronic properties of graphite/MoS2 heterostructures. *Journal of Physics and Chemistry of Solids* **2019**, *128*, 325-330.

207.     Hong, M.; Zhou, X.; Gao, N.; Jiang, S.; Xie, C.; Zhao, L.; Gao, Y.; Zhang, Z.; Yang, P.; Shi, Y., Identifying the non-identical outermost selenium atoms and invariable band gaps across the grain boundary of anisotropic rhenium diselenide. *ACS nano* **2018**, *12* (10), 10095-10103.

208.     Chen, X.; Lei, B.; Zhu, Y.; Zhou, J.; Liu, Z.; Ji, W.; Zhou, W., Pristine edge structures of T″-phase transition metal dichalcogenides (ReSe 2, ReS 2) atomic layers. *Nanoscale* **2020**, *12* (32), 17005-17012.

209.     Qiu, Z.; Trushin, M.; Fang, H.; Verzhbitskiy, I.; Gao, S.; Laksono, E.; Yang, M.; Lyu, P.; Li, J.; Su, J., Giant gate-tunable bandgap renormalization and excitonic effects in a 2D semiconductor. *Science advances* **2019**, *5* (7), eaaw2347.

210.     Plumadore, R.; Al Ezzi, M. M.; Adam, S.; Luican-Mayer, A., Moiré patterns in graphene–rhenium disulfide vertical heterostructures. *Journal of Applied Physics* **2020**, *128* (4), 044303.

211.     Pham, T. T.; Vancsó, P.; Szendrő, M.; Palotás, K.; Castelino, R.; Bouatou, M.; Chacon, C.; Henrard, L.; Lagoute, J.; Sporken, R., Higher-indexed Moiré patterns and surface states of MoTe2/graphene heterostructure grown by molecular beam epitaxy. *npj 2D Materials and Applications* **2022**, *6* (1), 48.

212.     Tong, Q.; Liu, F.; Xiao, J.; Yao, W., Skyrmions in the Moiré of van der Waals 2D Magnets. *Nano letters* **2018**, *18* (11), 7194-7199.

213.     Li, Q.; Zhang, C.-x.; Wang, D.; Chen, K.-Q.; Tang, L.-M., Giant valley splitting in a MoTe 2/MnSe 2 van der Waals heterostructure with room-temperature ferromagnetism. *Materials Advances* **2022**, *3* (6), 2927-2933.

214.     Norden, T.; Zhao, C.; Zhang, P.; Sabirianov, R.; Petrou, A.; Zeng, H., Giant valley splitting in monolayer WS2 by magnetic proximity effect. *Nature communications* **2019**, *10* (1), 4163.

215.     Kim, H. S.; Gye, G.; Lee, S.-H.; Wang, L.; Cheong, S.-W.; Yeom, H. W., Moire superstructure and dimensional crossover of 2D electronic states on nanoscale lead quantum films. *Scientific reports* **2017**, *7* (1), 12735.

216.     Kim, H. S.; Kim, S.; Kim, K.; Min, B. I.; Cho, Y.-H.; Wang, L.; Cheong, S.-W.; Yeom, H. W., Nanoscale superconducting honeycomb charge order in IrTe2. *Nano Letters* **2016**, *16* (7), 4260-4265.

217.     Dahal, A.; Batzill, M., Growth from behind: Intercalation-growth of two-dimensional FeO moiré structure underneath of metal-supported graphene. *Scientific Reports* **2015**, *5* (1), 1-9.

218.     Loginova, E.; Nie, S.; Thürmer, K.; Bartelt, N. C.; McCarty, K. F., Defects of graphene on Ir (111): Rotational domains and ridges. *Physical Review B* **2009**, *80* (8), 085430.

219.     Kwon, S.-Y.; Ciobanu, C. V.; Petrova, V.; Shenoy, V. B.; Bareno, J.; Gambin, V.; Petrov, I.; Kodambaka, S., Growth of semiconducting graphene on palladium. *Nano letters* **2009**, *9* (12), 3985-3990.

220.     Moritz, W.; Wang, B.; Bocquet, M.-L.; Brugger, T.; Greber, T.; Wintterlin, J.; Günther, S., Structure determination of the coincidence phase of graphene on Ru (0001). *Physical review letters* **2010**, *104* (13), 136102.

221.     Martoccia, D.; Willmott, P.; Brugger, T.; Björck, M.; Günther, S.; Schlepütz, C.; Cervellino, A.; Pauli, S.; Patterson, B.; Marchini, S., Graphene on Ru (0001): a 25× 25 supercell. *Physical Review Letters* **2008**, *101* (12), 126102.

222.     Gao, L.; Guest, J. R.; Guisinger, N. P., Epitaxial graphene on Cu (111). *Nano letters* **2010**, *10* (9), 3512-3516.

223.    Eom, D.; Prezzi, D.; Rim, K. T.; Zhou, H.; Lefenfeld, M.; Xiao, S.; Nuckolls, C.; Hybertsen, M. S.; Heinz, T. F.; Flynn, G. W., Structure and electronic properties of graphene nanoislands on Co (0001). *Nano letters* **2009,** *9* (8), 2844-2848.

224.    Sutter, P.; Sadowski, J. T.; Sutter, E., Graphene on Pt (111): Growth and substrate interaction. *Physical Review B* **2009,** *80* (24), 245411.

225.    Merino, P.; Svec, M.; Pinardi, A. L.; Otero, G.; Martin-Gago, J. A., Strain-driven moiré superstructures of epitaxial graphene on transition metal surfaces. *Acs Nano* **2011,** *5* (7), 5627-5634.

226.    Kim, Y.; Westphal, C.; Ynzunza, R.; Wang, Z.; Galloway, H.; Salmeron, M.; Van Hove, M.; Fadley, C., The growth of iron oxide films on Pt (111): a combined XPD, STM, and LEED study. *Surface Science* **1998,** *416* (1-2), 68-111.

227.    Weiss, W.; Ranke, W., Surface chemistry and catalysis on well-defined epitaxial iron-oxide layers. *Progress in Surface Science* **2002,** *70* (1-3), 1-151.

228.    Rienks, E. D.; Nilius, N.; Rust, H.-P.; Freund, H.-J., Surface potential of a polar oxide film: FeO on Pt (111). *Physical Review B* **2005,** *71* (24), 241404.

229.    Giordano, L.; Pacchioni, G.; Goniakowski, J.; Nilius, N.; Rienks, E. D.; Freund, H.-J., Charging of metal adatoms on ultrathin oxide films: Au and Pd on FeO/Pt (111). *Physical review letters* **2008,** *101* (2), 026102.

230.    Tian, X.; Kim, D. S.; Yang, S.; Ciccarino, C. J.; Gong, Y.; Yang, Y.; Yang, Y.; Duschatko, B.; Yuan, Y.; Ajayan, P. M., Correlating the three-dimensional atomic defects and electronic properties of two-dimensional transition metal dichalcogenides. *Nature materials* **2020,** *19* (8), 867-873.

231.    Sørensen, S. G.; Fuchtbauer, H. G.; Tuxen, A. K.; Walton, A. S.; Lauritsen, J. V., Structure and electronic properties of in situ synthesized single-layer MoS2 on a gold surface. *ACS nano* **2014,** *8* (7), 6788-6796.

232.    Murata, Y.; Nie, S.; Ebnonnasir, A.; Starodub, E.; Kappes, B.; McCarty, K.; Ciobanu, C.; Kodambaka, S., Growth structure and work function of bilayer graphene on Pd (111). *Physical Review B* **2012,** *85* (20), 205443.

233.    Yasuda, S.; Takahashi, R.; Osaka, R.; Kumagai, R.; Miyata, Y.; Okada, S.; Hayamizu, Y.; Murakoshi, K., Out-of-Plane Strain Induced in a Moiré Superstructure of Monolayer MoS2 and MoSe2 on Au (111). *Small* **2017,** *13* (31), 1700748.

234.    Tang, B.; Che, B.; Xu, M.; Ang, Z. P.; Di, J.; Gao, H.-J.; Yang, H.; Zhou, J.; Liu, Z., Recent advances in synthesis and study of 2D twisted transition metal dichalcogenide bilayers. *Small Structures* **2021,** *2* (5), 2000153.

235.    Puretzky, A. A.; Liang, L.; Li, X.; Xiao, K.; Sumpter, B. G.; Meunier, V.; Geohegan, D. B., Twisted MoSe2 bilayers with variable local stacking and interlayer coupling revealed by low-frequency raman spectroscopy. *ACS nano* **2016,** *10* (2), 2736-2744.

236.    Huang, S.; Ling, X.; Liang, L.; Kong, J.; Terrones, H.; Meunier, V.; Dresselhaus, M. S., Probing the interlayer coupling of twisted bilayer MoS2 using photoluminescence spectroscopy. *Nano letters* **2014,** *14* (10), 5500-5508.

237.    Fan, S.; Vu, Q. A.; Tran, M. D.; Adhikari, S.; Lee, Y. H., Transfer assembly for two-dimensional van der Waals heterostructures. *2D Materials* **2020,** *7* (2), 022005.

238.    Chen, X. D.; Xin, W.; Jiang, W. S.; Liu, Z. B.; Chen, Y.; Tian, J. G., High-Precision Twist-Controlled Bilayer and Trilayer Graphene. *Advanced Materials* **2016,** *28* (13), 2563-2570.

239.    Tan, Z.; Yin, J.; Chen, C.; Wang, H.; Lin, L.; Sun, L.; Wu, J.; Sun, X.; Yang, H.; Chen, Y., Building large-domain twisted bilayer graphene with van Hove singularity. *ACS nano* **2016,** *10* (7), 6725-6730.

240.    Wang, K.; Huang, B.; Tian, M.; Ceballos, F.; Lin, M. W.; Mahjouri-Samani, M.; Boulesbaa, A.; Puretzky, A. A.; Rouleau, C. M.; Yoon, M.; Zhao, H.; Xiao, K.; Duscher, G.; Geohegan, D. B., Interlayer Coupling in Twisted WSe2/WS2 Bilayer Heterostructures Revealed by Optical Spectroscopy. *ACS Nano* **2016,** *10* (7), 6612-22.

241.    Chung, T.-F.; Xu, Y.; Chen, Y. P., Transport measurements in twisted bilayer graphene: Electron-phonon coupling and Landau level crossing. *Physical review B* **2018,** *98* (3), 035425.

242.    Xu, Y.; Ray, A.; Shao, Y.-T.; Jiang, S.; Lee, K.; Weber, D.; Goldberger, J. E.; Watanabe, K.; Taniguchi, T.; Muller, D. A., Coexisting ferromagnetic–antiferromagnetic state in twisted bilayer CrI3. *Nature Nanotechnology* **2022,** *17* (2), 143-147.

243.    Jin, C.; Regan, E. C.; Wang, D.; Iqbal Bakti Utama, M.; Yang, C.-S.; Cain, J.; Qin, Y.; Shen, Y.; Zheng, Z.; Watanabe, K., Identification of spin, valley and moiré quasi-angular momentum of interlayer excitons. *Nature Physics* **2019,** *15* (11), 1140-1144.

244.    Castellanos-Gomez, A.; Buscema, M.; Molenaar, R.; Singh, V.; Janssen, L.; van der Zant, H. S. J.; Steele, G. A., Deterministic transfer of two-dimensional materials by all-dry viscoelastic stamping. *2D Materials* **2014,** *1* (1).

245.    Meitl, M. A.; Zhu, Z.-T.; Kumar, V.; Lee, K. J.; Feng, X.; Huang, Y. Y.; Adesida, I.; Nuzzo, R. G.; Rogers, J. A., Transfer printing by kinetic control of adhesion to an elastomeric stamp. *Nature Materials* **2005,** *5* (1), 33-38.

246.    Zhang, L.; Yan, H.; Sun, X.; Dong, M.; Yildirim, T.; Wang, B.; Wen, B.; Neupane, G. P.; Sharma, A.; Zhu, Y., Modulated interlayer charge transfer dynamics in a monolayer TMD/metal junction. *Nanoscale* **2019,** *11* (2), 418-425.

247.    Puretzky, A. A.; Liang, L.; Li, X.; Xiao, K.; Sumpter, B. G.; Meunier, V.; Geohegan, D. B., Twisted MoSe(2) Bilayers with Variable Local Stacking and Interlayer Coupling Revealed by Low-Frequency Raman Spectroscopy. *ACS Nano* **2016,** *10* (2), 2736-44.

248.    Ma, X.; Liu, Q.; Xu, D.; Zhu, Y.; Kim, S.; Cui, Y.; Zhong, L.; Liu, M., Capillary-Force-Assisted Clean-Stamp Transfer of Two-Dimensional Materials. *Nano Lett* **2017,** *17* (11), 6961-6967.

249.    Tao, L.; Li, H.; Gao, Y.; Chen, Z.; Wang, L.; Deng, Y.; Zhang, J.; Xu, J.-B., Deterministic and Etching-Free Transfer of Large-Scale 2D Layered Materials for Constructing Interlayer Coupled van der Waals Heterostructures. *Advanced Materials Technologies* **2018,** *3* (5).

250.    Zhang, Y.; Lee, Y.; Zhang, W.; Song, D.; Lee, K.; Zhao, Y.; Hao, H.; Zhao, Z.; Wang, S.; Kim, K., Deterministic Fabrication of Twisted Van Der Waals Structures. *Advanced Functional Materials* **2023,** *33* (23), 2212210.

251.    Yan, Z.; Liu, Y.; Ju, L.; Peng, Z.; Lin, J.; Wang, G.; Zhou, H.; Xiang, C.; Samuel, E. L. G.; Kittrell, C.; Artyukhov, V. I.; Wang, F.; Yakobson, B. I.; Tour, J. M., Large Hexagonal Bi- and Trilayer Graphene Single Crystals with Varied Interlayer Rotations. *Angewandte Chemie International Edition* **2014,** *53* (6), 1565-1569.

252.    Gao, Z.; Zhang, Q.; Naylor, C. H.; Kim, Y.; Abidi, I. H.; Ping, J.; Ducos, P.; Zauberman, J.; Zhao, M.-Q.; Rappe, A. M.; Luo, Z.; Ren, L.; Johnson, A. T. C., Crystalline Bilayer Graphene with Preferential Stacking from Ni–Cu Gradient Alloy. *ACS Nano* **2018,** *12* (3), 2275-2282.

253.    Cho, H.; Park, Y.; Kim, S.; Ahn, T.; Kim, T.-H.; Choi, H. C., Specific stacking angles of bilayer graphene grown on atomic-flat and -stepped Cu surfaces. *npj 2D Materials and Applications* **2020,** *4* (1), 35.

254.    Yin, J.; Wang, H.; Peng, H.; Tan, Z.; Liao, L.; Lin, L.; Sun, X.; Koh, A. L.; Chen, Y.; Peng, H., Selectively enhanced photocurrent generation in twisted bilayer graphene with van Hove singularity. *Nature communications* **2016,** *7* (1), 10699.

255.    Lin, L.; Sun, L.; Zhang, J.; Sun, J.; Koh, A. L.; Peng, H.; Liu, Z., Rapid Growth of Large Single-Crystalline Graphene via Second Passivation and Multistage Carbon Supply. *Adv Mater* **2016,** *28* (23), 4671-7.

256.    Sun, L.; Wang, Z.; Wang, Y.; Zhao, L.; Li, Y.; Chen, B.; Huang, S.; Zhang, S.; Wang, W.; Pei, D., Hetero-site nucleation for growing twisted bilayer graphene with a wide range of twist angles. *Nature communications* **2021,** *12* (1), 2391.

257.    Zhang, X.; Nan, H.; Xiao, S.; Wan, X.; Gu, X.; Du, A.; Ni, Z.; Ostrikov, K. K., Transition metal dichalcogenides bilayer single crystals by reverse-flow chemical vapor epitaxy. *Nat Commun* **2019,** *10* (1), 598.

258.    Zheng, S.; Sun, L.; Zhou, X.; Liu, F.; Liu, Z.; Shen, Z.; Fan, H. J., Coupling and Interlayer Exciton in Twist-Stacked WS2Bilayers. *Advanced Optical Materials* **2015,** *3* (11), 1600-1605.

259.    Zhang, D.; Zeng, Z.; Tong, Q.; Jiang, Y.; Chen, S.; Zheng, B.; Qu, J.; Li, F.; Zheng, W.; Jiang, F.; Zhao, H.; Huang, L.; Braun, K.; Meixner, A. J.; Wang, X.; Pan, A., Near-Unity Polarization of Valley-Dependent Second-Harmonic Generation in Stacked TMDC Layers and Heterostructures at Room Temperature. *Adv Mater* **2020,** *32* (29), e1908061.

260.    Bachmatiuk, A.; Abelin, R. F.; Quang, H. T.; Trzebicka, B.; Eckert, J.; Rummeli, M. H., Chemical vapor deposition of twisted bilayer and few-layer MoSe(2) over SiO(x) substrates. *Nanotechnology* **2014,** *25* (36), 365603.

261.     Mandyam, S. V.; Zhao, M. Q.; Masih Das, P.; Zhang, Q.; Price, C. C.; Gao, Z.; Shenoy, V. B.; Drndic, M.; Johnson, A. T. C., Controlled Growth of Large-Area Bilayer Tungsten Diselenides with Lateral P-N Junctions. *ACS Nano* **2019,** *13* (9), 10490-10498.

262.     Yuan, L.; Zheng, B.; Kunstmann, J.; Brumme, T.; Kuc, A. B.; Ma, C.; Deng, S.; Blach, D.; Pan, A.; Huang, L., Twist-angle-dependent interlayer exciton diffusion in WS2-WSe2 heterobilayers. *Nat Mater* **2020,** *19* (6), 617-623.

263.     Li, Z.; Tabataba-Vakili, F.; Zhao, S.; Rupp, A.; Bilgin, I.; Herdegen, Z.; März, B.; Watanabe, K.; Taniguchi, T.; Schleder, G. R.; Baimuratov, A. S.; Kaxiras, E.; Müller-Caspary, K.; Högele, A., Lattice Reconstruction in MoSe2–WSe2 Heterobilayers Synthesized by Chemical Vapor Deposition. *Nano Letters* **2023,** *23* (10), 4160-4166.

264.     Yan, A.; Velasco, J., Jr.; Kahn, S.; Watanabe, K.; Taniguchi, T.; Wang, F.; Crommie, M. F.; Zettl, A., Direct Growth of Single- and Few-Layer MoS2 on h-BN with Preferred Relative Rotation Angles. *Nano Letters* **2015,** *15* (10), 6324-6331.

265.     Ribeiro-Palau, R.; Zhang, C.; Watanabe, K.; Taniguchi, T.; Hone, J.; Dean, C. R., Twistable electronics with dynamically rotatable heterostructures. *Science* **2018,** *361* (6403), 690-693.

266.     Yang, Y.; Li, J.; Yin, J.; Xu, S.; Mullan, C.; Taniguchi, T.; Watanabe, K.; Geim, A. K.; Novoselov, K. S.; Mishchenko, A., In situ manipulation of van der Waals heterostructures for twistronics. *Science advances* **2020,** *6* (49), eabd3655.

267.     Hu, C.; Wu, T.; Huang, X.; Dong, Y.; Chen, J.; Zhang, Z.; Lyu, B.; Ma, S.; Watanabe, K.; Taniguchi, T.; Xie, G.; Li, X.; Liang, Q.; Shi, Z., In-situ twistable bilayer graphene. *Scientific Reports* **2022,** *12* (1), 204.

268.     Carozo, V.; Almeida, C.; Fragneaud, B.; Bedê, P.; Moutinho, M.; Ribeiro-Soares, J.; Andrade, N.; Souza Filho, A.; Matos, M.; Wang, B., Resonance effects on the Raman spectra of graphene superlattices. *Physical Review B* **2013,** *88* (8), 085401.

269.     Chen, H.; Zhang, X.-L.; Zhang, Y.-Y.; Wang, D.; Bao, D.-L.; Que, Y.; Xiao, W.; Du, S.; Ouyang, M.; Pantelides, S. T., Atomically precise, custom-design origami graphene nanostructures. *Science* **2019,** *365* (6457), 1036-1040.

270.     Wang, B.; Huang, M.; Kim, N. Y.; Cunning, B. V.; Huang, Y.; Qu, D.; Chen, X.; Jin, S.; Biswal, M.; Zhang, X., Controlled folding of single crystal graphene. *Nano letters* **2017,** *17* (3), 1467-1473.

271.     Du, X.; Lee, Y.; Zhang, Y.; Yu, T.; Kim, K.; Liu, N., Electronically weak coupled bilayer MoS2 at various twist angles via folding. *ACS Applied Materials & Interfaces* **2021,** *13* (19), 22819-22827.

272.     Li, B.; Wei, J.; Jin, C.; Si, K.; Meng, L.; Wang, X.; Jia, Y.; He, Q.; Zhang, P.; Wang, J.; Gong, Y., Twisted Bilayer Graphene Induced by Intercalation. *Nano Letters* **2023,** *23* (12), 5475-5481.

273.     Kerelsky, A.; Rubio-Verdú, C.; Xian, L.; Kennes, D. M.; Halbertal, D.; Finney, N.; Song, L.; Turkel, S.; Wang, L.; Watanabe, K.; Taniguchi, T.; Hone, J.; Dean, C.; Basov, D. N.; Rubio, A.; Pasupathy, A. N., Moiréless correlations in ABCA graphene. *Proceedings of the National Academy of Sciences* **2021,** *118* (4), e2017366118.

274.     Halbertal, D.; Finney, N. R.; Sunku, S. S.; Kerelsky, A.; Rubio-Verdu, C.; Shabani, S.; Xian, L.; Carr, S.; Chen, S.; Zhang, C.; Wang, L.; Gonzalez-Acevedo, D.; McLeod, A. S.; Rhodes, D.; Watanabe, K.; Taniguchi, T.; Kaxiras, E.; Dean, C. R.; Hone, J. C.; Pasupathy, A. N.; Kennes, D. M.; Rubio, A.; Basov, D. N., Moire metrology of energy landscapes in van der Waals heterostructures. *Nat Commun* **2021,** *12* (1), 242.

275.     Hämäläinen, S. K.; Boneschanscher, M. P.; Jacobse, P. H.; Swart, I.; Pussi, K.; Moritz, W.; Lahtinen, J.; Liljeroth, P.; Sainio, J., Structure and local variations of the graphene moire on Ir(111). *Physical Review B* **2013,** *88* (20), 201406.

276.     Chiodini, S.; Kerfoot, J.; Venturi, G.; Mignuzzi, S.; Alexeev, E. M.; Teixeira Rosa, B.; Tongay, S.; Taniguchi, T.; Watanabe, K.; Ferrari, A. C.; Ambrosio, A., Moire Modulation of Van Der Waals Potential in Twisted Hexagonal Boron Nitride. *ACS Nano* **2022,** *16* (5), 7589-7604.

277.     Garcia, R.; Gómez, C. J.; Martinez, N. F.; Patil, S.; Dietz, C.; Magerle, R., Identification of Nanoscale Dissipation Processes by Dynamic Atomic Force Microscopy. *Physical Review Letters* **2006,** *97* (1), 016103.

278.	Wang, X.; Yasuda, K.; Zhang, Y.; Liu, S.; Watanabe, K.; Taniguchi, T.; Hone, J.; Fu, L.; Jarillo-Herrero, P., Interfacial ferroelectricity in rhombohedral-stacked bilayer transition metal dichalcogenides. *Nat Nanotechnol* **2022,** *17* (4), 367-371.

279.	McGilly, L. J.; Kerelsky, A.; Finney, N. R.; Shapovalov, K.; Shih, E. M.; Ghiotto, A.; Zeng, Y.; Moore, S. L.; Wu, W.; Bai, Y.; Watanabe, K.; Taniguchi, T.; Stengel, M.; Zhou, L.; Hone, J.; Zhu, X.; Basov, D. N.; Dean, C.; Dreyer, C. E.; Pasupathy, A. N., Visualization of moire superlattices. *Nat Nanotechnol* **2020,** *15* (7), 580-584.

280.	Deb, S.; Cao, W.; Raab, N.; Watanabe, K.; Taniguchi, T.; Goldstein, M.; Kronik, L.; Urbakh, M.; Hod, O.; Ben Shalom, M., Cumulative polarization in conductive interfacial ferroelectrics. *Nature* **2022,** *612* (7940), 465-469.

281.	Gadelha, A. C.; Ohlberg, D. A. A.; Rabelo, C.; Neto, E. G. S.; Vasconcelos, T. L.; Campos, J. L.; Lemos, J. S.; Ornelas, V.; Miranda, D.; Nadas, R.; Santana, F. C.; Watanabe, K.; Taniguchi, T.; van Troeye, B.; Lamparski, M.; Meunier, V.; Nguyen, V. H.; Paszko, D.; Charlier, J. C.; Campos, L. C.; Cancado, L. G.; Medeiros-Ribeiro, G.; Jorio, A., Localization of lattice dynamics in low-angle twisted bilayer graphene. *Nature* **2021,** *590* (7846), 405-409.

282.	Luo, Y.; Engelke, R.; Mattheakis, M.; Tamagnone, M.; Carr, S.; Watanabe, K.; Taniguchi, T.; Kaxiras, E.; Kim, P.; Wilson, W. L., In situ nanoscale imaging of moiré superlattices in twisted van der Waals heterostructures. *Nature Communications* **2020,** *11* (1), 4209.

283.	Sunku, S. S.; Ni, G. X.; Jiang, B. Y.; Yoo, H.; Sternbach, A.; McLeod, A. S.; Stauber, T.; Xiong, L.; Taniguchi, T.; Watanabe, K.; Kim, P.; Fogler, M. M.; Basov, D. N., Photonic crystals for nano-light in moiré graphene superlattices. *Science* **2018,** *362* (6419), 1153-1156.

284.	Moore, S. L.; Ciccarino, C. J.; Halbertal, D.; McGilly, L. J.; Finney, N. R.; Yao, K.; Shao, Y.; Ni, G.; Sternbach, A.; Telford, E. J.; Kim, B. S.; Rossi, S. E.; Watanabe, K.; Taniguchi, T.; Pasupathy, A. N.; Dean, C. R.; Hone, J.; Schuck, P. J.; Narang, P.; Basov, D. N., Nanoscale lattice dynamics in hexagonal boron nitride moire superlattices. *Nat Commun* **2021,** *12* (1), 5741.

285.	Hesp, N. C. H.; Torre, I.; Barcons-Ruiz, D.; Herzig Sheinfux, H.; Watanabe, K.; Taniguchi, T.; Krishna Kumar, R.; Koppens, F. H. L., Nano-imaging photoresponse in a moiré unit cell of minimally twisted bilayer graphene. *Nature Communications* **2021,** *12* (1), 1640.

286.	Yao C.; Ma, Y., Superconducting materials: Challenges and opportunities for large-scale applications. *iScience* **2021,** *24* (6), 102541.

287.	Pantaleón, P. A.; Jimeno-Pozo, A.; Sainz-Cruz, H.; Phong, V. T.; Cea, T.; Guinea, F., Superconductivity and correlated phases in non-twisted bilayer and trilayer graphene. *Nature Reviews Physics* **2023,** *5* (5), 304-315.

288.	Fischer, A.; Goodwin, Z. A.; Mostofi, A. A.; Lischner, J.; Kennes, D. M.; Klebl, L., Unconventional superconductivity in magic-angle twisted trilayer graphene. *npj Quantum Materials* **2022,** *7* (1), 5.

289.	Wong, D.; Wang, Y.; Jung, J.; Pezzini, S.; DaSilva, A. M.; Tsai, H.-Z.; Jung, H. S.; Khajeh, R.; Kim, Y.; Lee, J.; Kahn, S.; Tollabimazraehno, S.; Rasool, H.; Watanabe, K.; Taniguchi, T.; Zettl, A.; Adam, S.; MacDonald, A. H.; Crommie, M. F., Local spectroscopy of moiré-induced electronic structure in gate-tunable twisted bilayer graphene. *Physical Review B* **2015,** *92* (15).

290.	Pei, C.; Zhang, J.; Wang, Q.; Zhao, Y.; Gao, L.; Gong, C.; Tian, S.; Luo, R.; Li, M.; Yang, W.; Lu, Z. Y.; Lei, H.; Liu, K.; Qi, Y., Pressure-induced superconductivity at 32 K in MoB(2). *Natl Sci Rev* **2023,** *10* (5), nwad034.

291.	Sun, L.; Lin, L.; Zhang, J.; Wang, H.; Peng, H.; Liu, Z., Visualizing fast growth of large single-crystalline graphene by tunable isotopic carbon source. *Nano Research* **2016,** *10* (2), 355-363.

292.	Tilak, N.; Lai, X.; Wu, S.; Zhang, Z.; Xu, M.; Ribeiro, R. d. A.; Canfield, P. C.; Andrei, E. Y., Flat band carrier confinement in magic-angle twisted bilayer graphene. *Nature communications* **2021,** *12* (1), 4180.

293.	Lisi, S.; Lu, X.; Benschop, T.; de Jong, T. A.; Stepanov, P.; Duran, J. R.; Margot, F.; Cucchi, I.; Cappelli, E.; Hunter, A., Observation of flat bands in twisted bilayer graphene. *Nature Physics* **2021,** *17* (2), 189-193.

294.	Li, X.-F.; Sun, R.-X.; Wang, S.-Y.; Li, X.; Liu, Z.-B.; Tian, J.-G., Recent Advances in Moiré Superlattice Structures of Twisted Bilayer and Multilayer Graphene. *Chinese Physics Letters* **2022,** *39* (3).

295.     Xiao, Y.; Liu, J.; Fu, L., Moiré is More: Access to New Properties of Two-Dimensional Layered Materials. *Matter* **2020**, *3* (4), 1142-1161.

296.     Morell, E. S.; Correa, J.; Vargas, P.; Pacheco, M.; Barticevic, Z., Flat bands in slightly twisted bilayer graphene: Tight-binding calculations. *Physical Review B* **2010**, *82* (12), 121407.

297.     Sherkunov, Y.; Betouras, J. J., Electronic phases in twisted bilayer graphene at magic angles as a result of Van Hove singularities and interactions. *Physical Review B* **2018**, *98* (20), 205151.

298.     Liao, L.; Wang, H.; Peng, H.; Yin, J.; Koh, A. L.; Chen, Y.; Xie, Q.; Peng, H.; Liu, Z., van Hove singularity enhanced photochemical reactivity of twisted bilayer graphene. *Nano letters* **2015**, *15* (8), 5585-5589.

299.     Moon, P.; Son, Y.-W.; Koshino, M., Optical absorption of twisted bilayer graphene with interlayer potential asymmetry. *Physical Review B* **2014**, *90* (15), 155427.

300.     Ma, Z.; Li, S.; Zheng, Y.-W.; Xiao, M.-M.; Jiang, H.; Gao, J.-H.; Xie, X., Topological flat bands in twisted trilayer graphene. *Science Bulletin* **2021**, *66* (1), 18-22.

301.     Oh, M.; Nuckolls, K. P.; Wong, D.; Lee, R. L.; Liu, X.; Watanabe, K.; Taniguchi, T.; Yazdani, A., Evidence for unconventional superconductivity in twisted bilayer graphene. *Nature* **2021**, *600* (7888), 240-245.

302.     Pierce, A. T.; Xie, Y.; Park, J. M.; Khalaf, E.; Lee, S. H.; Cao, Y.; Parker, D. E.; Forrester, P. R.; Chen, S.; Watanabe, K.; Taniguchi, T.; Vishwanath, A.; Jarillo-Herrero, P.; Yacoby, A., Unconventional sequence of correlated Chern insulators in magic-angle twisted bilayer graphene. *Nature Physics* **2021**, *17* (11), 1210-1215.

303.     Jin, C.; Kim, J.; Utama, M. I. B.; Regan, E. C.; Kleemann, H.; Cai, H.; Shen, Y.; Shinner, M. J.; Sengupta, A.; Watanabe, K., Imaging of pure spin-valley diffusion current in WS2-WSe2 heterostructures. *Science* **2018**, *360* (6391), 893-896.

304.     Yu, H.; Wang, Y.; Tong, Q.; Xu, X.; Yao, W., Anomalous light cones and valley optical selection rules of interlayer excitons in twisted heterobilayers. *Physical review letters* **2015**, *115* (18), 187002.

305.     Van der Donck, M.; Peeters, F., Interlayer excitons in transition metal dichalcogenide heterostructures. *Physical Review B* **2018**, *98* (11), 115104.

306.     Khan, A. R.; Liu, B.; Lü, T.; Zhang, L.; Sharma, A.; Zhu, Y.; Ma, W.; Lu, Y., Direct measurement of folding angle and strain vector in atomically thin WS2 using second-harmonic generation. *ACS nano* **2020**, *14* (11), 15806-15815.

307.     Khan, A. R.; Liu, B.; Zhang, L.; Zhu, Y.; He, X.; Zhang, L.; Lü, T.; Lu, Y., Extraordinary temperature dependent second harmonic generation in atomically thin layers of transition-metal dichalcogenides. *Advanced Optical Materials* **2020**, *8* (17), 2000441.

308.     Khan, A. R.; Zhang, L.; Ishfaq, K.; Ikram, A.; Yildrim, T.; Liu, B.; Rahman, S.; Lu, Y., Optical harmonic generation in 2d materials. *Advanced Functional Materials* **2022**, *32* (3), 2105259.

309.     Janisch, C.; Wang, Y.; Ma, D.; Mehta, N.; Elías, A. L.; Perea-López, N.; Terrones, M.; Crespi, V.; Liu, Z., Extraordinary second harmonic generation in tungsten disulfide monolayers. *Scientific reports* **2014**, *4* (1), 5530.

310.     Shinde, S. M.; Dhakal, K. P.; Chen, X.; Yun, W. S.; Lee, J.; Kim, H.; Ahn, J.-H., Stacking-controllable interlayer coupling and symmetric configuration of multilayered MoS2. *NPG Asia Materials* **2018**, *10* (2), e468-e468.

311.     Zeng, Z.; Sun, X.; Zhang, D.; Zheng, W.; Fan, X.; He, M.; Xu, T.; Sun, L.; Wang, X.; Pan, A., Controlled vapor growth and nonlinear optical applications of large-area 3R phase WS2 and WSe2 atomic layers. *Advanced Functional Materials* **2019**, *29* (11), 1806874.

312.     Rosa, H. G.; Junpeng, L.; Gomes, L. C.; Rodrigues, M. J.; Haur, S. C.; Gomes, J. C., Second-Harmonic spectroscopy for defects engineering monitoring in transition metal dichalcogenides. *Advanced Optical Materials* **2018**, *6* (5), 1701327.

313.     Kim, C.-J.; Brown, L.; Graham, M. W.; Hovden, R.; Havener, R. W.; McEuen, P. L.; Muller, D. A.; Park, J., Stacking order dependent second harmonic generation and topological defects in h-BN bilayers. *Nano letters* **2013**, *13* (11), 5660-5665.

314.     Karvonen, L.; Säynätjoki, A.; Huttunen, M. J.; Autere, A.; Amirsolaimani, B.; Li, S.; Norwood, R. A.; Peyghambarian, N.; Lipsanen, H.; Eda, G., Rapid visualization of grain boundaries in monolayer MoS2 by multiphoton microscopy. *Nature communications* **2017**, *8* (1), 15714.

315.    Yin, X.; Ye, Z.; Chenet, D. A.; Ye, Y.; O'Brien, K.; Hone, J. C.; Zhang, X., Edge nonlinear optics on a MoS2 atomic monolayer. *Science* **2014,** *344* (6183), 488-490.

316.    Shan, Y.; Li, Y.; Huang, D.; Tong, Q.; Yao, W.; Liu, W.-T.; Wu, S., Stacking symmetry governed second harmonic generation in graphene trilayers. *Science advances* **2018,** *4* (6), eaat0074.

317.    Song, Y.; Tian, R.; Yang, J.; Yin, R.; Zhao, J.; Gan, X., Second harmonic generation in atomically thin MoTe2. *Advanced Optical Materials* **2018,** *6* (17), 1701334.

318.    Xiao, J.; Zhu, H.; Wang, Y.; Feng, W.; Hu, Y.; Dasgupta, A.; Han, Y.; Wang, Y.; Muller, D. A.; Martin, L. W., Intrinsic two-dimensional ferroelectricity with dipole locking. *Physical review letters* **2018,** *120* (22), 227601.

319.    Zhou, Y.; Wu, D.; Zhu, Y.; Cho, Y.; He, Q.; Yang, X.; Herrera, K.; Chu, Z.; Han, Y.; Downer, M. C., Out-of-plane piezoelectricity and ferroelectricity in layered α-In2Se3 nanoflakes. *Nano letters* **2017,** *17* (9), 5508-5513.

320.    Zhu, H.; Wang, Y.; Xiao, J.; Liu, M.; Xiong, S.; Wong, Z. J.; Ye, Z.; Ye, Y.; Yin, X.; Zhang, X., Observation of piezoelectricity in free-standing monolayer MoS2. *Nature nanotechnology* **2015,** *10* (2), 151-155.

321.    Mennel, L.; Furchi, M. M.; Wachter, S.; Paur, M.; Polyushkin, D. K.; Mueller, T., Optical imaging of strain in two-dimensional crystals. *Nature communications* **2018,** *9* (1), 516.

322.    Beach, K.; Lucking, M. C.; Terrones, H., Strain dependence of second harmonic generation in transition metal dichalcogenide monolayers and the fine structure of the C exciton. *Physical Review B* **2020,** *101* (15), 155431.

323.    Liang, J.; Wang, J.; Zhang, Z.; Su, Y.; Guo, Y.; Qiao, R.; Song, P.; Gao, P.; Zhao, Y.; Jiao, Q., Universal imaging of full strain tensor in 2D crystals with third-harmonic generation. *Advanced Materials* **2019,** *31* (19), 1808160.

324.    Le, C. T.; Clark, D. J.; Ullah, F.; Jang, J. I.; Senthilkumar, V.; Sim, Y.; Seong, M.-J.; Chung, K.-H.; Kim, J. W.; Park, S., Impact of selenium doping on resonant second-harmonic generation in monolayer MoS2. *Acs Photonics* **2017,** *4* (1), 38-44.

325.    Yu, H.; Talukdar, D.; Xu, W.; Khurgin, J. B.; Xiong, Q., Charge-induced second-harmonic generation in bilayer WSe2. *Nano Letters* **2015,** *15* (8), 5653-5657.

326.    Margulis, V. A.; Muryumin, E.; Gaiduk, E., Electric-field-induced optical second-harmonic generation in doped graphene. *Solid State Communications* **2016,** *246*, 76-81.

327.    LaComb, R.; Nadiarnykh, O.; Townsend, S. S.; Campagnola, P. J., Phase matching considerations in second harmonic generation from tissues: effects on emission directionality, conversion efficiency and observed morphology. *Optics communications* **2008,** *281* (7), 1823-1832.

328.    Cai, L.; Gorbach, A. V.; Wang, Y.; Hu, H.; Ding, W., Highly efficient broadband second harmonic generation mediated by mode hybridization and nonlinearity patterning in compact fiber-integrated lithium niobate nano-waveguides. *Scientific Reports* **2018,** *8* (1), 12478.

329.    Zhang, J.; Cassan, E.; Gao, D.; Zhang, X., Highly efficient phase-matched second harmonic generation using an asymmetric plasmonic slot waveguide configuration in hybrid polymer-silicon photonics. *Optics Express* **2013,** *21* (12), 14876-14887.

330.    Jiang, T.; Liu, H.; Huang, D.; Zhang, S.; Li, Y.; Gong, X.; Shen, Y.-R.; Liu, W.-T.; Wu, S., Valley and band structure engineering of folded MoS2 bilayers. *Nature nanotechnology* **2014,** *9* (10), 825-829.

331.    Paradisanos, I.; Raven, A. M. S.; Amand, T.; Robert, C.; Renucci, P.; Watanabe, K.; Taniguchi, T.; Gerber, I. C.; Marie, X.; Urbaszek, B., Second harmonic generation control in twisted bilayers of transition metal dichalcogenides. *Physical Review B* **2022,** *105* (11), 115420.

332.    Wu, F.; Lovorn, T.; MacDonald, A., Theory of optical absorption by interlayer excitons in transition metal dichalcogenide heterobilayers. *Physical Review B* **2018,** *97* (3), 035306.

333.    Wu, F.; Lovorn, T.; MacDonald, A. H., Topological exciton bands in moiré heterojunctions. *Physical review letters* **2017,** *118* (14), 147401.

334.    Wei, G.; Czaplewski, D. A.; Lenferink, E. J.; Stanev, T. K.; Jung, I. W.; Stern, N. P., Size-tunable lateral confinement in monolayer semiconductors. *Scientific reports* **2017,** *7* (1), 1-8.

335.    Ren, S.; Tan, Q.; Zhang, J., Review on the quantum emitters in two-dimensional materials. *Journal of Semiconductors* **2019,** *40* (7), 071903.

336.     He, Y.-M.; Clark, G.; Schaibley, J. R.; He, Y.; Chen, M.-C.; Wei, Y.-J.; Ding, X.; Zhang, Q.; Yao, W.; Xu, X., Single quantum emitters in monolayer semiconductors. *Nature nanotechnology* **2015,** *10* (6), 497-502.

337.     Koperski, M.; Nogajewski, K.; Arora, A.; Cherkez, V.; Mallet, P.; Veuillen, J.-Y.; Marcus, J.; Kossacki, P.; Potemski, M., Single photon emitters in exfoliated WSe2 structures. *Nature nanotechnology* **2015,** *10* (6), 503-506.

338.     Tran, T. T.; Bray, K.; Ford, M. J.; Toth, M.; Aharonovich, I., Quantum emission from hexagonal boron nitride monolayers. *Nature nanotechnology* **2016,** *11* (1), 37-41.

339.     Zheng, H.; Xie, X.; Li, S.; Chen, S.; He, J.; Liu, Z.; Wang, C.-T.; Wang, J.-T.; Duan, X.; Lu, Y., Observation of Robust Interlayer Biexcitons in Twisted Homobilayer Superlattices. **2023.**

340.     Ko, J. S.; Jang, C. W.; Lee, W. J.; Kim, J. K.; Kim, H. K.; Liu, B.; Lu, Y.; Crosse, J.; Moon, P.; Kim, S., Blue-shifted and strongly-enhanced light emission in transition-metal dichalcogenide twisted heterobilayers. *npj 2D Materials and Applications* **2022,** *6* (1), 36.

341.     Nagler, P.; Plechinger, G.; Ballottin, M. V.; Mitioglu, A.; Meier, S.; Paradiso, N.; Strunk, C.; Chernikov, A.; Christianen, P. C.; Schüller, C., Interlayer exciton dynamics in a dichalcogenide monolayer heterostructure. *2D Materials* **2017,** *4* (2), 025112.

342.     Laikhtman, B.; Rapaport, R., Exciton correlations in coupled quantum wells and their luminescence blue shift. *Physical Review B* **2009,** *80* (19), 195313.

343.     Stern, M.; Umansky, V.; Bar-Joseph, I., Exciton liquid in coupled quantum wells. *Science* **2014,** *343* (6166), 55-57.

344.     Lahaye, T.; Menotti, C.; Santos, L.; Lewenstein, M.; Pfau, T., The physics of dipolar bosonic quantum gases. *Reports on Progress in Physics* **2009,** *72* (12), 126401.

345.     Astrakharchik, G.; Boronat, J.; Kurbakov, I.; Lozovik, Y. E., Quantum phase transition in a two-dimensional system of dipoles. *Physical review letters* **2007,** *98* (6), 060405.

346.     Liu, E.; Barré, E.; van Baren, J.; Wilson, M.; Taniguchi, T.; Watanabe, K.; Cui, Y.-T.; Gabor, N. M.; Heinz, T. F.; Chang, Y.-C., Signatures of moiré trions in WSe2/MoSe2 heterobilayers. *Nature* **2021,** *594* (7861), 46-50.

347.     Wang, J.; Ardelean, J.; Bai, Y.; Steinhoff, A.; Florian, M.; Jahnke, F.; Xu, X.; Kira, M.; Hone, J.; Zhu, X.-Y., Optical generation of high carrier densities in 2D semiconductor heterobilayers. *Science Advances* **2019,** *5* (9), eaax0145.

348.     Andrei, E. Y.; MacDonald, A. H., Graphene bilayers with a twist. *Nature materials* **2020,** *19* (12), 1265-1275.

349.     Li, Z.; Wang, T.; Miao, S.; Li, Y.; Lu, Z.; Jin, C.; Lian, Z.; Meng, Y.; Blei, M.; Taniguchi, T., Phonon-exciton Interactions in WSe2 under a quantizing magnetic field. *Nature communications* **2020,** *11* (1), 3104.

350.     Stier, A. V.; Wilson, N. P.; Velizhanin, K. A.; Kono, J.; Xu, X.; Crooker, S. A., Magnetooptics of exciton Rydberg states in a monolayer semiconductor. *Physical review letters* **2018,** *120* (5), 057405.

351.     Wu, F.; Lovorn, T.; Tutuc, E.; MacDonald, A. H., Hubbard model physics in transition metal dichalcogenide moiré bands. *Physical review letters* **2018,** *121* (2), 026402.

352.     Brem, S.; Linderälv, C.; Erhart, P.; Malic, E., Tunable phases of moiré excitons in van der Waals heterostructures. *Nano letters* **2020,** *20* (12), 8534-8540.

353.     Jauregui, L. A.; Joe, A. Y.; Pistunova, K.; Wild, D. S.; High, A. A.; Zhou, Y.; Scuri, G.; De Greve, K.; Sushko, A.; Yu, C.-H., Electrical control of interlayer exciton dynamics in atomically thin heterostructures. *Science* **2019,** *366* (6467), 870-875.

354.     Sun, Z.; Ciarrocchi, A.; Tagarelli, F.; Gonzalez Marin, J. F.; Watanabe, K.; Taniguchi, T.; Kis, A., Excitonic transport driven by repulsive dipolar interaction in a van der Waals heterostructure. *Nature photonics* **2022,** *16* (1), 79-85.

355.     Ziegler, J. D.; Zipfel, J.; Meisinger, B.; Menahem, M.; Zhu, X.; Taniguchi, T.; Watanabe, K.; Yaffe, O.; Egger, D. A.; Chernikov, A., Fast and anomalous exciton diffusion in two-dimensional hybrid perovskites. *Nano Letters* **2020,** *20* (9), 6674-6681.

356.     Fang, H.; Battaglia, C.; Carraro, C.; Nemsak, S.; Ozdol, B.; Kang, J. S.; Bechtel, H. A.; Desai, S. B.; Kronast, F.; Unal, A. A., Strong interlayer coupling in van der Waals heterostructures built from single-layer chalcogenides. *Proceedings of the National Academy of Sciences* **2014,** *111* (17), 6198-6202.

357.     Li, Z.; Li, Y.; Han, T.; Wang, X.; Yu, Y.; Tay, B.; Liu, Z.; Fang, Z., Tailoring MoS2 exciton–plasmon interaction by optical spin–orbit coupling. *ACS nano* **2017,** *11* (2), 1165-1171.

358.     Park, K.-D.; Jiang, T.; Clark, G.; Xu, X.; Raschke, M. B., Radiative control of dark excitons at room temperature by nano-optical antenna-tip Purcell effect. *Nature nanotechnology* **2018,** *13* (1), 59-64.

359.     Akselrod, G. M.; Ming, T.; Argyropoulos, C.; Hoang, T. B.; Lin, Y.; Ling, X.; Smith, D. R.; Kong, J.; Mikkelsen, M. H., Leveraging nanocavity harmonics for control of optical processes in 2D semiconductors. *Nano Letters* **2015,** *15* (5), 3578-3584.

360.     Luo, Y.; Shepard, G. D.; Ardelean, J. V.; Rhodes, D. A.; Kim, B.; Barmak, K.; Hone, J. C.; Strauf, S., Deterministic coupling of site-controlled quantum emitters in monolayer WSe2 to plasmonic nanocavities. *Nature nanotechnology* **2018,** *13* (12), 1137-1142.

361.     Dufferwiel, S.; Lyons, T.; Solnyshkov, D.; Trichet, A.; Catanzaro, A.; Withers, F.; Malpuech, G.; Smith, J.; Novoselov, K.; Skolnick, M., Valley coherent exciton-polaritons in a monolayer semiconductor. *Nature communications* **2018,** *9* (1), 4797.

362.     Chen, Y.; Miao, S.; Wang, T.; Zhong, D.; Saxena, A.; Chow, C.; Whitehead, J.; Gerace, D.; Xu, X.; Shi, S.-F., Metasurface integrated monolayer exciton polariton. *Nano Letters* **2020,** *20* (7), 5292-5300.

363.     Tran, T. N.; Kim, S.; White, S. J.; Nguyen, M. A. P.; Xiao, L.; Strauf, S.; Yang, T.; Aharonovich, I.; Xu, Z. Q., Enhanced emission from interlayer excitons coupled to plasmonic gap cavities. *Small* **2021,** *17* (45), 2103994.

364.     Förg, M.; Colombier, L.; Patel, R. K.; Lindlau, J.; Mohite, A. D.; Yamaguchi, H.; Glazov, M. M.; Hunger, D.; Högele, A., Cavity-control of interlayer excitons in van der Waals heterostructures. *Nature communications* **2019,** *10* (1), 3697.

365.     Latini, S.; Ronca, E.; De Giovannini, U.; Hübener, H.; Rubio, A., Cavity control of excitons in two-dimensional materials. *Nano letters* **2019,** *19* (6), 3473-3479.

366.     Zhang, L.; Wu, F.; Hou, S.; Zhang, Z.; Chou, Y.-H.; Watanabe, K.; Taniguchi, T.; Forrest, S. R.; Deng, H., Van der Waals heterostructure polaritons with moiré-induced nonlinearity. *Nature* **2021,** *591* (7848), 61-65.

367.     Rahman, S.; Lu, Y., Nano-engineering and nano-manufacturing in 2D materials: marvels of nanotechnology. *Nanoscale Horizons* **2022,** *7* (8), 849-872.

368.     Lin, M.-L.; Tan, Q.-H.; Wu, J.-B.; Chen, X.-S.; Wang, J.-H.; Pan, Y.-H.; Zhang, X.; Cong, X.; Zhang, J.; Ji, W.; Hu, P.-A.; Liu, K.-H.; Tan, P.-H., Moiré Phonons in Twisted Bilayer MoS2. *ACS Nano* **2018,** *12* (8), 8770-8780.

369.     Huang, S.; Liang, L.; Ling, X.; Puretzky, A. A.; Geohegan, D. B.; Sumpter, B. G.; Kong, J.; Meunier, V.; Dresselhaus, M. S., Low-Frequency Interlayer Raman Modes to Probe Interface of Twisted Bilayer MoS2. *Nano Letters* **2016,** *16* (2), 1435-1444.

370.     Zhang, X.; Qiao, X.-F.; Shi, W.; Wu, J.-B.; Jiang, D.-S.; Tan, P.-H., Phonon and Raman scattering of two-dimensional transition metal dichalcogenides from monolayer, multilayer to bulk material. *Chemical Society Reviews* **2015,** *44* (9), 2757-2785.

371.     Lin, K.-Q.; Holler, J.; Bauer, J. M.; Parzefall, P.; Scheuck, M.; Peng, B.; Korn, T.; Bange, S.; Lupton, J. M.; Schüller, C., Large-Scale Mapping of Moiré Superlattices by Hyperspectral Raman Imaging. *Advanced Materials* **2021,** *33* (34), 2008333.

372.     Mak, K. F.; Lee, C.; Hone, J.; Shan, J.; Heinz, T. F., Atomically thin MoS 2: a new direct-gap semiconductor. *Physical review letters* **2010,** *105* (13), 136805.

373.     Verble, J. L.; Wieting, T. J., Lattice Mode Degeneracy in $MoS_2$ and Other Layer Compounds. *Physical Review Letters* **1970,** *25* (6), 362-365.

374.     Parzefall, P.; Holler, J.; Scheuck, M.; Beer, A.; Lin, K.-Q.; Peng, B.; Monserrat, B.; Nagler, P.; Kempf, M.; Korn, T.; Schüller, C., Moiré phonons in twisted $MoSe_2/WSe_2$ heterobilayers and their correlation with interlayer excitons. *2D Materials* **2021,** *8* (3), 035030.

375.     Wu, L.; Cong, C.; Shang, J.; Yang, W.; Chen, Y.; Zhou, J.; Ai, W.; Wang, Y.; Feng, S.; Zhang, H.; Liu, Z.; Yu, T., Raman scattering investigation of twisted WS2/MoS2 heterostructures: interlayer mechanical coupling versus charge transfer. *Nano Research* **2021,** *14* (7), 2215-2223.

376.     Parzefall, P.; Holler, J.; Scheuck, M.; Beer, A.; Lin, K.-Q.; Peng, B.; Monserrat, B.; Nagler, P.; Kempf, M.; Korn, T.; Schüller, C., Moiré phonons in twisted MoSe2–WSe2 heterobilayers and their correlation with interlayer excitons. *2D Materials* **2021,** *8* (3), 035030.

377. Lee, J.-U.; Woo, S.; Park, J.; Park, H. C.; Son, Y.-W.; Cheong, H., Strain-shear coupling in bilayer MoS2. *Nature Communications* **2017,** *8* (1), 1370.

378. Wang, Y.; Cong, C.; Qiu, C.; Yu, T., Raman Spectroscopy Study of Lattice Vibration and Crystallographic Orientation of Monolayer MoS2 under Uniaxial Strain. *Small* **2013,** *9* (17), 2857-2861.

379. Yu, L.; Deng, M.; Zhang, J. L.; Borghardt, S.; Kardynal, B.; Vučković, J.; Heinz, T. F., Site-controlled quantum emitters in monolayer MoSe2. *Nano Letters* **2021,** *21* (6), 2376-2381.

380. Sim, K.; Chen, S.; Li, Z.; Rao, Z.; Liu, J.; Lu, Y.; Jang, S.; Ershad, F.; Chen, J.; Xiao, J., Three-dimensional curvy electronics created using conformal additive stamp printing. *Nature Electronics* **2019,** *2* (10), 471-479.

381. Okada, M.; Kutana, A.; Kureishi, Y.; Kobayashi, Y.; Saito, Y.; Saito, T.; Watanabe, K.; Taniguchi, T.; Gupta, S.; Miyata, Y.; Yakobson, B. I.; Shinohara, H.; Kitaura, R., Direct and Indirect Interlayer Excitons in a van der Waals Heterostructure of hBN/WS2/MoS2/hBN. *ACS Nano* **2018,** *12* (3), 2498-2505.

382. Rahman, S.; Sun, X.; Zhu, Y.; Lu, Y., Extraordinary Phonon Displacement and Giant Resonance Raman Enhancement in WSe2/WS2 Moiré Heterostructures. *ACS Nano* **2022,** *16* (12), 21505-21517.

383. Basov, D.; Fogler, M.; García de Abajo, F., Polaritons in van der Waals materials. *Science* **2016,** *354* (6309), aag1992.

384. Chaves, A.; Azadani, J. G.; Alsalman, H.; Da Costa, D.; Frisenda, R.; Chaves, A.; Song, S. H.; Kim, Y. D.; He, D.; Zhou, J., Bandgap engineering of two-dimensional semiconductor materials. *npj 2D Materials and Applications* **2020,** *4* (1), 1-21.

385. Low, T.; Chaves, A.; Caldwell, J. D.; Kumar, A.; Fang, N. X.; Avouris, P.; Heinz, T. F.; Guinea, F.; Martin-Moreno, L.; Koppens, F., Polaritons in layered two-dimensional materials. *Nature materials* **2017,** *16* (2), 182-194.

386. Ma, W.; Shabbir, B.; Ou, Q.; Dong, Y.; Chen, H.; Li, P.; Zhang, X.; Lu, Y.; Bao, Q., Anisotropic polaritons in van der Waals materials. *InfoMat* **2020,** *2* (5), 777-790.

387. Xiao, S.; Zhu, X.; Li, B.-H.; Mortensen, N. A., Graphene-plasmon polaritons: From fundamental properties to potential applications. *Frontiers of Physics* **2016,** *11*, 1-13.

388. Schäfer, C.; Baranov, D. G., Chiral Polaritonics: Analytical Solutions, Intuition, and Use. *The Journal of Physical Chemistry Letters* **2023,** *14* (15), 3777-3784.

389. Wu, Y.; Duan, J.; Ma, W.; Ou, Q.; Li, P.; Alonso-González, P.; Caldwell, J. D.; Bao, Q., Manipulating polaritons at the extreme scale in van der Waals materials. *Nature Reviews Physics* **2022,** *4* (9), 578-594.

390. Muir, J. B.; Levinsen, J.; Earl, S. K.; Conway, M. A.; Cole, J. H.; Wurdack, M.; Mishra, R.; David, J.; Estrecho, E.; Lu, Y., Exciton-polaron interactions in monolayer WS $_2$. *arXiv preprint arXiv:2206.12007* **2022**.

391. Fitzgerald, J. M.; Thompson, J. J.; Malic, E., Twist angle tuning of moiré exciton polaritons in van der Waals heterostructures. *Nano Letters* **2022,** *22* (11), 4468-4474.

392. Yu, H.; Yao, W., Electrically tunable topological transport of moiré polaritons. *Science Bulletin* **2020,** *65* (18), 1555-1562.

393. Togan, E.; Lim, H.-T.; Faelt, S.; Wegscheider, W.; Imamoglu, A., Enhanced interactions between dipolar polaritons. *Physical review letters* **2018,** *121* (22), 227402.

394. Tsintzos, S.; Tzimis, A.; Stavrinidis, G.; Trifonov, A.; Hatzopoulos, Z.; Baumberg, J.; Ohadi, H.; Savvidis, P., Electrical tuning of nonlinearities in exciton-polariton condensates. *Physical Review Letters* **2018,** *121* (3), 037401.

395. Hu, G.; Ou, Q.; Si, G.; Wu, Y.; Wu, J.; Dai, Z.; Krasnok, A.; Mazor, Y.; Zhang, Q.; Bao, Q.; Qiu, C.-W.; Alù, A., Topological polaritons and photonic magic angles in twisted α-MoO3 bilayers. *Nature* **2020,** *582* (7811), 209-213.

396. Zheng, Z.; Sun, F.; Huang, W.; Jiang, J.; Zhan, R.; Ke, Y.; Chen, H.; Deng, S., Phonon polaritons in twisted double-layers of hyperbolic van der Waals crystals. *Nano letters* **2020,** *20* (7), 5301-5308.

397. Duan, J.; Capote-Robayna, N.; Taboada-Gutiérrez, J.; Álvarez-Pérez, G.; Prieto, I.; Martín-Sánchez, J.; Nikitin, A. Y.; Alonso-González, P., Twisted nano-optics: manipulating light at the nanoscale with twisted phonon polaritonic slabs. *Nano Letters* **2020,** *20* (7), 5323-5329.

398.    Hu, G.; Krasnok, A.; Mazor, Y.; Qiu, C.-W.; Alù, A., Moiré hyperbolic metasurfaces. *Nano letters* **2020,** *20* (5), 3217-3224.

399.    Jiang, L.; Shi, Z.; Zeng, B.; Wang, S.; Kang, J.-H.; Joshi, T.; Jin, C.; Ju, L.; Kim, J.; Lyu, T., Soliton-dependent plasmon reflection at bilayer graphene domain walls. *Nature materials* **2016,** *15* (8), 840-844.

400.    Huang, T.; Tu, X.; Shen, C.; Zheng, B.; Wang, J.; Wang, H.; Khaliji, K.; Park, S. H.; Liu, Z.; Yang, T., Observation of chiral and slow plasmons in twisted bilayer graphene. *Nature* **2022,** *605* (7908), 63-68.

401.    Peng, J.; Tang, G.; Wang, L.; Macêdo, R.; Chen, H.; Ren, J., Twist-induced near-field thermal switch using nonreciprocal surface magnon-polaritons. *ACS Photonics* **2021,** *8* (8), 2183-2189.

402.    Mellado, P., Intrinsic topological magnons in arrays of magnetic dipoles. *Scientific Reports* **2022,** *12* (1), 1420.

403.    Gubbiotti, G., *Three-dimensional magnonics: layered, micro-and nanostructures*. CRC Press: 2019.

404.    Lisiecki, F.; Rychły, J.; Kuświk, P.; Głowiński, H.; Kłos, J. W.; Groß, F.; Bykova, I.; Weigand, M.; Zelent, M.; Goering, E. J., Reprogrammability and scalability of magnonic Fibonacci quasicrystals. *Physical Review Applied* **2019,** *11* (5), 054003.

405.    Wang, Z. K.; Zhang, V. L.; Lim, H. S.; Ng, S. C.; Kuok, M. H.; Jain, S.; Adeyeye, A. O., Nanostructured magnonic crystals with size-tunable bandgaps. *ACS nano* **2010,** *4* (2), 643-648.

406.    Sluka, V.; Schneider, T.; Gallardo, R. A.; Kákay, A.; Weigand, M.; Warnatz, T.; Mattheis, R.; Roldán-Molina, A.; Landeros, P.; Tiberkevich, V., Emission and propagation of 1D and 2D spin waves with nanoscale wavelengths in anisotropic spin textures. *Nature nanotechnology* **2019,** *14* (4), 328-333.

407.    Rana, B.; Fukuma, Y.; Miura, K.; Takahashi, H.; Otani, Y., Excitation of coherent propagating spin waves in ultrathin CoFeB film by voltage-controlled magnetic anisotropy. *Applied Physics Letters* **2017,** *111* (5), 052404.

408.    Verba, R.; Tiberkevich, V.; Krivorotov, I.; Slavin, A., Parametric excitation of spin waves by voltage-controlled magnetic anisotropy. *Physical Review Applied* **2014,** *1* (4), 044006.

409.    Bozhko, D. A.; Serga, A. A.; Clausen, P.; Vasyuchka, V. I.; Heussner, F.; Melkov, G. A.; Pomyalov, A.; L'vov, V. S.; Hillebrands, B., Supercurrent in a room-temperature Bose–Einstein magnon condensate. *Nature Physics* **2016,** *12* (11), 1057-1062.

410.    Li, X.; Labanowski, D.; Salahuddin, S.; Lynch, C. S., Spin wave generation by surface acoustic waves. *Journal of Applied Physics* **2017,** *122* (4), 043904.

411.    Panda, S.; Mondal, S.; Sinha, J.; Choudhury, S.; Barman, A., All-optical detection of interfacial spin transparency from spin pumping in β-Ta/CoFeB thin films. *Science advances* **2019,** *5* (4), eaav7200.

412.    Pal, S.; Klos, J.; Das, K.; Hellwig, O.; Gruszecki, P.; Krawczyk, M.; Barman, A., Optically induced spin wave dynamics in [Co/Pd] 8 antidot lattices with perpendicular magnetic anisotropy. *Applied Physics Letters* **2014,** *105* (16), 162408.

413.    Pan, S.; Mondal, S.; Zelent, M.; Szwierz, R.; Pal, S.; Hellwig, O.; Krawczyk, M.; Barman, A., Edge localization of spin waves in antidot multilayers with perpendicular magnetic anisotropy. *Physical Review B* **2020,** *101* (1), 014403.

414.    Chen, T.; Dumas, R. K.; Eklund, A.; Muduli, P. K.; Houshang, A.; Awad, A. A.; Dürrenfeld, P.; Malm, B. G.; Rusu, A.; Åkerman, J., Spin-torque and spin-Hall nano-oscillators. *Proceedings of the IEEE* **2016,** *104* (10), 1919-1945.

415.    Garcia-Sanchez, F.; Borys, P.; Soucaille, R.; Adam, J.-P.; Stamps, R. L.; Kim, J.-V., Narrow magnonic waveguides based on domain walls. *Physical review letters* **2015,** *114* (24), 247206.

416.    Graczyk, P.; Kłos, J.; Krawczyk, M., Broadband magnetoelastic coupling in magnonic-phononic crystals for high-frequency nanoscale spin-wave generation. *Physical Review B* **2017,** *95* (10), 104425.

417.    Schneider, M.; Brächer, T.; Breitbach, D.; Lauer, V.; Pirro, P.; Bozhko, D. A.; Musiienko-Shmarova, H. Y.; Heinz, B.; Wang, Q.; Meyer, T., Bose–Einstein condensation of quasiparticles by rapid cooling. *Nature Nanotechnology* **2020,** *15* (6), 457-461.

418.    Onose, Y.; Ideue, T.; Katsura, H.; Shiomi, Y.; Nagaosa, N.; Tokura, Y., Observation of the magnon Hall effect. *Science* **2010,** *329* (5989), 297-299.

419.     Hirschberger, M.; Krizan, J. W.; Cava, R.; Ong, N., Large thermal Hall conductivity of neutral spin excitations in a frustrated quantum magnet. *Science* **2015,** *348* (6230), 106-109.

420.     Yao, W.; Li, C.; Wang, L.; Xue, S.; Dan, Y.; Iida, K.; Kamazawa, K.; Li, K.; Fang, C.; Li, Y., Topological spin excitations in a three-dimensional antiferromagnet. *Nature Physics* **2018,** *14* (10), 1011-1015.

421.     Li, Y.-H.; Cheng, R., Moiré magnons in twisted bilayer magnets with collinear order. *Physical Review B* **2020,** *102* (9), 094404.

422.     Wang, H.; Madami, M.; Chen, J.; Jia, H.; Zhang, Y.; Yuan, R.; Wang, Y.; He, W.; Sheng, L.; Zhang, Y., Observation of spin-wave moiré edge and cavity modes in twisted magnetic lattices. *Physical Review X* **2023,** *13* (2), 021016.

423.     Chen, J.; Zeng, L.; Wang, H.; Madami, M.; Gubbiotti, G.; Liu, S.; Zhang, J.; Wang, Z.; Jiang, W.; Zhang, Y., Magic-angle magnonic nanocavity in a magnetic moiré superlattice. *Physical Review B* **2022,** *105* (9), 094445.

424.     Wang, C.; Gao, Y.; Lv, H.; Xu, X.; Xiao, D., Stacking domain wall magnons in twisted van der Waals magnets. *Physical Review Letters* **2020,** *125* (24), 247201.

425.     Andersen, U. L.; Neergaard-Nielsen, J. S.; van Loock, P.; Furusawa, A., Hybrid discrete- and continuous-variable quantum information. *Nature Physics* **2015,** *11* (9), 713-719.

426.     Hong, C.-K.; Ou, Z.-Y.; Mandel, L., Measurement of subpicosecond time intervals between two photons by interference. *Physical review letters* **1987,** *59* (18), 2044.

427.     Meyer-Scott, E.; Silberhorn, C.; Migdall, A., Single-photon sources: Approaching the ideal through multiplexing. *Review of Scientific Instruments* **2020,** *91* (4), 041101.

428.     Kimble, H. J.; Dagenais, M.; Mandel, L., Photon antibunching in resonance fluorescence. *Physical Review Letters* **1977,** *39* (11), 691.

429.     Aharonovich, I.; Englund, D.; Toth, M., Solid-state single-photon emitters. *Nature photonics* **2016,** *10* (10), 631-641.

430.     Kaneda, F.; Kwiat, P. G., High-efficiency single-photon generation via large-scale active time multiplexing. *Science advances* **2019,** *5* (10), eaaw8586.

431.     Wang, H.; Qin, J.; Ding, X.; Chen, M.-C.; Chen, S.; You, X.; He, Y.-M.; Jiang, X.; You, L.; Wang, Z., Boson sampling with 20 input photons and a 60-mode interferometer in a 1 0 14-dimensional hilbert space. *Physical review letters* **2019,** *123* (25), 250503.

432.     Kumar, S.; Brotóns-Gisbert, M.; Al-Khuzheyri, R.; Branny, A.; Ballesteros-Garcia, G.; Sánchez-Royo, J. F.; Gerardot, B. D., Resonant laser spectroscopy of localized excitons in monolayer WSe 2. *Optica* **2016,** *3* (8), 882-886.

433.     Palacios-Berraquero, C.; Barbone, M.; Kara, D. M.; Chen, X.; Goykhman, I.; Yoon, D.; Ott, A. K.; Beitner, J.; Watanabe, K.; Taniguchi, T.; Ferrari, A. C.; Atatüre, M., Atomically thin quantum light-emitting diodes. *Nature Communications* **2016,** *7* (1), 12978.

434.     Vogl, T.; Lecamwasam, R.; Buchler, B. C.; Lu, Y.; Lam, P. K., Compact cavity-enhanced single-photon generation with hexagonal boron nitride. *ACS Photonics* **2019,** *6* (8), 1955-1962.

435.     Peyskens, F.; Chakraborty, C.; Muneeb, M.; Van Thourhout, D.; Englund, D., Integration of single photon emitters in 2D layered materials with a silicon nitride photonic chip. *Nature communications* **2019,** *10* (1), 4435.

436.     Gao, T.; von Helversen, M.; Antón-Solanas, C.; Schneider, C.; Heindel, T., Atomically-thin single-photon sources for quantum communication. *npj 2D Materials and Applications* **2023,** *7* (1), 4.

437.     Srivastava, A.; Sidler, M.; Allain, A. V.; Lembke, D. S.; Kis, A.; Imamoğlu, A., Optically active quantum dots in monolayer WSe2. *Nature nanotechnology* **2015,** *10* (6), 491-496.

438.     Chakraborty, C.; Kinnischtzke, L.; Goodfellow, K. M.; Beams, R.; Vamivakas, A. N., Voltage-controlled quantum light from an atomically thin semiconductor. *Nature nanotechnology* **2015,** *10* (6), 507-511.

439.     Palacios-Berraquero, C.; Palacios-Berraquero, C., Atomically-thin quantum light emitting diodes. *Quantum confined excitons in 2-dimensional materials* **2018,** 71-89.

440.     Abobeih, M. H.; Cramer, J.; Bakker, M. A.; Kalb, N.; Markham, M.; Twitchen, D. J.; Taminiau, T. H., One-second coherence for a single electron spin coupled to a multi-qubit nuclear-spin environment. *Nature communications* **2018,** *9* (1), 2552.

441.    Nagy, R.; Niethammer, M.; Widmann, M.; Chen, Y.-C.; Udvarhelyi, P.; Bonato, C.; Hassan, J. U.; Karhu, R.; Ivanov, I. G.; Son, N. T., High-fidelity spin and optical control of single silicon-vacancy centres in silicon carbide. *Nature communications* **2019**, *10* (1), 1-8.

442.    Klein, J.; Sigl, L.; Gyger, S.; Barthelmi, K.; Florian, M.; Rey, S.; Taniguchi, T.; Watanabe, K.; Jahnke, F.; Kastl, C., Engineering the luminescence and generation of individual defect emitters in atomically thin MoS2. *ACS Photonics* **2021**, *8* (2), 669-677.

443.    Barthelmi, K.; Klein, J.; Hötger, A.; Sigl, L.; Sigger, F.; Mitterreiter, E.; Rey, S.; Gyger, S.; Lorke, M.; Florian, M., Atomistic defects as single-photon emitters in atomically thin MoS2. *Applied Physics Letters* **2020**, *117* (7).

444.    Zhong, M.; Hedges, M. P.; Ahlefeldt, R. L.; Bartholomew, J. G.; Beavan, S. E.; Wittig, S. M.; Longdell, J. J.; Sellars, M. J., Optically addressable nuclear spins in a solid with a six-hour coherence time. *Nature* **2015**, *517* (7533), 177-180.

445.    Zhao, H.; Pettes, M. T.; Zheng, Y.; Htoon, H., Site-controlled telecom-wavelength single-photon emitters in atomically-thin MoTe2. *Nature communications* **2021**, *12* (1), 6753.

446.    Tran, T. T.; Zachreson, C.; Berhane, A. M.; Bray, K.; Sandstrom, R. G.; Li, L. H.; Taniguchi, T.; Watanabe, K.; Aharonovich, I.; Toth, M., Quantum emission from defects in single-crystalline hexagonal boron nitride. *Physical review applied* **2016**, *5* (3), 034005.

447.    Tran, T. T.; Elbadawi, C.; Totonjian, D.; Lobo, C. J.; Grosso, G.; Moon, H.; Englund, D. R.; Ford, M. J.; Aharonovich, I.; Toth, M., Robust multicolor single photon emission from point defects in hexagonal boron nitride. *ACS nano* **2016**, *10* (8), 7331-7338.

448.    Ziegler, J.; Klaiss, R.; Blaikie, A.; Miller, D.; Horowitz, V. R.; Alemán, B. J., Deterministic quantum emitter formation in hexagonal boron nitride via controlled edge creation. *Nano letters* **2019**, *19* (3), 2121-2127.

449.    Vogl, T.; Sripathy, K.; Sharma, A.; Reddy, P.; Sullivan, J.; Machacek, J. R.; Zhang, L.; Karouta, F.; Buchler, B. C.; Doherty, M. W., Radiation tolerance of two-dimensional material-based devices for space applications. *Nature communications* **2019**, *10* (1), 1202.

450.    Grosso, G.; Moon, H.; Lienhard, B.; Ali, S.; Efetov, D. K.; Furchi, M. M.; Jarillo-Herrero, P.; Ford, M. J.; Aharonovich, I.; Englund, D., Tunable and high-purity room temperature single-photon emission from atomic defects in hexagonal boron nitride. *Nature Communications* **2017**, *8* (1), 705.

451.    White, S. J. U.; Yang, T.; Dontschuk, N.; Li, C.; Xu, Z.-Q.; Kianinia, M.; Stacey, A.; Toth, M.; Aharonovich, I., Electrical control of quantum emitters in a Van der Waals heterostructure. *Light: Science & Applications* **2022**, *11* (1), 186.

452.    Bourrellier, R.; Meuret, S.; Tararan, A.; Stéphan, O.; Kociak, M.; Tizei, L. H.; Zobelli, A., Bright UV single photon emission at point defects in h-BN. *Nano letters* **2016**, *16* (7), 4317-4321.

453.    Meuret, S.; Tizei, L.; Cazimajou, T.; Bourrellier, R.; Chang, H.; Treussart, F.; Kociak, M., Photon bunching in cathodoluminescence. *Physical review letters* **2015**, *114* (19), 197401.

454.    Gottscholl, A.; Kianinia, M.; Soltamov, V.; Orlinskii, S.; Mamin, G.; Bradac, C.; Kasper, C.; Krambrock, K.; Sperlich, A.; Toth, M., Initialization and read-out of intrinsic spin defects in a van der Waals crystal at room temperature. *Nature materials* **2020**, *19* (5), 540-545.

455.    Meng, Y.; Feng, J.; Han, S.; Xu, Z.; Mao, W.; Zhang, T.; Kim, J. S.; Roh, I.; Zhao, Y.; Kim, D.-H.; Yang, Y.; Lee, J.-W.; Yang, L.; Qiu, C.-W.; Bae, S.-H., Photonic van der Waals integration from 2D materials to 3D nanomembranes. *Nature Reviews Materials* **2023**, *8* (8), 498-517.

456.    Montblanch, A. R.-P.; Kara, D. M.; Paradisanos, I.; Purser, C. M.; Feuer, M. S.; Alexeev, E. M.; Stefan, L.; Qin, Y.; Blei, M.; Wang, G., Confinement of long-lived interlayer excitons in WS2/WSe2 heterostructures. *Communications Physics* **2021**, *4* (1), 119.

457.    Shanks, D. N.; Mahdikhanysarvejahany, F.; Muccianti, C.; Alfrey, A.; Koehler, M. R.; Mandrus, D. G.; Taniguchi, T.; Watanabe, K.; Yu, H.; LeRoy, B. J., Nanoscale trapping of interlayer excitons in a 2D semiconductor heterostructure. *Nano letters* **2021**, *21* (13), 5641-5647.

458.    Wang, W.; Ma, X., Strain-induced trapping of indirect excitons in MoSe2/WSe2 heterostructures. *ACS Photonics* **2020**, *7* (9), 2460-2467.

459.    Parzefall, M.; Szabó, Á.; Taniguchi, T.; Watanabe, K.; Luisier, M.; Novotny, L., Light from van der Waals quantum tunneling devices. *Nature communications* **2019**, *10* (1), 292.

460.    Wang, Z.; Kalathingal, V.; Hoang, T. X.; Chu, H.-S.; Nijhuis, C. A., Optical Anisotropy in van der Waals materials: Impact on Direct Excitation of Plasmons and Photons by Quantum Tunneling. *Light: Science & Applications* **2021**, *10* (1), 230.

461. Gould, M.; Schmidgall, E. R.; Dadgostar, S.; Hatami, F.; Fu, K.-M. C., Efficient extraction of zero-phonon-line photons from single nitrogen-vacancy centers in an integrated gap-on-diamond platform. *Physical Review Applied* **2016**, *6* (1), 011001.

462. Bachor, H.-A.; Ralph, T. C., *A guide to experiments in quantum optics*. John Wiley & Sons: 2019.

463. Li, X.; Scully, R. A.; Shayan, K.; Luo, Y.; Strauf, S., Near-unity light collection efficiency from quantum emitters in boron nitride by coupling to metallo-dielectric antennas. *ACS nano* **2019**, *13* (6), 6992-6997.

464. Fröch, J. E.; Kim, S.; Mendelson, N.; Kianinia, M.; Toth, M.; Aharonovich, I., Coupling hexagonal boron nitride quantum emitters to photonic crystal cavities. *ACS nano* **2020**, *14* (6), 7085-7091.

465. Tran, T. T.; Wang, D.; Xu, Z.-Q.; Yang, A.; Toth, M.; Odom, T. W.; Aharonovich, I., Deterministic coupling of quantum emitters in 2D materials to plasmonic nanocavity arrays. *Nano letters* **2017**, *17* (4), 2634-2639.

466. Chejanovsky, N.; Mukherjee, A.; Geng, J.; Chen, Y.-C.; Kim, Y.; Denisenko, A.; Finkler, A.; Taniguchi, T.; Watanabe, K.; Dasari, D. B. R., Single-spin resonance in a van der Waals embedded paramagnetic defect. *Nature materials* **2021**, *20* (8), 1079-1084.

467. Flatten, L.; Weng, L.; Branny, A.; Johnson, S.; Dolan, P.; Trichet, A.; Gerardot, B. D.; Smith, J., Microcavity enhanced single photon emission from two-dimensional WSe2. *Applied Physics Letters* **2018**, *112* (19).

468. Blauth, M.; Jurgensen, M.; Vest, G.; Hartwig, O.; Prechtl, M.; Cerne, J.; Finley, J. J.; Kaniber, M., Coupling single photons from discrete quantum emitters in WSe2 to lithographically defined plasmonic slot waveguides. *Nano letters* **2018**, *18* (11), 6812-6819.

469. Tonndorf, P.; Del Pozo-Zamudio, O.; Gruhler, N.; Kern, J.; Schmidt, R.; Dmitriev, A. I.; Bakhtinov, A. P.; Tartakovskii, A. I.; Pernice, W.; Michaelis de Vasconcellos, S., On-chip waveguide coupling of a layered semiconductor single-photon source. *Nano letters* **2017**, *17* (9), 5446-5451.

470. Errando-Herranz, C.; Schöll, E.; Picard, R.; Laini, M.; Gyger, S.; Elshaari, A. W.; Branny, A.; Wennberg, U.; Barbat, S.; Renaud, T., Resonance fluorescence from waveguide-coupled, strain-localized, two-dimensional quantum emitters. *ACS photonics* **2021**, *8* (4), 1069-1076.

471. Kok, P.; Munro, W. J.; Nemoto, K.; Ralph, T. C.; Dowling, J. P.; Milburn, G. J., Linear optical quantum computing with photonic qubits. *Reviews of modern physics* **2007**, *79* (1), 135.

472. Wehner, S.; Elkouss, D.; Hanson, R., Quantum internet: A vision for the road ahead. *Science* **2018**, *362* (6412), eaam9288.

473. Pirandola, S.; Braunstein, S. L., Physics: Unite to build a quantum Internet. *Nature* **2016**, *532* (7598), 169-171.

474. Wei, S. H.; Jing, B.; Zhang, X. Y.; Liao, J. Y.; Yuan, C. Z.; Fan, B. Y.; Lyu, C.; Zhou, D. L.; Wang, Y.; Deng, G. W., Towards real-world quantum networks: a review. *Laser & Photonics Reviews* **2022**, *16* (3), 2100219.

475. Azuma, K.; Economou, S. E.; Elkouss, D.; Hilaire, P.; Jiang, L.; Lo, H.-K.; Tzitrin, I., Quantum repeaters: From quantum networks to the quantum internet. *arXiv preprint arXiv:2212.10820* **2022**.

476. Turunen, M.; Brotons-Gisbert, M.; Dai, Y.; Wang, Y.; Scerri, E.; Bonato, C.; Jöns, K. D.; Sun, Z.; Gerardot, B. D., Quantum photonics with layered 2D materials. *Nature Reviews Physics* **2022**, *4* (4), 219-236.

477. Montblanch, A. R. P.; Barbone, M.; Aharonovich, I.; Atatüre, M.; Ferrari, A. C., Layered materials as a platform for quantum technologies. *Nature Nanotechnology* **2023**, *18* (6), 555-571.

478. Dastidar, M. G.; Thekkooden, I.; Nayak, P. K.; Bhallamudi, V. P., Quantum emitters and detectors based on 2D van der Waals materials. *Nanoscale* **2022**, *14* (14), 5289-5313.

479. Luo, W.; Cao, L.; Shi, Y.; Wan, L.; Zhang, H.; Li, S.; Chen, G.; Li, Y.; Li, S.; Wang, Y.; Sun, S.; Karim, M. F.; Cai, H.; Kwek, L. C.; Liu, A. Q., Recent progress in quantum photonic chips for quantum communication and internet. *Light: Science & Applications* **2023**, *12* (1), 175.

480. Bao, J.; Fu, Z.; Pramanik, T.; Mao, J.; Chi, Y.; Cao, Y.; Zhai, C.; Mao, Y.; Dai, T.; Chen, X.; Jia, X.; Zhao, L.; Zheng, Y.; Tang, B.; Li, Z.; Luo, J.; Wang, W.; Yang, Y.; Peng, Y.; Liu, D.; Dai, D.; He, Q.; Muthali, A. L.; Oxenløwe, L. K.; Vigliar, C.; Paesani, S.; Hou, H.; Santagati, R.; Silverstone,

J. W.; Laing, A.; Thompson, M. G.; O'Brien, J. L.; Ding, Y.; Gong, Q.; Wang, J., Very-large-scale integrated quantum graph photonics. *Nature Photonics* **2023,** *17* (7), 573-581.

481.    Pelucchi, E.; Fagas, G.; Aharonovich, I.; Englund, D.; Figueroa, E.; Gong, Q.; Hannes, H.; Liu, J.; Lu, C.-Y.; Matsuda, N.; Pan, J.-W.; Schreck, F.; Sciarrino, F.; Silberhorn, C.; Wang, J.; Jöns, K. D., The potential and global outlook of integrated photonics for quantum technologies. *Nature Reviews Physics* **2022,** *4* (3), 194-208.

482.    Wang, J.; Sciarrino, F.; Laing, A.; Thompson, M. G., Integrated photonic quantum technologies. *Nature Photonics* **2020,** *14* (5), 273-284.

483.    Madsen, L. S.; Laudenbach, F.; Askarani, M. F.; Rortais, F.; Vincent, T.; Bulmer, J. F. F.; Miatto, F. M.; Neuhaus, L.; Helt, L. G.; Collins, M. J.; Lita, A. E.; Gerrits, T.; Nam, S. W.; Vaidya, V. D.; Menotti, M.; Dhand, I.; Vernon, Z.; Quesada, N.; Lavoie, J., Quantum computational advantage with a programmable photonic processor. *Nature* **2022,** *606* (7912), 75-81.

484.    Arrazola, J. M.; Bergholm, V.; Brádler, K.; Bromley, T. R.; Collins, M. J.; Dhand, I.; Fumagalli, A.; Gerrits, T.; Goussev, A.; Helt, L. G.; Hundal, J.; Isacsson, T.; Israel, R. B.; Izaac, J.; Jahangiri, S.; Janik, R.; Killoran, N.; Kumar, S. P.; Lavoie, J.; Lita, A. E.; Mahler, D. H.; Menotti, M.; Morrison, B.; Nam, S. W.; Neuhaus, L.; Qi, H. Y.; Quesada, N.; Repingon, A.; Sabapathy, K. K.; Schuld, M.; Su, D.; Swinarton, J.; Száva, A.; Tan, K.; Tan, P.; Vaidya, V. D.; Vernon, Z.; Zabaneh, Z.; Zhang, Y., Quantum circuits with many photons on a programmable nanophotonic chip. *Nature* **2021,** *591* (7848), 54-60.

485.    Giordani, T.; Flamini, F.; Pompili, M.; Viggianiello, N.; Spagnolo, N.; Crespi, A.; Osellame, R.; Wiebe, N.; Walschaers, M.; Buchleitner, A.; Sciarrino, F., Experimental statistical signature of many-body quantum interference. *Nature Photonics* **2018,** *12* (3), 173-178.

486.    Raussendorf, R.; Briegel, H. J., A One-Way Quantum Computer. *Physical Review Letters* **2001,** *86* (22), 5188-5191.

487.    Luo, K.-H.; Brauner, S.; Eigner, C.; Sharapova, P. R.; Ricken, R.; Meier, T.; Herrmann, H.; Silberhorn, C., Nonlinear integrated quantum electro-optic circuits. *Science Advances* **2019,** *5* (1), eaat1451.

488.    Kim, J.; Jo, H.; Hong, H.; Kim, M. H.; Kim, J. M.; Lee, J.-K.; Heo, W. D.; Kim, J., Actin remodelling factors control ciliogenesis by regulating YAP/TAZ activity and vesicle trafficking. *Nature Communications* **2015,** *6* (1), 6781.

489.    Li, Z.-D.; Zhang, R.; Yin, X.-F.; Liu, L.-Z.; Hu, Y.; Fang, Y.-Q.; Fei, Y.-Y.; Jiang, X.; Zhang, J.; Li, L.; Liu, N.-L.; Xu, F.; Chen, Y.-A.; Pan, J.-W., Experimental quantum repeater without quantum memory. *Nature Photonics* **2019,** *13* (9), 644-648.

490.    Xu, X.; Trovatello, C.; Mooshammer, F.; Shao, Y.; Zhang, S.; Yao, K.; Basov, D. N.; Cerullo, G.; Schuck, P. J., Towards compact phase-matched and waveguided nonlinear optics in atomically layered semiconductors. *Nature Photonics* **2022,** *16* (10), 698-706.

491.    Leifgen, M.; Schröder, T.; Gädeke, F.; Riemann, R.; Métillon, V.; Neu, E.; Hepp, C.; Arend, C.; Becher, C.; Lauritsen, K., Evaluation of nitrogen-and silicon-vacancy defect centres as single photon sources in quantum key distribution. *New journal of physics* **2014,** *16* (2), 023021.

492.    Waks, E.; Inoue, K.; Santori, C.; Fattal, D.; Vuckovic, J.; Solomon, G. S.; Yamamoto, Y., Quantum cryptography with a photon turnstile. *Nature* **2002,** *420* (6917), 762-762.

493.    Nilsson, J.; Stevenson, R. M.; Chan, K. H. A.; Skiba-Szymanska, J.; Lucamarini, M.; Ward, M. B.; Bennett, A. J.; Salter, C. L.; Farrer, I.; Ritchie, D. A.; Shields, A. J., Quantum teleportation using a light-emitting diode. *Nature Photonics* **2013,** *7* (4), 311-315.

494.    Humphreys, P. C.; Kalb, N.; Morits, J. P. J.; Schouten, R. N.; Vermeulen, R. F. L.; Twitchen, D. J.; Markham, M.; Hanson, R., Deterministic delivery of remote entanglement on a quantum network. *Nature* **2018,** *558* (7709), 268-273.

495.    Anderson, M.; Müller, T.; Huwer, J.; Skiba-Szymanska, J.; Krysa, A.; Stevenson, R.; Heffernan, J.; Ritchie, D.; Shields, A., Quantum teleportation using highly coherent emission from telecom C-band quantum dots. *npj Quantum Information* **2020,** *6* (1), 14.

496.    Kalb, N.; Reiserer, A. A.; Humphreys, P. C.; Bakermans, J. J. W.; Kamerling, S. J.; Nickerson, N. H.; Benjamin, S. C.; Twitchen, D. J.; Markham, M.; Hanson, R., Entanglement distillation between solid-state quantum network nodes. *Science* **2017,** *356* (6341), 928-932.

497.    Pompili, M.; Hermans, S. L. N.; Baier, S.; Beukers, H. K. C.; Humphreys, P. C.; Schouten, R. N.; Vermeulen, R. F. L.; Tiggelman, M. J.; dos Santos Martins, L.; Dirkse, B.; Wehner, S.; Hanson,

R., Realization of a multinode quantum network of remote solid-state qubits. *Science* **2021,** *372* (6539), 259-264.

498. Hermans, S. L. N.; Pompili, M.; Beukers, H. K. C.; Baier, S.; Borregaard, J.; Hanson, R., Qubit teleportation between non-neighbouring nodes in a quantum network. *Nature* **2022,** *605* (7911), 663-668.

499. Oh, S.-H.; Altug, H.; Jin, X.; Low, T.; Koester, S. J.; Ivanov, A. P.; Edel, J. B.; Avouris, P.; Strano, M. S., Nanophotonic biosensors harnessing van der Waals materials. *Nature Communications* **2021,** *12* (1), 3824.

500. Lee, G.-H.; Moon, H.; Kim, H.; Lee, G. H.; Kwon, W.; Yoo, S.; Myung, D.; Yun, S. H.; Bao, Z.; Hahn, S. K., Multifunctional materials for implantable and wearable photonic healthcare devices. *Nature Reviews Materials* **2020,** *5* (2), 149-165.

501. Sim, K.; Chen, S.; Li, Z.; Rao, Z.; Liu, J.; Lu, Y.; Jang, S.; Ershad, F.; Chen, J.; Xiao, J.; Yu, C., Three-dimensional curvy electronics created using conformal additive stamp printing. *Nature Electronics* **2019,** *2* (10), 471-479.

502. Hou, H.-L.; Anichini, C.; Samorì, P.; Criado, A.; Prato, M., 2D Van der Waals Heterostructures for Chemical Sensing. *Advanced Functional Materials* **2022,** *32* (49), 2207065.

503. Ashida, Y.; İmamoğlu, A.; Demler, E., Cavity Quantum Electrodynamics with Hyperbolic van der Waals Materials. *Physical Review Letters* **2023,** *130* (21), 216901.

504. Rodrigo, D.; Limaj, O.; Janner, D.; Etezadi, D.; García de Abajo, F. J.; Pruneri, V.; Altug, H., Mid-infrared plasmonic biosensing with graphene. *Science* **2015,** *349* (6244), 165-168.

505. Wan, N. H.; Lu, T.-J.; Chen, K. C.; Walsh, M. P.; Trusheim, M. E.; De Santis, L.; Bersin, E. A.; Harris, I. B.; Mouradian, S. L.; Christen, I. R.; Bielejec, E. S.; Englund, D., Large-scale integration of artificial atoms in hybrid photonic circuits. *Nature* **2020,** *583* (7815), 226-231.

506. Yoon, J.; Jo, S.; Chun, I. S.; Jung, I.; Kim, H.-S.; Meitl, M.; Menard, E.; Li, X.; Coleman, J. J.; Paik, U.; Rogers, J. A., GaAs photovoltaics and optoelectronics using releasable multilayer epitaxial assemblies. *Nature* **2010,** *465* (7296), 329-333.

507. Zheng, W.; Liu, X.; Xie, J.; Lu, G.; Zhang, J., Emerging van der Waals junctions based on TMDs materials for advanced gas sensors. *Coordination Chemistry Reviews* **2021,** *447*, 214151.

508. Farmer, D. B.; Avouris, P.; Li, Y.; Heinz, T. F.; Han, S.-J., Ultrasensitive Plasmonic Detection of Molecules with Graphene. *ACS Photonics* **2016,** *3* (4), 553-557.

509. Hu, H.; Yang, X.; Guo, X.; Khaliji, K.; Biswas, S. R.; García de Abajo, F. J.; Low, T.; Sun, Z.; Dai, Q., Gas identification with graphene plasmons. *Nature Communications* **2019,** *10* (1), 1131.

510. Zhao, Y.; Wu, G.; Hung, K.-M.; Cho, J.; Choi, M.; Ó Coileáin, C.; Duesberg, G. S.; Ren, X.-K.; Chang, C.-R.; Wu, H.-C., Field Effect Transistor Gas Sensors Based on Mechanically Exfoliated Van der Waals Materials. *ACS Applied Materials & Interfaces* **2023,** *15* (13), 17335-17343.

511. Liu, L.; Xu, N. S.; Ke, Y.; Chen, H.; Zhang, Y.; Deng, S., Sensing by Surface Work Function Modulation: High Performance Gas Sensing using van der Waals Stacked Bipolar Junction Transistor. *Sensors and Actuators B: Chemical* **2021,** *345*, 130340.

512. Lee, J.; Lee, W.; Kim, G.-Y.; Choi, Y.-B.; Park, J.; Jang, S.; Gu, G.; Choi, S.-Y.; Cho, G. Y.; Lee, G.-H., Twisted van der Waals Josephson Junction Based on a High-T c Superconductor. *Nano Letters* **2021,** *21* (24), 10469-10477.

513. Yan, W.; Fuh, H.-R.; Lv, Y.; Chen, K.-Q.; Tsai, T.-Y.; Wu, Y.-R.; Shieh, T.-H.; Hung, K.-M.; Li, J.; Zhang, D.; Ó Coileáin, C.; Arora, S. K.; Wang, Z.; Jiang, Z.; Chang, C.-R.; Wu, H.-C., Giant gauge factor of Van der Waals material based strain sensors. *Nature Communications* **2021,** *12* (1), 2018.

514. Zhang, Y.-n.; Song, Z.-y.; Qiao, D.; Li, X.-h.; Guang, Z.; Li, S.-p.; Zhou, L.-b.; Chen, X.-h., 2D van der Waals materials for ultrafast pulsed fiber lasers: review and prospect. *Nanotechnology* **2022,** *33* (8), 082003.

515. Jin, C.; Ma, E. Y.; Karni, O.; Regan, E. C.; Wang, F.; Heinz, T. F., Ultrafast dynamics in van der Waals heterostructures. *Nature Nanotechnology* **2018,** *13* (11), 994-1003.

516. Zhu, J.; Shang, W.; Ning, J.; Wang, D.; Zhang, J.; Hao, Y., Twisted angle modulated structural property, electronic structure and carrier transport of MoS2/AlN(0001) mixed-dimensional van der Waals heterostructure. *Applied Surface Science* **2021,** *563*, 150330.

517.	Xia, Z.; Song, H.; Kim, M.; Zhou, M.; Chang, T.-H.; Liu, D.; Yin, X.; Xiong, K.; Mi, H.; Wang, X.; Xia, F.; Yu, Z.; Ma, Z.; Gan, Q., Single-crystalline germanium nanomembrane photodetectors on foreign nanocavities. *Science Advances* **2017,** *3* (7), e1602783.

518.	Liu, S.; Shah, D. S.; Kramer-Bottiglio, R., Highly stretchable multilayer electronic circuits using biphasic gallium-indium. *Nature Materials* **2021,** *20* (6), 851-858.

519.	Wang, C.; You, L.; Cobden, D.; Wang, J., Towards two-dimensional van der Waals ferroelectrics. *Nature Materials* **2023,** *22* (5), 542-552.

520.	Mao, X.-R.; Shao, Z.-K.; Luan, H.-Y.; Wang, S.-L.; Ma, R.-M., Magic-angle lasers in nanostructured moiré superlattice. *Nature Nanotechnology* **2021,** *16* (10), 1099-1105.

521.	Paik, E. Y.; Zhang, L.; Burg, G. W.; Gogna, R.; Tutuc, E.; Deng, H., Interlayer exciton laser of extended spatial coherence in atomically thin heterostructures. *Nature* **2019,** *576* (7785), 80-84.

522.	Liu, Y.; Fang, H.; Rasmita, A.; Zhou, Y.; Li, J.; Yu, T.; Xiong, Q.; Zheludev, N.; Liu, J.; Gao, W., Room temperature nanocavity laser with interlayer excitons in 2D heterostructures. *Science Advances* **2019,** *5* (4), eaav4506.

523.	Xin, W.; Chen, X. D.; Liu, Z. B.; Jiang, W. S.; Gao, X. G.; Jiang, X. Q.; Chen, Y.; Tian, J. G., Photovoltage Enhancement in Twisted-Bilayer Graphene Using Surface Plasmon Resonance. *Advanced Optical Materials* **2016,** *4* (11), 1703-1710.

524.	Deng, B.; Ma, C.; Wang, Q.; Yuan, S.; Watanabe, K.; Taniguchi, T.; Zhang, F.; Xia, F., Strong mid-infrared photoresponse in small-twist-angle bilayer graphene. *Nature Photonics* **2020,** *14* (9), 549-553.

525.	Zou, Z.; Li, D.; Liang, J.; Zhang, X.; Liu, H.; Zhu, C.; Yang, X.; Li, L.; Zheng, B.; Sun, X., Epitaxial synthesis of ultrathin β-In 2 Se 3/MoS 2 heterostructures with high visible/near-infrared photoresponse. *Nanoscale* **2020,** *12* (11), 6480-6488.

526.	Xin, W.; Jiang, H. B.; Li, X. K.; Zhou, X. F.; Lu, J. L.; Yang, J. J.; Guo, C.; Liu, Z. B.; Tian, J. G., Photoinduced Orientation-Dependent Interlayer Carrier Transportation in Cross-Stacked Black Phosphorus van der Waals Junctions. *Advanced Materials Interfaces* **2018,** *5* (20), 1800964.

527.	Xiong, Y.; Wang, Y.; Zhu, R.; Xu, H.; Wu, C.; Chen, J.; Ma, Y.; Liu, Y.; Chen, Y.; Watanabe, K., Twisted black phosphorus–based van der Waals stacks for fiber-integrated polarimeters. *Science Advances* **2022,** *8* (18), eabo0375.

528.	Srivastava, P. K.; Hassan, Y.; de Sousa, D. J.; Gebredingle, Y.; Joe, M.; Ali, F.; Zheng, Y.; Yoo, W. J.; Ghosh, S.; Teherani, J. T., Resonant tunnelling diodes based on twisted black phosphorus homostructures. *Nature Electronics* **2021,** *4* (4), 269-276.

529.	Otteneder, M.; Hubmann, S.; Lu, X.; Kozlov, D. A.; Golub, L. E.; Watanabe, K.; Taniguchi, T.; Efetov, D. K.; Ganichev, S. D., Terahertz photogalvanics in twisted bilayer graphene close to the second magic angle. *Nano Letters* **2020,** *20* (10), 7152-7158.

530.	Xin, W.; Chen, X.-D.; Liu, Z.-B.; Jiang, W.-S.; Gao, X.-G.; Jiang, X.-Q.; Chen, Y.; Tian, J.-G., Photovoltage Enhancement in Twisted-Bilayer Graphene Using Surface Plasmon Resonance. *Advanced Optical Materials* **2016,** *4* (11), 1703-1710.

531.	Chen, X. D.; Xin, W.; Jiang, W. S.; Liu, Z. B.; Chen, Y.; Tian, J. G., High-Precision Twist-Controlled Bilayer and Trilayer Graphene. *Adv Mater* **2016,** *28* (13), 2563-70.

532.	al., S. W. e., Multiple hot-carrier collection in photo-excitedgraphene Moiré superlattices. *Sci. Adv.* **2016,** *2* (e160000), 1-6.

533.	Xin, W.; Li, X. K.; He, X. L.; Su, B. W.; Jiang, X. Q.; Huang, K. X.; Zhou, X. F.; Liu, Z. B.; Tian, J. G., Black-Phosphorus-Based Orientation-Induced Diodes. *Adv Mater* **2018,** *30* (2).

534.	Xie, L.; Wang, L.; Zhao, W.; Liu, S.; Huang, W.; Zhao, Q., WS(2) moire superlattices derived from mechanical flexibility for hydrogen evolution reaction. *Nat Commun* **2021,** *12* (1), 5070.

535.	Jiang, Z.; Zhou, W.; Hong, A.; Guo, M.; Luo, X.; Yuan, C., MoS2 Moiré Superlattice for Hydrogen Evolution Reaction. *ACS Energy Letters* **2019,** *4* (12), 2830-2835.

536.	al., E. C. e., Correlated insulating and superconducting states intwisted bilayer graphene below the magic angle. *Sci. Adv.* **2019,** *5* (eaaw9770), 1-5.

537.	Hao, Z.; Zimmerman, A. M.; Ledwith, P.; Khalaf, E.; Najafabadi, D. H.; Watanabe, K.; Taniguchi, T.; Vishwanath, A.; Kim, P., Electric field-tunable superconductivity in alternating-twist magic-angle trilayer graphene. *Science* **2021,** *371* (6534), 1133-1138.

538.	Wang, Y.; Lan, Y.; Song, Q.; Vogelbacher, F.; Xu, T.; Zhan, Y.; Li, M.; Sha, W. E. I.; Song, Y., Colorful Efficient Moire-Perovskite Solar Cells. *Adv Mater* **2021,** *33* (15), e2008091.

539.     Alexeev, E. M.; Ruiz-Tijerina, D. A.; Danovich, M.; Hamer, M. J.; Terry, D. J.; Nayak, P. K.; Ahn, S.; Pak, S.; Lee, J.; Sohn, J. I.; Molas, M. R.; Koperski, M.; Watanabe, K.; Taniguchi, T.; Novoselov, K. S.; Gorbachev, R. V.; Shin, H. S.; Fal'ko, V. I.; Tartakovskii, A. I., Resonantly hybridized excitons in moire superlattices in van der Waals heterostructures. *Nature* **2019,** *567* (7746), 81-86.

540.     Mao, X. R.; Shao, Z. K.; Luan, H. Y.; Wang, S. L.; Ma, R. M., Magic-angle lasers in nanostructured moire superlattice. *Nat Nanotechnol* **2021,** *16* (10), 1099-1105.

541.     Patel, H.; Huang, L.; Kim, C. J.; Park, J.; Graham, M. W., Stacking angle-tunable photoluminescence from interlayer exciton states in twisted bilayer graphene. *Nat Commun* **2019,** *10* (1), 1445.

542.     Chen, M.; Lin, X.; Dinh, T. H.; Zheng, Z.; Shen, J.; Ma, Q.; Chen, H.; Jarillo-Herrero, P.; Dai, S., Configurable phonon polaritons in twisted α-MoO3. *Nature materials* **2020,** *19* (12), 1307-1311.

543.     Su, C.; Zhang, F.; Kahn, S.; Shevitski, B.; Jiang, J.; Dai, C.; Ungar, A.; Park, J.-H.; Watanabe, K.; Taniguchi, T., Tuning color centers at a twisted interface. *arXiv preprint arXiv:2108.04747* **2021**.

544.     Gupta, N.; Walia, S.; Mogera, U.; Kulkarni, G. U., Twist-dependent Raman and electron diffraction correlations in twisted multilayer graphene. *The Journal of Physical Chemistry Letters* **2020,** *11* (8), 2797-2803.

545.     Farrar, L. S.; Nevill, A.; Lim, Z. J.; Balakrishnan, G.; Dale, S.; Bending, S. J., Superconducting quantum interference in twisted van der Waals heterostructures. *Nano Letters* **2021,** *21* (16), 6725-6731.

546.     Ghader, D., Magnon magic angles and tunable Hall conductivity in 2D twisted ferromagnetic bilayers. *Scientific Reports* **2020,** *10* (1), 15069.

547.     Castellanos-Gomez, A.; Duan, X.; Fei, Z.; Gutierrez, H. R.; Huang, Y.; Huang, X.; Quereda, J.; Qian, Q.; Sutter, E.; Sutter, P., Van der Waals heterostructures. *Nature Reviews Methods Primers* **2022,** *2* (1), 58.

548.     Cho, S.; Kim, S.; Kim, J. H.; Zhao, J.; Seok, J.; Keum, D. H.; Baik, J.; Choe, D.-H.; Chang, K. J.; Suenaga, K.; Kim, S. W.; Lee, Y. H.; Yang, H., Phase patterning for ohmic homojunction contact in MoTe2. *Science* **2015,** *349* (6248), 625-628.

549.     Li, J.; Yang, X.; Liu, Y.; Huang, B.; Wu, R.; Zhang, Z.; Zhao, B.; Ma, H.; Dang, W.; Wei, Z.; Wang, K.; Lin, Z.; Yan, X.; Sun, M.; Li, B.; Pan, X.; Luo, J.; Zhang, G.; Liu, Y.; Huang, Y.; Duan, X.; Duan, X., General synthesis of two-dimensional van der Waals heterostructure arrays. *Nature* **2020,** *579* (7799), 368-374.

550.     Zhang, Y.; Yin, L.; Chu, J.; Shifa, T. A.; Xia, J.; Wang, F.; Wen, Y.; Zhan, X.; Wang, Z.; He, J., Edge-Epitaxial Growth of 2D NbS2-WS2 Lateral Metal-Semiconductor Heterostructures. *Advanced Materials* **2018,** *30* (40), 1803665.

551.     Qin, B.; Saeed, M. Z.; Li, Q.; Zhu, M.; Feng, Y.; Zhou, Z.; Fang, J.; Hossain, M.; Zhang, Z.; Zhou, Y., General low-temperature growth of two-dimensional nanosheets from layered and nonlayered materials. *Nature Communications* **2023,** *14* (1), 304.

552.     Kim, H.-U.; Kim, M.; Jin, Y.; Hyeon, Y.; Kim, K. S.; An, B.-S.; Yang, C.-W.; Kanade, V.; Moon, J.-Y.; Yeom, G. Y., Low-temperature wafer-scale growth of MoS2-graphene heterostructures. *Applied Surface Science* **2019,** *470*, 129-134.

553.     Van der Donck, M.; Conti, S.; Perali, A.; Hamilton, A. R.; Partoens, B.; Peeters, F. M.; Neilson, D., Three-dimensional electron-hole superfluidity in a superlattice close to room temperature. *Physical Review B* **2020,** *102* (6), 060503.

554.     Ren, H.; Wan, Z.; Duan, X., Van der Waals superlattices. *National Science Review* **2021,** *9* (5).

555.     Zhao, B.; Wan, Z.; Liu, Y.; Xu, J.; Yang, X.; Shen, D.; Zhang, Z.; Guo, C.; Qian, Q.; Li, J.; Wu, R.; Lin, Z.; Yan, X.; Li, B.; Zhang, Z.; Ma, H.; Li, B.; Chen, X.; Qiao, Y.; Shakir, I.; Almutairi, Z.; Wei, F.; Zhang, Y.; Pan, X.; Huang, Y.; Ping, Y.; Duan, X.; Duan, X., High-order superlattices by rolling up van der Waals heterostructures. *Nature* **2021,** *591* (7850), 385-390.

556.     Gibertini, M.; Koperski, M.; Morpurgo, A. F.; Novoselov, K. S., Magnetic 2D materials and heterostructures. *Nature Nanotechnology* **2019,** *14* (5), 408-419.

557.     Huang, M.; Wu, Z.; Hu, J.; Cai, X.; Li, E.; An, L.; Feng, X.; Ye, Z.; Lin, N.; Law, K. T.; Wang, N., Giant nonlinear Hall effect in twisted bilayer WSe2. *National Science Review* **2022,** *10* (4).

558.     Yan, X.; Liu, C.; Gadre, C. A.; Gu, L.; Aoki, T.; Lovejoy, T. C.; Dellby, N.; Krivanek, O. L.; Schlom, D. G.; Wu, R., Single-defect phonons imaged by electron microscopy. *Nature* **2021,** *589* (7840), 65-69.

559.     López, L. E. P.; Rosławska, A.; Scheurer, F.; Berciaud, S.; Schull, G., Tip-induced excitonic luminescence nanoscopy of an atomically resolved van der Waals heterostructure. *Nature Materials* **2023,** *22* (4), 482-488.

560.     Li, H.; Xiang, Z.; Naik, M. H.; Kim, W.; Li, Z.; Sailus, R.; Banerjee, R.; Taniguchi, T.; Watanabe, K.; Tongay, S., Imaging Moir\'e Excited States with Photocurrent Tunneling Microscopy. *arXiv preprint arXiv:2306.00859* **2023**.

561.     Nayak, G.; Lisi, S.; Liu, W.; Jakubczyk, T.; Stepanov, P.; Donatini, F.; Watanabe, K.; Taniguchi, T.; Bid, A.; Kasprzak, J., Cathodoluminescence enhancement and quenching in type-I van der Waals heterostructures: Cleanliness of the interfaces and defect creation. *Physical Review Materials* **2019,** *3* (11), 114001.

562.     Lee, H. Y.; Al Ezzi, M. M.; Raghuvanshi, N.; Chung, J. Y.; Watanabe, K.; Taniguchi, T.; Garaj, S.; Adam, S.; Gradecak, S., Tunable optical properties of thin films controlled by the interface twist angle. *Nano Letters* **2021,** *21* (7), 2832-2839.

563.     Bai, Y.; Zhou, L.; Wang, J.; Wu, W.; McGilly, L. J.; Halbertal, D.; Lo, C. F. B.; Liu, F.; Ardelean, J.; Rivera, P.; Finney, N. R.; Yang, X.-C.; Basov, D. N.; Yao, W.; Xu, X.; Hone, J.; Pasupathy, A. N.; Zhu, X. Y., Excitons in strain-induced one-dimensional moiré potentials at transition metal dichalcogenide heterojunctions. *Nature Materials* **2020,** *19* (10), 1068-1073.

564.     Bao, X.; Wu, X.; Ke, Y.; Wu, K.; Jiang, C.; Wu, B.; Li, J.; Yue, S.; Zhang, S.; Shi, J., Giant out-of-plane exciton emission enhancement in two-dimensional indium selenide via a plasmonic nanocavity. *Nano Letters* **2023,** *23* (9), 3716-3723.